\documentclass{ucbthesis}

\usepackage[style=authoryear, backend=biber, natbib=true, maxcitenames=2, maxbibnames=5, uniquename=false]{biblatex}

\usepackage{rotating}   % provides sidewaystable and        
\usepackage{subfig}
\usepackage{amsmath}	% Advanced maths commands
\usepackage{amssymb}	% Extra maths symbols
\usepackage{aas_macros}
\usepackage{xcolor}
\usepackage{hyperref}

\hypersetup{
    colorlinks=true, %set true if you want colored links
    linktoc=all,     %set to all if you want both sections and subsections linked
    linkcolor=black,  %choose some color if you want links to stand out
    citecolor=blue,
}

\usepackage{indentfirst}

\newcommand{\matr}[1]{\mathsf{\mathbf{#1}}}

\begin{document}

% Declarations for Front Matter

\title{Instrumentation for Radio Interferometers with Antennas on a Regular Grid}
\author{Deepthi Bhavana Gorthi}
\degreesemester{Spring}
\degreeyear{2021}
\degree{Doctor of Philosophy}
\chair{Professor Aaron Parsons}
\othermembers{Professor Jessica Lu \\
  Professor Uros Seljak}
% For a co-chair who is subordinate to the \chair listed above
% \cochair{Professor Benedict Francis Pope}
% For two co-chairs of equal standing (do not use \chair with this one)
% \cochairs{Professor Richard Francis Sony}{Professor Benedict Francis Pope}
\numberofmembers{3}
% Previous degrees are no longer to be listed on the title page.
% \prevdegrees{B.A. (University of Northern South Dakota at Hoople) 1978 \\
%   M.S. (Ed's School of Quantum Mechanics and Muffler Repair) 1989}
\field{Astrophysics}
% Designated Emphasis -- this is optional, and rare
% \emphasis{Colloidal Telemetry}
% This is optional, and rare
% \jointinstitution{University of Western Maryland}
% This is optional (default is Berkeley)
% \campus{Berkeley}

% For a masters thesis, replace the above \documentclass line with
% \documentclass[masters]{ucbthesis}
% This affects the title and approval pages, which by default calls this
% document a "dissertation", not a "thesis".

\maketitle
% Delete (or comment out) the \approvalpage line for the final version.
%\approvalpage
\copyrightpage

\begin{abstract}

In the past two decades, a rebirth of interest in low-frequency radio astronomy, for 21\,cm tomography of the Epoch of Reionization, has given rise to a new class of radio interferometers with $N \gg 100$ antennas. The availability of low-noise receivers that do not require cryogenic cooling has driven down the cost of antennas, making it affordable to build sensitivity with numerous small antennas rather than traditional large dish structures. However, the computational- and storage-costs of such radio arrays, determined by the $\mathcal{O}(N^2)$ scaling of visibility products that need to be computed for calibration and imaging, become proportional to the cost of the array itself and drive up the overall cost of the radio telescope.

When antennas in the array are built on a regular grid, direct-imaging methods based on spatial Fourier transforms of the array can be exploited to avoid computing the intermediate visibility matrices that drive the unfavorable scaling. However, such methods rely on the availability of calibrated antenna voltages which are themselves difficult to obtain without using visibility matrices. In this thesis, I explore two real-time calibration strategies that can operate on subsets of visibility matrices, which can be computed without compromising on the $\mathcal{O}(N\log{N})$ scaling of direct-imaging systems.

For more general radio interferometer layouts, baseline-dependent averaging with fringe stopping can be used to decrease the data rate of visibility products. While the computational cost is nearly unchanged, this technique can decrease the data volume of cross-correlation products, making it more tenable to store, process, and calibrate the output of the correlator. In this thesis, I describe the entire signal processing pipeline built for the Hydrogen Epoch of Reionization Array (HERA), which is currently being commissioned for detecting and characterizing the power spectrum of neutral hydrogen in the redshift range $5 < z < 28$. The HERA correlator implements both fringe stopping and baseline dependent averaging to bring down the data rate from nearly 1\,Tbps to 15\,Gbps.

\end{abstract}

\begin{frontmatter}

\begin{dedication}
\null\vfil
\begin{center}
To my parents and brother,\\\vspace{12pt}
For tolerating my obsession with space and being my first audience.

\end{center}
\vfil\null
\end{dedication}

% You can delete the \clearpage lines if you don't want these to start on separate pages.

\tableofcontents
\clearpage
\listoffigures
\clearpage
\listoftables

\begin{acknowledgements}

There are many people and institutions that I am indebted to for this opportunity and experience. I am grateful to the University of California, Berkeley for providing us a platform for research and pursuit of intellectual curiosity. Thank you members of the South African Radio Astronomy Observatory, for always making trip to the site so pleasant and accommodating us in your office. I also want to thank the National Science Foundation and the Gordon and Betty More Foundation for supporting my research with HERA.

I would sincerely like to thank Dr.~Aaron Parsons for being my advisor and guide throughout my course. Aaron, your vision for radio astronomy and EoR science, and your understanding of instrumentation and its pitfalls are truly inspiring. Thank you for helping me appreciate the nuances of building a correlator system, for teaching me to be prepared for most eventualities during field deployments and for letting me participate in pushing the boundaries of radio astronomy just a little bit.

Dr.~Jack Hickish, thank you for being my mentor, friend, confidant and a fantastic teacher. Without your patient tutorials that sometimes ran late into the night and into weekends and, possibly, to the brink of your patience, I would never have been able to complete my degree. Thank you for challenging me and letting me fail and learn at my own pace. Thank you for letting me grow as a graduate student and as an engineer, I really appreciate the increasing amount of responsibility you gave me with the correlator. Lastly, thank you for entertaining my (crazy) theories about radio astronomy, life, universe and everything!

I am extremely grateful to Dr.~Dan Werthimer for his guidance and mentorship throughout my time in Berkeley. Dan, you and Mary Kate made me feel at home right from the first day I was in Berkeley. I have immensely benefited from your open-hearted good will and ability to bring different people together. Thank you for indulging my projections of various career paths and guiding me in the right direction at every step. Dave Macmahon, thank you for teaching me how to debug in a structured manner. Designing and programming with you is a lot of fun. Jonathon Kocz, thank you for motivating me to apply to postdoctoral positions and encouraging me about my prospects after graduation. 

I am grateful to many members of the Department of Astronomy for their encouragement and help. Thank you, Dr.~Eugene Chiang and Dr.~Eliot Quataert, for the invaluable support that you both gave me in the capacity of Chair. Dr.~Dan Weisz, thank you for talking to me about my dissertation work and long term goals, for encouraging me to pursue my interests, and for giving me the courage to endure the hard parts of graduate school. Dr.~Chung-Pei Ma, thank you for being my academic advisor and helping me develop perspective on my thesis work. You gave me crucial guidance and support during a critical time in my PhD. Dr.~Jessica Lu, thank you for chairing my qualifying exam committee and for helping me through some sticky situations in graduate school.

I would like to thank International House and the Allan \& Kathleen Gateway Fellowship for the opportunity to live in a wonderful community of students from all around the world and from multiple disciplines. I have made some life-long friends here and will always cherish the memories of playing mafia late into the night on many a Friday. The astronomy graduate student community, BADGrads, has been fantastic company and solace throughout my time in Berkeley.

Lastly, I owe a huge debt of gratitude to Saundra Albers and Alex Lyons for keeping me sane through this time. I am extremely grateful to my parents, Dr. Rajendra Prasad and Dr. Padmavathi Choudeswari for their unwavering support throughout graduate school. Your encouragement and support during the hard parts of graduate school kept me going and motivated me to bring it to a finish. Kaustav Majhi, thank you for giving me company through the highs and lows of graduate student life. This would not have been possible without you.

\end{acknowledgements}

\end{frontmatter}

\pagestyle{headings}

% (Optional) \part{First Part}

% Introduction
\chapter{Introduction}
\label{chp:Intro}

%The tools used to probe nature determine the quality of laws deduced about it. 
In a field like astronomy, where one is only a passive observer, advances in instrumentation can enable observations in new phase spaces, reveal new objects and new physics. In this thesis, I attempt to push the boundary of radio interferometry in a direction that will enable sensitive experiments like the detection of the weak neutral hydrogen signal from a period in the Universe called the Epoch of Reionization. In this chapter, I lay out the framework and context for understanding the rest of this thesis, motivating the scientific goals of my instrumentation work and the foundations of radio interferometry on which it is based. 

\section{Epoch of Reionization}
\label{chp:Intro:sec:EoR}

The Big Bang Model is currently our best theoretical understanding of the origin and evolution of our Universe, and is amply supported by observational evidence. The model postulates that everything within our observable Universe originated from a hot, dense singularity which expanded and cooled over time. Observations of the Cosmic Microwave Background Radiation (CMBR;~\citealt{Penzias_and_Wilson_1965}), which consists of photons from the first optically observable moment in the history of our Universe, form the principal proof for this model of our Universe \citep{Dicke_et_al_1965}.

About 380,000 years after the Universe started expanding from a singularity, it cooled sufficiently enough for the formation of neutral hydrogen from free electrons and protons, during an event called Recombination. This relatively quick capture of free electrons released photons from Thompson scattering and decoupled them from the charged baryonic particles, allowing them to stream unhindered toward us, forming the CMBR. Observing the footprint of the Universe at recombination gave us valuable information about it-- the mass, composition and age of the Universe, proportions of dark matter, dark energy, photons and observable matter and the curvature being a few of them. Another key piece of information that is evident from the CMBR is that the Universe used to be remarkably smooth, i.e. to 1 part in $\sim$10,000 the temperature in one direction was the same as in any other direction~\citep{Mather_et_al_1990}.

\begin{figure}
    \centering
    \includegraphics[width=\linewidth]{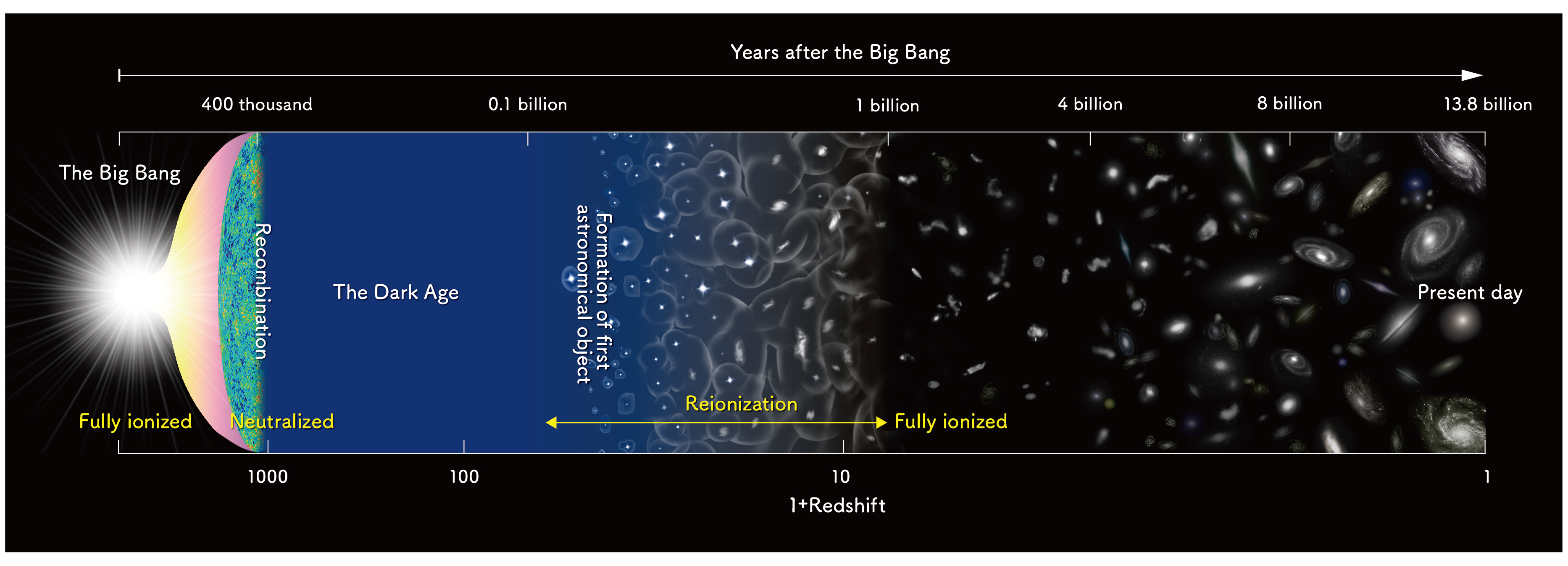}
    \caption{Artist's impression of the evolution of the Universe in the standard cosmological paradigm. The first luminous structures are theoretically expected to have formed at a definite time, and for a finite period called the Epoch of Reionization. Observations of this time span could answer many questions about the evolution of our Universe from a homogeneous isotropic medium at recombination to the hierarchical structures we see today. (Credit: NAOJ)}
    \label{fig:Universe_evolution}
\end{figure}

The Universe we observe today is unsmooth or non-isotropic on small scales ($\lesssim$100\,Mpc), with clusters of galaxies forming filaments and sheets in a cosmic web~\citep{Dodelson_2003, Springel_et_al_2006}. As depicted in Figure~\ref{fig:Universe_evolution}, observable baryonic matter must have undergone structure formation in the intervening years between recombination and today. Our current understanding of structure formation is largely incomplete and has been pieced together from observations of the CMBR and the local Universe~\citep{Zaroubi_2013}. A distinct event that can shed more light on the evolution of the Universe is the period of the formation of the first luminous structures, consisting of stars, galaxies and quasars (henceforth collectively referred to as stars). The astrophysics of star-formation indicates that the first luminous sources must have ionized the neutral hydrogen formed during Recombination, leading to the term Epoch of Reionization (EoR) for this era of cosmic dawn~\citep{Loeb_and_Furlanetto_2013}. Observations of the EoR promise to answer many questions-- what phenomena triggered and governed the formation of the first stars? How much ionizing radiation did they produce? Were the first luminous structures dwarf galaxies or quasars or other astrophysical objects? Did the high density regions get ionized first or the low density regions? What is the temperature evolution of the inter-galactic medium? and much more. 

Despite the pivotal role of EoR in the history of our Universe, observational evidence that can constrain the details of this period is scarce. The strongest constraints currently available come from two key probes-- the Gunn-Peterson~\citep{Gunn_and_Peterson_1965} trough in the spectra of distant quasars and observations of the CMBR~\citep{Plank_Collaboration_2016}, which are discussed in Sections~\ref{chp:Intro:sec:EoR:subsec:LyA} and~\ref{chp:Intro:sec:EoR:subsec:CMBR} respectively.

There are also a variety of other indirect probes that are not discussed here, but are worth mentioning. The width of low column density absorption lines in the Ly-$\alpha$ forest can be used to model the temperature of the inter-galactic medium (IGM) to high redshifts~\citep{Hui_and_Haiman_2003, Theuns_Schaye_et_al_2002}. These estimations~\citep{Schaye_et_al_2000, Theuns_Zaroubi_et_al_2002, Zaldarriaga_2002, Lidz_et_al_2010} indicate the universe was too hot by redshift $\sim$6 for reionization to occur at $z \gtrsim 10$ \citep{Bolton_et_al_2010}, but are dependent on the assumed cooling function for the IGM at later redshifts. 

Lyman-break galaxies can be used to model the number of ionizing photons per baryon and, hence, to map out the ionizing emissivity as a function of redshift~\citep{Bolton_and_Haehnelt_2007}. These observations~\citep{Bouwens_et_al_2005, Oesch_et_al_2010, Bunker_et_al_2010, McLure_et_al_2010, Bouwens_et_al_2011} indicate that the amount of ionizing radiation emitted by high-redshift galaxies is insufficient for reionization to have occurred primarily due to galaxies~\citep{Bolton_and_Haehnelt_2007, Calverley_et_al_2011}. However, the number of galaxies as a function of luminosity (or mass) is poorly constrained and can affect this conclusion.

Surveys of other galaxies, like Ly-$\alpha$ emitters~\citep{Ouchi_et_al_2009}, at high redshifts provide some indication about the relative contribution of dwarf galaxies and quasars to the reionization process~\citep{Bouwens_et_al_2017} but these also suffer from modelling uncertainties and contamination from foreground cool stars or interloper galaxies~\citep{Ellis_2008}. High redshift quasars~\citep{Mortlock_et_al_2011}, metal abundances at high redshifts~\citep{Rudie_et_al_2012}, and GRBs~\citep{Bromm_and_Loeb_2006} are other probes to reionization but provide limited constraints on the EoR.

\subsection{\texorpdfstring{Ly-$\alpha$}{Ly-A} forest probes}
\label{chp:Intro:sec:EoR:subsec:LyA}

The spectra of distant quasars are an evidence that the universe contains more neutral hydrogen at high redshifts than in the local Universe. Quasars (quasi-stellar objects; QSOs) are distant, extremely luminous sources that have broad spectrum emission ranging from X-rays to radio on the electromagnetic spectrum. Their high luminosity, combined with cosmic redshifts, can be used to probe the inter-galactic medium (IGM) along the line-of-sight between the quasar and us~\citep{Frank_King_Raine_2002}.

\begin{figure}
    \centering
    \includegraphics[width=0.75\linewidth]{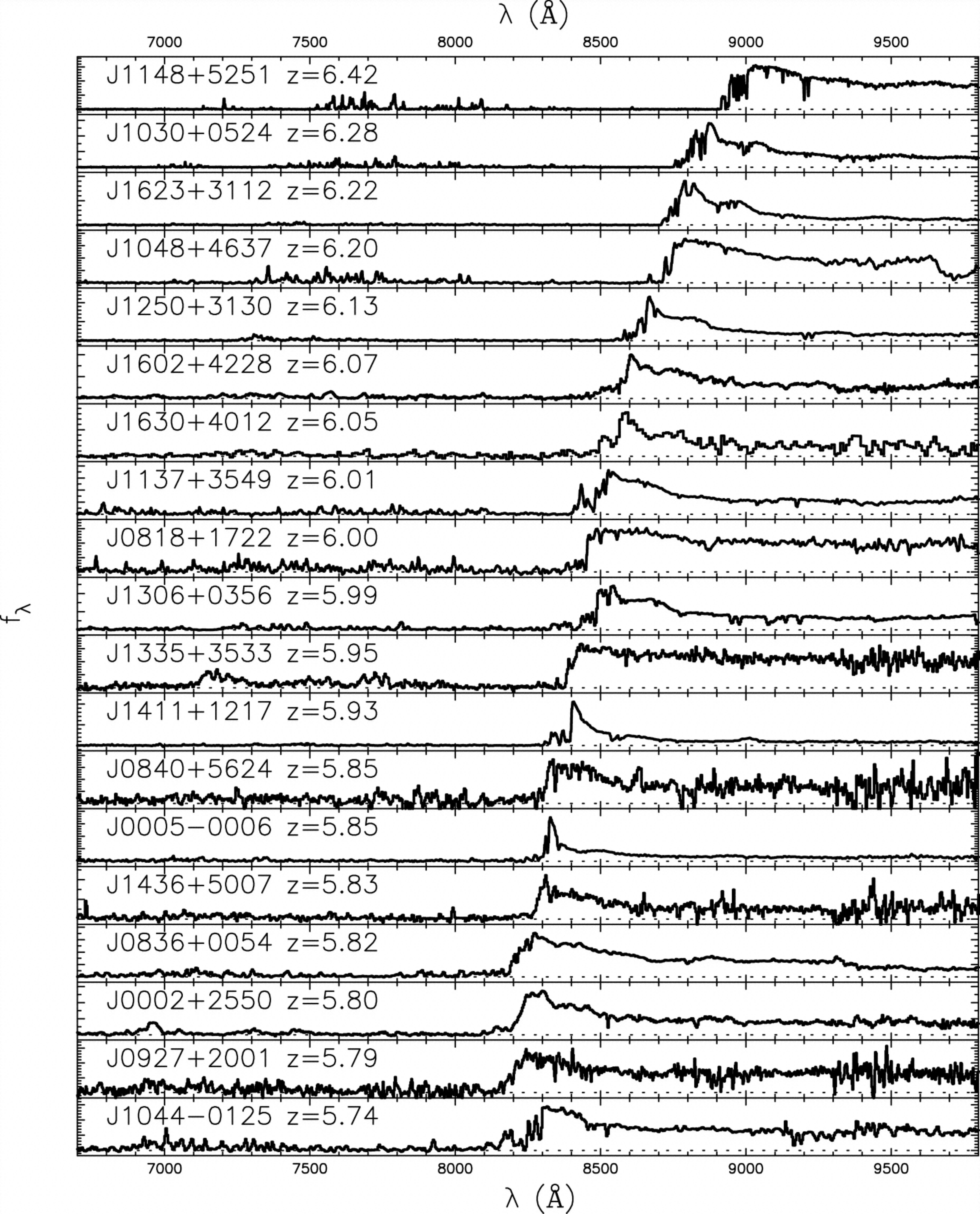}
    \caption{Spectra of QSOs located in the redshift range of the epoch of reionization, borrowed from \citet{Fan_et_al_2006}. The Gunn-Peterson trough, evident in the spectra at redshifts of $z\sim 6$ gradually disappears and is replaced by the Ly-$\alpha$ forest at lower redshifts, indicating the presence of large quantities of neutral hydrogen at high redshifts that disappear over time.}
    \label{fig:quasar_spectra}
\end{figure}

Within their rest-frame UV-optical band, quasar spectra show an emission peak around the Ly-$\alpha$ wavelength of 121.6\,nm which corresponds to the electronic transition from $n=2$ to $n=1$ orbital in a neutral hydrogen atom~\citep{VandenBerk_et_al_2001, Fan_et_al_2006}. The constant expansion of the Universe redshifts this ultraviolet wavelength to $\sim$1000\,nm, in the near-infrared, based on the distance to the quasar. Larger the redshift, larger the wavelength at which the peak emission is detected. This trend in evident in Figure~\ref{fig:quasar_spectra} which shows the spectra of quasars in the redshift range $6.42 \geq z \geq 5.74$. 

Neutral hydrogen in the IGM between the quasar and us, can absorb some of this emission at its own rest-frame Ly-$\alpha$ wavelength. For example, a quasar located at $z=6$ emits Ly-$\alpha$ photons at its rest-frame 121.6\,nm which corresponds to 850\,nm at Earth. If a pocket of neutral hydrogen is located at $z=5.90$, it can absorb some of this radiation and convert ground state hydrogen to the excited state. This absorption occurs at the rest-frame Ly-$\alpha$ wavelength at $z=5.90$ which corresponds to a lower wavelength of $\lambda=$ 839\,nm when observed from Earth. 

Multiple such clouds of neutral hydrogen in the IGM absorb portions of the emitted radiation blueward of the Ly-$\alpha$ emission peak, creating a series of absorption lines called the Ly-$\alpha$ forest~\citep{Lynds_1970}. When the density of neutral hydrogen is high, the absorption lines are hard to distinguish and instead appear as an absorption trough called the Gunn-Peterson trough~\citep{Gunn_and_Peterson_1965}, after the scientists who first characterized it from observations. The higher density of absorbers at redshifts of $z\sim$ 6 is larger than expected from passive evolution of the density due to expansion of the Universe. This increased density is attributed to a larger fraction of neutral hydrogen in the IGM of the Universe at these times~\citep{Fan_et_al_2003, Fan_et_al_2006}.

While the presence of the Gunn-Peterson trough in the spectra of high redshift quasars indicates a larger fraction of neutral hydrogen in the Universe, it does not constrain the value of that fraction very well. Theoretically, it can be shown that the optical depth to Ly-$\alpha$ emission gets saturated when the ionized fraction of the Universe is only on the order of $10^{-4}$~\citep[Section 2.1]{Zaroubi_2013}. Hence, Ly-$\alpha$ forest and the Gunn-Peterson trough can only probe the tail end of the reionization process. Moreover, quasars at high redshifts are far and few. Line-of-sight variations~\citep{Mesinger_and_Furlanetto_2009} and model dependent mechanisms like the radius of the ionization bubble around the quasar~\citep{Wyithe_and_Loeb_jul2004, Wyithe_and_Loeb_nov2004}, can significantly change the inferred neutral hydrogen fraction~\citep{Wyithe_and_Loeb_feb2004}.

\subsection{Evidence from CMBR}
\label{chp:Intro:sec:EoR:subsec:CMBR}

The power spectrum of the CMB gives valuable information about the time and duration of the EoR. The CMB is evidence that Recombination, or formation of neutral hydrogen took place around $z$$\sim$1100. If free electrons and protons were not bound-up in neutral hydrogen atoms, Thomson scattering from these charged particles would have damped the CMB until the Universe cooled to a temperature where photons could decouple. A similar argument can be used to determine the redshift at which reionization of the Universe could have occurred since the free-electrons created by reionization introduce Thomson scattering at later times.

The column density of free electrons along the line-of-sight to the CMB can be modelled in terms of an optical depth. The Thomson scattering optical depth ($\tau$) has a one-to-one correspondence to the redshift of reionization ($z_{re}$), under the assumption that the Universe reionized instantaneously~\citep{Griffths_Barbosa_Liddle_1999, Venkatesan_Aparna_2000}. Assuming a constant profile of electron density, as a function of redshift, a higher $\tau$ imples a higher $z_{re}$ and hence an earlier onset of star formation. The Wilkinson Microwave Anisotropy Probe (WMAP) was the first experiment to measure the optical depth to CMB and placed it at $\tau = 0.085$ which was significantly higher than what other probes indicated~\citep{Hinshaw_et_al_2013}. However, the latest results from the Planck mission which use higher sensitivity polarization measurements, report $\tau = 0.051$ and an instantaneous redshift of reionization $z_{re} = 7.68 \pm 0.79$ which is compatible with other probes~\citep{Plank_Collaboration_2016}.

The CMB can also be used to constrain the evolution of the neutral hydrogen fraction as a function of redshift, using the kinetic Sunyaev-Zel'dovich effect~\citep{Sunyaev_and_Zeldovich_1980}. This is based on the idea that ionized electrons, post Recombination can interact with CMB photons and scatter them into our line-of-sight. This causes anisotropies in the CMB power spectrum, but on scales different from the primordial anisotropies imprinted during Recombination. During EoR, reionization was ``patchy", creating low-amplitude spatial structure in the CMB radiation. Measurements of the kSZ power spectrum, by the South Pole Telescope, place the duration of reionization at $\Delta z < 7.9$~\citep{Zahn_et_al_2012} These estimates, however, depend on the model of the redshift evolution of electron density assumed and the optical depth to CMB measurements.

% \paragraph{}
% In recent years, some authors~\citep{Sobacchi_Mesinger_Greig_2016, Park_et_al_2019, Qin_et_al_2021} have tried combining observations of CMB, Ly-$\alpha$, galaxy surveys and simulated neutral hydrogen power spectra to obtain better limits.

\section{Neutral Hydrogen}
\label{chp:Intro:sec:HI}

A promising probe of cosmic dawn is the redshifted 21\,cm signal from neutral hydrogen (HI) emitted by the spin-flip transition of the lone electron in a hydrogen atom~\citep{Hogan_and_Rees_1979, Scott_and_Rees_1990, Madau_Meiksin_Rees_1997}. The electron-proton pair in a neutral hydrogen atom can be in a spin anti-aligned state (singlet state, $1_{0}S_{1/2}$) or in a spin aligned state (triplet state, $1_{1}S_{1/2}$). At Recombination, when neutral hydrogen formed from free electrons and protons, about a fourth of the atoms formed in the singlet state and the rest in the triplet state, proportional to their degeneracy ratio. Neither alignment was particularly favored relative to its degeneracy, since the energy difference between the triplet and singlet state is much smaller than the thermal energy of atoms at Recombination. 

However, the hydrogen atom is slightly more stable in the singlet state than in the triplet state. A direct transition from the triplet to the singlet state is ``forbidden" by dipole selection rules of spectroscopy~\citep{Harris_and_Bertolucci_1980}, and has an extremely small rate of $3 \times 10^{-15}\,\mathrm{s}^{-1}$ or about one transition in a few million years. Despite this, it is still observable in the Universe due to the sheer quantities of neutral hydrogen available~\citep{Field_1958}. 

The spontaneous spin-flip transition releases a small quanta of energy corresponding to a photon at a wavelength of 21\,cm or 1.4\,GHz. This falls in the radio portion of the electromagnetic spectrum, making it observable by Earth-based radio telescopes. Warm neutral hydrogen in the outer rims of our galaxy~\citep{Ewen_and_Purcell_1951} and other neighbouring galaxies can be observed through this weak emission line to establish rotation velocities~\citep{Muller_and_Oort_1951}. Observations of the rotation velocity profile of our galaxy even led to one of the first speculations of the existence of dark matter~\citep{Rubin_and_Ford_1970}. Most observations of neutral hydrogen are so far limited to the local Universe. The highest redshift at which we currently have a direct detection of neutral hydrogen is $z$$\sim$1~\citep{Chang_et_al_2010, Masui_et_al_2013}.

Recently, there has been a rebirth of interest in 21\,cm studies due to its potential in uncovering the physics that governed the dark ages and cosmic dawn. Observations of neutral hydrogen in the early Universe can be characterized as emission/absorption profiles in contrast to a background radio source like the CMB emission. Occasionally, a bright radio source, like a radio-loud quasar, can be used as the background source to probe the ``21\,cm forest"~\citep{Carilli_Gnedin_Owen_2002} in a similar manner as the Ly-$\alpha$ forest discussed above in Section~\ref{chp:Intro:sec:EoR:subsec:LyA}. In this text, I will focus on the former, general probe which contrasts the 21\,cm emission with the CMB. 

\subsection{Physics of \texorpdfstring{21\,cm}{21 cm}}
\label{chp:Intro:sec:HI:subsec:21cm}

The intensity of emission\,($I_{\nu}$) at the large wavelengths of the spin-flip transition can be modelled in the Rayleigh-Jeans limit of black-body radiation as~\citep[from][]{Rybicki_and_Lightman_1979}:
\begin{equation}
    I_{\nu} \equiv 2k_B \frac{\nu^2}{c^2} T_b
\end{equation}
\noindent
where $k_B$ is the Boltzmann constant, $\nu$ is the frequency of observation, $c$ is the speed of light and $T_b$ is a brightness temperature that can be used to characterize the intensity of emission. Henceforth, I will refer to all radio intensities in terms of their brightness temperature, for example, the intensity of the CMB radiation will be represented in terms of its brightness temperature $T_{\mathrm{CMB}}$ and so on.

The brightness temperature contrast of the 21\,cm signal with the background can be modelled as:
\begin{equation}
\label{eq:brightness_temperature}
    \delta T_b = \frac{(T_S - T_{\mathrm{CMB}})}{1+z} \left(1 - e^{-\tau_{\nu}}\right)
\end{equation}
\noindent
where $z$ is the redshift at which the contrast is being measured, $\tau_{\nu}$ is the optical depth along the line of sight and $T_{S}$ is the spin temperature of neutral hydrogen. 

The spin temperature represents the ratio of atoms in the triplet state\,($n_1$) to the singlet state\,($n_0$):
\begin{equation}
    \frac{n_1}{n_0} = \frac{g_1}{g_0}\;\mathrm{exp}\left(-\frac{T_{*}}{T_S}\right) = 3 \; \mathrm{exp}\left(-\frac{T_{*}}{T_S}\right)
\end{equation}
\noindent
where $(g_1/g_0)$ is the ratio of the statistical degeneracy of each state, and $T_{*} \equiv hc/k_B\lambda_{\mathrm{21cm}} \approx 0.068\mathrm{K}$.
The key point to note from Equation~\ref{eq:brightness_temperature} is that only deviations between the spin temperature and the background CMB can be detected. If the spin temperature is equal to the background CMB temperature, that period in the reionization history will not be visible to 21\,cm probes~\citep{Field_1958, Field_1959a, Field_1959b, Hogan_and_Rees_1979}.

\subsection{Spin Temperature}
\label{chp:Intro:sec:HI:subsec:Ts}

Since the 21\,cm transition can take millions of years to occur spontaneously, other mechanisms can effectively couple with neutral hydrogen creating local deviations in the intensity of emission~\citep{Field_1958}. In the early Universe, i.e. through the period of dark ages, cosmic dawn into the EoR, three mechanisms predominantly rule the spin temperature:
\begin{enumerate}[(i)]
    \item Emission/Absorption of background CMB photons.
    \item Collisions with hydrogen atoms or electrons.
    \item Resonant scattering with Ly-$\alpha$ photons (Wouthuysen-Field effect).
\end{enumerate}

The effect of either of these factors on the spin temperature can be quantified as~\citep[see][]{Field_1958}:
\begin{equation}
\label{eq:spin_temperature}
    T_S^{-1} = \frac{T_{\gamma}^{-1} + x_{\alpha}T_{\alpha}^{-1} + x_{c}T_{K}^{-1}}{1 + x_{\alpha} + x_{c}}
\end{equation}
\noindent
where $T_{\gamma}$ is the intensity of the background photons and can be set to the CMB brightness temperature\,($T_{\mathrm{CMB}}$). $T_{\alpha}$ is the wavelength of the radiation at which coupling with Ly-$\alpha$ photons occurs (expressed as a temperature in a similar way to $T_{*}$). $T_K$ is the kinetic temperature of gas (the only ``conventional" temperature in the equation). The coefficients $x_{\alpha}$ and $x_{c}$ determine the coupling to Ly-$\alpha$ photons and gas temperature respectively. If $x_{\alpha} \gg x_{c}$ and $x_{\alpha} \gg 1$, the coupling with Ly-$\alpha$ photons determines the spin temperature and the other two effects can be neglected; a similar argument can be made for $x_c$. $x_{\alpha} + x_{c} < 1$ would result in the spin temperature relaxing to the background CMB temperature.

Collisions between a neutral hydrogen atom, and another hydrogen atom or electron or proton can induce spin-flip transitions in the hydrogen atom and drive a spin temperature that is different from the background CMB~\citep{Field_1958, furlanetto_et_al_2006}. The exact physics of this coupling can be parameterized through the collisional cross-section between either of the three combinations of interactions~\citep{Allison_and_Dalgarno_1969, Liszt_2001, Smith_1966, Wild_1952, Zygelman_2005}. In this text I do not present the exact details of this coupling because they are likely more important for studying the dark ages than the EoR. The period, starting with the formation of the first stars until the completion of reionization, was likely more dominated by coupling with resonant Ly-$\alpha$ photons through the Wouthuysen-Field effect~\citep{Wouthuysen_1952, Field_1958} and to the kinetic gas temperature which increases due to X-ray heating~\citep{furlanetto_et_al_2006}. 

\subsubsection{Wouthuysen-Field Effect}
\label{chp:Intro:sec:HI:subsec:Ts:subsubsec:WF}

\begin{figure}
    \centering
    \includegraphics[width=0.75\linewidth]{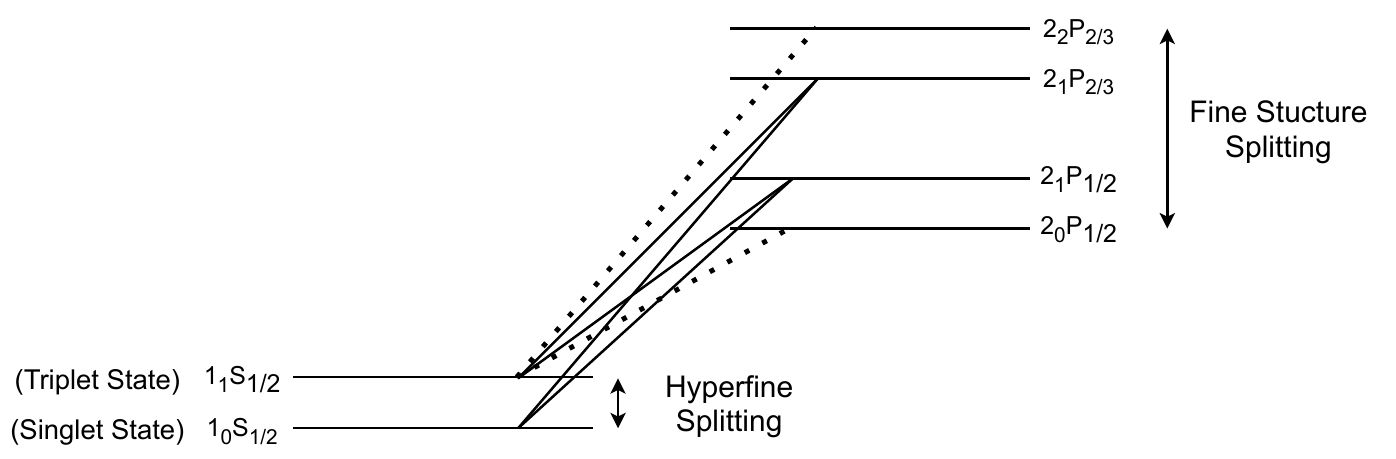}
    \caption{Electronic transitions in a hydrogen atom that produce the Wouthuysen-Field effect, or a change in the spin-state of the atom due to Ly-$\alpha$ coupling (solid lines). The dotted transitions are permissible but do not result in a change in spin state. The quantum orbital notation $\mathrm{n}_{\mathrm{F}}\mathrm{L}_{\mathrm{J}}$ is explained in the text.}
    \label{fig:wf_effect}
\end{figure}

The Wouthuysen-Field effect (WF effect;~\citealt{Wouthuysen_1952, Field_1958}) is mechanism by which a spin-flip transition is induced in the hydrogen atom by a Ly-$\alpha$ photon. Say, a hydrogen atom in the singlet state is excited by a Ly-$\alpha$ photon, moving the electron to one of two central p-orbital hyperfine states. A subsequent de-excitation of this atom can put the atom in the triplet state, causing a change in the spin of the electron. This change in spin, caused by the coupling with the Ly-$\alpha$ radiation field, would have dominated the spin temperature evolution in the period when the first stars formed.

Figure~\ref{fig:wf_effect} illustrates this mechanism methodically. The quantum state of the electron in the hydrogen atom can be depicted in the notation $\mathrm{n}_{\mathrm{F}}\mathrm{L}_{\mathrm{J}}$ where $\mathrm{n}$ is the principal quantum number and $\mathrm{L}$ is the electron orbital angular momentum denoted by the azimuthal quantum number\,($l$) that can take values $0,1,2,$\,etc. corresponding to s,p,d,\,etc. orbitals. $\mathrm{J}$ is the electron total angular momentum given by $\left|\mathrm{L} + \mathrm{S}_e\right|$ where $\mathrm{S}_e$ is the electron spin, and $\mathrm{F}$ is the total angular momentum of the hydrogen atom given by $\left|\mathrm{L} + \mathrm{S}_e + \mathrm{S}_p\right|$ where $\mathrm{S}_p$ is the spin of the proton in the nucleus of the hydrogen atom. The splitting the energy levels of the p-orbital (or fine structure splitting) occurs due to coupling of the electron spin with the magnetic field generated by the electron's orbit around the nucleus. The hyperfine splitting within each level is caused by coupling between the spin of the nucleus and the magnetic field from the electron's movement.

Dipole selection rules for electronic transitions prohibit the direct transition from $1_{1}\mathrm{S}_{1/2}$ to $1_{0}\mathrm{S}_{1/2}$ (or the ``forbidden" spin-flip transition), which allows the WF effect to dominate in the presence of Ly-$\alpha$ photons. Conservation of total angular momentum dictates that the transition of an electron from the ground singlet state cannot occur to either $2_0\mathrm{P}_{1/2}$ or $2_2\mathrm{P}_{2/3}$. The two possible states allowed are the central 2P orbitals with hyperfine splitting. However, de-excitation leading to the release of a ly-$\alpha$ photon, can place the hydrogen atom in the singlet or the triplet state. If the result is a hydrogen atom in the triplet state, a spin-flip has occurred.

The actual physics of this coupling is much more nuanced. The radiation field around the Ly-$\alpha$ line (denoted by color temperature\,$T_{\alpha}$ in Equation~\ref{eq:spin_temperature}) is affected by photons red-shifting into the Ly-$\alpha$ resonant frequencies, scattering with hydrogen atoms, spin-exchanges and the local gas temperature~\citep{Rybicki_2006, Hirata_2006}. That is, the WF effect couples the spin temperature of neutral hydrogen to the color temperature of the Ly-$\alpha$ radiation field, that is set to the kinetic gas temperature by Doppler broadening. Both, the total intensity of Ly-$\alpha$ radiation and the exact shape of the photon distribution close to the line center, determine the extent of coupling and hence the effect on the spin temperature observed during the EoR. 

\subsubsection{X-ray Heating}
\label{chp:Intro:sec:HI:subsec:Ts:subsubsec:Xray}

In the period immediately following the formation of first stars, the spin temperature coupling to Ly-$\alpha$ photons saturates and the coupling to gas temperature is expected to take over. Heating of the IGM is expected to increase the spin temperature and make it visible in emission against the CMB. There are many speculated sources for this heating-- shocks from the gas collapsing against the Hubble flow~\citep{Furlanetto_and_Loeb_2004}, Ly-$\alpha$ photons scattering off of hydrogen atoms imparting them with momentum~\citep{Madau_Meiksin_Rees_1997}, Compton up-scattering of CMB photons (which dominates heating through the dark ages, \citealt{Naoz_and_Barkana_2005}), and other exotic heating mechanisms like dark matter annihilation~\citep{Furlanetto_Oh_Pierpaoli_2006}. However, many authors~\citep{Madau_and_Efstathiou_1999, Venkatesan_Giroux_Shull_2001, Chen_and_Miralda_2004, Pritchard_and_Furlanetto_2007, Zaroubi_et_al_2007} agree that the dominant heating mechanism must be from X-ray photons. The source of this X-ray emission is likely high mass X-ray binaries that formed after the death of the first stars~\citep{Grimm_Gilfanov_Sunyaev_2003}.

X-rays have a much larger mean free path than Ly-$\alpha$ photons due to a low collisional cross-section. This allows them to penetrate the IGM more effectively and increase the gas kinetic temperature. The mean free path of X-ray photons can be characterized as~\citep[see][]{furlanetto_et_al_2006}:
\begin{equation}
    \lambda_X \propto \bar{x}_{\mathrm{HI}} \, (1+z)^{-2} \, E^3
\end{equation}
\noindent
where $x_{HI}$ is the neutral hydrogen fraction, and $E$ is the energy of the X-ray photon. The strong dependence of the mean free path on energy indicates that the heating is dominated by soft X-rays, which show structure on small scales. Hard X-rays contribute to a more uniform heating of the IGM~\citep{Pritchard_and_Loeb_2012}. X-rays heat the IGM predominantly through photo-ionization of hydrogen which releases high-velocity electrons. These photo-electrons subsequently dissipate heat to their surroundings, or lose energy by causing secondary ionizations or atomic excitations. The exact rate of heating can be estimated by computing the relative distribution of X-ray flux density to these mechanisms~\citep{Shull_1985, Furlanetto_and_Stoever_2010, Valdes_Evoli_Ferrara_2010}. The total X-ray flux density itself, is predominantly generated by high-mass X-ray binaries (HMXBs;~\citealt{Grimm_Gilfanov_Sunyaev_2003}). The population density of HMXBs at high redshifts can be extrapolated from observations of the local Universe, where HXMBs track the star formation rate (SFR) due to their relatively small lifespan~\citep{Lehmer_et_al_2010, Mineo_Gilfanov_Sunyaev_2011}. However, such conclusions should be regarded carefully since the ratio of HMXBs to SFR could vary with redshift~\citep{Dijkstra_et_al_2012}. The luminosity, number density and steepness of X-ray spectra at the redshifts of reionization are currently speculative, making these projections variable~\citep{Fialkov_Barkana_Visbal_2014}.

\subsection{Global Signal Model}
\label{chp:Intro:sec:HI:subsec:GSM}

Currently, we only have a notional model of the evolution of spin temperature as a function of redshift. There are large uncertainties in theoretical models and in the inferences from observations. In this section I will present a qualitative picture of the evolution of the sky-averaged 21\,cm signal that broadly sets the expectations in the signal threshold and detectability.

\begin{figure}
    \centering
    \includegraphics[width=\linewidth]{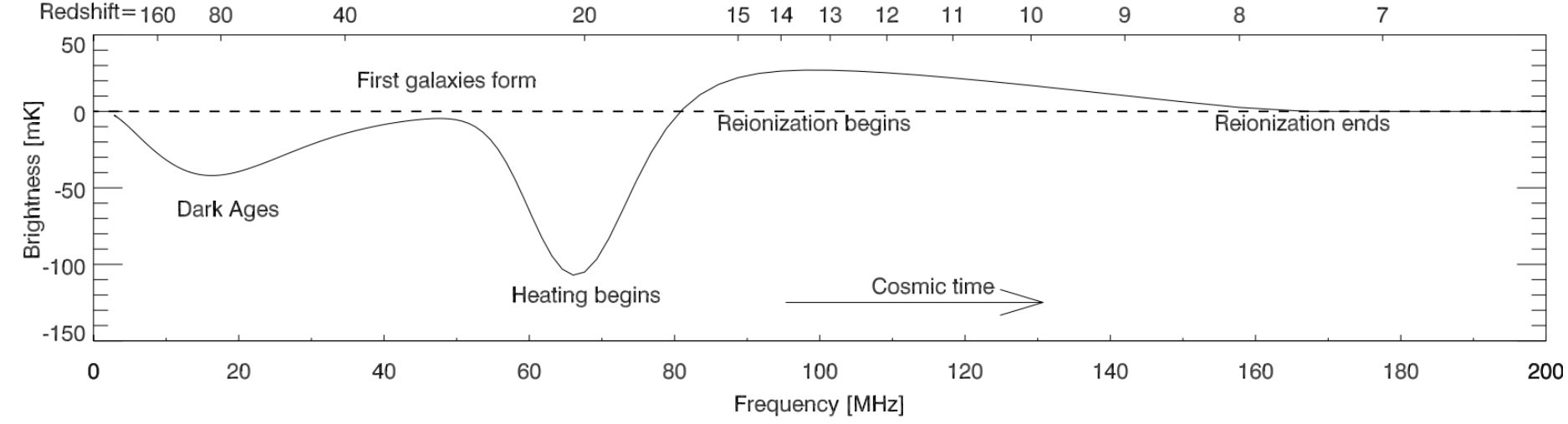}
    \caption{Evolution of the brightness temperature of the sky-averaged 21\,cm signal, taken from \citet[Figure 1]{Pritchard_and_Loeb_2012}.}
    \label{fig:global_signal}
\end{figure}

The evolution of the brightness temperature (Equation~\ref{eq:brightness_temperature}) of the 21\,cm signal can be split into distinct regimes based on the parameter that the spin temperature is most coupled to. In reality, the evolution between these distinct phases is gradual and our lack of understanding in the astrophysics of the early Universe could greatly mean that the order of these phases is different or that altogether different particles influence the evolution in these times~\citep{Bowman_et_al_2018, Barkana_et_al_2018}. Figure~\ref{fig:global_signal} shows the tentative prediction in the evolution of the brightness temperature and the different phases are explained below, following \citet{Pritchard_and_Loeb_2012}.

\paragraph{$\mathbf{1000 \gtrsim z \gtrsim 200}$} 
Compton scattering with the residual free electrons from recombination causes the spin temperature to be coupled to the CMB photons. This leads to $T_{S} \approx T_{\gamma}$ and a $\delta T_b \approx 0$ (not shown in the figure).

\paragraph{$\mathbf{200 \gtrsim z \gtrsim z_d}$} 
When the electron density is too low for effective Compton scattering, the spin temperature couples to the kinetic gas temperature. Adiabatic cooling of the gas leads to a decrease in the gas temperature as $(1+z)^2$, which comes from relating the temperature evolution of a gas with an adiabatic index of $\gamma = 5/3$ to the cosmic expansion. Collisional coupling with the gas leads to a spin temperature that falls relative to the background, making the 21\,cm signal detectable as an absorption feature until a redshift in the dark ages $z_d$.

\paragraph{$\mathbf{z_d \gtrsim z \gtrsim z_{*}}$} 
Around $z_d$, a redshift set by cosmological parameters, coupling between the spin temperature and the gas temperature diminishes as the gas density falls with the expansion of the Universe. This leads to radiative coupling with background CMB (whose temperature gradient with redshift is slower than gas) bringing up the spin temperature to the background CMB temperature, until the first stars and galaxies start forming at $z_{*}$.

\paragraph{$\mathbf{z_{*} \gtrsim z \gtrsim z_{\alpha}}$} 
This phase is dominated by spatial inhomogenities created by the first stars. The high cross-section of $Ly$-$\alpha$ photons, emitted by the first stars, results in ionization ``bubbles" that drive the spatial fluctuations in brightness temperature (not shown in the figure). During this period, the spin temperature in some areas is coupled to the CMB while in Ly-$\alpha$ photon rich environments, it is coupled to this radiation via the Wouthuysen-Field effect. This coupling mechanism to resonant Ly-$\alpha$ photons remains in place until the neutral hydrogen gets ionized at $z$$\sim$6. The global signature, which does not probe the spatial structure, is expected to be observed in absorption against the CMB background until the onset of X-ray heating at $z_{\alpha}$.

\paragraph{$\mathbf{z_{\alpha} \gtrsim z \gtrsim z_{h}}$} Coupling with Ly-$\alpha$ saturates as star formation occurs, leading to a redshift\,($z_{\alpha}$) where the spin temperature is coupled to the gas temperature. X-ray heating from high mass X-ray binaries, that form on the death of the first stars, drives up the spin temperature in this phase. While the exact astrophysics during this time determines the amount of X-ray heating, it is likely that some regions will be visible in emission against the CMB background around redshift $z_{h}$.

\paragraph{$\mathbf{z_{h} \gtrsim z_{r}}$} As the first sources start reionizing the neutral hydrogen, the ionization fluctuations become more important than the gas temperature and $T_{S}$ can be neglected in Equation~\ref{eq:brightness_temperature} until the end of reionization at $z_{r}$.

\paragraph{$\mathbf{z_{r} \gtrsim z}$} As neutral hydrogen is depleted, the 21\,cm signal disappears, and any residual signal only originates from isolated pockets.

\begin{figure}
    \centering
    \includegraphics[width=0.55\textwidth]{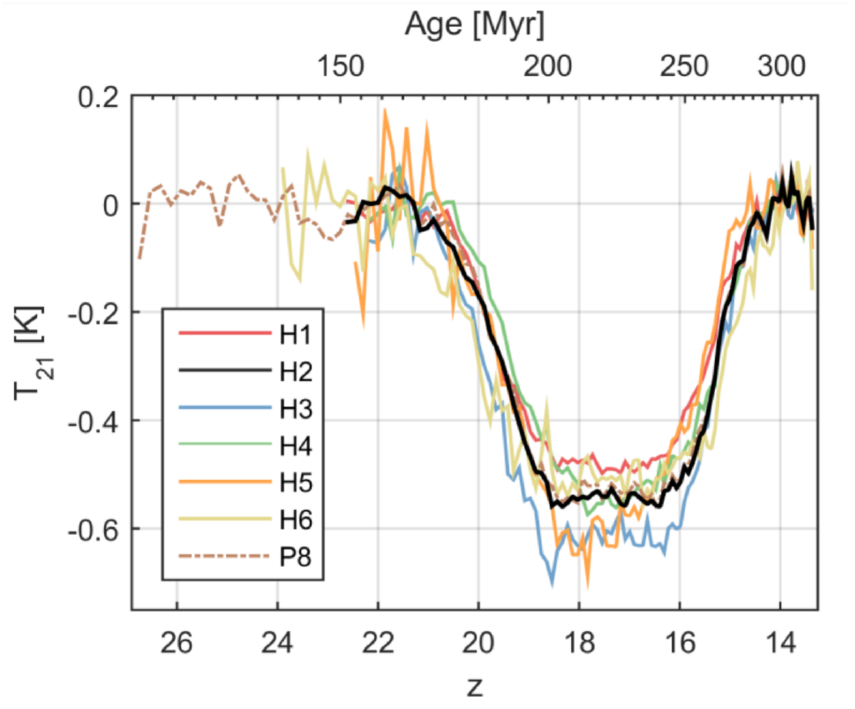}
    \caption{Results from the Experiment to Detect the Global EoR Signature (EDGES;~\citealt{Bowman_et_al_2018}) showing an absorption profile that is $\sim$50\% higher in amplitude than theoretically possible. The different lines show the best-fit profile obtained using different hardware configurations.}
    \label{fig:EDGES_result}
\end{figure}

\paragraph{}
As mentioned, this is a cursory picture of the evolution of the 21\,cm signal and there are major knobs in this theoretical model that can be adjusted by direct observations. However, measuring the global signal is onerous. The signal is faint compared to foregrounds (a problem that affects power spectrum measurements as well, Section~\ref{chp:Intro:sec:Pspec:subsec:Foregrounds}) and instrument systematics can be hard to remove. Measurements of the global signal, performed by ~\citet{Bowman_et_al_2018} (shown in Figure~\ref{fig:EDGES_result}) indicate that the theoretical models could be off by at least $\sim$50\% in the predicted $T_{\mathrm{CMB}}/T_S$ ratio. They suggest that this deep absorption profile could be caused by either a high Ly-$\alpha$ photon flux, which saturates the spin temperature and recouples it to the adiabatically cooling gas, or by more exotic scenarios like dark matter-baryon interactions~\citep{Barkana_et_al_2018}. However, \citet{Hills_et_al_2018} throw this result into question by observing that the foreground modelling used to derive the above result yields unphysical foreground emission parameters and that the profile obtained is not a unique solution for the given data.

Observations of the global 21\,cm signal can reveal a lot about the physics governing the formation of the first stars and subsequent heating of the IGM. As evident from the EDGES experiment, it has the potential to uncover exotic heating and cooling mechanisms involving non-baryonic particles. However, the rich statistical properties of reionization are encoded in the spatial fluctuations that the global signal does not capture. The spatial fluctuations can be characterized by a power spectrum, whose formalism is discussed in the next section.

\subsection{Power Spectrum}
\label{chp:Intro:sec:HI:subsec:Pspec}

The global signal is a sky-averaged (or large angular scale-averaged) quantity that can be interpreted as the zeroth order approximation of the full 21\,cm power spectrum. The actual evolution of structure in spin temperature is expected to be highly inhomogeneous, resulting from the interaction of gas and the various radiation fields present during this time. The spatial variations in the redshifted 21\,cm signal from neutral hydrogen can be statistically described in the form of a power spectrum. While images of the EoR can capture the details of the reionization process most accurately, a power spectrum can reasonably capture all the physics that governed this period. Power spectrum measurements can be used to constrain or estimate cosmological parameters, and also to distinguish between various theoretical reionization models. Most of the current-generation experiments that are targeting EoR observations aim to measure the power spectrum of this signal within a redshift range of interest.

Mathematically\footnote{Equations in this section have been adapted from \citet{Parsons_et_al_2012a, Liu_and_Shaw_2019}.}, a power spectrum represents the Fourier transform of a 3D correlation function. Say, $T(\mathbf{r})$ is the EoR brightness temperature measured at various 3D vector locations\,($\mathbf{r}$) in the Universe. The spatial correlation of this signal can be computed as a function of separation\,($\mathbf{x})$ as:
\begin{equation}
\label{eq:correlation_function}
    \xi(\mathbf{x}) = \left<T(\mathbf{r})\;T(\mathbf{r}-\mathbf{x}) \right>
\end{equation}
\noindent
where the angular brackets\,$\left<\right>$ denote an ensemble average over multiple locations\,$\mathbf{r}$. Then the power spectrum of this 3D map of the brightness temperature is given by:
\begin{equation}
\label{eq:power_spectrum_1}
    P(\mathbf{k}) = \int \xi(\mathbf{x})\;e^{-i\mathbf{k}\cdot\mathbf{x}}\; \mathrm{d}^3\mathbf{x}
\end{equation}
\noindent
where $\mathbf{k}$ is a measure of the spatial scale, represented as a Fourier mode or comoving wave-vector. This formulation of the power spectrum is useful to develop the intuition that a power spectrum probes the position space correlations in data. For a field that is homogeneous and isotropic with no large-scale structure, like the Universe through the dark ages, the power spectrum can be imagined roughly as a straight line with more power at large k-modes (small spatial scales) and less power at small k-modes (large spatial scales). For practical measurements of the power spectrum, it is useful to define it in terms of a Fourier transform of the brightness temperature field:
\begin{equation}
\label{eq:tb_fft}
    \widetilde{T}(\mathbf{k}) = \int\limits_{-\infty}^{\infty} T(\mathbf{r})\; e^{-i\mathbf{k}\cdot\mathbf{r}}\;\mathrm{d}^3\mathbf{r}
\end{equation}

The power spectrum is then defined by the correlation in various modes:
\begin{equation}
\label{eq:power_spectrum_2}
    \left<\widetilde{T}(\mathbf{k})\; \widetilde{T}(\mathbf{k}')\right> = (2\pi)^3 \delta^{D}(\mathbf{k}-\mathbf{k'}) P(\mathbf{k})
\end{equation}
\noindent
where $\delta^{D}$ is the Dirac delta function that is non-zero only when the argument of the function is zero. This allows the approximation:
\begin{equation}
\label{eq:power_spectrum_3}
    P(\mathbf{k}) \approx \frac{1}{V}\; \left<\left|\widetilde{T}^{\mathrm{obs}}(\mathbf{k})\right|^2\right>
\end{equation}
\noindent
where the total volume, $(2\pi)^3$ in Equation~\ref{eq:power_spectrum_2}  is replaced by the volume of the survey\,($V$). The complication in this practical recipe for computing the power spectrum is the ensemble average, which requires measurements of the brightness temperature field from numerous surveys, each with volume $V$ and probing regions that are governed by the same underlying statistical processes. This is nearly impossible for any survey. Power spectrum measurements of various parameters in the early Universe, like the dark matter power spectrum or the CMB angular power spectrum, exploit the isotropy in the Universe to perform an ensemble average over wave-vectors with the same wavenumber. That is, by recognizing that vector $\mathbf{k}$ can be replaced by the spherically averaged scalar $k = |k|$, the ensemble average can be replaced by an average in direction.

\begin{figure}
     \centering
     \subfloat[][]{\includegraphics[width=0.43\textwidth]{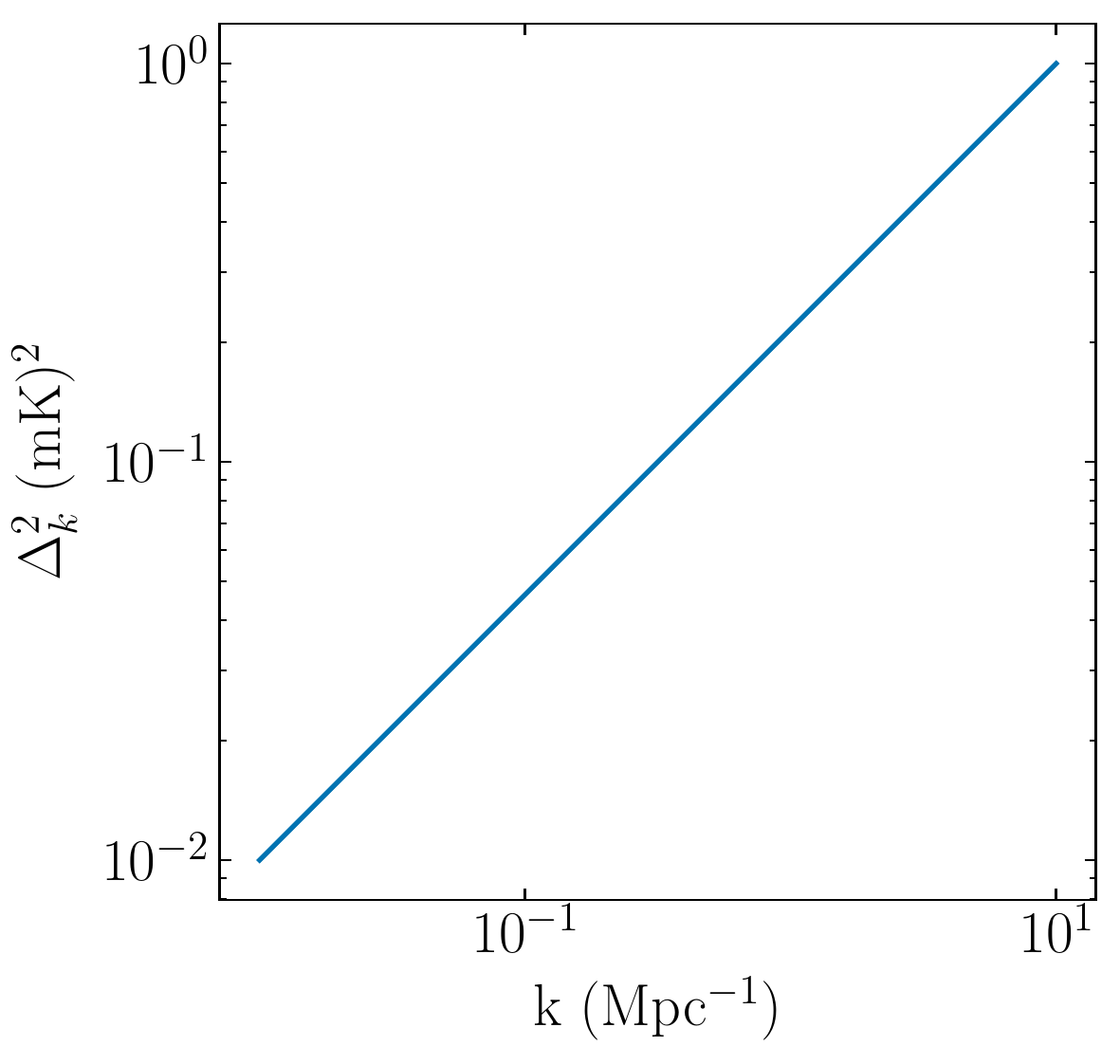}}
     \hspace{1.5em}
     \subfloat[][]{\includegraphics[width=0.5\textwidth]{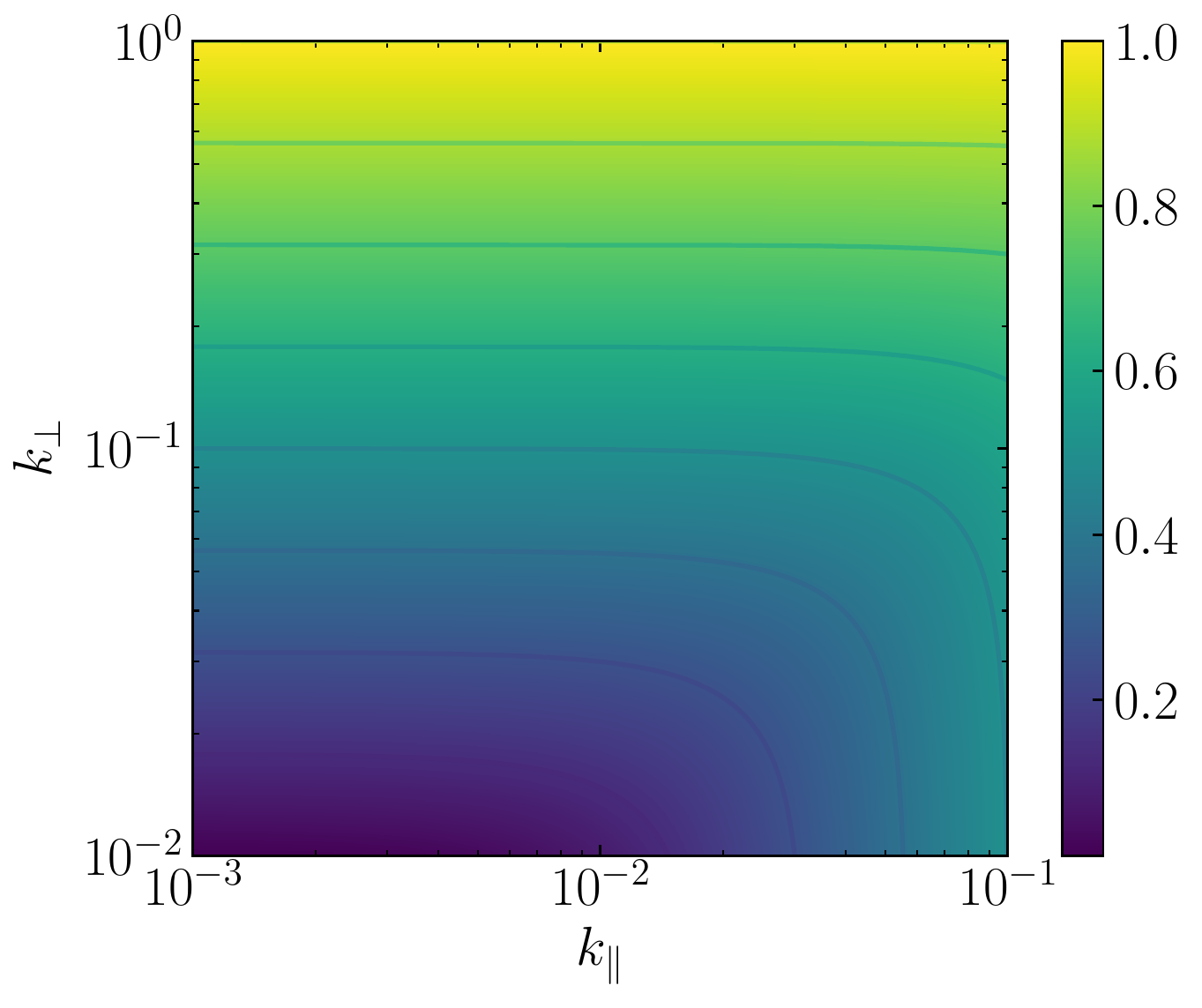}}
     \caption{Figure illustrating the relationship between spherically averaged coordinates and cylindrical coordinates, given by $k = \sqrt{k_{\bot}^2 + k_{\|}^2}$.  Panel\,(a): A simple mock power spectrum plotted against spherically averaged k-modes. Panel\,(b): The same power spectrum profile, plotted as a color-map against cylindrically averaged coordinates.}
     \label{fig:mock_power_spectrum}
\end{figure}

In estimating the power spectrum of the EoR, it is useful to maintain a distinction between the Fourier modes along the line-of-sight\,($k_{\|}$) and perpendicular to the line-of-sight\,($\mathbf{k}_{\bot}$). This is because the instrumental response and systematics that govern both these modes are very different. While the radio telescope's field-of-view and resolution determine the $\mathbf{k}_{\bot}$ modes, the frequency resolution and bandwidth determine the $k_{\|}$ modes. In the flat-sky limit, the cylindrical power spectrum can be estimated by averaging over rings of radius $k_{\bot}$ located at $\pm k_{\|}$ to obtain $P(k_{\bot},k_{\|})$ for small scales. In the cylindrical coordinate space, Equations~\ref{eq:tb_fft} and~\ref{eq:power_spectrum_3} can be rewritten as:
\begin{align}
\label{eq:tb_fft_cylindrical}
    \widetilde{T}(\mathbf{k}_{\bot}, k_{\|}) &= \int\limits_{-\infty}^{\infty} T(\mathbf{r}_{\bot}, r_{\|}) \, e^{-i\left(\mathbf{k}_{\bot}\cdot \mathbf{r}_{\bot} + k_{\|}r_{\|}\right)} \, \mathrm{d}^2 \mathbf{r}_{\bot} \, \mathrm{d}r_{\|} \\[1em]
\label{eq:power_spectrum_cylindrical}
    P(\mathbf{k}_{\bot}, k_{\|}) &\approx \frac{1}{V} \left<\left|\widetilde{T}^{\mathrm{obs}}(\mathbf{k}_{\bot}, k_{\|})\right|^2\right>
\end{align}

Figure~\ref{fig:mock_power_spectrum} shows a mock power spectrum, that could represent the 21\,cm signal from the early Universe before reionization. It shows the power spectrum in both spherically averaged k-modes and in cylindrical coordinates for illustration. The quantity plotted on both panels is called the \textit{dimensionless} power spectrum, rather unhelpfully since it has the units of $(\mathrm{mK})^2$. The dimensionless power spectrum is related to the definition we have been using so far as:
\begin{equation}
\label{eq:dimensionless_power_spectrum}
    \Delta^2(k) = \frac{k^3}{2\pi^2}P(k)
\end{equation}
\noindent
and can be intuitively understood as the power per $\log{k}$ bin. 

Measuring the 21\,cm power spectrum from the EoR is arduous. The signal is extremely faint compared to foreground sources, by roughly five orders of magnitude, making it hard to detect (Section~\ref{chp:Intro:sec:Pspec:subsec:Foregrounds}). In addition, radio telescopes that can measure the redshifted signal have a chromatic beam response (Section~\ref{chp:Intro:sec:Interf:subsec:Chromaticity}) which couples the foregrounds to the EoR signal. This coupling make it difficult to model and filter the foregrounds without filtering the EoR signal itself. To understand this effect, the challenge it poses to characterizing the EoR power spectrum, and the potential solution to this problem, it is necessary to detour briefly and discuss radio interferometry.

\section{Radio Interferometry}
\label{chp:Intro:sec:Interf}

In general, telescopes used for observing astronomical objects in the radio frequencies (10\,MHz - 30\,GHz) can be built as a single dish or in multiple dishes. Single dish telescopes operate similarly to their optical counterparts, and small dishes are often used as antennas in an interferometer. Telescopes which operate using multiple antennas are called interferometers and are the main topic of this section. Both these telescopes were used in the past for EoR observations. The Green Bank telescope in West Virginia~\citep{Prestage_et_al_2009} and the Parkes telescope in Australia~\citep{Staveley-Smith_1996} are both single dish structures, while the Murchison Wide-field Array (MWA;~\citealt{Tingay_et_al_2013}) in Western Australia, the Giant Metrewave Radio Telescope (GMRT;~\citealt{Kapahi_and_Ananthakrishnan_1995}) in India, the Owens Valley Radio Observatory Long Wavelength Array (OVRO-LWA), the Donald C. Backer Precision Array for Probing the Epoch of Reionization in South Africa (PAPER;~\citealt{Parsons_et_al_2010}) and the Low Frequency Array (LOFAR;~\citealt{VanHaarlem_et_al_2013}), in addition to other  upcoming EoR experiments, are all interferometers. As evident, interferometers are preferable for probing this weak signal. This is because of they offer more flexibility in dealing with foreground systematics, have lower cost per collecting area, and can be tailored to make a detection in specific power spectrum modes. In this section, I will present the theory of interferometry, the visibility equation that drives interferometric measurements and beam chromaticity that results in foreground coupling. \citet{tms2017} is an excellent reference for a detailed review of radio interferometry. 

\subsection{Radio Antennas}
\label{chp:Intro:sec:Interf:subsec:Antennas}

A radio antenna is a device that is capable of measuring the intensity of radio waves, originating from some source (astronomical or terrestrial), by generating a proportional voltage difference or current in the receiving element. Though not strictly necessary, radio antennas usually have a parabolic reflector around the receiver to increase the collecting area for radiation and improve signal-to-noise at the receiver or feed. Radio antennas differ from typical optical telescopes in their ability to probe the electric field; optical antennas can only measure the total intensity at the focal point via a photographic plate or a digital charged compact device (CCD). Another difference is that a radio telescope with a single parabolic reflector usually operates in a diffraction-limited regime, unlike its optical counterpart which is usually seeing-limited due to the Earth's atmosphere. 

To theoretically understand the beam response of a single radio antenna, let us consider a hypothetical single dimensional telescope. The beam response of the reflector in this case can be approximated to the diffraction pattern generated by a single slit of classical wave optics. The intensity due to the electric field pattern created by this reflector at the focal point is given by the single slit diffraction equation:
\begin{equation}
    I(\theta) = I_0\; \mathrm{sinc}^2 \left(\frac{\pi w \sin\theta}{\lambda}\right)
\end{equation}
\noindent
where $I_0$ is the intensity of the source, $\theta$ is a small angle around the focal point, $w$ is the width of the slit and $\lambda$ is the wavelength at which the telescope operates. 

For a two dimensional parabolic reflector, the electric field at the focal point is given by the Airy pattern, or the Fourier transform of a circular aperture. This can mathematically be represented as:
\begin{equation}
\label{eq:airy_disk}
    I(\theta) = I_0 \left[ \frac{2\mathrm{J}_1(x)}{x} \right]^2 ; \quad x = \frac{2\pi D \sin\theta}{\lambda}
\end{equation}
\noindent
where $\mathrm{J}_1 (x)$ is a Bessel function of the first kind of order one, $D$ is the diameter of the dish and $\lambda$ is the wavelength of observation. This equation underlies the Rayleigh criterion for the resolution of an image constructed by a single dish antenna:
\begin{equation}
    \theta = 1.22 \; \frac{\lambda}{D}
\end{equation}

That is, the larger the diameter of the single dish telescope, higher the resolution of the image that one can build using it. This is the driving factor behind building large telescopes like Arecibo~\citep{Castleberg_and_Xilouris_1997} (RIP), the Green Bank telescope, Five-hundred-meter Aperture Spherical radio Telescope (FAST;~\citealt{Nan_2006}) etc. However, the mechanical costs of building a large structure are empirically known to scale roughly as $D^{2.7}$, making it prohibitively expensive to build single dish telescopes beyond a point. Interferometers are a cheaper solution to building resolution since the effective diameter is the distance between two antennas, which can be made arbitrarily large with minimal cost.

\subsection{Two-element Interferometer}
\label{chp:Intro:sec:Interf:subsec:Twoelement}

Radio interferometers consist of multiple antennas (with possible rare exceptions), with the number of antennas being anywhere between a few tens to a few hundreds. Each of these antennas typically has a reflector, albeit much smaller in diameter that a single dish telescope. The signal from the different antennas is usually combined to form a single radio image, with a resolution equivalent to the span of the array, either in real-time or in post-processing. Interferometers are so called because the signal from any two antennas in the array can be vector added (or scalar multiplied) to construct an interference pattern. In fact, imaging with interferometers is performed by computing the interference pattern generated by all antenna pairs in the array.

\begin{figure}
    \centering
    \includegraphics[width=0.45\textwidth]{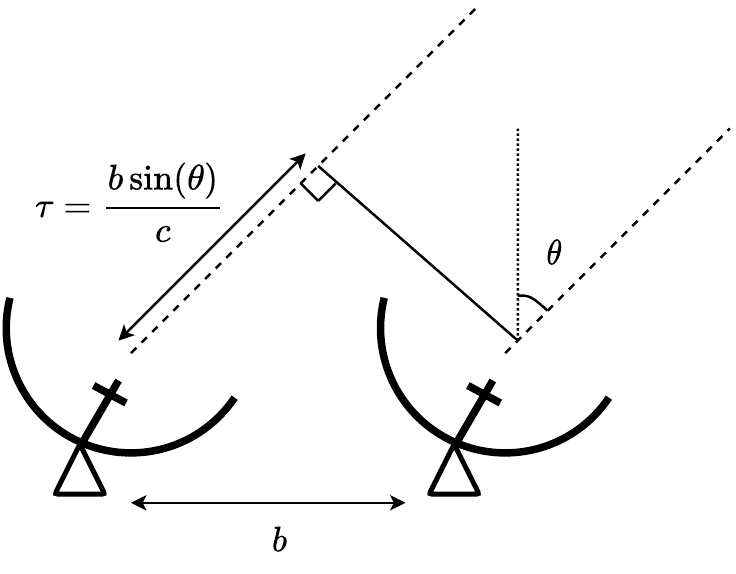}
    \caption{In a simple two element interferometer, the small difference in the arrival time of a signal between the two antennas generates interferometric fringes in the combined response.}
    \label{fig:two_element_interferometer}
\end{figure}

In the simplest case, let us consider a two-element interferometer like the one shown in Figure~\ref{fig:two_element_interferometer}. The analysis of this setup is similar to that of a Young's double slit experiment in the optical domain. The distance between the antennas, or length of the baseline vector\,($b$), creates a small path difference which leads to a time delay in the arrival of the signal:
\begin{equation}
\label{eq:time_delay}
    \tau = \frac{b\,\sin(\theta)}{c}
\end{equation}
\noindent
where $c$ is the speed of light and $\theta$ is the relative angle between the source location and zenith. This can be see from simple trigonometry on the right-angled triangle formed by the two rays of light and the baseline vector.

This delay in arrival time can also be viewed as a phase difference in the electric field probed by the two antennas. That is, if the electric field probed by the antenna on the right can be modelled as $\mathbf{E}_r = \mathbf{E}_0 \sin{(\omega t)}$, then the electric field probed by the antenna on the left can be represented as $\mathbf{E}_l = \mathbf{E}_0 \sin{(\omega t + \phi)}$ where the phase difference\,($\phi$) is given by:
\begin{equation}
\label{eq:phase_difference}
    \phi = {2\pi}\,\frac{\mathbf{b}\cdot\mathbf{\hat{s}}}{\lambda}
\end{equation}
\noindent
Here, the path difference has been formalized as a vector product between the baseline vector and the direction of the source, measured from the horizon rather than the zenith. Note that the phase difference is necessarily defined only for a particular wavelength.

Using the above definitions for the electric field probed by each antenna, and substituting the phase difference we can derive the intensity of the combined signal as:
\begin{align}
\label{eq:fringes1}
    I &\propto \left(\mathbf{E}_r + \mathbf{E}_l\right)^2 \\
\label{eq:fringes2}
    &=  I_0 \, \left[ 1 + \cos\left(\frac{2\pi \mathbf{b}\cdot\mathbf{\hat{s}}}{\lambda}\right) \right]
\end{align}

The $\cos^2$ response in the signal, as function of the projected baseline vector in direction of the source, is characteristic of interferometers and represents the primary difference from single dish telescopes. As the source moves in the sky (due to the Earth's rotation), the projection of the baseline vector in the direction of the source continuously changes causing the characteristic \textit{interferometric fringe pattern}.

\subsection{Visibility Equation}
\label{chp:Intro:sec:Interf:subsec:Vis}

Formalizing the above derivation, the response of a two-element interferometer can written as:
\begin{equation}
\label{eq:visibility}
    \mathcal{V}(\mathbf{b},\lambda) = \int\limits_{\mathrm{sky}} I(\mathbf{\hat{s}}, \lambda) \, B(\mathbf{\hat{s}}, \lambda) \, e^{-2\pi i (\mathbf{b}\cdot\mathbf{\hat{s}})/\lambda}\; \mathrm{d}\Omega
\end{equation}
\noindent 
where $\mathcal{V}(\mathbf{b},\lambda)$ is the \textit{visibility} measured by the pair of antennas separated by baseline vector $\mathbf{b}$, at the wavelength $\lambda$. The intensity of the single source\,($I_0$) in Equation~\ref{eq:fringes1} is replaced by the intensity of the sky in all directions, which is sampled by the primary beam of each antenna, given by $B(\mathbf{\hat{s}},\lambda)$. The primary beam is theoretically equivalent to the Bessel function described in Equation~\ref{eq:airy_disk}, but can vary for different reflector designs. The exponential term represents the interferometric fringe pattern in the visibility field, previously encoded by the cosine term.

Under a flat-sky approximation, the vectors in Equation~\ref{eq:visibility} can be decomposed into $x,y$ components. The baseline vector can be represented by its components $(b_x,b_y)$ or the wavelength normalized baseline coordinates $u \equiv b_x/\lambda$ and $v \equiv b_y/\lambda$, and the direction of the source is represented by the coordinates $(l,m)$ where $l \equiv \sin{\theta_x}$ and $m \equiv \sin{\theta_y}$. Using these terms the visibility equation can be written as:
\begin{equation}
\label{eq:visibility_uv}
    \mathcal{V}(u,v) = \int\frac{\mathrm{d}l\,\mathrm{d}m}{\sqrt{1- l^2 - m^2}} I(l,m,\lambda)\; B(l,m,\lambda)\; e^{-2\pi i (ul + vm)}
\end{equation}

From the above equation you can see that the visibilities measured by different pairs of antennas in a array, with different baseline vectors, can all be mapped to a single $(u,v)$ coordinate space; usually called the \textit{uv-plane} in radio astronomy. Another interesting feature of interferometers that is evident from the above equation is that the uv-coordinates are the Fourier equivalent of the sky coordinates $(l,m)$. That is, the visibilities measured by an interferometer sample the spatial Fourier transform of the sky. 

This Fourier relationship between the uv-plane and the sky drives the design of interferometers that have a primary science goal of imaging. In traditional radio astronomy, interferometers were designed to maximize uv-coverage or sample as many unique points on the uv-plane as possible. The Giant Meterwave Radio Telescope (GMRT;~\citealt{Kapahi_and_Ananthakrishnan_1995}), Atacama Large Millimeter Array (ALMA;~\citealt{Escoffier_et_al_2007}), the Very Large Array (VLA;~\citealt{vla_1980}), etc. have all been designed with antennas pairs at unique distances so as to measure different uv-samples. 

\subsubsection{UV Plane and the Synthesized Beam}
\label{chp:Intro:sec:Interf:subsec:Vis:subsubsec:uv}

\begin{figure}
\centering
\subfloat[]{\label{main:a}\includegraphics[width=0.6\textwidth]{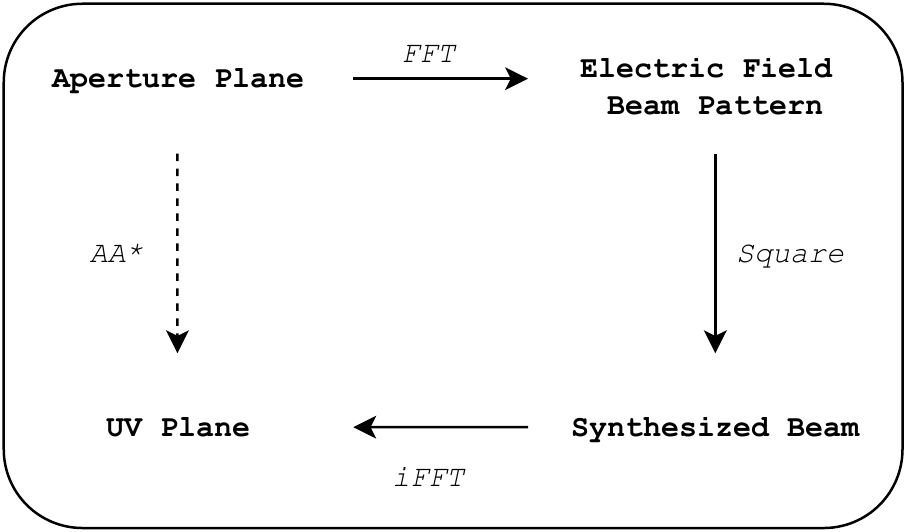}}\par\medskip
\begin{minipage}{.5\linewidth}
\centering
\subfloat[Single Dish Telescope]{\label{main:b}\includegraphics[width=\textwidth]{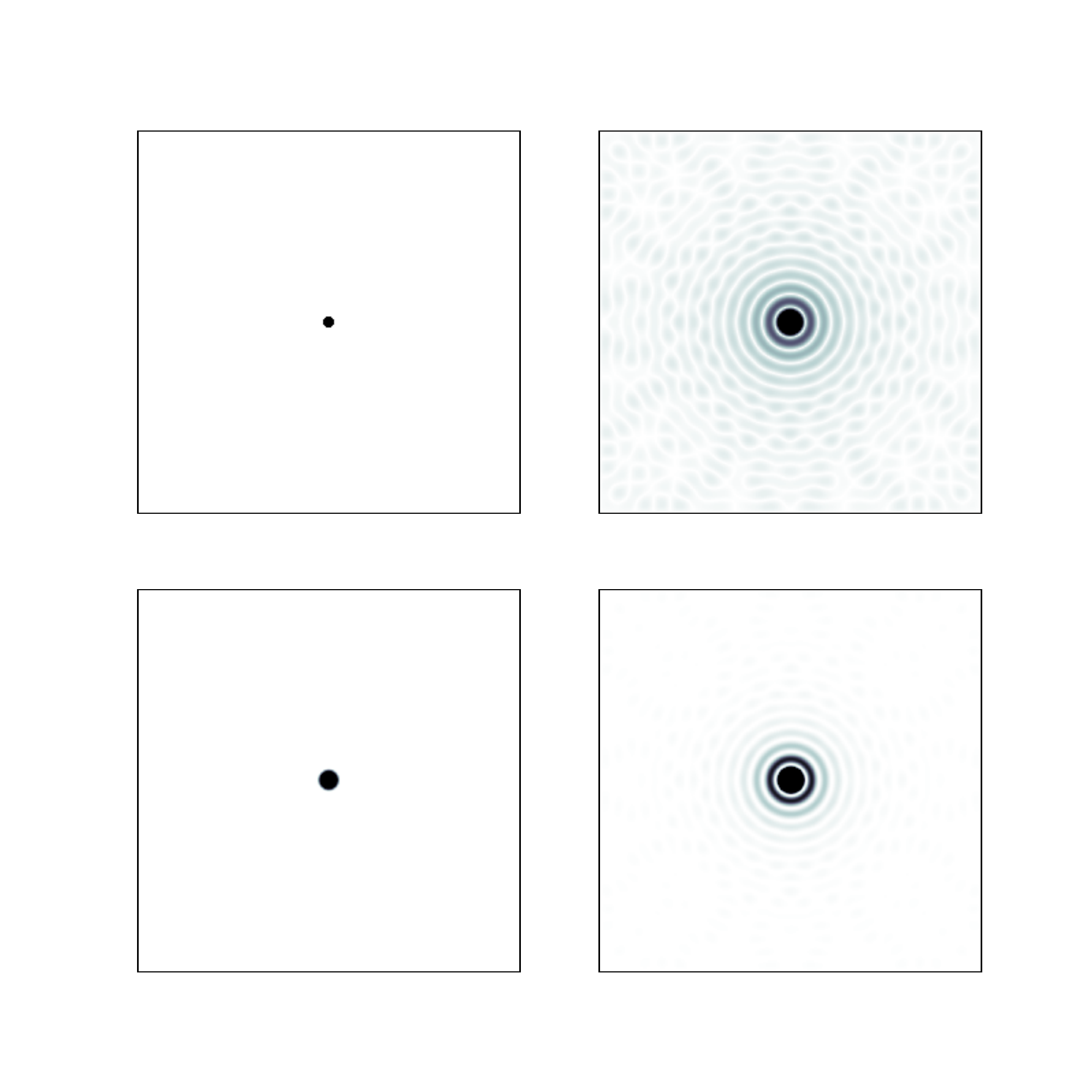}}
\end{minipage}%
\begin{minipage}{.5\linewidth}
\centering
\subfloat[Interferometer]{\label{main:c}\includegraphics[width=\textwidth]{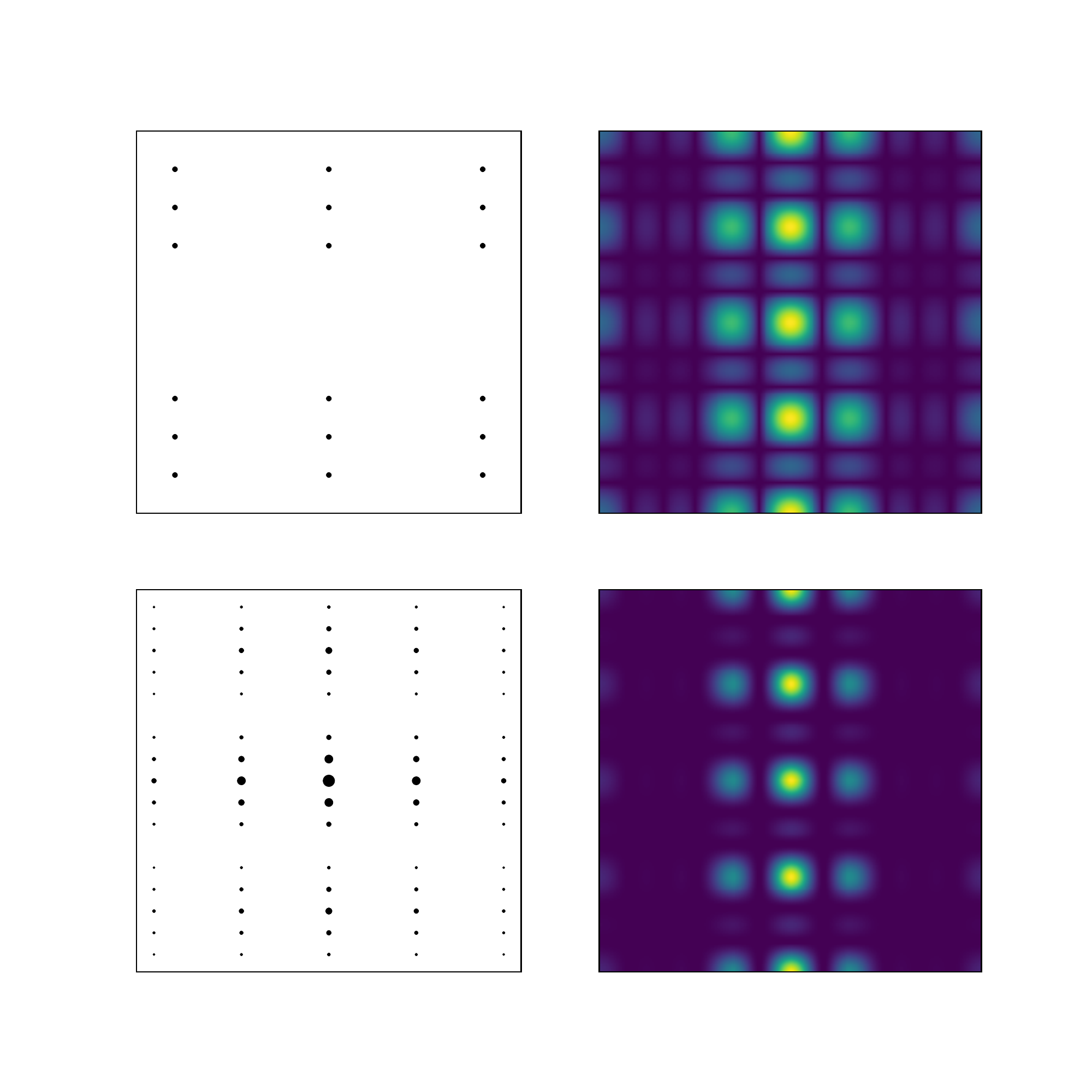}}
\end{minipage}

\caption{Panel\,(a): Relationship between the aperture plane, the electric field beam pattern of the reflector, the synthesized beam (primary beam) of an interferometer (single dish), and the uv-plane of the telescope. Panels\,(b) and\,(c): Clockwise from the top left is the aperture plane with the antenna(s) pointing to the zenith, the electric field beam pattern, the primary beam and the uv-plane. The uv-plane of a single dish telescope is continuously sampled up to a cut-off boundary. The uv-plane of an interferometer can extend to a larger boundary, but is often more sparsely sampled than a single dish.}
\label{fig:uvplane}
\end{figure}

Another way of understanding the relationship between the antenna layout and the uv-plane is via the synthesized beam of the interferometer. We previously discussed that the electric field probed by a single dish can be represented by the Airy pattern, given by Equation~\ref{eq:airy_disk}. This is the Fourier transform of a circular aperture with a diameter equal to the diameter of the single dish telescope. When there are multiple antennas, the electric field probed by the combination of all elements is equivalent to the spatial Fourier transform of their layout in the field, which is sometimes called the aperture plane. The synthesized beam of the interferometer, obtained by squaring the electric field beam pattern, acts as a convolution kernel to the true sky signal and is analogous to the point-spread function of an optical telescope. For clarity:
\begin{equation}
\label{eq:synthesized_beam}
    I^{\mathrm{meas}}(\lambda) = B(\lambda) * I(\lambda)
\end{equation}
\noindent
where $B$ and $I$ are the synthesized beam and the true sky signal (similar to the definitions in Equations~\ref{eq:visibility} and \ref{eq:visibility_uv}), and $I^{\mathrm{meas}}$ is the measured intensity. Using the convolution theorem, the above equation can also be written as:
\begin{equation}
\label{eq:synthesize_beam_fft}
    \tilde{I}^{\mathrm{meas}}(\lambda) = \tilde{B}(\lambda) \tilde{I}(\lambda)
\end{equation}
\noindent
where the tilde denotes a Fourier transform of the underlying parameter. This equation implies that the Fourier transform of the synthesized beam of the interferometer, acts as sampling function to the Fourier transform of the true sky signal. In other words, Equation~\ref{eq:synthesize_beam_fft} represents the same relation shown in Equation \ref{eq:visibility_uv}, or the uv-plane is equivalent to the Fourier transform of the synthesized beam of the telescope.

Figure~\ref{fig:uvplane} shows the relationship between the aperture plane, the electric field probed by the interferometer, the synthesized beam and the uv-plane. For single dish telescopes, this relationship is almost trivial. However, it is interesting to note that the uv-plane of a single dish is continuously sampled up to a sharp cut-off boundary (evident in the figure as a solid circle). The uv-plane sampled by an interferometer is completely a function of the layout of antennas in the field. Interferometers with unique antenna spacings can have a large uv-coverage, though it is often more sparsely sampled than a single dish telescope. The larger extent of uv-sampling that interferometers typically provide, leads to a higher resolution in images than what single dish antennas can provide.

In Panel\,(c) of Figure~\ref{fig:uvplane} the interferometer plotted has multiple pairs of antennas with the same baseline vector. All these antenna pairs probe the same uv-mode (size of the markers in the uv-plane of Panel\,(c) shows the number of times that uv-mode is measured), compromising the uv-coverage. However, antennas with such layouts are favourable for EoR experiments where building sensitivity in a few uv-modes helps in improving the signal-to-noise ratio in measurements, which is critical for detecting the weak cosmological signal. Experiments like the Donald C. Backer Precision Array for Probing the Epoch of Reionization (PAPER;~\citealt{Parsons_et_al_2010}), the Murchison Widefield Array (MWA;~\citealt{Tingay_et_al_2013}), the Canadian Hydrogen Intensity Mapping Experiment (CHIME;~\citealt{Newburgh_et_al_2014}), Hydrogen Intensity and Real-time Analysis eXperiment (HIRAX;~\citealt{Hirax_2016}) and the Tianlai Telescope~\citep{Xu_et_al_2015, Xu_et_al_2015, Das_et_al_2018} have numerous antennas spaced equidistantly (or in a grid-like layout), to obtain multiple measurements of the same set of uv-modes.

\subsection{Beam Chromaticity}
\label{chp:Intro:sec:Interf:subsec:Chromaticity}

\begin{figure}
    \centering
    \includegraphics[width=\textwidth]{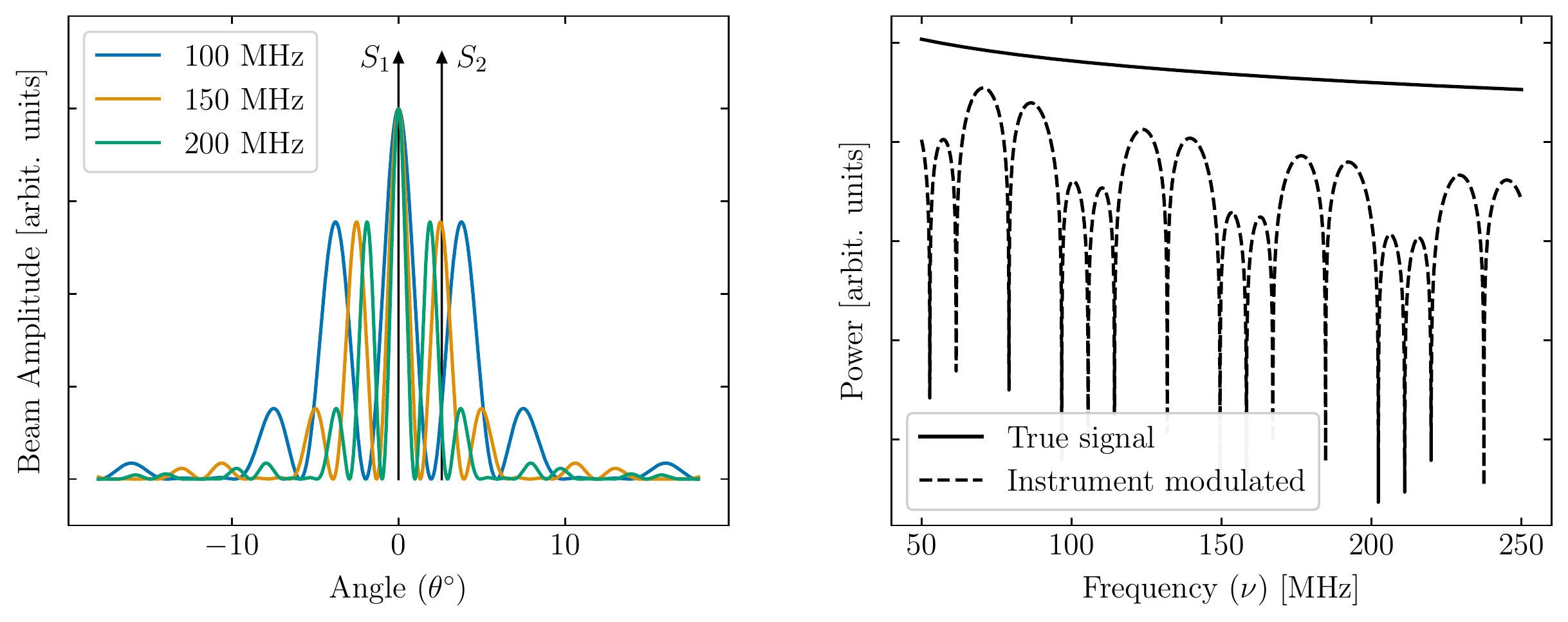}
    \caption{Panel\,(a): Synthesized beam of a two-element interferometer with 14.6\,m diameter dishes spaced three times the diameter apart. $S_1$ and $S_2$ are two instantaneous source directions. Panel\,(b): Spectral distortion in a smooth-spectrum foreground located at $S_2$, caused by the chromatic beam of an interferometer.}
    \label{fig:beam_chromaticity}
\end{figure}

In Section~\ref{chp:Intro:sec:Pspec:subsec:Foregrounds}, we will discuss the idea that bright foregrounds, distorted by a chromatic instrument response, contaminate the EoR signal and make it difficult to detect. This phenomenon, often referred to as mode-mixing, presents a formidable challenge to the separation of foregrounds from the cosmological 21\,cm signal, and warrants a closer look. Panel\,(a) of Figure~\ref{fig:beam_chromaticity} shows the synthesized beam of a two-element interferometer operating at frequencies that can target the redshifted 21\,cm signal. The antennas are moderately sized for an EoR experiment, with 14.6\,m diameter dishes, and spaced 43.8\,m apart or three times the diameter. Sources that are off-zenith, like direction $S_2$ in the figure, are picked up at different amplitudes over the bandwidth of the telescope. Panel\,(b) shows the typical modulation that a chromatic interferometer beam causes to a smooth-spectrum foreground that is off-zenith. 

The change in the antenna beam response as a function of frequency, originates from the wavelength dependence in the argument of Equation~\ref{eq:fringes2}. Just like the interferometric fringe pattern is caused by a changing projection of the baseline vector in the direction of the source, in exactly the same way, a changing wavelength of observation also creates fringes in the observed signal, manifesting as beam chromaticity. This beam chromaticity is similar to the chromaticity in the primary beam of a telescope, even though Equations~\ref{eq:visibility} and~\ref{eq:visibility_uv} factor them separately. The primary beam of a single dish telescope also changes as function of frequency, given by the wavelength dependence in Equation~\ref{eq:airy_disk}. This is similar to chromatic aberration exhibited by optical mirrors and lens; although the larger fractional bandwidth of low frequency radio telescopes, compared to optical telescopes, leads to a higher degree of chromaticity in the radio beam. In the overall synthesized beam of a radio telescope, the frequency dependence of the fringe pattern is usually a more dominant effect than the primary beam chromaticity because the uv-range probed by a single dish is often smaller. \footnote{For compact radio interferometers, like the layouts being designed for EoR experiments, the antenna diameter and distance between antennas are comparable. In such cases, the chromatic effect of the primary beam and the fringe modulation could be comparable for the shorter baselines.}. 

\begin{figure}
    \centering
    \includegraphics[width=0.65\textwidth]{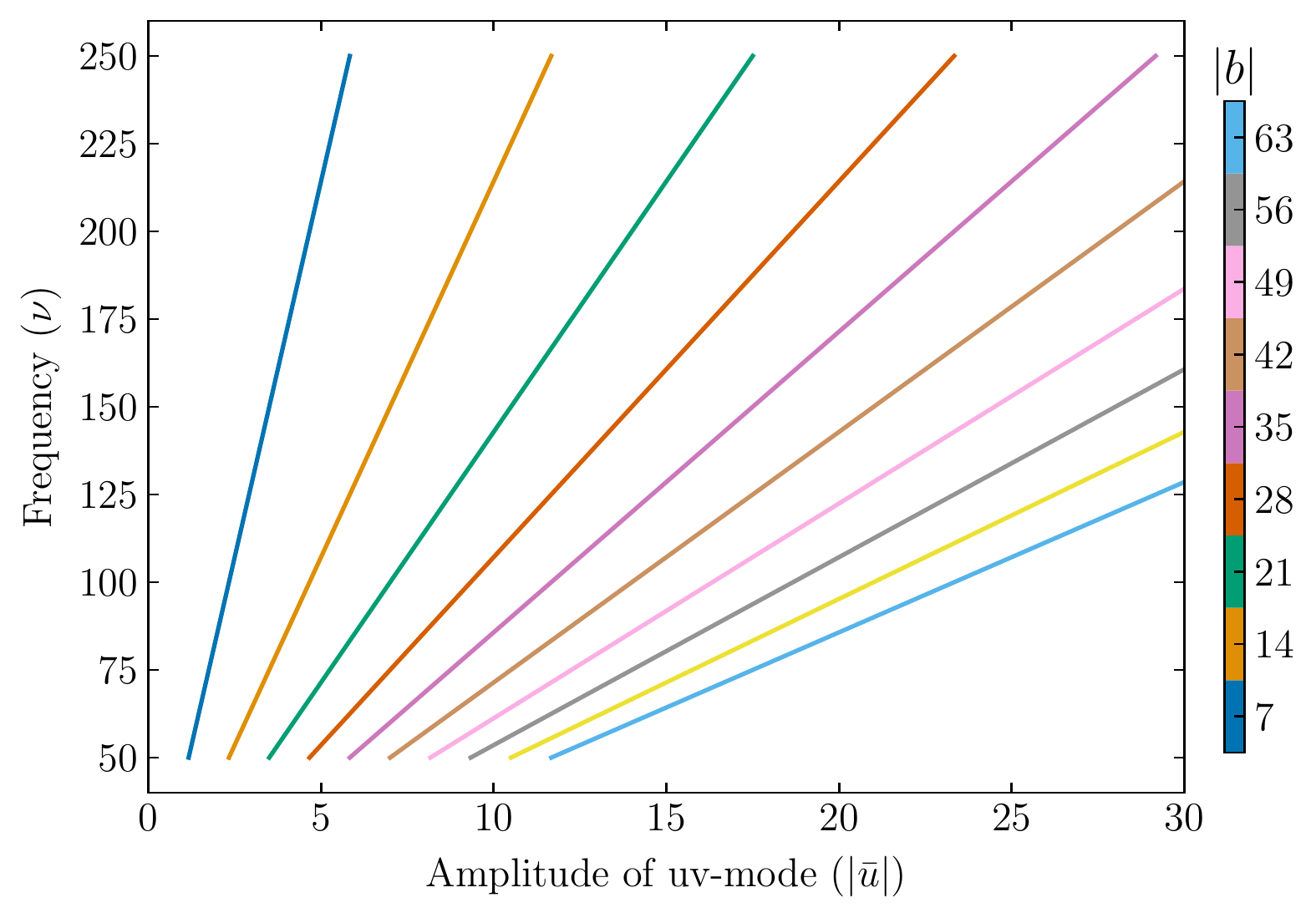}
    \caption{The $(u,v)$ coordinate sampled by a baseline vector as a function of frequency. The slope of the linear relationship is determined by the length of the baseline vector.}
    \label{fig:uv_scaling}
\end{figure}

\subsubsection{Mode Mixing}
\label{chp:Intro:sec:Interf:subsec:Chromaticity:subsubsec:Modemixing}

The frequency modulation caused by interferometers can also understood through Equations~\ref{eq:visibility} and~\ref{eq:visibility_uv}, where exponential term depends on the wavelength/ frequency of observation. The visibility measured by a pair of antennas samples different points on the uv-plane as a function of frequency. This feature of interferometers is routinely used in radio imaging to increase uv-coverage of a particular layout, through a technique called frequency synthesis. However, in power spectrum measurements this results in mode-mixing of high amplitude foregrounds and becomes a nuisance. To understand the mechanism through which high amplitude foregrounds contaminate the EoR signal, and how they can be filtered (discussed in the next section), it is useful to explicitly write the relationship between baseline length and $(u,v)$ coordinates in terms of frequency:
\begin{align}
    u = b_x \frac{\nu}{c}; &\quad v = b_y \frac{\nu}{c} \\[1em]
    |\mathbf{u}| &= |\mathbf{b}| \frac{\nu}{c}
\end{align}
\noindent
That is, the sampled uv-mode and the frequency of observation have a linear relationship, with a slope given by the length of the baseline vector between the pair of antennas measuring that visibility. 

%\citet{Parsons_et_al_2012} show that the visibilities measured by short baselines, which have a small slope in the $(|u|, \nu)$ plane, can be Fourier transformed and interpreted directly as modes of the EoR power spectrum. This technique, called the \textit{delay transform}, is also useful to understand the mechanism through which foregrounds couple with the EoR signal and how they can be filtered. We will discuss this in the next section.

\section{Measuring the \texorpdfstring{21\,cm}{21 cm} Power Spectrum}
\label{chp:Intro:sec:Pspec}

Probing the EoR, for the purpose of characterizing the spatial fluctuations caused by ionization bubbles, requires addressing at least three important analysis choices: (a) power spectrum estimators (b) foreground filtering techniques, and (c) calibration methods. These three axes of analysing data from a telescope are loosely interlinked-- the power spectrum estimator used determines the kind of foreground filtering technique that is most efficient and drives the calibration methodology for maximizing the signal-to-noise ratio. In the following sections, I will
outline the delay transform proposed by \citet{Parsons_et_al_2012} as a power spectrum estimator and qualitatively describe a few foreground filtering techniques. Calibration is briefly discussed in Section~\ref{chp:Intro:sec:FFTCorr:subsec:Calibration} and forms the focus of Chapter~\ref{chp:Redredcal}.

\subsection{Delay Transform}
\label{chp:Intro:sec:Pspec:subsec:DelayTransform}

In the previous section, we discussed the visibility equation and showed how the visibility measured by a baseline samples a mode in the Fourier transform of the sky. That is, each point on the uv-plane, sampled by a visibility, corresponds to a spatial Fourier mode of the sky. Interferometers designed for imaging aim to maximize uv-coverage and measure as many modes as possible, and apply an inverse Fourier transform to the uv-plane to form radio images of the sky. For power spectrum measurements, uv-modes can be directly mapped to $k_{\bot}$ modes in the cylindrical coordinate space. To see this relationship, let us rewrite Equation~\ref{eq:visibility_uv} in terms of the brightness temperature of neutral hydrogen~\citep[adapted from][]{Liu_and_Shaw_2019}. Neglecting the primary beam term briefly, and using frequency instead of wavelength we get:
\begin{equation}
    \mathcal{V}(u,v) = \frac{2k_B}{\lambda^2} \, \int T(l,m,\nu) \, e^{-2\pi i (ul + vm)} \mathrm{d}l \, \mathrm{d}m
\end{equation}
\noindent
which is very similar in form to the Fourier transform of the brightness temperature field in Equation~\ref{eq:tb_fft_cylindrical}. The difference is that the visibilities represent only a partial Fourier transform, corresponding to the $\mathbf{r}_{\bot}$ coordinates. The result of the above integration can be written as a hybrid parameter $\widetilde{T}(\mathbf{u}, \nu)$, where $\mathbf{u}$ is a vector representing the coordinates $(u,v)$ and is directly related to the power spectrum coordinate $k_{\bot}$ via cosmological constants.

Due to the expansion of the Universe, and a sharp line profile of the 21\,cm emission, every frequency bin within the bandwidth of the experiment corresponds to a unique slice of the Universe along our line-of-sight. The other cylindrical coordinate of the power spectrum, $k_{\|}$, represents perturbations along the line-of-sight and can be measured by computing a Fourier transform of visibilities along the frequency axis. However, note that a naive Fourier transform of $\widetilde{T}(\mathbf{u}, \nu)$ along the frequency axis is often insufficient, because of the frequency dependence of $\mathbf{u}$ in the visibility measured by a single baseline. As shown in Figure~\ref{fig:uv_scaling}, the diameter of the uv-ring sampled by a single visibility, changes as a function of frequency, coupling the frequency axis and $\mathbf{k}_{\bot}$ modes. A true Fourier transform along the frequency axis typically requires combining information from multiple baselines.

Figure~\ref{fig:uv_scaling} also shows that the `amount' of mixing between $\mathbf{u}$ and frequency depends on the length of the baseline. Within a finite bandwidth, longer baselines sweep a wider area in the uv-plane than short baselines. In the limit that the ring spanned by a visibility in the uv-plane is infinitesimally small, a Fourier transform along the frequency axis will yield uncorrupted $k_{\|}$ modes. Visibilities measured by short baselines, which have the least width on the uv-plane, can be Fourier transformed along the frequency axis and used as approximate probes of the power spectrum. Mathematically, this can be written as:
\begin{equation}
\label{eq:delay_transform}
    \widetilde{V}(u,v,\eta) = \int I(l, m, \nu) \, B(l, m, \nu) e^{-2\pi i (ul + vm + \nu\eta)} \, \mathrm{d}l\,\mathrm{d}m\,\mathrm{d}\nu
\end{equation}
\noindent
where $\eta$ is the Fourier equivalent of $\nu$. It has the dimensions of time, and is equal to $\tau$ at the geometric delay between these two antennas. The distinction between $\eta$ and $\tau$ is an important one. In Figure~\ref{fig:uv_scaling}, $\eta$ can be see as the Fourier equivalent of frequency, while $\tau$ is the Fourier equivalent of the sloped line representing a given baseline. For short baselines, they are roughly approximate, and can be mapped to $k_{\|}$ through multiplicative cosmological constants. The exact consequences of this approximation are dealt with by~\citet{Parsons_et_al_2012}.

The advantage of the delay transform over other power spectrum estimators is that it is a per-baseline approach. Visibilities do not have to be combined across baselines, making the calibration requirements less stringent than for map-making techniques. The visibility measured by a particular baseline, for one integration time sample can be Fourier transformed and squared to obtain the power in a given $(k_{\bot}, k_{\|})$ bin. Averaging over multiple measurements of the same mode helps build the sensitivity required to make a detection. In the original formulation, the delay transform did not use the information encoded in the cross-correlation of visibilities measured by baselines of different lengths, although this can be done~\citep{Zhang_Liu_Parsons_2018}. When observations are sample-variance limited, ignoring cross-correlation between baselines leads to a decrease in noise as $1/\sqrt{N_t}$ instead of $1/N_t$~\citep[Section 11.2]{Liu_and_Shaw_2019}. Overall however, the advantages of using a per-baseline approach out-weight the disadvantages, especially in the presence of practical issues like radio frequency interference, malfunctioning baselines, etc.

\subsection{Foreground Contamination}
\label{chp:Intro:sec:Pspec:subsec:Foregrounds}

\begin{figure}
    \centering
    \includegraphics[width=\textwidth]{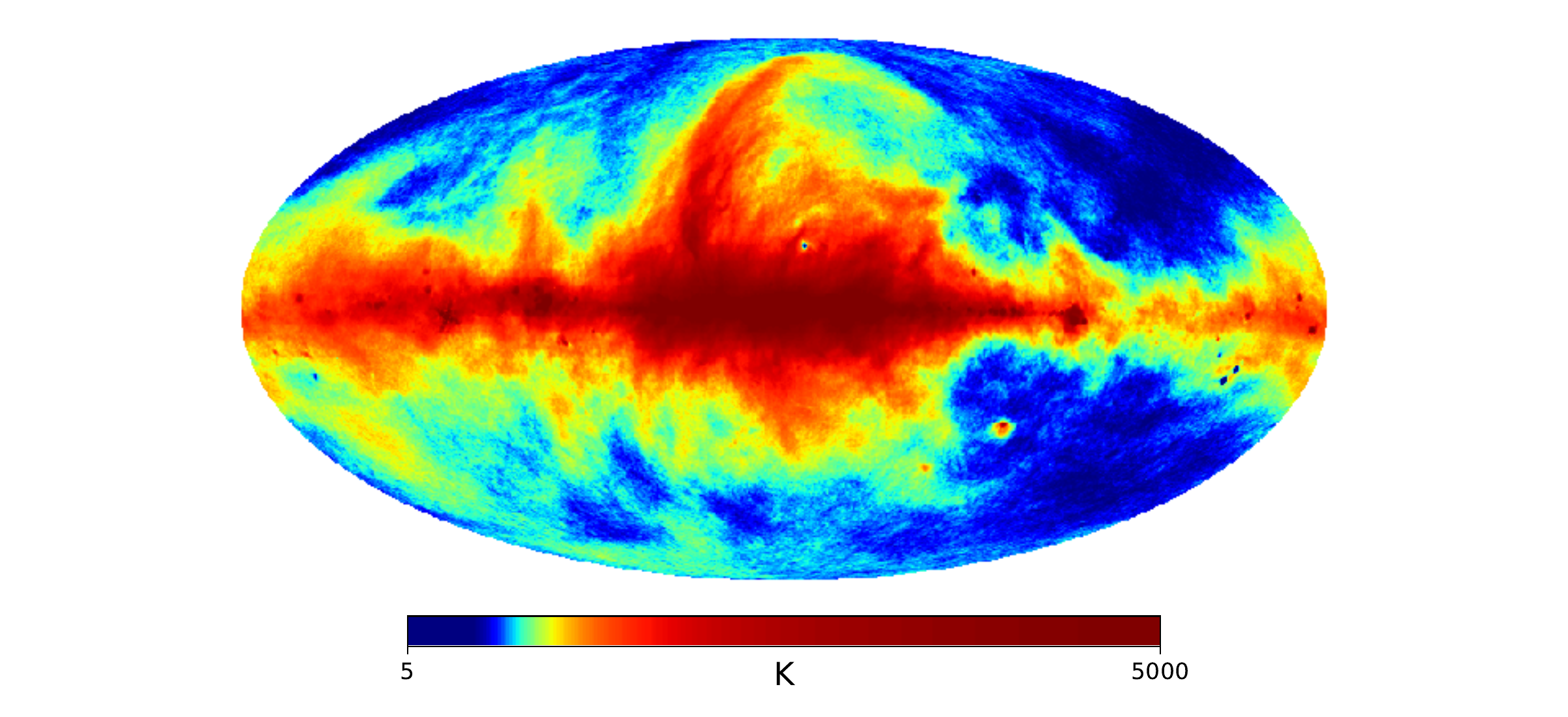}
    \caption{Galactic synchrotron emission at 150\,MHz~\citep{Zheng_et_al_2017b}. The colorbar shows the brightness temperature of the emission in units of K.}
    \label{fig:foregrounds}
\end{figure}

A daunting challenge in the detection of the cosmological 21\,cm signal is separating the contamination from foreground sources that locally produce radiation at the same wavelengths as redshifted neutral hydrogen. Figure~\ref{fig:foregrounds} shows the brightness temperature of galactic emission at 150\,MHz, which is on the order of 100\,K~\citep{Zheng_et_al_2017b}. The galactic emission at these frequencies is dominated by synchrotron radiation from relativistic free electrons. In general, synchrotron emission is higher at lower frequencies, following a power law trend in its power spectrum set by the power law energy distribution of relativistic electrons~\citep{Asakimori_et_al_1998}. The signal from redshifted neutral hydrogen in a similar frequency range is expected to be on the order of 10\,mK to 0.1\,mK at lower frequencies~\citep{Mesinger_et_al_2011}. This nearly five orders of magnitude difference in the foreground contamination and the cosmological signal of interest poses a huge challenge to the effort of 21\,cm tomography.

Most current-generation experiments that target EoR use some combination of the following three methods for foreground filtering:

\subsubsection{Foreground Modelling}

Under the assumption that the foreground signal is dominated by synchrotron radiation, the smooth power law slope in its spectrum can be used to model its emission. Experiments in the past have used parametric fits like orthogonal polynomials to model the foreground emission pixel-by-pixel in radio images taken~\citep{Di_Matteo_et_al_2002, Santos_et_al_2005}. However, as we saw in Section~\ref{chp:Intro:sec:Interf:subsec:Chromaticity}, interferometers are inherently chromatic and do not sample the same pixel on the uv-plane at all frequencies. This leads to sparse sampling of pixels that correspond to long baselines; sometimes data for a given pixel is not even available at all frequencies. Hence, smooth spectral foregrounds may appear unsmooth in an image, leading to poor foreground rejection when modelling only for smooth foregrounds.

An alternative approach that some experiments have taken is to use non-parametric functional forms to model the foregrounds. Rather than fitting the data to some choice of basis vectors (like polynomials), a template for the foregrounds is generated from the data itself by using the idea that foregrounds have slower-varying spectra than the cosmological signal. Wp smoothing~\citep{Harker_et_al_2009} and Gaussian Process Regression (GPR;~\citealt{Rasmussen_and_Williams_2006}) are two such techniques.

\subsubsection{Foreground Subtraction}

A simpler, data-driven solution that some experiments adopt is to represent the data in some basis where the foreground signal is occupied by only a few components and project out those components from the data. For example, a principal component analysis on the data could restrict most of the foreground signal to only the dominant modes since foregrounds are much higher in amplitude than the cosmological signal~\citep{Liu_et_al_2012}. Rejection of bright point-sources from the images is also a similar idea, though it requires forward-modelling from a pre-existing catalog of point-source intensities at the given frequency~\citep{Bernardi_et_al_2011, Sullivan_et_al_2012}. Foreground subtraction, or mode projection, has the advantage that the instrumental corruption of smooth-spectral foregrounds is accounted for, but the nuances in the exact basis used and the number of modes rejected determine the effectiveness of the technique and the amount of signal-loss incurred. The Karhunen-Lo\`eve transform~\citep{Shaw_et_al_2014}, Independent Component Analysis (ICA;~\citealt{Chapman_et_al_2012}), and Generalized Morphological Component Analysis (GMCA;~\citealt{Chapman_et_al_2013}) are all various modelling schemes for foreground subtraction.

\subsubsection{Foreground Avoidance}

\begin{figure}
    \centering
    \includegraphics[width=\textwidth]{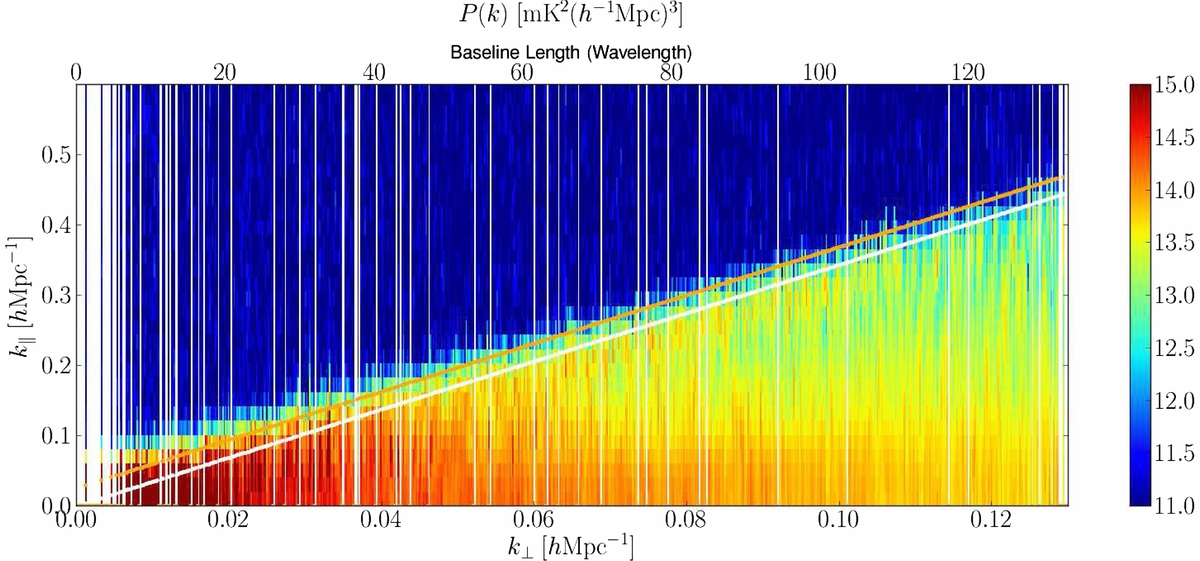}
    \caption{Smooth-spectral foregrounds, modified by a chromatic instrumental response, are contained to a wedge-like portion in the cylindrical coordinate phase space of the power spectrum. This figure, borrowed from~\citet{Pober_et_al_2013}, shows the power spectrum obtained from a 4\,hr integration using PAPER antennas. The white line shows the horizon limit and the orange line is 50\,ns beyond.}
    \label{fig:wedge}
\end{figure}

An idea that developed over the last decade is to avoid foregrounds rather than model them or subtract them from data. Unfortunately, unlike CMB experiments where the cosmological signal dominates in regions away from the Galactic plane, the 21\,cm foreground emission has a high amplitude every where in the sky (see Figure~\ref{fig:foregrounds}). Foregrounds cannot be avoided by simply observing in selected regions of the sky. Rather, they can be avoided in the power spectrum phase space where the smooth-spectral nature of the foregrounds and the chromaticity of the telescope, sequester them to a wedge-like shape in the cylindrical coordinate space~\cite{Datta_et_al_2010, Morales_et_al_2012, Parsons_et_al_2012, Vedantham_et_al_2012, Trott_et_al_2012, Hazelton_et_al_2013, Pober_et_al_2013, Thyagarajan_et_al_2013} as shown in Figure~\ref{fig:wedge} taken from \citet{Pober_et_al_2013}.

\begin{figure}
    \centering
    \includegraphics[width=0.95\textwidth]{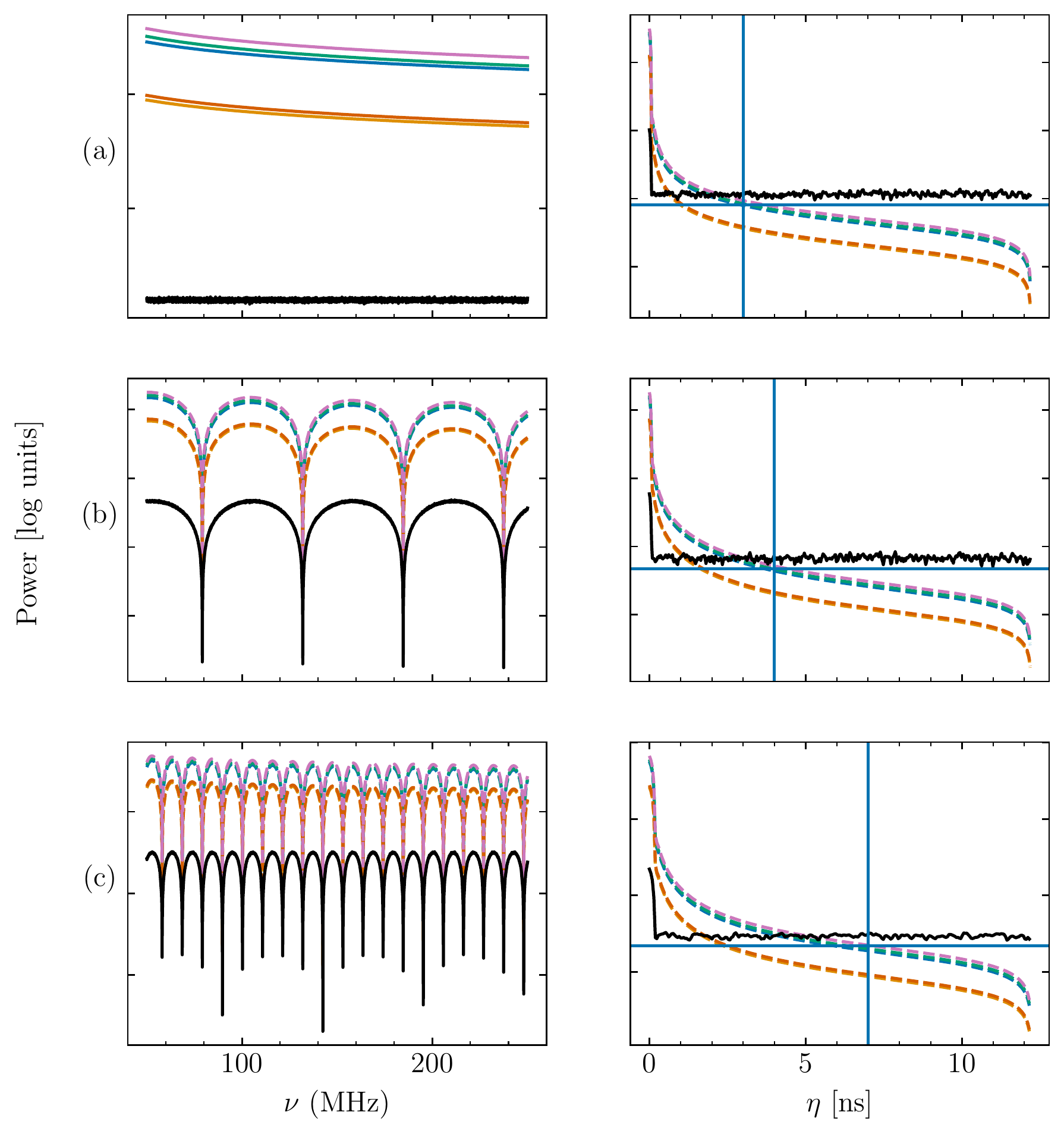}
    \caption{Modulation of smooth-spectrum foregrounds by a chromatic instrument response increases the delay cut-off as a function of baseline length. Panel\,(a) shows simulated smooth-spectrum foregrounds signals (colored lines) and a noise-like EoR signal (black) in frequency and delay domain without any instrument modulation. The delay cut-off in this case is determined by the horizon limit. Panel\,(b) shows the same signals when modulated by a short baseline. Beam chromaticity increases the delay cut-off where EoR dominates. Panel\,(c) shows the same, with the typical modulation caused by a long baseline.}
    \label{fig:delay_spectra}
\end{figure}

This containment of foregrounds can be understood by reinterpreting the delay transform in a slightly different way, for local astronomical sources. To understand this, let us rewrite Equation~\ref{eq:delay_transform} in terms of time delay:
\begin{equation}
    \widetilde{\mathcal{V}}_{\mathbf{b}}(\tau) = \int I(l,m,\nu)\, B(l,m,\nu)\, e^{-2\pi i \nu (\tau_g - \tau)}\; \mathrm{d}l\, \mathrm{d}m\, \mathrm{d}\nu
\end{equation}
\noindent
where $\tau_g$ is the geometric delay in arrival time between two antennas, given by $\mathbf{b}\cdot\mathbf{\hat{s}}/c$. To develop the intuition for what this measures, let us briefly ignore the frequency dependence of the sky signal and the primary beam. The above equation can then be simplified as:
\begin{equation}
    \widetilde{\mathcal{V}}_{\mathbf{b}}(\tau) \sim \int I(l,m)\; B(l,m)\; \delta^{D}(\tau_g - \tau)\; \mathrm{d}l\, \mathrm{d}m
\end{equation}
\noindent
with $\delta^{D}(x)$ representing the Dirac-delta function that is only defined when the argument is zero. The above equation shows that the delay transform is useful for selecting rings of constant delay on the sky.

Observe that the horizon forms a natural upper limit to the delay at which a foreground source can be observed. Foreground avoidance takes advantage of this upper limit, and uses modes that correspond to higher delays for probing the EoR signal. Reintroducing the frequency dependence of the primary beam and the sky complicates this interpretation by only a little. The chromaticity of the beam acts as a convolution kernel to the signal in delay space, smearing out the footprint of a foreground source at a given delay. This increases the upper limit of delay placed by the horizon, but the cut-off still remains. The foreground wedge originates from the changing width of the convolution kernel. Since beam chromaticity is smaller for short baselines and increases as a function of baseline length, the convolution kernel smears out long baseline more. That is, mode-mixing is higher in long baselines than short baselines. 

Mode-mixing can also be understood in a different way. Figure~\ref{fig:delay_spectra} shows simulated foreground and EoR signals, modulated by the chromatic instrument response of two different baseline lengths. Panel\,(a) shows the expected conversion from frequency domain to $\eta$-domain or delay space. The simulated foreground signals have a smooth frequency response, with spectral indices between $-2.5 < \alpha -2.3$. These nearly flat spectra have concentrated power in only a few modes in the delay-space. Beyond these modes, the EoR signal is expected to dominate. The $\eta$ cut-off, beyond which the EoR signal is more prominent, is dictated by the spectral indices. In Panel\,(b) the signals have been modulated by the chromatic beam response of a typical `short' baseline. For this simulation, we assumed HERA-like 14.6\,m diameter dishes, spaced 14.6\,m apart. The beam chromaticity includes a primary beam response that is modelled as a Bessel function of the first kind (Equation~\ref{eq:airy_disk}) in addition to interferometric fringes. As evident from the right side figure, this modulation pushes the delay cut-off to higher $\eta$. Panel\,(c) shows the modulation caused by a `moderate' baseline in the HERA layout, with the antennas spaced 73\,m apart. Since the chromaticity is larger for this baseline length, the cut-off is pushed to higher a $\eta$ value. This increasing $\eta$ (probe of $k_{\|}$) as a function of baseline length (probe of $k_{\bot}$), creates the foreground wedge.

\subsection{Experiments and Current Status}
\label{chp:Intro:sec:Pspec:subsec:Experiments}

\begin{figure}
    \centering
    \includegraphics[width=0.65\textwidth]{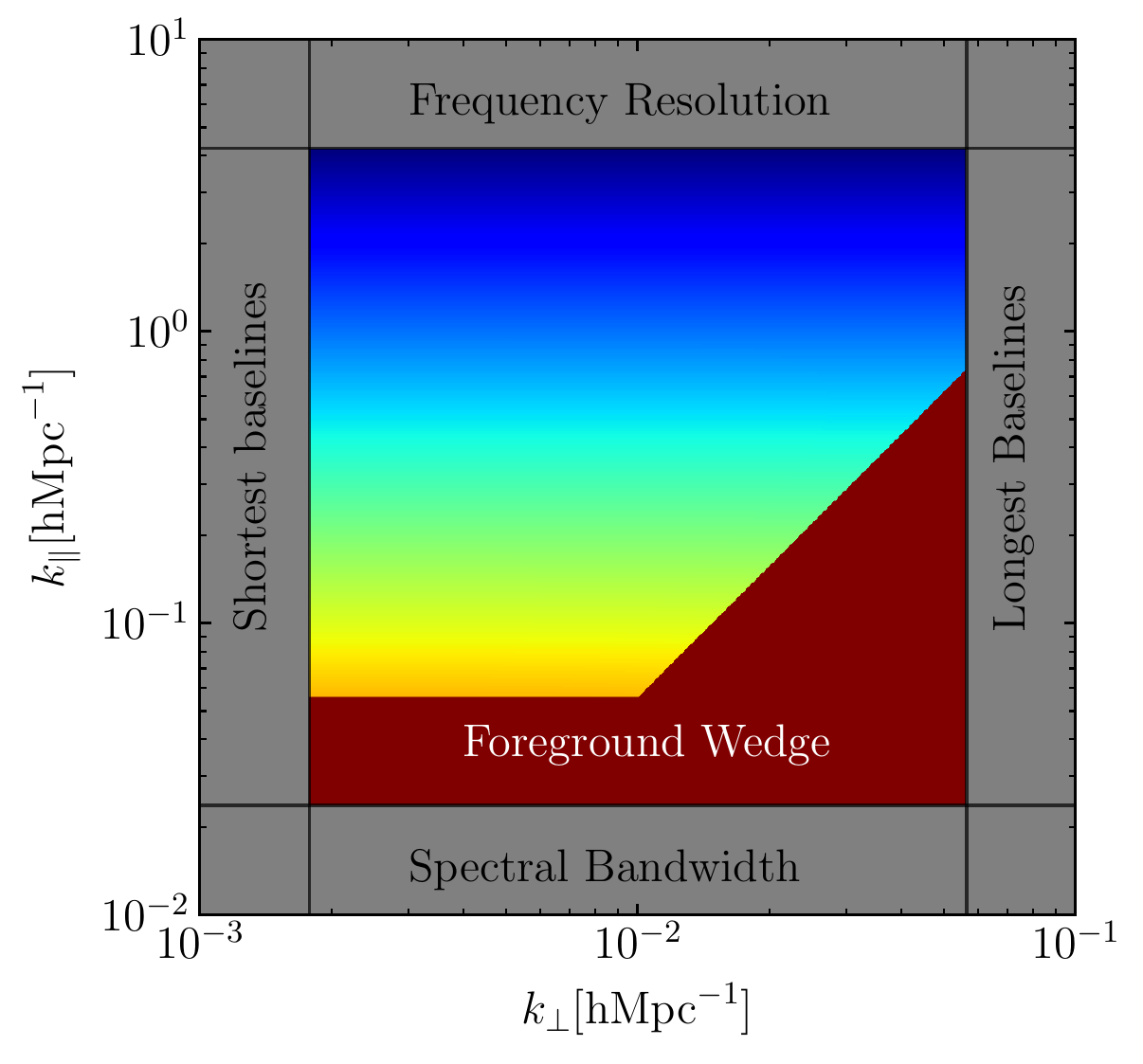}
    \caption{Limits to the power spectrum modes probed by a survey, based on the design parameters, shown for a typical EoR experiment. The foreground wedge is contaminated by the instrument coupling with bright synchrotron foregrounds, and prevents extraction of the EoR signal (see text). The remaining portion of the phase-space is expected to be dominated by the EoR signal, although heavily suppressed by noise.}
    \label{fig:eor_survey}
\end{figure}

Combining these ideas, a comprehensive view of the modes that can be probed by a realistic survey can be formed. Figure~\ref{fig:eor_survey} shows the typical $(k_{\bot}, k_{\|})$ modes that can be probed by a survey. The lowest angular scales (highest $k_{\bot}$) are limited by the maximum resolution of the telescope, which is set by the length of the longest baselines. In reality, this limit is some-what reduced by the sparse uv-coverage resulting from the lack of a large number of baselines of these lengths. The smallest angular scales are set by the length of the shortest baselines, or in the case that the layout is a filled aperture, by the field-of-view of each antenna. 

On the $k_{\|}$ axis, the upper limit is set by the resolution of each frequency channel. Typically, for radio interferometers at low frequencies, frequency resolution can be made arbitrarily large by the signal processing backend. The angular resolution on the other hand, is driven by the size of array and is more expensive to build. This often makes it easier to probe large-$k_{\|}$ modes as compared to large-$k_{\bot}$ modes. The lower limit of $k_{\|}$ modes is set by the bandwidth of the experiment. For telescopes with a large bandwidth, lower $k_{\|}$ modes can be probed. However, this limit is actually set by the period over which a certain phase of reionization lasts.

\begin{figure}
    \centering
    \includegraphics[width=\textwidth]{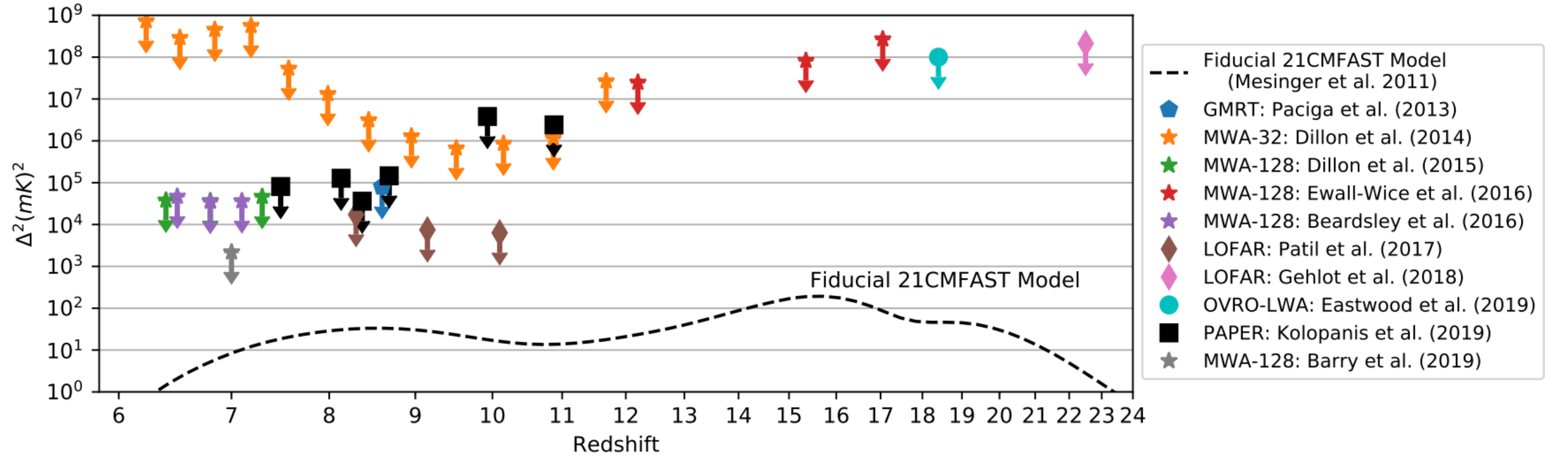}
    \caption{Upper limits on the reionization power spectrum set by various experiments. Figure borrowed from~\citet{Liu_and_Shaw_2019}.}
    \label{fig:reionization_limits}
\end{figure}

Despite the rich literature surrounding EoR experiments and methodologies for power spectrum estimation, none of the experiments performed so far have been successful at detecting the cosmological 21\,cm power spectrum. However, many experiments have been successful at placing upper limits to the EoR signal and these limits are constantly being improved by new analyses and more sensitive interferometers. Figure~\ref{fig:reionization_limits} shows the upper limits computed by different surveys and the fiducial signal theoretically predicted using the semi-analytical 21CM FAST code~\citep{Mesinger_et_al_2011}.  

Early experiments performed using the Giant Meterwave Radio Telescope (GMRT;~\citealt{Pen_et_al_2009}), the Parkes Radio telescope~\citep{Anderson_et_al_2018}, the Green Bank Telescope~\citep{Chang_et_al_2010, Masui_et_al_2013} and the Owens Valley Radio Observatory Low Wavelength Array (OVRO-LWA;~\citealt{Eastwood_et_al_2019}) used foreground subtraction techniques to establish their limits. The Murchison Widefield Array (MWA;~\citealt{Tingay_et_al_2013}) was built to utilize a hybrid approach involving bright point source subtraction and foreground avoidance. Multiple approaches have been used to analyse that data to obtain upper limits in various k-modes~\citep{Dillon_et_al_2014, Dillon_et_al_2015, Ewall-Wice_et_al_2016, Beardsley_et_al_2016, Barry_et_al_2019}, and as such should not be compared on the same scale. However, they have been included in the same plot for completeness. The Low Frequency Array (LOFAR;~\citealt{Chapman_et_al_2012, Patil_et_al_2017, Gehlot_et_al_2018}) survey used a combination of GMCA and GPR for foreground rejection to obtain their limits in addition to sophisticated direction-dependent calibration algorithms to obtain beam models. The Donald C. Backer Precision Array for Probing the Epoch of Reionization (PAPER;~\citealt{Parsons_et_al_2010, Ali_et_al_2015, Kolopanis_et_al_2019}) exclusively used foreground avoidance as a demonstration of the technique. 

Using the lessons learnt from these experiments, new arrays are being commissioned to make a detection of the 21\,cm signal. Next-generation experiments include the Hydrogen Epoch of Reionization Array (HERA;~\citealt{Deboer_et_al_2017}), the Hydrogen Intensity and Real-time Analysis eXperiment (HIRAX;~\citealt{Hirax_2016}), the Tianlai experiment in China~\citep{Chen_2011, Xu_et_al_2015, Das_et_al_2018} and the Square Kilometer Array (SKA;~\citealt{Santos_et_al_2005}). These telescopes have the sensitivity and collecting area required to make a detection and hold the promise of opening up neutral hydrogen as a cosmological probe to the physics governing the formation of the first stars.

\section{Correlators}
\label{chp:Intro:sec:Corr}

Moving from the theoretical realm to the physical, radio telescopes require signal processing backends that can convert raw, analog antenna voltages into a data product that can be stored, calibrated, imaged or converted into a power spectrum. A major portion of this thesis is concerning the design (Chapter~\ref{chp:HERACorr}) and calibration (Chapter~\ref{chp:Redredcal}) of such instruments, required to achieve the science goals discussed so far.

A correlator is a signal processing backend, that computes the visibility measured by each antenna pair in the array, within narrow spectral bins to avoid decorrelation. While most interferometers employ a correlator backend to compute visibilities, some interferometers have beamforming pipelines that directly store images of the source being observed. For wide-field surveys like EoR experiments, correlators are ubiquitous because visibilities allow precision calibration that is necessary for detecting the EoR. 

Equation~\ref{eq:fringes1} suggests a practical way of computing visibilities with a correlator, and was the design of the very first interferometers built. However, the summation between electric fields results in a DC-offset which also encodes unnecessary signals like the Galactic background, thermal noise from the ground and noise generated by the receivers, in addition to the signal of interest. Expanding Equation~\ref{eq:fringes1} we get:
\begin{equation}
    I \propto \left|\mathbf{E}_r\right|^2 + \left|\mathbf{E}_l\right|^2 + 2 \mathbf{E}_r \cdot \mathbf{E}_l
\end{equation}
\noindent
By computing only the final cross-correlation product, the DC-offset in total intensity can be avoided while capturing the interferometric fringe pattern. Due to the frequency dependence of visibilities, computing only one product for the entire bandwidth of the telescope results in signal decorrelation. To mitigate this, the cross-correlation products are often computed in narrow spectral bins. The set of cross-correlation products that comprise of all antenna pairs in the array, and all frequency channels that span the bandwidth of observation, is called the \textit{visibility matrix}. The correlator is an instrument which computes the visibility matrix for every time sample that is observed by the telescope. In this section, I will discuss the two traditional architectures in which correlators are built and hardware platform choices for them.

\subsection{Design Architectures}
\label{chp:Intro:sec:Corr:subsec:Design}

To compute the visibility matrix, the correlator needs to perform two distinct operations: (a) compute the signal power in narrow spectral bins and (b) compute cross-correlation products of antenna pairs. This can be mathematically represented as:
\begin{equation}
\label{eq:fxcorr}
    V_{ij}(\nu) = \mathcal{F}\left({v}_i\right) \cdot \mathcal{F}\left({v}_j\right)
\end{equation}
\noindent
where $v_i$ is the voltage time stream generated by the $i^{\mathrm{th}}$ antenna in the array, $\mathcal{F}$ indicates a spectral Fourier transform of the parameter in brackets, the asterisk indicates conjugation and the left hand side of the equation represents an entry in the visibility matrix. Applying the convolution theorem to the above equation we can write:
\begin{equation}
\label{eq:xfcorr}
    V_{ij}(\nu) = \mathcal{F} (v_i \star v_j)
\end{equation}
\noindent
where the star-symbol indicates a convolution operation between the time-domain voltages of both antennas. Correlators can be built in two different architectures following Equations~\ref{eq:fxcorr} and \ref{eq:xfcorr}.

\subsubsection{FX Correlators}

Most current-generation radio telescopes employ correlators built in the FX architecture. Following the order of operations in Equation~\ref{eq:fxcorr}, the first F-stage converts the voltage time-domain signal of each antenna to frequency-domain spectra through a Fourier transform. The second X-stage computes cross-correlation products within each frequency channel of the first stage to generate the visibility matrix. Computationally, the F-stage performs $\mathcal{O}(NF\log{F})$ operations where $N$ is the number of antennas in the array and $F$ is the number of frequency channels computed in the Fourier transform. The X-stage is often the more computationally intensive step, performing $\mathcal{O}(FN^2)$ operations on every time sample. There are many radio telescopes that employ correlators of this design, like the Murchison Widefield Array~\citep{Ord_et_al_2015}, the Large-Aperture Experiment to Detect the Dark Ages~\citep{Kocz_et_al_2015, Price_et_al_2018}, the Canadian Hydrogen Intensity Mapping Experiment~\citep{Bandura_et_al_2016}, the Karoo Array Telescope~\citep{Foley_et_al_2016}, Expanded Owens Valley Solar Array~\citep{Nita_et_al_2016}, the Sub-Millimeter Array~\citep{Primiani_et_al_2016}, the upgraded Giant Metrewave Radio Telescope~\citep{Reddy_et_al_2017}, the Arcminute Microkelvin Imager~\citep{Hickish_et_al_2018} and the Deep Synoptic Array~\citep{Kocz_et_al_2019} to name a few.

\subsubsection{XF Correlators}

This is a relatively older design of correlators, that follows the order of operations in Equation~\ref{eq:xfcorr}. The XF correlator design is more conducive to analog systems where the convolution can be implemented through a series of `lag' operations on the time-domain voltages followed by multiplication through analog mixers. The time lag is introduced in the signal path of each antenna via varying cable lengths, or in the case of completely digital systems via shift registers. The convolution output can be time-averaged, unlike the intermediate product of an FX correlator. The final F-stage operation performs digitization (if the signal is still analog) and channelization of these cross-correlation products. For digital XF correlators, only the first stage is computationally intensive requiring $\mathcal{O}(FN^2)$ operations where $F$ is the number of frequency channels or `lags' in the convolution operation. The compute in the second stage can be ignored since it occurs only once per integration time period. Before digital electronics became affordable, most large arrays built their correlator in the XF architecture. The Very Large Array~\citep{Napier_Thompson_Ekers_1983} and its extended version~\citep{Perley_et_al_2009}, the Atacama Large Millimeter Array\footnote{The new ALMA correlator is built in a hybrid FXF architecture that is not discussed here~\citep{Escoffier_et_al_2007}.}~\citep{Escoffier_et_al_2000} and the Institut de Radio Astronomie Millim\'{e}trique in France~\citep{Guilloteau_and_Lucas_2000} employ lag correlators. 

\subsection{Hardware Platforms}
\label{chp:Intro:sec:Corr:subsec:Hardware}

Within the electronics industry, there are four major hardware platforms commonly identified for general-purpose computational operations-- Field Programmable Gate Arrays (FPGAs), Graphics Processing Units (GPUs), Central Processing Units (CPUs) and Application Specific Integrated Circuits (ASICs). Each of these platforms have their own pros and cons, making some of them more suitable for certain stages of the correlation operation. 

Before delving into the details of each platform, it is useful to define two parameters, which serve as a common ground for comparison: (a) input bandwidth and (b) computational capacity. Input bandwidth is the rate at which data can be ingested by a platform, usually measured in units of bits-per-second\,(bps). Commercially available platforms today have input bandwidths in the range of 1--20\,Tbps. The definition of computational capacity is slightly different for each platform, with ASICs and FPGAs using the number of multiply-accumulate (MAC) operations per second as a measure, and CPUs and GPUs using the number of floating-point operations per second (FLOPs). For the purpose of this text, I use the rough equality of two MAC operations comprising a floating-point operation to compare between these platforms.

\subsubsection{Field Programmable Gate Arrays (FPGAs)}

FPGAs are integrated-circuits that can be programmed to perform a given set of operations. The `program' or signal processing design, is often written in a Hardware Description Language (HDL) and converted to a gate-level map before being programmed into the FPGA fabric. FPGAs are excellent for repetitive operations that need to be performed in parallel on many inputs while maintaining accurate clock precision. For example, FPGAs form a good choice for the F-stage of an FX correlator since the voltage signal of all antennas need to be digitized and channelized with accurate timestamps on each spectrum.

Wide-band radio telescopes that need to be sampled at high sampling rates of $\sim$100\,Msps--5\,Gsps, generate a high input bandwidth which FPGAs can accommodate. The rate at which data can be ingested by an FPGA is determined by the number of I/O lanes it provides and the clock rate. Most commercial FPGAs operate with clock rates between 10--300\,MHz and can accommodate much high input bandwidths than either GPUs or CPUs. The compute capacity of an FPGA depends on the number of DSP slices and Look-Up Tables (LUTs) available. There are a wide variety of FPGA platforms available commercially, with compute capacities anywhere from a few GFLOPs to tens of TFLOPs. 

While FPGAs have a high input bandwidth and some of them offer high compute capacity, their primary limitation is the amount of RAM available on-chip. In general, the total amount of block RAM available on an FPGA can vary between a few KB to tens of MB. High-end FPGAs offer up to 8\,GB of block RAM, but these chips are usually too expensive for academic radio astronomy applications. Considering practical limitations, a significant disadvantage of FPGAs is that they are not easy to program and operate, usually requiring specialized personnel like FPGA engineers. Moreover, they cannot be used as stand-alone chips and need to be integrated into a larger circuit containing other devices necessary to operate FPGAs like ADCs, frequency synthesizers, off-chip storage, Ethernet ports etc.  

Many telescopes have developed their own FPGA boards for processing. For example, the ICE FPGA boards~\citep{Bandura_et_al_2016} developed for CHIME and HIRAX, the UniBoard and Uniboard$^2$~\citep{Szomuru_2010, Schoonderbeek_et_al_2019}, developed for the Low Frequency Array and other related projects and the Tunable Filterbank Cards~\citep{Escoffier_et_al_2007} developed for ALMA. The Collaboration for Astronomy Signal Processing and Electronics Research\footnote{\url{https://casper.berkeley.edu/}} (CASPER), has designed numerous FPGA boards and made them publicly available for any experiment that wants to use them. The Reconfigurable Open Architecture Computing Hardware board (ROACH, ROACH-2), Square Kilometre Array Reconfigurable Application Board (SKARAB) and the Smart Network ADC Processor (SNAP, SNAP-2) power many radio telescope facilities across the world \citep{Parsons_et_al_2008, Hickish_et_al_2016}. 

In addition to hardware, CASPER also provides software tools for designing DSP layouts in the interactive MATLAB$^{\tiny\text{\textregistered}}$ Simulink$^{\text{\copyright}}$ environment, and backend toolflow libraries\footnote{\url{https://github.com/casper-astro/mlib_devel}} to compile the design into a programmable bitfile. They also provide a light-weight Trivial File Transfer Protocol (TFTP) based command interface with the FPGA for programming, monitoring and changing  run-time configurable block RAM contents. This Python software package, called \texttt{casperfpga}\footnote{\url{https://github.com/casper-astro/casperfpga}}, is also publicly available on Github.

\subsubsection{Graphics Processing Units (GPUs)}

GPUs are specialized electronic hardware, originally designed to handle graphics manipulation and output. Their ability to highly-parallelize computation makes them versatile, and they are often used as accelerator cards to speed-up operations otherwise performed on a CPU~\citep{Barsdell_et_al_2011, Barsdell_et_al_2012}. Thanks to the gaming industry that heavily employs GPU cards, commercial GPUs are more affordable than FPGAs of a comparable compute capacity. Moreover, there are well-documented software libraries which can be used to program GPUs, like the CUDA library\footnote{\url{https://developer.nvidia.com/about-cuda}} for Nvidia chips, and the open-source OpenCL\footnote{\url{https://www.khronos.org/opencl/}} library that supports many commercial GPU chips. 

All these feature make GPUs a attractive option for the compute-intensive X-stage of an FX correlator~\citep{Kocz_et_al_2014}. The only disadvantages of GPUs, from the perspective of radio astronomy applications, are their low input bandwidth and high power consumption~\citep{Price_et_al_2014} which increases the maintenance cost of a telescope. The low input bandwidth is set by the number of PCIe links connecting to the motherboard of the server, with each link providing $\sim$32\,GBps input bandwidth for PCIe 3.0 protocol. Newer protocols like PCIe 5.0 offer higher speeds and might soon make input bandwidth of GPUs comparable to that of FPGAs. \citet{Clark_LaPlante_Greenhill} developed a software package, called \texttt{xGPU}\footnote{\url{https://github.com/GPU-correlators/xGPU}}, that can be used nearly as-is to perform antenna cross-correlations on Nvidia GPU chips. \citet{Denman_et_al_2015} developed a similar software package for AMD GPUs based on OpenCL.

\subsubsection{Central Processing Units (CPUs)}

CPUs are the most ubiquitous hardware platforms for general-purpose computations. While they are extremely versatile in the operations they can handle, most CPUs are not optimized to handle large input bandwidth or compute-intensive tasks. In most FX correlator designs, CPUs are used to issue control commands, process Ethernet packets, manage visibility matrix data before it is written to disk and for writing GPU code or designing FPGA layouts. DiFX~\citep{Deller_et_al_2007, Deller_et_al_2011} is a correlator completely built on CPUs, for VLBI cross-correlations. 

\subsubsection{Application Specific Integrated Circuits (ASICs)}

ASICs are integrated circuits designed for a specific application. They need to be designed by hardware engineers, prototyped and tested before being manufactured into chips at a fabrication facility. The overhead cost of developing ASICs is usually much higher than that of FPGAs or GPUs, especially when including the cost of hiring specialized engineers. However, ASICs out-perform FPGAs, GPUs and CPUs in run-time efficiency, power consumption and space occupied and are preferable for large-budget observatories. The second F-stage of the ALMA correlator~\citep{Escoffier_et_al_2007} is built on custom-designed ASICs.

\begin{figure}
    \centering
    \includegraphics[width=0.65\textwidth]{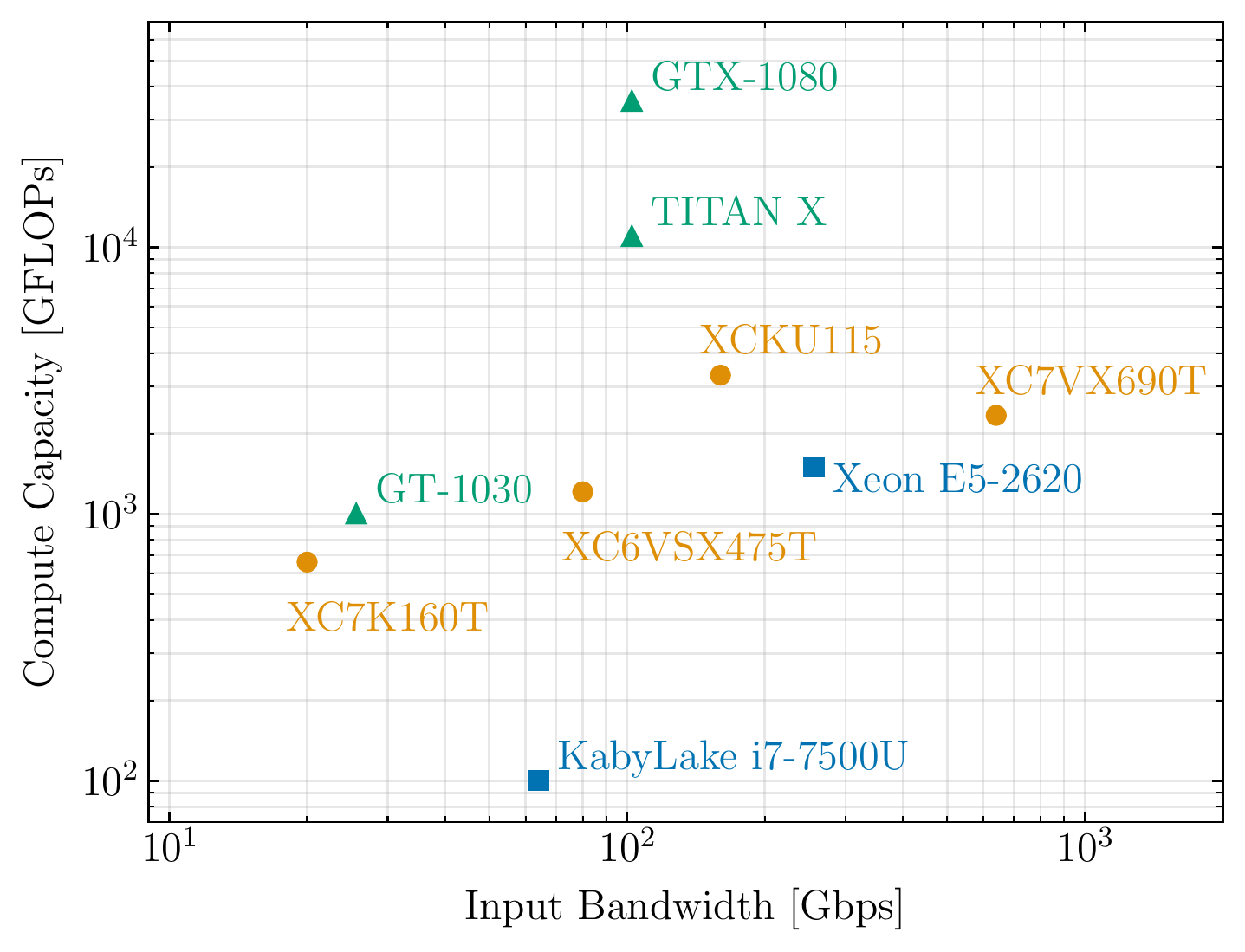}
    \caption{Input bandwidth and compute capacity for a few commercially available hardware platforms. Orange circles represent four FPGAs chips used in CASPER FPGA boards, the blue squares show two Intel CPUs and the green triangles are Nvidia GPUs.}
    \label{fig:platform_comparision}
\end{figure}

Figure~\ref{fig:platform_comparision} shows the comparison between a few commercially available hardware platforms. The FPGA chips shown in the figure are available as CASPER hardware. In the ascending order of input bandwidth these are the Kintex-7 (SNAP), Virtex-6 (ROACH-2), Kintex UltraScale (SNAP2) and the Virtex-7 (SKARAB). Though the plot is far from exhaustive, it is easy to see that GPUs generally offer higher compute capacity, and that FPGAs are more suitable for high input bandwidth operations.

\section{FFT Correlators}
\label{chp:Intro:sec:FFTCorr}

Chapter~\ref{chp:Redredcal} of this thesis, discusses calibration methods for a more recently proposed correlator architecture, the FFT correlator. This section tries to motivate this new correlator design and lays out the theoretical framework required to interpret the research presented there. 

The correlator architectures laid out in the previous section, have been used to build the signal processing system of many telescopes around the world. However, both FX- and XF-architectures require computational resources that scale as $\mathcal{O}(N^2)$ with the number of antennas. Recently, there has been a renewed interest in correlator architectures which require computational resources that scale less steeply with array size, for low-frequency radio astronomy applications that require a large collecting area. The Hydrogen Epoch of Reionization Array (HERA; \citealt{Deboer_et_al_2017}), the Canadian Hydrogen Intensity Mapping Experiment (CHIME; \citealt{bandura_et_al_2014, Newburgh_et_al_2014}), the Murchison Widefield Array (MWA;~\citealt{Tingay_et_al_2013}), Low Frequency Array (LOFAR; \citealt{VanHaarlem_et_al_2013}) and MITEoR \citep{Zheng_et_al_2014, Zheng_et_al_2017a} are all built with relatively cheap antennas that can scale to large-N arrays. At the low radio frequencies that these telescopes operate at, the signal chain can also be relatively inexpensive because cryogenic cooling of receivers is not essential. Receivers are sky-noise dominated at low radio frequencies \citep{Ellingson_2005}, decreasing the need to lower thermal-noise. If the correlator architecture can also scale up to large-N arrays, it will be more cost-efficient to build the collecting area required through numerous small antennas.

\begin{figure}
    \centering \includegraphics[width=\linewidth]{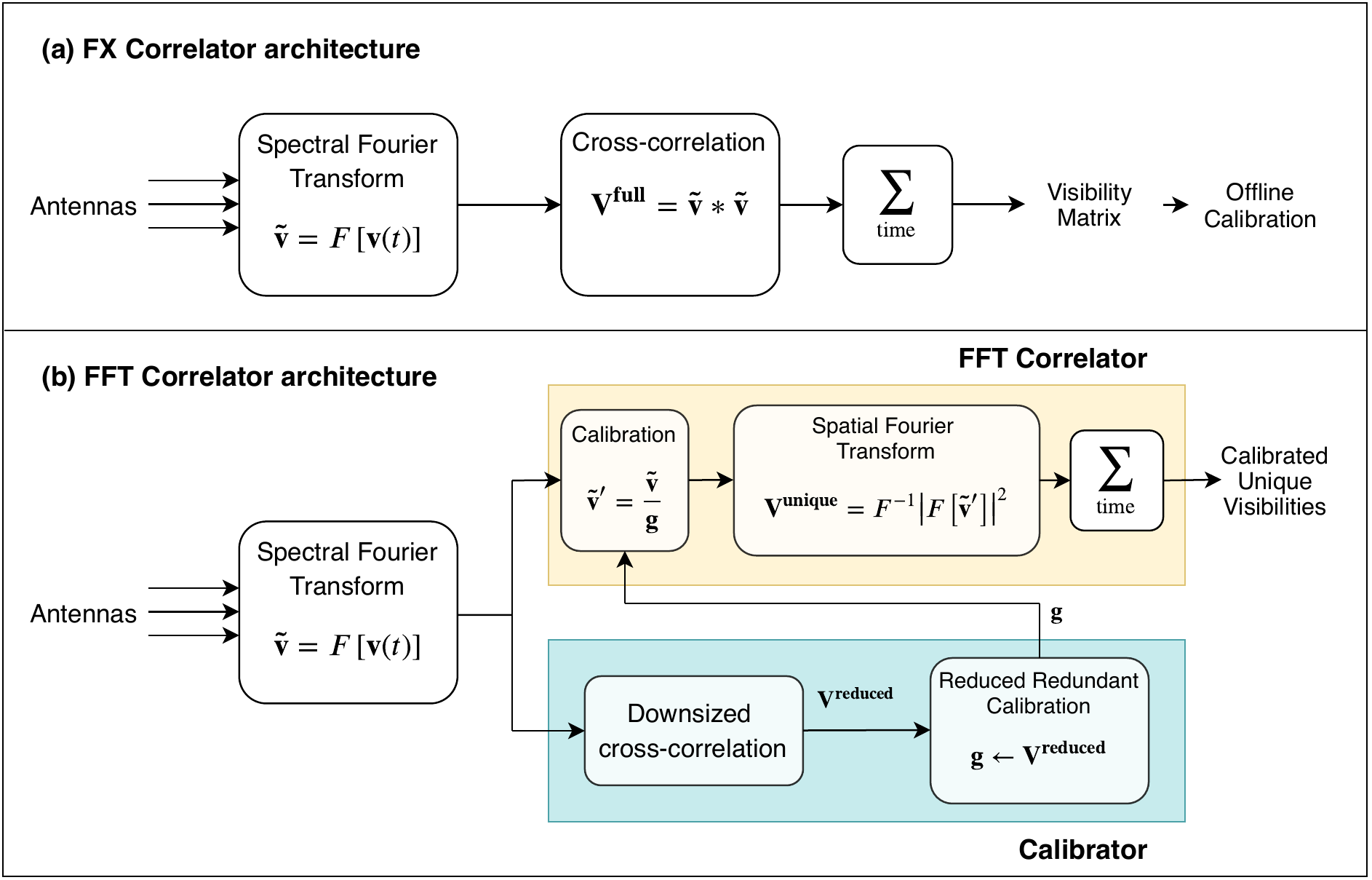}
    \caption{Panels (a) and (b) show the correlator architectures for a traditional FX-correlator and an FFT-correlator respectively. For either architecture, the first stage Fourier transforms the voltage measured by each antenna $\mathbf{v}(t)$ to obtain a spectrum $\mathbf{\tilde{v}}$, and the second stage computes visibilities of all antenna pairs. For an FX-correlator, the visibility matrix $\mathbf{V^{full}}$ is computed by cross-correlating the signal from every antenna with every other antenna in the array. For an FFT-correlator, the visibility matrix $\mathbf{V^{unique}}$ is computed by a spatial Fourier transform on the calibrated antenna voltages (yellow box). The calibrator (blue box) operates in parallel to the FFT-correlator and computes per-antenna gains for calibrating antenna voltages. The antenna gains are computed by performing one of the two reduced redundant-baseline calibration schemes described in this paper, on a smaller visibility matrix $\mathbf{V^{reduced}}$ computed for the purpose of calibration.
  \label{fig:systemlayout}}
\end{figure}

In a traditional FX correlator architecture, the signal from every antenna is cross-correlated with the signal from every other antenna in the array. Panel (a) of Figure~\ref{fig:systemlayout} shows this architecture. The first stage performs a spectral Fourier transform, computing a spectrum of the time-varying voltage signal from antennas. The second stage computes the cross-correlation of all antenna pairs producing a time-integrated visibility matrix. The computational resources required to generate the visibility matrix and, to store and process the output data products scale as $\mathcal{O}(N^2)$ with the number of antennas in the array. For large-N arrays, this cost can dominate the entire cost of the array and has been one of the limiting factors for interferometers built in the previous decade.

For interferometers with antennas on a regular grid, \citet{daishido_et_al_1991, Tegmark_and_Zaldarriaga_2009, Tegmark_and_Zaldarriaga_2010} have proposed FFT-correlators or FFT imagers as a potential solution to this steep scaling in cost- and computational-resources. Instead of cross-correlating antenna pairs, an FFT-correlator produces visibilities through a spatial Fourier transform. If the visibilities of redundant baselines, produced by an FX-correlator, can be averaged, these two methods are equivalent by the convolution theorem \citep[see][]{Tegmark_and_Zaldarriaga_2009, Tegmark_and_Zaldarriaga_2010}. By the nature of the Fast Fourier transform algorithm \citep{cooley_tukey65} FFT-correlators only scale as $\mathcal{O}(N\log{N})$, decreasing the number of computations performed in the correlator.

\subsection{Theoretical Motivation}
\label{chp:Intro:sec:FFTCorr:subsec:Theory}

The equivalence in the output of FX correlators and FFT correlators can be seen by rewriting Equations~\ref{eq:visibility_uv} and~\ref{eq:fxcorr} for a linear array where the uv-plane is reduced to a single line.
\begin{equation}
    \mathcal{V}(u) = \int v(x) v^{*}(x + u) \, \mathrm{d}x \equiv (v \star v) (u)
\end{equation}
\noindent
In other words, this equation encodes the relationship between the aperture plane and the uv-plane as shown in Figure~\ref{fig:uvplane}. Applying the convolution theorem to the above equation we can write:
\begin{equation}
    \mathcal{V}(u) = \mathcal{F}^{-1}\left[\mathcal{F}(v \star v^{*})\right] = \mathcal{F}^{-1}\left|\mathcal{F}(v)\right|^2
\end{equation}
\noindent
The last equality shows that visibility matrices can be generated by computing a spatial Fourier transform of the voltage signal measured by every antenna in the array, squaring it to obtain the amplitude and inverse Fourier transforming the result. Note that this equivalence holds for all array layouts, not just redundant arrays. However, this approach is more computationally efficient for redundant arrays than non-redundant layouts.

\subsection{Calibration}
\label{chp:Intro:sec:FFTCorr:subsec:Calibration}

An important difference between an FX-correlator and an FFT-correlator is that the latter does not preserve the full visibility matrix. The spatial Fourier transform averages redundant visibilities, which are expected to be the same, in principle, since they are visibilities measured by antenna pairs with the same displacement vector. However, in practice, redundant visibilities are different due to differences in the signal chain, structure of the dish, varying cable-lengths etc. If antenna voltages are not calibrated, the spatial Fourier transform could result in averaging dissimilar visibilities, making post-processing correction impossible as well. Hence, antenna gain- and phase-calibration, prior to the spatial Fourier transform, is essential to avoid signal-loss in the FFT-correlator.

An FFT-correlator that implements the design proposed by \citet{Tegmark_and_Zaldarriaga_2009}, is the one built by \citet{Foster_et_al_2014} on the BEST-2 array at Medicina, Italy. They demonstrated that the visibilities produced by the FFT-correlator and the redundantly-averaged visibilities of an FX-correlator are similar when all the antennas are calibrated before the spatial Fourier transform. However, they used a traditional FX-correlator working in parallel to generate all the visibilities required for point-source calibration. This is not a scalable solution for calibrating large-N arrays since building an FX-correlator may not be viable. 

A more generic alternative to the FFT-correlator, discussed by \citet{Thyagarajan_et_al_2017}, is a direct-imaging-correlator called the E-field Parallel Imaging Correlator (EPIC) that has now been deployed on the LWA \citep{Kent_2019}. It works like a Modular Optimal Frequency Fourier (MOFF; \citealt{Morales_2011}) correlator, where antenna voltages are gridded before a spatial Fourier transform produces an electric-field image. Unlike the FFT-Correlator, EPIC can also be implemented on non-redundant arrays, including arrays where the antenna beams are non-identical. For highly redundant arrays with identical antenna beams, EPIC becomes equivalent to the FFT-correlator. \citet{Beardsley_2017} propose an iterative sky-based calibration algorithm, EPICal, for such a correlator that does not require generating real-time visibility products and scales as $\mathcal{O}(N)$. However, EPICal requires prior knowledge of antenna beams which can be difficult to model or measure in situ at low radio frequencies. Moreover, the lack of accurate diffuse-sky models that also account for polarisation at these frequencies could make it harder to decouple the sky-signal from beam models. Here we choose to discuss only redundant array layouts where the redundancy can be exploited for calibration.

An ideal calibration scheme for FFT-correlators, must be capable of minimising the scatter in redundant visibilities because any residual scatter will become additional noise on the visibility returned by the correlator. Additionally, the calibration scheme must produce an output that can be applied to antenna voltages.  Redundant-baseline calibration \citep{wieringa_1992,Liu_et_al_2010, noorishad_et_al_2012, marthi_and_chengalur_2014}, that has been used to calibrate the Donald C. Backer Precision Array for Probing the Epoch of Reionization (PAPER; \citealt{Parsons_et_al_2010, Ali_et_al_2015, Kolopanis_et_al_2019}), LOFAR \citep{noorishad_et_al_2012}, MITEoR \citep{Zheng_et_al_2014} and HERA \citep{Dillon_et_al_2020}, results in complex antenna gains that can be applied to antenna voltages. The multiplicative antenna gains are computed by solving a system of equations that minimise the scatter in calibrated redundant visibilities.

A known caveat of redundant-baseline calibration is that it can only yield relative antenna gains \citep{Liu_et_al_2010, Dillon_et_al_2018} i.e, the equations can constrain the ratio of antenna gains but cannot determine their actual value. The system of equations has a null space with four degenerate parameters including the absolute amplitude and the phase of antenna gains. However, this is not a problem for calibrating voltages for the purpose of FFT-correlation since it requires only relative calibration of antennas so that visibilities of redundant baselines can be averaged coherently. Absolute calibration, to determine the degenerate parameters, can still be performed offline with the visibilities generated by the FFT-correlator. 

Applications of redundant-baseline calibration, so far, had the full visibility matrix available for constructing the system of equations for calibration. However, redundant-baseline calibration does not inherently require all $N(N-1)/2$ visibilities measured at high signal-to-noise ratio (SNR). This paper explores two redundant-baseline calibration schemes that can use $\mathcal{O}(N\log{N})$ computational resources for generating visibilities for the purpose of calibration and are henceforth referred to as \textit{reduced redundant-baseline calibration} schemes.

The FFT-correlator architecture assumed in this paper is similar to the one proposed by \citet{Zheng_et_al_2014}. This is shown in Panel (b) of Figure~\ref{fig:systemlayout}. The first stage, computing a spectrum of antenna voltages, is similar to the FX-correlator architecture. The yellow boxed region shows the FFT-correlation where a spatial Fourier transform on calibrated voltages results in the time-integrated unique visibilities of the array. Notice that the FFT-correlator does not produce the full visibility matrix; the redundant baselines are averaged by the spatial Fourier transform. 

The blue boxed region in Figure~\ref{fig:systemlayout} shows the \textit{calibrator}, which is the main focus of Chapter~\ref{chp:Redredcal}. It performs two functions: (a) cross-correlate the baselines required for calibration in a manner similar to the second stage of an FX-correlator and (b) compute antenna gains by applying one of the two reduced redundant-calibration schemes on this set of visibilities. The computation- and resource-intensive stage of the calibrator is the first step of cross-correlating antenna pairs. The number of baselines that need to be cross-correlated in a given integration cycle determines the computational resources required by this stage of the calibrator. The reduced redundant-baseline calibration scheme employed by the calibrator dictates the set of visibilities that need to be cross-correlated and hence determines the size of the calibrator. Both the calibration schemes discussed in Chapter~\ref{chp:Redredcal} can be adapted to a calibrator that scales as $\mathcal{O}(N\log{N})$, keeping the size of the calibrator comparable to the size of the FFT-correlator.

\section{Hydrogen Epoch of Reionization Array}
\label{chp:Intro:sec:HERA}

The Hydrogen Epoch of Reionization Array (HERA;~\citealt{Deboer_et_al_2017}) is an upcoming experiment designed to detect and characterize the power spectrum of neutral hydrogen from the epoch of reionization (EoR). It is a low frequency radio interferometer consisting of 350 parabolic dishes built in a compact hexagonal grid layout, to maximize sensitivity in the power spectrum modes that have low foreground contamination. 

\begin{figure}
    \centering
    \includegraphics[width=\linewidth]{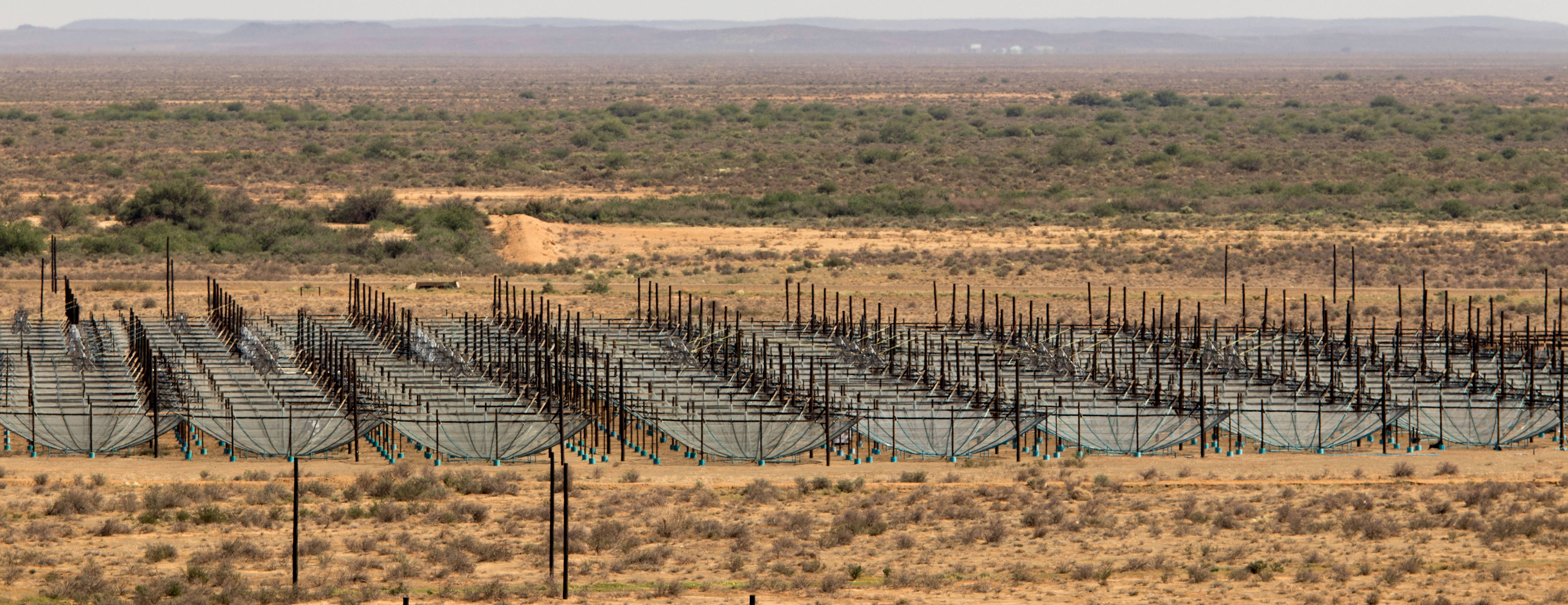}
    \caption{The Hydrogen Epoch of Reionization Array under construction in the South African Karoo Radio Astronomy Reserve. The antennas are 14.6\,m diameter non-tracking parabolic dishes supported by wooden poles and built out of PVC pipes. Each antenna has a Vivaldi feed that can observe in the frequency range 50--250\,MHz. The image shows 200 of the planned 350 antennas that are currently being commissioned for early science.}
    \label{fig:HERA}
\end{figure}

The layout, antenna design and analog signal chain of HERA have been designed to have a smooth spectral response in the frequency range 50--250\,MHz. While the inherent chromaticity of interferometers cannot be avoided, minimizing spurious spectral features due to a non-uniform bandpass or digital artefacts can greatly reduce the burden of modelling and rejecting foreground contamination. 

HERA is currently being built in the Karoo Radio Astronomy Preserve in South Africa. Roughly 200 of the planned 350 antennas have been constructed and deployed with the relevant signal processing backend. These antennas are being commissioned for pipeline robustness, testing reliability of the signal processing chain and for obtaining early science results. Figure~\ref{fig:HERA} shows the current status of the interferometer.

The primary science goal of HERA is characterizing the power spectrum of neutral hydrogen from cosmic dawn through EoR. HERA targets the redshift range $4.7 < z< 27.4$ which encompasses the period of X-ray heating, formation of the first stars and subsequent reionization of the Universe. Science results from HERA will help determine what objects composed the early luminous structures, place constrains on cosmological parameters like the optical depth to CMB which directly affects CMB lensing experiments that are attempting to establish gravitational waves in the primordial Universe, determine the amount of X-ray radiation produced by early stars and determine the redshift boundaries of the different phases of neutral hydrogen evolution that we outlined in Section~\ref{chp:Intro:sec:HI:subsec:GSM}.

The most important advance, made early in the last decade, that motivated the construction of a new telescope is the containment of high-amplitude foregrounds to a wedge-like shape in the cylindrical coordinate space of the power spectrum. Section~\ref{chp:Intro:sec:Pspec:subsec:DelayTransform} outlines the delay transform and which allows the visibilities of `short' baselines to be used as a proxy for estimating modes of the power spectrum. The delay transform applied to visibilities is also a way of foreground filtering (Section~\ref{chp:Intro:sec:Pspec:subsec:Foregrounds}), since it naturally contains the foreground contamination to a wedge-like region in cylindrical power spectrum coordinates.

\subsubsection{Design Specifications}

The design of HERA is built on the lessons learnt from previous EoR experiments like PAPER, MWA, GMRT and LOFAR. It is purpose-designed for detecting and characterizing the power spectrum of EoR, unlike other general-purpose radio observatories. It provides a substantial increase in sensitivity over previous experiments, with a large collecting area of nearly $54,000\,\mathrm{m}^2$, and holds the promise of making a robust detection. Table~\ref{tab:HERA} shows the design specifications for HERA-350.

\begin{table}
    \centering
    \begin{tabular}{cc|cc}
    \hline
        Specification &   & Observational Parameter & \\
    \hline
        Antenna Diameter &  14.6\,m &  Primary beam   &   $9.54^{\circ}$  \\
        Shortest Baseline & 14.6\,m & Largest scale & $7.8^{\circ}$ \\
        Number of antennas  &  350 & &      \\
        Outrigger antennas  &  30  & &      \\
        Longest Baseline (core)  &  292\,m   &  Synthesized beam (core)  &  $25'$  \\
        Longest Baseline (outrigger) &  876\,m  &   Synthesized beam (outrigger) & $11'$ \\ 
        Frequency range          &  50--250\,MHz      &   Redshift range  &  $4.7 < z < 27.4$ \\
        Number of channels       &  8192        \\ 
        Frequency resolution     &  24.4\,kHz         &   LoS Comoving Resolution & 1.7\,Mpc (at z=8.5) \\
        \hline
    \end{tabular}
    \caption{HERA design specifications and observational parameters for EoR science}
    \label{tab:HERA}
\end{table}

\begin{figure}
    \centering
    \includegraphics[width=0.65\textwidth]{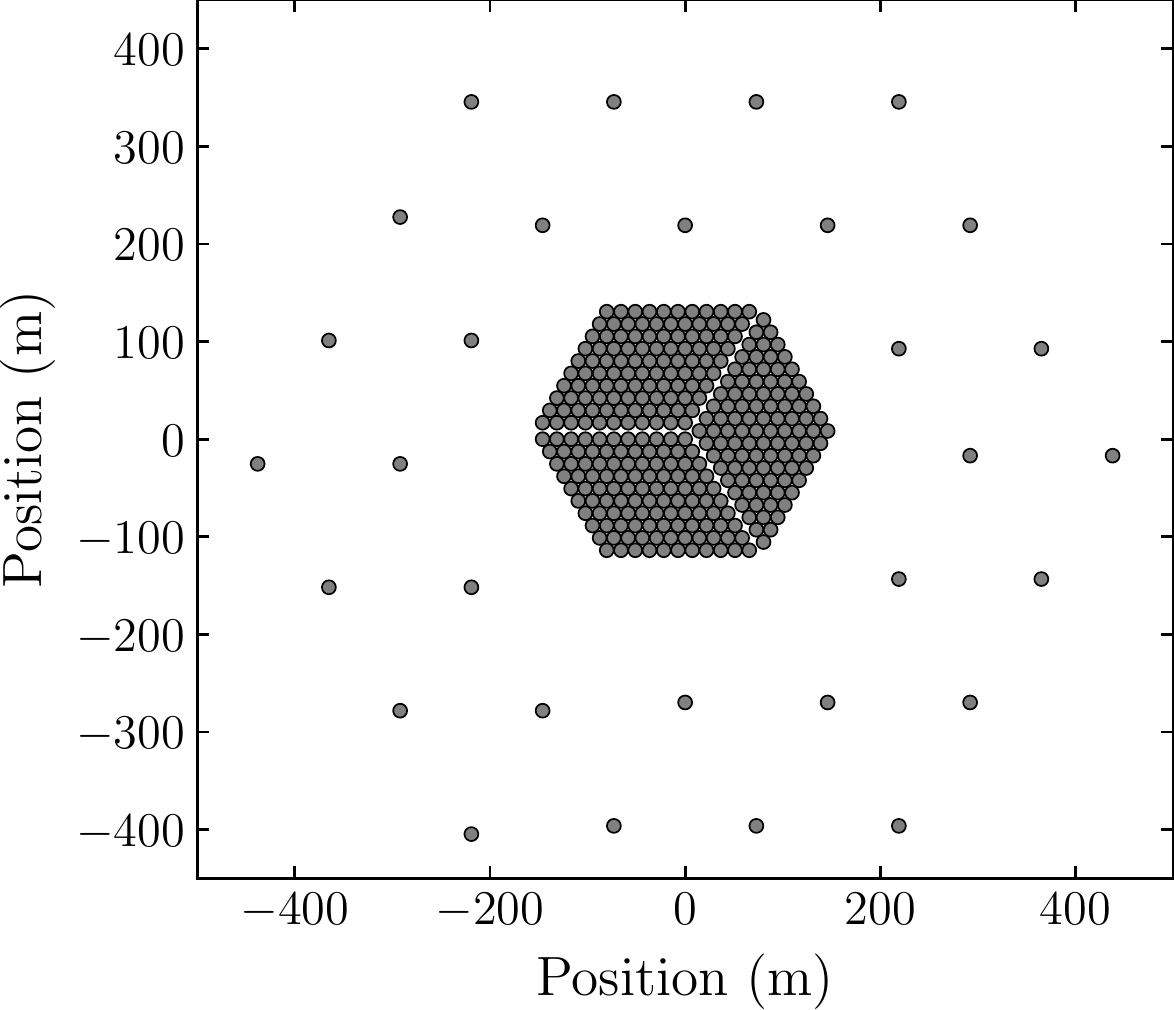}
    \caption{Layout of HERA with 350 antennas~\citep{Dillon_and_Parsons_2016}. The core consists of 320 antennas built in a hexagonal grid layout, spanning a diameter of nearly 300\,m. The split-core configuration and the 30 outrigger antennas improve the uv-coverage of the array, which is useful for foreground calibration and imaging the EoR.}
    \label{fig:HERA_layout}
\end{figure}

HERA is a filled-aperture interferometer with a core diameter of nearly 300\,m (see Figure~\ref{fig:HERA_layout}). This makes it almost as large as Arecibo in terms of collecting area. It is built in a redundant configuration, with antennas located on hexagonal grid points~\citep{Dillon_and_Parsons_2016}. The antenna diameter is same as the shortest baseline length, meaning that the antennas are built edge-to-edge to achieve a large collecting area. This also maximizes the number of short baselines, which are critical to the delay transform approach of measuring the power spectrum. The large number of short baselines also allows numerous instantaneous measurements of the same power spectrum mode, allowing a faster increase in the signal-to-noise ratio than afforded by time alone.

The redundant configuration also enables the use of redundant-baseline calibration for computing antenna gains. At the low frequencies where HERA operates, the sky signal is dominated by diffuse emission from the galaxy. The lack of high precision models of this emission in addition to instrument chromaticity makes sky-based calibration difficult. Redundant-baseline calibration, which relies on the redundancy of the array and is nearly independent of the sky signal, is an attractive alternative. However, the requirement of redundancy is a strict one. Antenna positions errors, beam point errors and beam width can all result in poor calibration solutions, and manifest as mode-mixing in the power spectrum. \citet{Dillon_et_al_2020} quantify the redundancy in the HERA layout and show how instrument systematics like cross-talk and antenna sidelobes can affect calibration solutions.

A disadvantage of a completely redundant configuration is the sparse uv-coverage obtained for the number of elements built. While the primary science goal of HERA is power spectrum estimation, it theoretically has the sensitivity required to image the EoR, at least at some redshifts. Moreover, a good uv-coverage can enable precision imaging of foregrounds which is useful for subtracting bright point-sources and for secondary sky-based calibration. To mitigate this issue, \citet{Dillon_and_Parsons_2016} suggest a split-core configuration as shown in Figure~\ref{fig:HERA_layout}. This layout maintains the redundancy required for calibration and power spectrum estimation, while improving uv-coverage. The 30 outrigger antennas serve the same purpose, and additionally, increase the resolution of the telescope.

The dish structure and feeds have been designed to enable observations in the wide spectral bandwidth range of 50--250\,MHz. This enables HERA to target observations from cosmic dawn, when neutral hydrogen was dominated by X-ray heating from the first stars, through the end of the reionization when filling fraction of ionized hydrogen is nearly one. The frequency resolution of the signal processing backend, set by fringe decorrelation on the largest baselines, informs the resolution along the line-of-sight. 

The antenna diameter and the total number of antennas have been optimized for cost, considering the total collecting area required to make a detection. The antennas are built from readily available materials like wire mesh, PVC pipes, concrete and wooden poles to minimize the cost per antenna. The Vivaldi feeds are suspended by line, borrowed from boating suspension cables.

\section{Thesis Layout}
\label{chp:Intro:sec:Layout}

The field of 21\,cm cosmology holds the promise of uncovering the physics of early star formation, properties of the IGM, amount of X-ray radiation in galaxies and determining the drivers of cosmic reionization. Opening up this era in the evolution of our Universe requires well-designed radio interferometers that have a large collecting area, precision control over antenna beams, and numerous short baselines. Backed by the low cost of non-cryogenic receivers, this has propelled a new class of radio interferometers with a large number of antennas ($N \gg 100$) and wide bandwidths (and relatively low budgets). The digital signal processing requirements for such telescopes are enormous compared to a previous generation of radio observatories with the number of antennas on the order of 10s.

Often, theoretical limits on the cosmological parameters, that can be probed by an experiment, do not account for instrument induced errors like malfunctioning antennas, rounding errors, asynchronous clock domains in digital hardware etc. Instruments are expected to function perfectly and deliver accurate data products that reflect the measured signal within noise limits. This requirement is more stringent for precise experiments aimed at detecting the EoR signal, by sequestering foreground contamination to a tight region in the power spectrum phase space. Instrument systematics can widen the foreground wedge, and digital errors can affect signals which need to be integrated for days.

This thesis addresses the digital signal processing requirements of one such current-generation EoR experiment, HERA, and paves the way for next-generation larger telescopes. While the material presented here targets EoR experiments, it is in general applicable to large-N radio interferometers that may be aimed at FRB observations, dark energy surveys, primordial inflation field power spectrum or surveys of radio transients. 

Chapter~\ref{chp:Swcorr} describes a raw voltage recorder pipeline that I built to collect voltage data from PAPER and HERA antennas, for the purpose of verifying the feasibility of an FFT correlator. The exercise of building this instrument helped me appreciate the nuances of designing and deploying real-time systems. However, a big learning through this exercise was that not all data lead to tangible outcomes. While I did show that the output of FFT correlators matches that of an FX design when antennas are calibrated, the data was insufficient to compare various calibration schemes. 

Chapter~\ref{chp:Redredcal} motivates two calibration schemes that can work with FFT correlators. Both schemes are built on the idea that FFT correlators are more suitable for redundant arrays, and exploit this redundancy to estimate antenna gains. I place theoretical limits on the precision of the obtained calibration solutions and show that noise in the measurement dominates the result in both cases. Lastly, I compared both these schemes and suggest which experiments either would be suitable for, based on science-goals and amount of computational resources available.

Chapter~\ref{chp:HERACorr} lays out the lessons learned in the practical exercise of building an FX correlator for the Hydrogen Epoch of Reionization Array. This correlator is built on a hybrid platform, using custom-built FPGA boards for the F-stage and GPUs for the X-stage. General-purpose off-the-shelf Ethernet switches connect the two systems. The entire system processes data in real-time at roughly 1\,Tbps and is one of the largest correlators currently deployed on a telescope.

Finally, Chapter~\ref{chp:Conclusion} presents the conclusions of this dissertation work, with an outlook towards the next-generation telescopes. Overall, this thesis tries to motivate that instrumentation is an integral part of radio astronomy, and an aspect as important as the data analysis process required to produce the correct results.

% Software Voltage Recorder
\chapter{Software FFT Correlator}
\label{chp:Swcorr}

Section~\ref{chp:Intro:sec:FFTCorr} outlines the design of this recently proposed correlator architecture and motivates its cost- and computational-resource advantageous for redundant arrays like the Hydrogen Epoch of Reionization Array (HERA; Section~\ref{chp:Intro:sec:HERA};~\citealt{Deboer_et_al_2017}). FFT correlators~\citep{Tegmark_and_Zaldarriaga_2009, Tegmark_and_Zaldarriaga_2010} rely on the redundancy of a radio interferometer i.e., at a given time and frequency the cross-correlation products of all antenna pairs with the same baseline vector (redundant baselines) should be nearly equal. This is an important criterion for the proper functioning of FFT correlators because they only output the average visibility measured by all the redundant baselines in the array. If redundant baselines have dissimilar visibilities, FFT correlators can result in signal loss. The dissimilarity in redundant visibilities can arise from differences in the analog signal chain, the digital signal processing or individual receiver noise. However, most of these effects can be grouped into per-antenna complex calibration parameters which calibration algorithms can reasonably estimate~\citep{Liu_et_al_2010, Dillon_et_al_2018}.

The motivation to build a software-based FFT correlator was to prototype an FFT correlator for maximally-redundant arrays, like HERA, and experiment with different calibration schemes that can yield a result closest to that produced by a traditional $\mathcal{O}(N^2)$ correlator. Specifically, we wanted to explore the effect of two unrelated sources of signal loss in an FFT correlator-- (a) non-redundancy in baselines that are designed to be redundant and (b) calibration errors that can arise from not utilizing the full visibility matrix of $\sim$$N^2/2$ visibility products.

Non-redundancy in baselines, that have been designed to be redundant, arise from practical issues during the construction of the array~\citep{Orosz_et_al_2018}. For example, antennas in the field cannot be laid out to millimeter precision in x,y,z coordinates. There can be considerable (at least a few centimeters) error, both in the position of antennas and in the height of the antenna from a reference point in the ground due to non-uniform levelling of the antenna site prior to construction. For reference, a centimeter error in the East-West positioning of an antenna creates a $60^{\circ}$ phase error with a source on the horizon, for the shortest baseline of 14.6\,m at 250\,MHz frequency. In a similar way, there could be beam-size and beam-pointing errors that originate from improper suspension of the feed over the antenna dish or irregularities in the reflecting surface of the dish itself. 

Non-redundancy in the array is difficult to calibrate even with all the cross-correlation products produced by a traditional correlator, and are impossible to correct with just the output of the FFT correlator. For an array that is designed to be redundant, like HERA, it is important to quantify the error originating from imperfections in the field and find ways to minimize them in the field itself. Since these errors also have a significant impact on the science-goals of HERA, the software FFT correlator, which is faster and cheaper to build than a full correlator system, was an easy way to look for inherent redundancy in the HERA layout as it was being built.

The second motivation to build a software FFT correlator was to experiment with different calibration schemes. Most existing calibration algorithms rely on the availability of the full visibility matrix to compute the per-antenna complex calibration parameters. FFT correlators attempt to minimize cost by avoiding the computation of the full visibility matrix in arrays with many redundant baselines. If any or some of the existing calibration schemes could be adapted to run on fewer visibilities and have a scaling $\leq\,\mathcal{O}(N\log{N})$, it would be a good calibration methodology for FFT correlation. However, the effect of providing reduced information to a calibration algorithm could be larger calibration errors. Since calibration needs to be performed prior to FFT correlation, comparing the outcomes of multiple calibration schemes entails FFT correlating multiple sets of calibrated antenna voltages that are otherwise similar in time and frequency. A software correlator, running on a CPU, is easier to apply to multiple sets of data than a hardware-based system. 

The software FFT correlator was straight-forward to write in Python using the Numpy FFT algorithm. The more challenging part of the experiment was collecting raw voltage data from antennas, which forms the input to both the FFT-correlator and the calibration algorithms. At the time when this experiment was conceived and planned both, the Donald C. Backer Precision Array for Probing the Epoch of Reionization (PAPER;~\citealt{Parsons_et_al_2010}) antennas and HERA Phase-I antennas, operated with a bandwidth of 100\,MHz between 100-200\,MHz, requiring a sampling frequency of at least 200\,MHz. At 8\,bit resolution of each sample, the raw voltage stream of a single antenna produces a datarate of 1.5\,Gbps. Recording the raw voltages of even $\sim$10 antennas for an hour requires $\sim$10\,TB of disk space.

If the raw voltage data could be channelized into narrow spectral bins, the experiment could be performed with only a few channels of voltage data. That is, recording a fraction of the bandwidth that the antennas are capable of observing is sufficient for prototyping the calibration algorithms that can operate on an FFT correlator. This significantly decreases the data rate at which raw voltages need to be recorded to disk and the entire system could be built on a small portable server. After considering various factors which are laid out in the following sections, we decided to record the voltage data within 3 narrow frequency channels of 390\,kHz each, from a total of 12 antennas, and for a time period of 6 hours.

The experimental setup consists of two parts, a digital signal processing (DSP) pipeline that digitizes and channelizes the analog time-domain voltage stream, and a 1U server that can write the data to disk. The DSP pipeline is built on the Smart Network ADC Processor (SNAP) boards that host 250\,MHz ADC samplers, a Xilinx Kintex-7 FPGA and two 10\,GbE ports for sending data via Ethernet. The 1U Server is equipped with 10\,GbE NIC cards and two 6TB hard-drives for data collection and file writing. 

The layout of this chapter is as follows: Section~\ref{chp:Swcorr:sec:Fengine} discusses the FPGA-based channel selector design that is used to program the Kintex-7 FPGA. Section~\ref{chp:Swcorr:sec:Xengine} describes the CPU-based file writing pipeline that was written for this experiment. Section~\ref{chp:Swcorr:sec:observing} relates the adventures of deploying this correlator at two different sites and the quick decisions that had to be made in lieu of unforeseen circumstances. Finally, Section~\ref{chp:Swcorr:sec:Analysis} presents the results from the data collected from both PAPER and HERA antennas. Unfortunately, both data sets were insufficient to test the two objectives of this experiment. However, the lessons learnt while building, deploying and analysing this data proved valuable when I started working on the HERA correlator system that is much more complex.

\section{FPGA-based Channel Selector}
\label{chp:Swcorr:sec:Fengine}

The primary purpose of the first stage of the voltage recorder, is to digitize the analog voltage stream of each antenna, compute a spectrum over a well-defined time window and select a finite number of frequency channels for writing to disk. The selected frequency channels are built into UDP packets with unique time-stamps and sent to a server via a 10\,GbE port. An FPGA platform was chosen to build this system due to the precise time-stamping and time synchronization required across all the antennas.

\begin{sidewaysfigure}
    \centering
    \includegraphics[width=\linewidth]{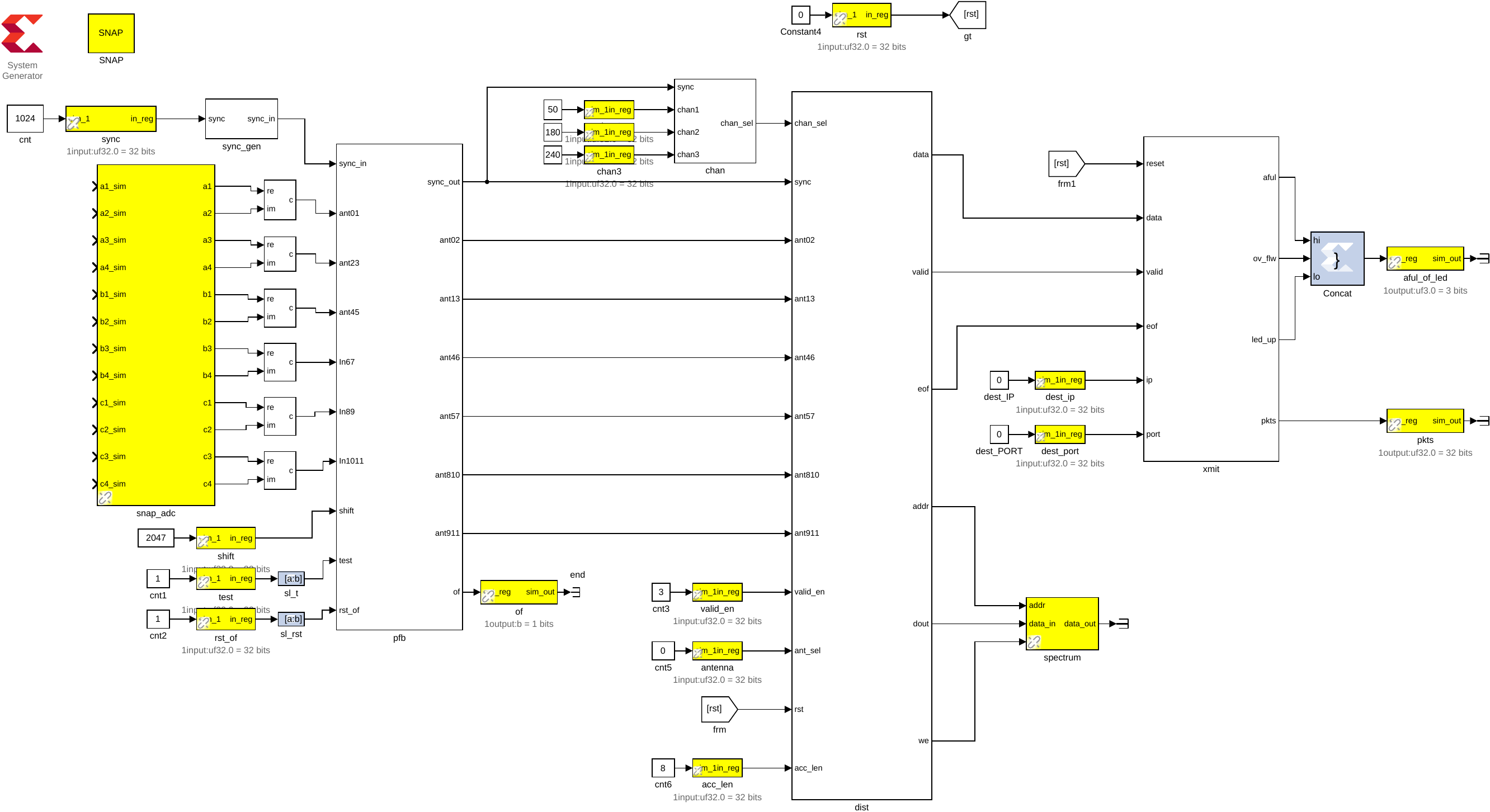}
    \caption{MATLAB$^{\tiny\text{\textregistered}}$ Simulink design that is used to program the Xilinx$^{\tiny\text{\textregistered}}$ Kintex-7 FPGA on-board the SNAP. The yellow colored blocks are CASPER-designed interfaces to hardware devices on the SNAP, like ADCs and 32-bit registers. The five main modules of the design are \texttt{snap\_adc}, \texttt{pfb}, \texttt{chan}, \texttt{dist} and \texttt{xmit}.}
    \label{fig:volt_recorder_full}
\end{sidewaysfigure}

At the time of this experiment, the Smart Network ADC Processor (SNAP) boards had just been designed and manufactured for deployment on the HERA Phase-II system. SNAP boards are capable of processing up to 12 inputs at a sampling rate of $\leq$\,250\,MHz using the on-board HMCAD1511 ADC samplers. This was the motivation for building the software correlator with just 12 antennas-- with a single SNAP board, synchronization issues between multiple boards, which can be cumbersome to calibrate with only limited frequency channels, can be avoided.

The Xilinx Kintex-7 FPGA on-board the SNAP was designed using MATLAB$^{\tiny\text{\textregistered}}$ Simulink$^{\text{\copyright}}$ and the System Generator for DSP$^{\text{\copyright}}$ module provided by Xilinx, which can compile the Simulink design into a valid Vivado$^{\tiny\text{\textregistered}}$ project that can be used to program the FPGA. Simulink-compatible modules to interface with the other hardware devices on the SNAP board, like ADCs, 10\,GbE ports, 32-bit registers, GPIO pins etc, were collectively designed by members of the Collaboration for Astronomy Signal Processing and Electronics Research (CASPER;~\citealt{Parsons_et_al_2008, Hickish_et_al_2016}). The Simulink design, shown in Figure~\ref{fig:volt_recorder_full}, outlines six major modules that implement the functionality required and are as follows:

\subsubsection{ADC Module (\texttt{snap\_adc})}
The SNAP ADC block is the interface to the HMCAD1511 ADC chips on the SNAP board and was designed by members of CASPER. A separate Python program, with code that is now integrated into \texttt{casperfpga}\footnote{\url{https://github.com/casper-astro/casperfpga/blob/master/src/snapadc.py}}, was used to calibrate the ADCs before every run.

\subsubsection{Polyphase Filterbank (\textbf{\texttt{pfb}})}

The Polyphase Filter Bank (PFB) module converts 1024 time-domain samples to 512 frequency-domain channels by performing a 1024-point FFT. In a traditional 1024-point FFT 1024 time samples are converted to 1024 frequency channels. However, since our input is always real-valued, the amplitude of the negative frequencies is mathematically guaranteed to be a mirror image of the positive frequencies, and can be discarded. Only 512 frequency channels with complex-valued amplitude, corresponding to the positive frequencies, are retained after the FFT.

\begin{figure}
    \centering
    \includegraphics[width=\textwidth]{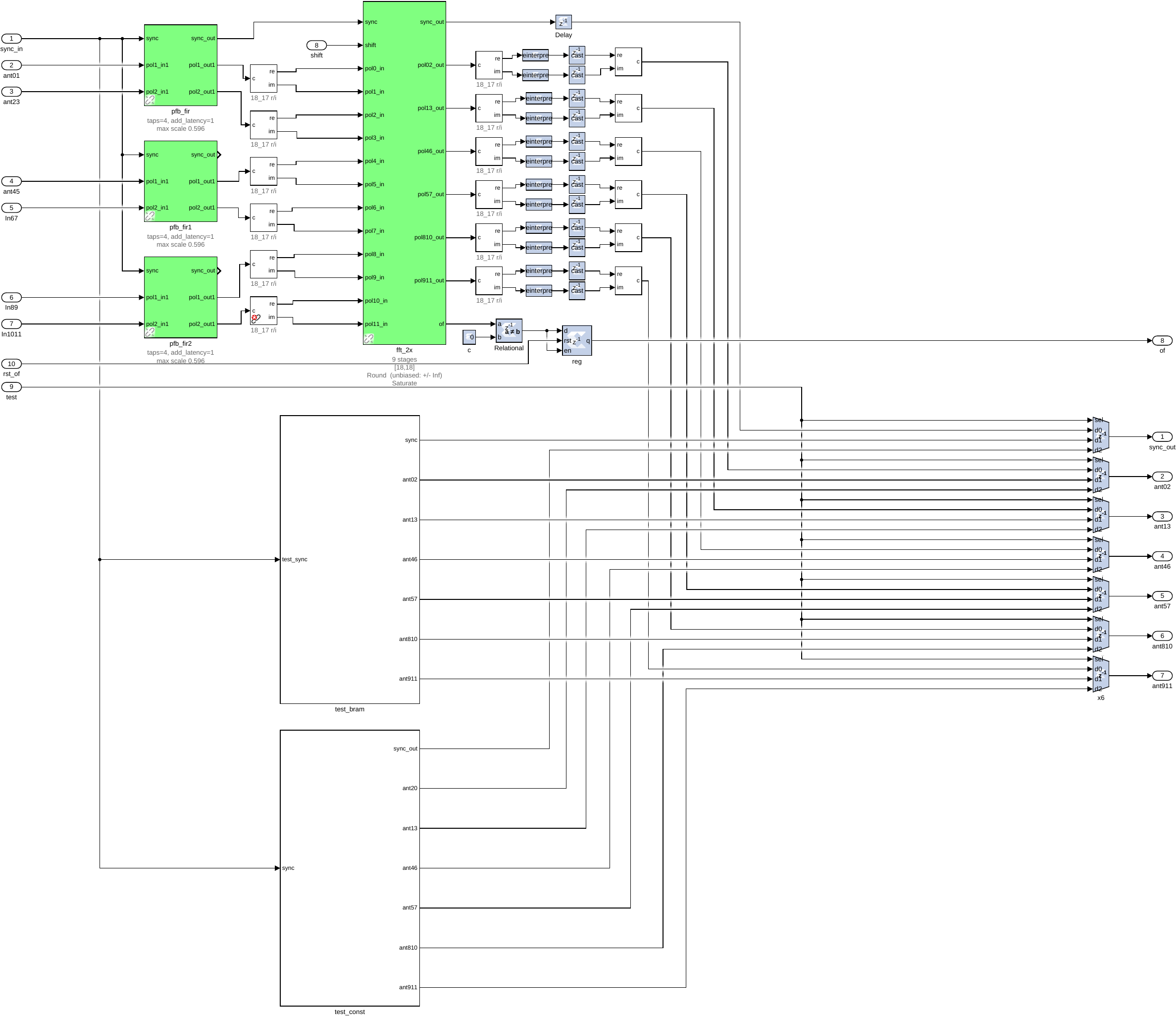}
    \caption{The Polyphase Filter Bank (\texttt{pfb}) module of the FPGA design. The green colored blocks are CASPER-designed modules to implement a 4-tap FIR filter with a Hamming windowing function, and a 1024-point FFT on each of the 12 inputs. The two blocks in parallel to the FFT block generate test vectors for checking the downstream functionality. \texttt{test\_bram} allows the user to set a custom vector in BRAM and \texttt{test\_const} sets the input of all antennas to a constant value.
    \label{fig:volt_recorder_pfb}}
\end{figure}

Prior to the 1024-point FFT, the signal is passed through a 4-tap Finite Impulse Response (FIR) filter with a Hamming window function. The FIR filter, effectively, multiplies 4096 ($4 \times 1024$) time-domain samples with a Hamming window function. By the convolution theorem, this is equivalent to convolving the frequency-domain spectrum by the Fourier transform of the Hamming window which is narrow with adjacent sidelobes suppressed down to $-40$\,dB. This decreases spectral leakage in the discrete Fourier transform and improves the sharpness of the spectral response.Both the FIR filter and the FFT module are available as CASPER-designed modules in the CASPER DSP library\footnote{\url{https://github.com/casper-astro/mlib_devel}} (see Figure~\ref{fig:volt_recorder_pfb}). 

The FIR filter module accepts complex-valued input and performs a linear additive-multiplicative operation which is the same for both the real and imaginary parts of the input. Both, multiplication with a real-valued function and summation of time samples, keep the real and imaginary parts of the input separate and prevent any mixing. This allows a user to concatenate two real-valued signals into a single complex-valued signal and pass it as input to the FIR filter. The resulting output can be re-split into real and imaginary parts to recover the processed version of the original two signals. In addition, each FIR filter module is capable of independently processing two complex-valued input streams. Hence, a total of three FIR filter modules are used to process the 12 inputs to the SNAP board.

The two blocks in parallel with the FFT module, \texttt{test\_bram} and \texttt{test\_const}, are designed to setup test vectors for checking the functionality of all the blocks downstream of the \texttt{pfb} module in the top layer i.e., \texttt{chan},  \texttt{dist} and  \texttt{xmit}. The \texttt{test\_bram} block allows the user to set custom test-vectors for each of the 12 outputs of the \texttt{pfb} block, by modifying the contents of BRAMs. This is useful for identifying bugs in parts of the design where antennas are treated in serial, for example, in building UDP packets for transporting via Ethernet. The \texttt{test\_const} block sets the input of all antennas to a frequency ramp. This is useful for testing the functionality of portions where antennas are treated in parallel but frequency channels are treated serially, for example, in channel selection.

\subsubsection{Channel Selector (\texttt{chan})}

The top layer of the design exposes three 32-bit registers which the user can use to set the three frequency channel numbers that need to be written to file. This is a user-defined parameter rather than a hardwired one because radio frequency interference (RFI) is site-specific and time variable. Moreover, this flexibility allows the user to change the three frequency channels being written to disk from one observing run to another, thus getting more bandwidth. 

The \texttt{chan} block takes the three user-input frequency channel numbers, which can be any number in the range [0, 512), and generates a mask spectrum that can be and-ed with the output of the \texttt{pfb} module to leave only the selected three channels with a non-zero value.

\subsubsection{Distribution and Reorder (\texttt{dist})}

This block performs two independent tasks-- (a) it masks the output of the \texttt{pfb} module with the template produced by \texttt{chan} to select only three frequency channel that need to be written to disk. Based on the \texttt{chan} module template, it also generates valid and end-of-file signals for the 10\,GbE module within the \texttt{xmit} block. (b) It vector accumulates the full spectrum of any one of the 12 antennas, based on user-input, and writes the output to a BRAM called \texttt{spectrum}. This functionality, which is completely independent of the selection and distribution of frequency channels to UDP packets, should ideally be designed as module of its own and labelled something meaningful like auto-correlation.

\subsubsection{Sync-Pulse Generator (\texttt{sync\_gen})}

This block generates the sync pulse which forms the heart-beat of the FPGA design. The sync pulse periodically synchronizes all the major blocks in the design and ensures that they are operating at the same pace. The period of the sync pulse is set to $9! \times 2^{10}$ to account for the periodicities of various blocks in the design.

\subsubsection{Packetize and Transmit (\texttt{xmit})}

This block primarily wraps a CASPER-module that is an interface to the 10\,Gb Ethernet port on the SNAP. The \texttt{dist} block generates the data, valid and end-of-file signals that drive the UDP packet transmission from this block and subsequently from the 10\,GbE port. The user can set the destination server IP address and port for sending the UDP packets via 32-bit software registers that are programmable.

\paragraph{}
In summary, the channel selector is an FPGA-based system that can digitize, channelize and transmit $3/512$ fraction of the input bandwidth. The output data rate of the entire system is 450\,Mbps. Each UDP packet output by the FPGA is 8072 bytes wide, with a 64 bit header (continuously incrementing counter) and 8064 bytes of data. The data consists of 3 frequency channels for each of 12 antennas and 56 time samples. The amplitude within a single frequency channel of one antenna is represented as a 32-bit complex number with a 16\_15 fixed point number representing each of the real and imaginary parts. Software to automate the observing runs and program the FPGA design has been written in Python and is publicly available\footnote{\url{https://github.com/dgorthi/SA-Data}}. 

\section{File Writing Pipeline}
\label{chp:Swcorr:sec:Xengine}

A single 1U server is used to collect the UDP packets output by the FPGA. The server contains two 10\,GbE Network Interface Cards (NICs) that can be connected directly to the Ethernet ports on the SNAP board. The server is also equipped with two 6\,TB hard drives to store the data collected during the observation run. A continuously running program\footnote{\url{https://github.com/dgorthi/acq_data}} on this server, written in the C language, is used to create and write the data files. 

Voltage data from antennas cannot be integrated in real-time because, unlike intensities, they represent the electric field observed by the antenna and cannot be scalar averaged. This implies that the data write speed has to keep up with the 56.25\,MBps output by the FPGA. Fortunately, most disk write speeds at the time of designing this system, were capable of writing data at 70\,MBps or higher which avoided us the hassle of setting up a disk RAID system where fractions of data are written to multiple disks to boost the write speed.

The program to collect UDP packets and write them to files is a multi-threaded pipeline that uses a shared ring buffer system to buffer the data in UDP packets before writing to disk. This is necessary because the UDP protocol for communication over Ethernet does not guarantee packet arrival order i.e., the order in which packets are sent and received could be different. This requires any UDP packet receiving application to re-order the packets based on the contents of the packet. To enable this ordering, every UDP packet sent by the FPGA is stamped with a unique, continuously incrementing, 64-bit header.

The program is also multi-threaded to enable two parallel processes for packet collection and data writing. It uses Open MP threads for parallelization since they are easier to implement than POSIX threads. Open MP threads are task based, easier to implement when fine control over the threads is not necessary and useful for scaling a given code. For instance, this program written in Open MP can be easily re-used to serve more Ethernet ports if two or more SNAP boards need to be setup up for the same experiment. Moreover, Open MP threads are portable and supported by multiple compilers across various platforms making the code platform independent.

The next few subsections describe the setup and layout of this program, the data structures used to format the shared ring buffers, how various packet statistics are monitored and the format of the files where the data is finally written.

\subsubsection{Shared Ring Buffer System}

Three shared memory segments labelled \texttt{curr}, \texttt{next}, \texttt{copy} are used to buffer the incoming UDP packets. Each buffer can hold the data from \texttt{NACC} packets, the default of which is set to 8192 packets. The buffers can be made larger or smaller by changing this macro at compile-time. Based on the 64-bit unique time stamp of a given UDP packet, it is placed into either the \texttt{curr} or \texttt{next} buffer. This system allows a UDP packet to be late by 16384 packets, or at 450\,Mbps, a little more than 36\,ms. If a UDP packet arrives later than this, it is discarded and marked as a packet lost.

When the \texttt{curr} buffer gets filled or the \texttt{next} buffer get half filled, the three labels are shifted cyclically so that the partially filled buffer becomes \texttt{curr}, the empty \texttt{copy} buffer become \texttt{next} and the filled \texttt{curr} buffer is now labelled \texttt{copy}. At this stage, packet statistics are computed, recorded to a log file and a parallel thread that has been waiting for a filled \texttt{copy} buffer is kicked into action. 

\paragraph{}
Three data structures (\texttt{struct}s in the C language) are used to keep track of various variables in this program. 

\paragraph{\texttt{SockPar}}
This structure contains all the variables that a thread requires to collect data from the Ethernet link and arrange it in time order. The variables stored by this struct are:

\begin{enumerate}
    \item fd: The socket file descriptor that enables connection to an Ethernet port.
    \item \texttt{sockaddr\_in} addr: The address of the SNAP board that is sending the UDP packets.
    \item ubuf: A temporary array to hold the UDP packet data before it is copied into the ring buffer.
    \item sbuf: A array of structs of type \texttt{SockBuf} which represent the ring buffer. The number of buffers in the ring are fixed to three (for \texttt{curr}, \texttt{next} and \texttt{copy}) but can by changed at compile-time.
    \item \texttt{curr}, \texttt{next}, \texttt{copy}: Shortcut pointers to the three buffers in sbuf.
    \item switch\_thresh: The threshold value that determines how long the program has to wait for \texttt{curr} to get filled. When this parameter times out, \texttt{curr} is emptied with the unreceived packets flagged.
    \item stat: This variable points to a struct of type SockStat, that is used to keep track of packet statistics like the number of packets received, lost and discarded. These stats are periodically printed to a log file for maintenance and debugging.
\end{enumerate}

\paragraph{\texttt{SockBuf}}
\noindent
This structure defines the memory format of each of the three \texttt{curr}, \texttt{next} and \texttt{copy} buffers. The contents of this structure are:

\begin{enumerate}
    \item data: A pointer to the data from UDP packets that is copied into the ring buffer. This shared memory is dynamically allocated at run-time, and can hold NACC number of UDP packets.
    \item flag: A pointer to the metadata of each buffer which contains information about what packets have been received and what packets have been lost. It holds NACC boolean values which are 0/1 corresponding to lost/received packets.
    \item start: Timestamp of the first packet in the buffer.
    \item stop: Timestamp of the last packet in the buffer. Packets received with a greater timestamp are passed on to the next buffer.
    \item idx: An index is assigned to each buffer in the ring to keep track of the buffer that has last been copied. This is one of 0/1/2 for each buffer.
    \item count: Counter to keep track of number of packets received in this buffer.
\end{enumerate}

\paragraph{\texttt{SockStat}}
\noindent
This structure keeps track of the statistics of packets received like the number of packets lost, discarded, copied, out-of-time etc. It computes both cumulative and differential stats i.e., stats from the beginning of program execution and individual stats for each buffer change. Variables that start with `d' hold the differential values, which are reset at every call to report statistics. The members of this structure are:

\begin{enumerate}
    \item start, dstart: These variables, of type \texttt{timeval}, hold the time at which the first packet was received.
    \item total, dtotal: Total number of packets sent by the FPGA, i.e. the total of received, lost and bad packets.
    \item got, dgot: Number of packets received at the right time i.e., packets whose data could be copied into \texttt{curr} or \texttt{next} buffers.
    \item bad, dbad: Packets that were received but could not be copied into the right location, so the packet either arrived too late or too early. NOTE: Too early should really not have been a red flag. In this case the buffers should update irrespective of threshold and filled state of \texttt{curr}.
    \item rate, drate: Estimate of the data rate in MBps.
    \item log\_rate: Frequency at which the statistics are being written to the log file.
\end{enumerate}

\subsubsection{Classes and Methods}

The code is broadly divided into two parts that are run parallelly: data collection from the Ethernet port (handled by function \texttt{acquire\_socket\_data}) and copying data into files (handled by function \texttt{transfer\_socket\_data}). Other functions support these two operations by setting up the environment and managing other variables. The tasks of the different functions in the code are outlined below.

\paragraph{\texttt{acquire\_socket\_data}}
\noindent
The function receives data from the socket and populates NACC number of UDP packets into the circular ring buffer. It checks the timestamp of the UDP packet, discards the 64-bit header and copies the data into the right location in the ring buffer. 

\paragraph{\texttt{init\_socket}}
\noindent
This function initializes the socket parameters, establishes a connection at the specified Ethernet port and allocates memory required to hold the ring buffers. It also initializes the \texttt{SockPar} struct variables for computing packet statistics. It is called once at the beginning of program execution.

\paragraph{\texttt{init\_sock\_stat}}
\noindent
This function sets-up the variables in struct \texttt{SockPar} and initializes them once at the beginning of program execution. 

\paragraph{\texttt{report\_sock\_stat}}
\noindent
This function is called every time NACC number of packets have been received by the program. It updates the variables in \texttt{SockPar} and prints the stats to a log file.

\paragraph{\texttt{transfer\_socket\_data}}
\noindent
This function polls for a filled \texttt{copy} buffer and writes data to disk when a buffer becomes available. It opens a new binary data file every 512MB of received packets and writes a metadata header and binary data to the file until it is filled. Once the file is 512MB large, corresponding to 8 completely filled buffers, the file is closed and a new file is opened.

\paragraph{}
In summary, the CPU-based file writing pipeline collects UDP packets that are sent out by the SNAP board and time-orders them using a shared ring-buffer system. The code is written to serve a general-purpose use of writing data to disk, at file write speeds up to 1\,Gbps. The files written are in raw binary format and require unpacking and reformatting before calibration or imaging.

\section{Observing Runs}
\label{chp:Swcorr:sec:observing}

\begin{figure}
    \centering
    \includegraphics{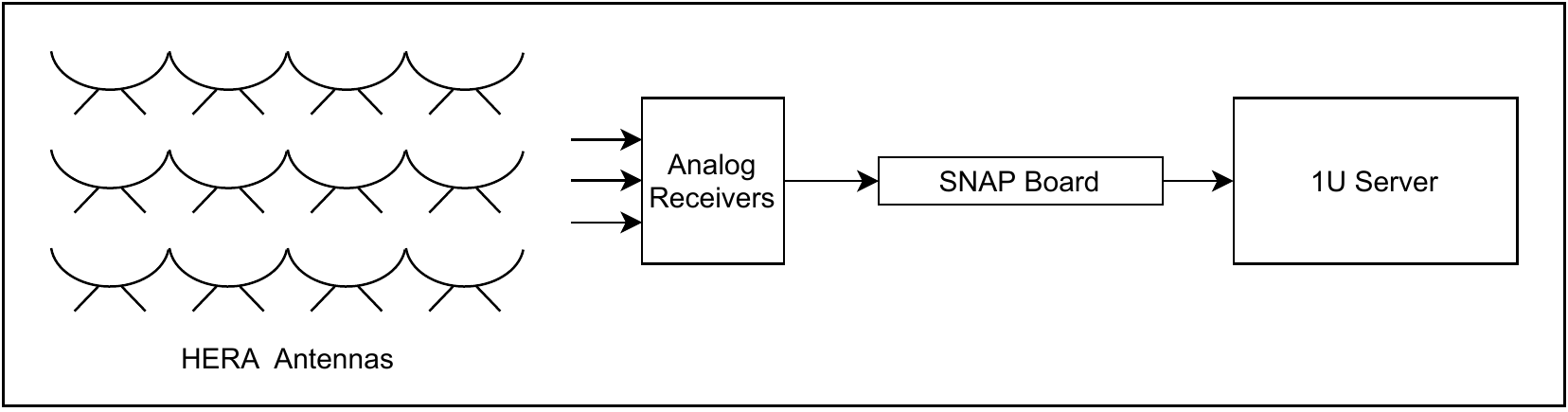}
    \caption{Setup of the experiment to record raw antenna voltages from 12 HERA antennas in a redundant layout. The voltages are recorded only in 3 of 512 frequency channels to keep the data rate and storage requirements small. The antenna voltages form the input to both the software FFT correlator and calibration algorithms that can be for an FFT correlator.}
    \label{fig:volt_recorder_setup}
\end{figure}

This experiment was performed in two locations: at the site of the Green Bank Telescope in West Virginia, where there were a few experimental antennas from the PAPER experiment, and on HERA Phase-I antennas in South Africa. The experimental setup at both these locations was exactly the same, and shown in Figure~\ref{fig:volt_recorder_setup} for the setup in South Africa.

The PAPER antennas at the site of the Green Bank Telescope were not in use for a couple of years so it was difficult to bring them to a functional state for the experiment. The RF cables that connected the dipole feeds to the receiver were matted by a thick layer of grass and it took a powerful weed-cutter to dig them out of the lush green undergrowth in the summer of West Virginia. The small wooden hut that was constructed to house a few receivers, analog electronics and servers for the PAPER antennas was locked and it took some bureaucracy-fu to get the keys to it. Unfortunately, linear power supplies and connectors to power and run the receivers were all missing from the hut and we needed the help of a GBT operator to locate spare supplies and had to re-install them ourselves. It was a huge lesson in field-deployments and patience to get the entire system working. After three days of hard effort we were able to get 4 antennas working and laid out in a straight line. We took 6 hours of data while the Sun was up and got good data with fringes showing up where expected.

The experiment with HERA antennas, a year later in South Africa, was equally challenging to setup. Due to a miscommunication between me, my advisor, the project manager and the team on site, I was uninformed that all the HERA antennas were in a transition state between HERA Phase-I and Phase-II. The feeds, receiver system, bandwidth of observation and cabling were all different between the two phases. By the time I reached the site, most of the antenna feeds had been dismantled and cabling had been removed in preparation for setting up Phase-II of the array. The site of the container (a modified RV), that held the electronics and servers of the previous correlator system, had also been changed so, unfortunately, rewiring the antennas was not a trivial process.

\begin{figure}
    \centering
    \includegraphics[width=0.9\linewidth]{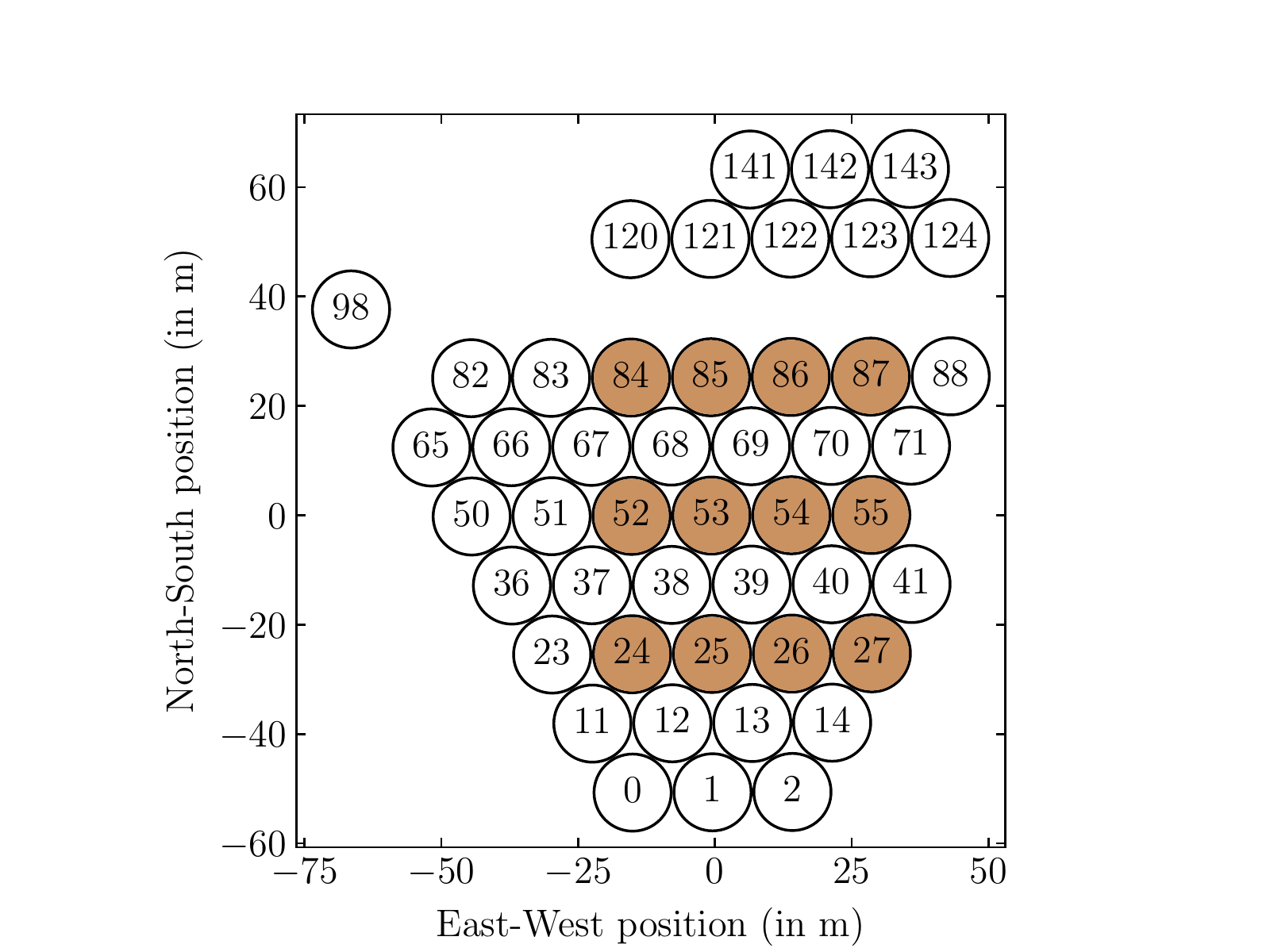}
    \caption{Position of antennas in HERA Phase-I layout. The antennas that are considered for the software correlator experiment are shaded in brown.}
    \label{fig:volt_recorder_layout}
\end{figure}

With the help of some on-site crew members, I moved cables for 12 antennas from the previous location of the container to the new location. Choosing which of the 52 antennas, constructed by that point, to include in my experiment was a tricky decision because the length of the cables available was fixed by the previous location of the container. The new location of the container was slightly further away, making the cable length too small for some antennas at the other extreme end. Moreover, for the purpose of testing various calibration schemes I required antennas in a redundant configuration. Finally, after many considerations and trials, the 12 antennas shown in Figure~\ref{fig:volt_recorder_layout} were selected for the experiment. 

\begin{figure}
    \centering
    \includegraphics[width=0.94\linewidth]{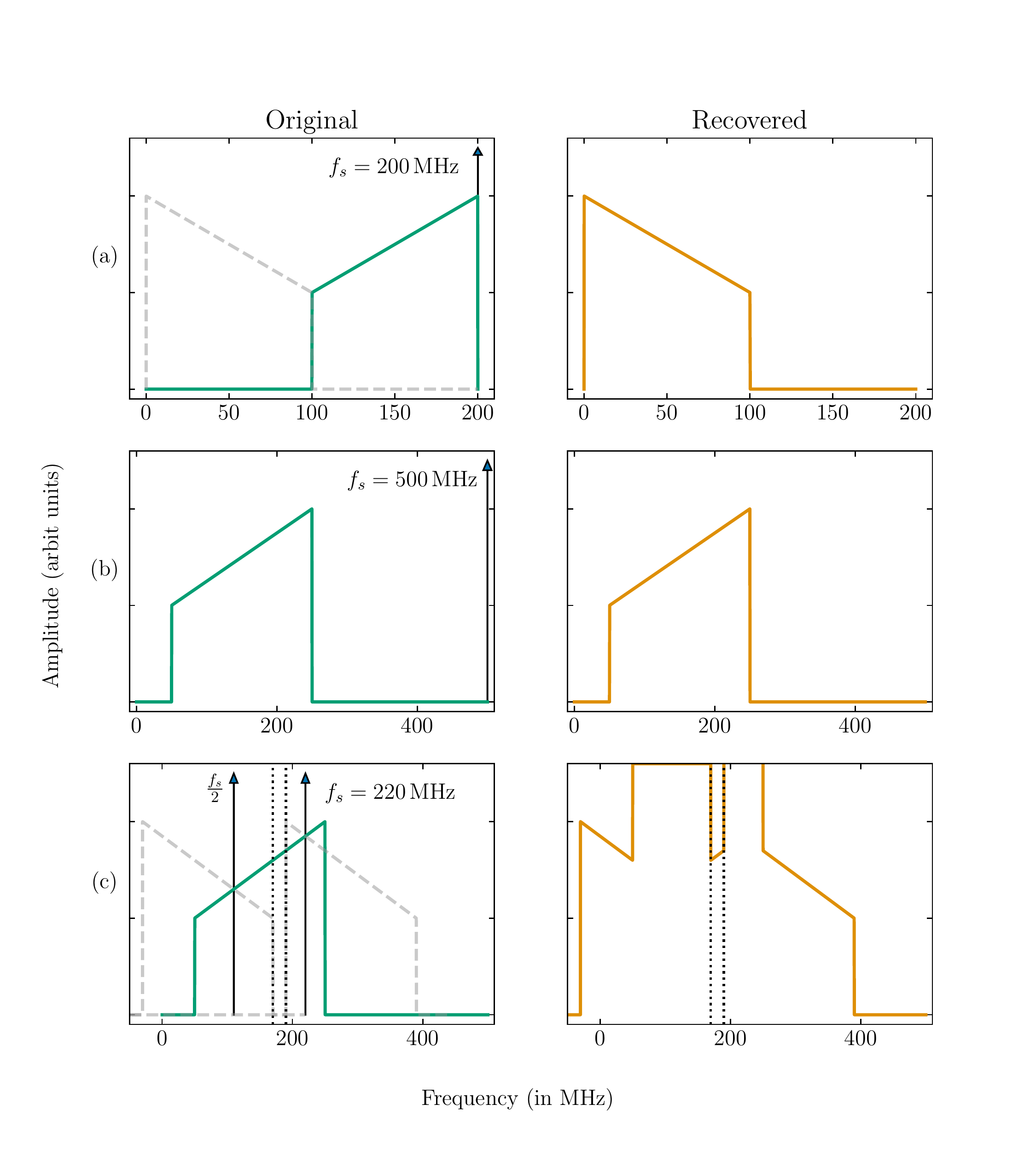}
    \caption{Bandpass that is recovered for a given sampling frequency, for HERA Phase-I and Phase-II receivers. Panels (a) show HERA Phase-I receivers with pass band between 100--200\,MHz. At a sampling frequency of 200\,MHz, this pass band can be recovered in the first Nyquist Zone. Panels (b) show Nyquist sampling of HERA Phase-II receivers. Panels (c) shows the aliasing introduced by sampling at 220\,MHz. Despite the top and bottom portions of the band being aliased, the region between 170--190\,MHz can be recovered without aliasing.}
    \label{fig:volt_recorder_aliasing}
\end{figure}

Despite the effort spent in getting the antennas working, unfortunately, the change in the bandwidth of the new receivers meant that a sampling frequency of 200\,MHz, as originally planned, would not be sufficient. The bandwidth of the receivers for HERA Phase-II is 50--250\,MHz, double that of HERA Phase-I which is 100--200\,MHz. The frequency generator that was available on-site could only synthesize up to 250\,MHz which meant that the Phase-II receivers could not be sampled without aliasing. HERA Phase-I receivers could be sampled in the second Nyquist window without aliasing at 200\,MHz.

The software FFT correlator requires only 3 frequency channels of unaliased data. Instead of abandoning the experiment at this stage, I decided to take data by intentionally aliasing only a portion of the bandwidth. At a sampling frequency of 220\,MHz, a portion of the bandwidth between 170--190\,MHz would fall under the second Nyquist window of 0-50\,MHz and and could be recovered (shown pictorially in Figure~\ref{fig:volt_recorder_aliasing}). I picked three channels in this narrow 20\,MHz band and collected data for 12 hours over two nights.

\section{Analysis and Results}
\label{chp:Swcorr:sec:Analysis}

\subsection{Data from Green Bank}
The four PAPER antennas, at the site of the Green Bank Telescope, were laid out in a single straight line in the East-West direction, with 45\,feet separation. The layout and spacing were chosen fairly arbitrarily, after taking into account the frequency of observation and the area of open space at the site. Data was taken for nearly three hours between 1.30\,pm and 5\,pm in local time (EST) while the Sun was up in the sky.

\begin{figure}
    \centering
    \includegraphics[width=\linewidth]{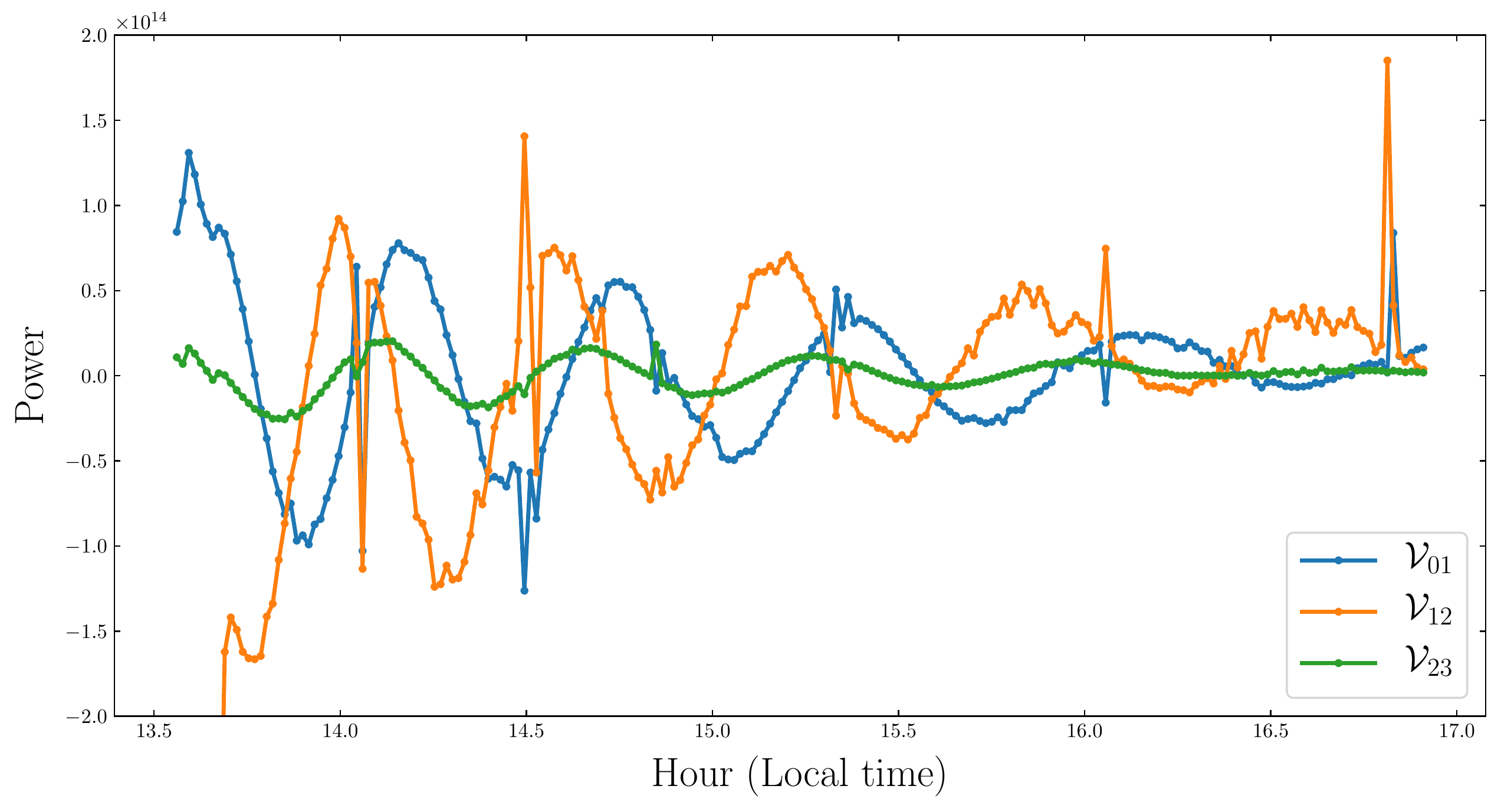}
    \caption{Uncalibrated raw visibilities measured by three antenna pairs with East-West baselines of length 45\,feet. The label $\mathcal{V}_{ij}$ indicates that the visibility function was constructed by cross-correlating antennas ($i,j$). These three visibilities measured by redundant baselines are expected to be the same but are different in practice due to differences in the signal chain.}
    \label{fig:GBT_uncalibrated}
\end{figure}

\begin{figure}
    \centering
    \includegraphics[width=\linewidth]{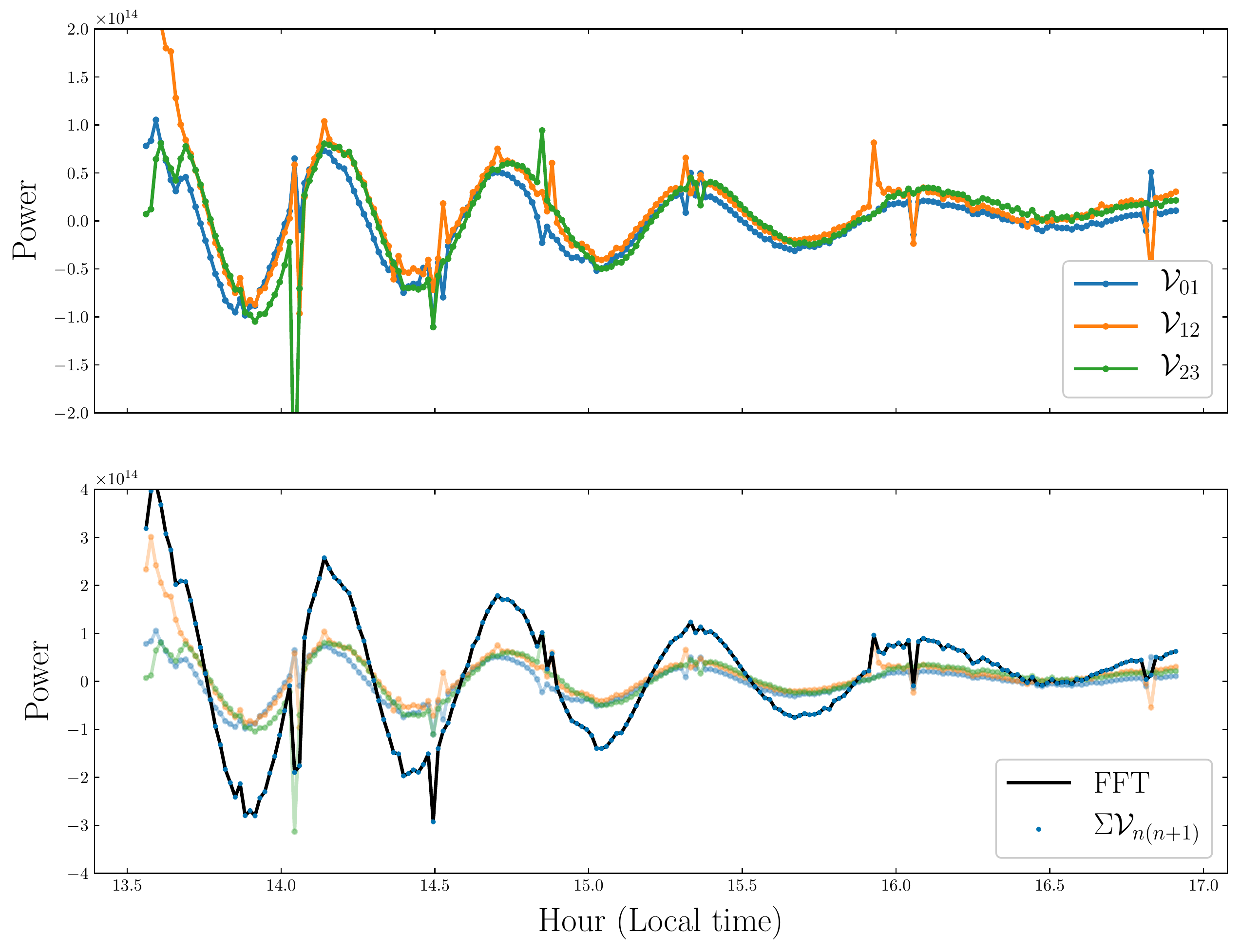}
    \caption{Top panel: Calibrated visibilities of three redundant baselines, measured using PAPER antennas, with time-independent calibration parameters applied to the measured values. Bottom panel: Comparison between the sum of the calibrated visibilities and the output of the software FFT correlator, when applied to calibrated antenna voltages. As expected from the convolution theorem, the two quantities match exactly.}
    \label{fig:GBT_const_calibrated}
\end{figure}

\begin{figure}
    \centering
    \includegraphics[width=\linewidth]{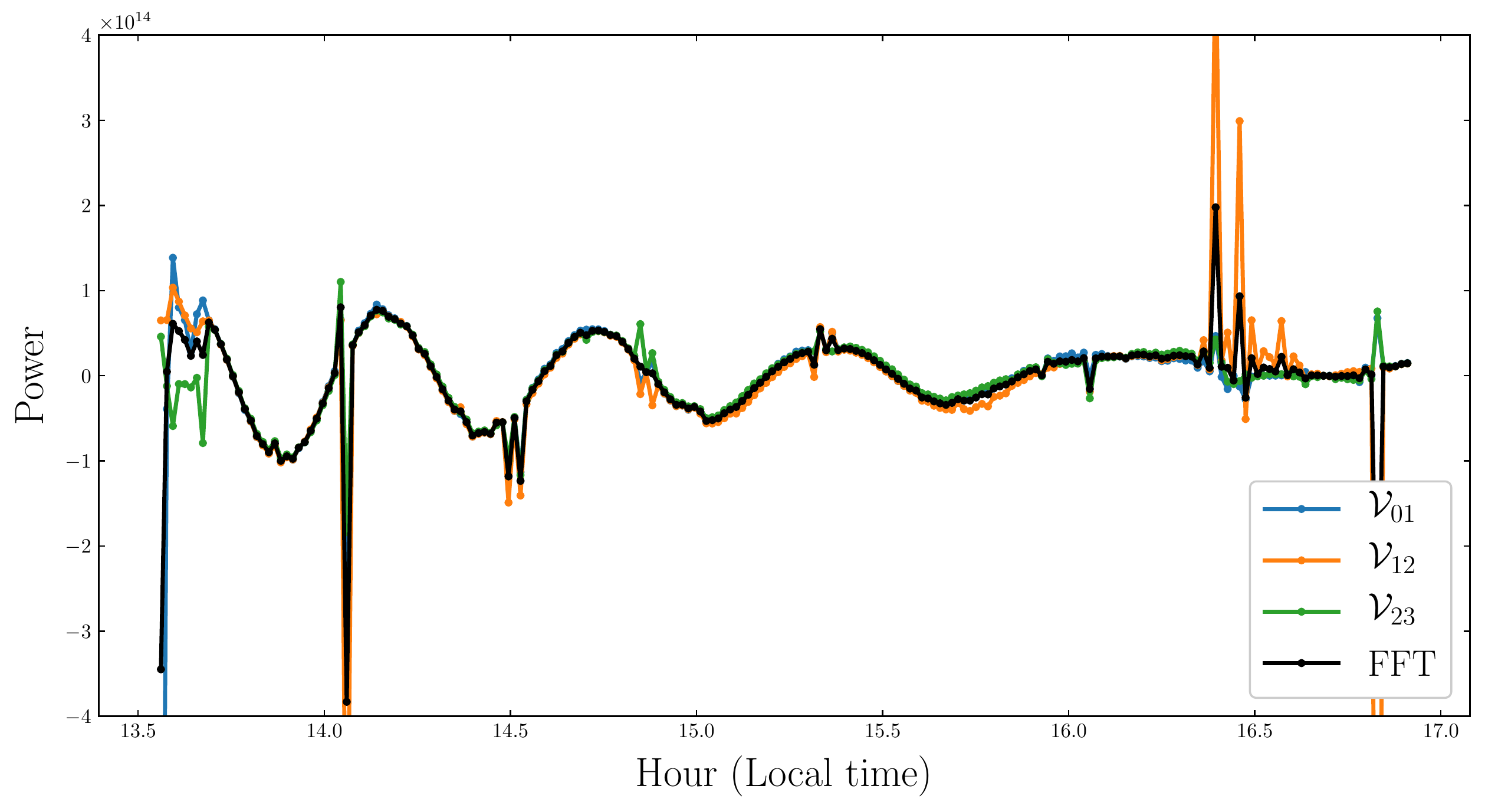}
    \caption{Time-dependent redundant-baseline calibration applied to the visibilities measured by PAPER antennas. Calibrating every integration period results in lower baseline-to-baseline variance, or higher redundancy, and hence reduces signal-loss in the FFT correlator.}
    \label{fig:GBT_timedep_calibrated}
\end{figure}

The Sun is a bright radio source, overshadowing other sources in the sky, and producing an interferometric response that is inline with a single bright point source in the beam of the telescope. The cross-correlation product of any two antennas, as a function of time, is a fringe pattern that has an amplitude which is dependent on the location of the point source in the sky. The fringe pattern, also called the visibility function, is given by:

\begin{equation}
\label{eq:visib}
    \mathcal{V}(u,v) = \iint B(l,m) \; \mathcal{I}(l,m) e^{-2\pi i (ul+vm)} \mathrm{d}l \mathrm{d}m
\end{equation}
\noindent
where $B$ is the antenna beam pattern and $\mathcal{I}$ is the intensity of the source(s) in the sky. When the Sun is directly overhead ($l=0, m=0$), the antenna beam pattern has the highest amplitude and correspondingly, the visibility function has the highest amplitude. As the Sun moves from the zenith to the horizon, the beam pattern tapers in amplitude, resulting in a decrease in amplitude in the visibility function. This change in amplitude with time, or location of the Sun in the sky, is evident in Figures~\ref{fig:GBT_uncalibrated}, \ref{fig:GBT_const_calibrated} and~\ref{fig:GBT_timedep_calibrated}.

The exponential term in the integral of Equation~\ref{eq:visib} creates sinusoidal fluctuations in the visibility function, called the \textit{fringe frequency}. The fringe frequency depends on the derivative of the time delay between the two antennas, and is predominantly dependent on the East-West component of the baseline when the source is at zenith and on the North-South component of the baseline when the source is closer to the horizon. The PAPER antennas used in this experiment were placed very close to an East-West spacing of 45\,feet. Hence the fringe frequency is highest when the Sun is at zenith, at the beginning of the experiment, and decreases as the Sun sets.

Figure~\ref{fig:GBT_uncalibrated} shows the cross-correlation products of antenna voltages, for the duration of the observation, for the three shortest baselines in the layout. The three baselines have the same East-West baseline vector of length 45\,feet and should, theoretically, measure the same visibility function. In practice, however, differences in the signal chain like the amplification produced by the low-noise amplifier, the length of the cable connecting the feed to the receiver, etc. create amplitude and phase differences in the measured visibilities. This is the reason why the raw-visibilities, prior to calibration, look extremely non-redundant. FFT correlation on these uncalibrated visibilities would result in signal-loss.

Figure~\ref{fig:GBT_const_calibrated} shows the result of applying constant, time-independent calibration parameters to the visibilities measured by the redundant baselines. The initial calibration parameters were set by-hand to match the phase of the visibilities and fine-tuned by iterating over these initial values. As evident from the figure, the redundant visibilities are closer in amplitude and phase after calibration. 

The output of the FFT correlator, for a particular baseline vector, is equivalent to the average of visibilities measured by all the antenna pairs separated by that baseline vector. This can be derived from the convolution theorem, applied to the layout of antennas~\citep{Tegmark_and_Zaldarriaga_2009, Tegmark_and_Zaldarriaga_2010}.

Mathematically, this can be represented as:
\begin{equation}
    \mathcal{V}^{\mathrm{FFT}}_{\alpha} = \frac{1}{N_{\alpha}} \sum_{(i,j) \in \alpha} \mathcal{V}_{ij}
\end{equation}
\noindent
where $\alpha$ represents a given baseline vector, $N_{\alpha}$ is the number of antenna pairs separated by that baseline vector, $\mathcal{V}^{\mathrm{FFT}}_{\alpha}$ is the output of the FFT correlator for that baseline vector and $\mathcal{V}_{ij}$ are the cross-correlation products of individual antenna pairs separated by that baseline vector. The bottom panel of Figure~\ref{fig:GBT_const_calibrated} shows this equivalence in the output of the FFT correlator and the sum of redundant visibilities for the three shortest baselines in the layout of this experiment. 

For experiments that are sensitive to the signal recovered, like detecting and characterizing the EoR power spectrum, constant time-independent calibration is not sufficient for recovering the original signal. Figure~\ref{fig:GBT_timedep_calibrated} shows the result of applying time-dependent redundant-baseline calibration to each data point i.e., to each integration bin of the visibility function. This produces a closer match between the three redundant visibilities, and in this case, the FFT correlator output matches the average visibility measured by these redundant baselines.

The experiment with PAPER antennas at the Green Bank Telescope was an exercise in understanding the operation of FFT correlators and the importance of time-dependent calibration. In this case, redundant-baseline calibration using the full visibility matrix was applied to the data for time-dependent calibration. However, the aim of constructing this experiment with 12 antennas was to test the applicability of calibration schemes that either use only a subset of the full visibility matrix or perform calibration less frequently. The data we collected using PAPER antennas was insufficient to test either of these calibration ideas.

\begin{figure}
    \centering
    \includegraphics[width=\linewidth]{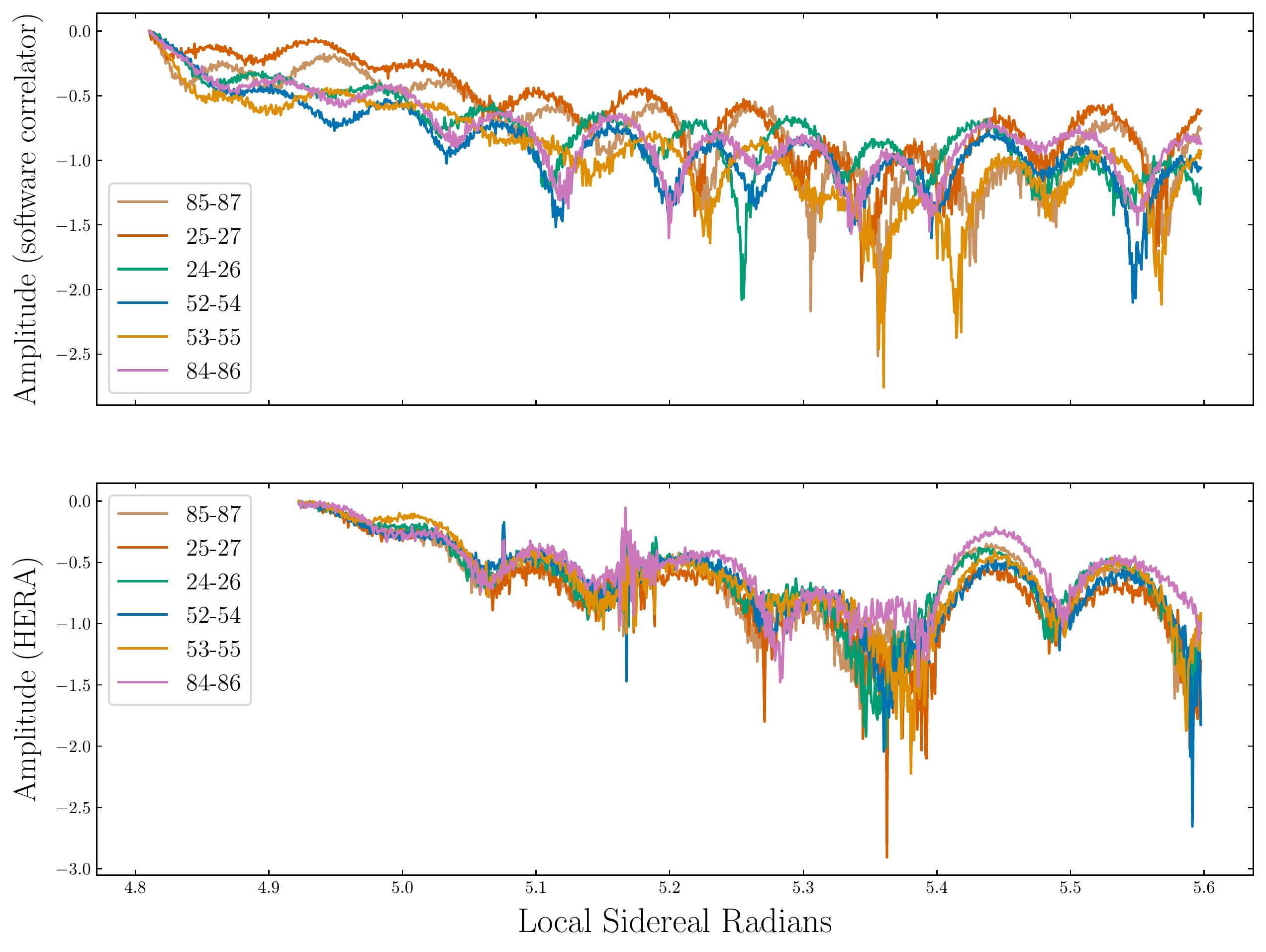}
    \caption{Top panel: Data taken using the custom setup for recording antenna voltages and cross-correlated on a software platform. The visibilities of these redundant baselines are expected to be similar but show non-redundancy. Bottom panel: Data taken using the HERA correlator system for the same set of antennas, on a different night, show redundancy.}
    \label{fig:SA_Data}
\end{figure}

\subsection{Data from South Africa}

In experiment with HERA antennas in South Africa, data was taken in three frequency channels within the range 170--190\,MHz at a sampling frequency of 220\,MHz which intentionally created aliasing in the bandwidth of the Phase-II receivers. Data was collected from the 12 antennas highlighted in Figure~\ref{fig:volt_recorder_layout}. On cross-correlating the data taken over nearly 9\,hours, redundant baselines were found to not show redundancy. 

The top panel of Figure~\ref{fig:SA_Data} shows the visibilities computed for the six East-West baselines with a single antenna separation. The visibilities of all these baselines are expected to similar in absolute amplitude $\left|\mathcal{V}_{i-j}\right|$, i.e. the fringe peaks and nulls are expected to match in time for the different baselines. Calibration, either time-dependent or constant, can only change the relative amplitude of the visibilities with respect to each other but cannot change the location of the null points in the fringes. Note that in Figures~\ref{fig:GBT_uncalibrated}, \ref{fig:GBT_const_calibrated} and~\ref{fig:GBT_timedep_calibrated} the y-axis shows the real-part of the visibility function $\mathcal{V}_{i-j}$ and not its absolute value. In this case, phase calibration could change the relative location of the fringe peaks and nulls.

The bottom panel of Figure~\ref{fig:SA_Data} shows the visibilities for the same set of baselines, computed by the HERA hardware correlator system on a different night. This data has been computed at the same frequency as the top panel, with the same frequency channel resolution. Since the HERA correlator system has four times the number of channels as the voltage recorder system, over the same bandwidth, this plot was generated by averaging the visibility function over four channels of the HERA correlator data. If the non-redundancy shown by the voltage recorder system were true non-redundancy, this would have been reflected in the HERA correlator data as well. However, this data shows the expected redundancy with nulls and peaks matching in time for all the redundant baselines.

This discrepancy between the data recorded by the HERA correlator system and the voltage recorder system exists in all the baselines between the 12 antennas. Moreover, it was found that the antennas involved in the baselines that are not matching are inconsistent between redundant sets. After much speculating and probing, the apparent non-redundancy in the voltage recorder system has been attributed to aliasing within the tail-ends of the bandpass filter, which can potentially corrupt the 170--190\,MHz region of the band.

Both the experiments, at Green Bank and South Africa, unfortunately, did not have an expected outcome. In both cases, the data recorded was not useful for testing calibration schemes that can work with an FFT correlator system. However, the experience of building an instrument and deploying it in the field was an extremely valuable one for me. It taught me to prepare for unexpected failure modes, that rigorous testing in the lab could save valuable time on site, and to take experimental failures lightheartedly.

% Reduced Redundant-Baseline Calibration
\chapter{Reduced Redundant-Baseline Calibration}
\label{chp:Redredcal}

In Section~\ref{chp:Intro:sec:FFTCorr}, I laid out the motivation for building FFT correlators for large-N redundant arrays. While this correlator architecture is an attractive solution to decreasing the cost scaling of the entire telescope, there are some practical considerations that need to be taken care of, like calibration of antenna elements prior to correlation. In this chapter, I will attempt to show that the gains estimated using either of two reduced redundant-baseline calibration schemes, low-cadence calibration or subset redundant calibration, are capable of minimising the scatter in the visibilities of redundant baselines. I lay out metrics for comparing the two reduced redundant-baseline calibration schemes and assessing their scalability to large-N arrays. Ultimately, I attempt to show that the calibrator design proposed in Section~\ref{chp:Intro:sec:FFTCorr:subsec:Calibration} makes self-contained $\mathcal{O}(N\log{N})$ correlators conceivable for future generation large-N arrays.

In the first reduced redundant-baseline calibration scheme, \textbf{low-cadence calibration}, the calibrator computes the full visibility matrix by cycling through baseline pairs. As an extreme example, if the computational resources allocated to the calibrator can only cross-correlate one antenna pair at a time, the low-cadence calibrator will generate the full visibility matrix by cycling through all the antenna pairs in the array. The full visibility matrix, constructed in this fashion, is then used for redundant-baseline calibration. While all the $\sim$$N^2/2$ visibilities are used for redundant-baseline calibration, they are computed by using only a small fraction of the resources of an $O(N^2)$ FX-correlator. By adjusting the integration time and the number of visibilities computed within each cycle, the computational resources required by a low-cadence calibrator can be limited to an $\mathcal{O}(N\log{N})$ scaling.

The second reduced redundant-baseline calibration scheme, \textbf{subset redundant calibration}\footnote{Shortened from subset redundant-baseline calibration}, is a generalisation of hierarchical redundant-baseline calibration described by \citet{Zheng_et_al_2014}. In subset redundant calibration, the calibrator computes a partial visibility matrix by cross-correlating only a limited set of antenna pairs. Redundant-baseline calibration is applied to this partial visibility matrix to estimate the antenna gains. For example, in highly redundant arrays, since the shortest baselines involve all the antennas in the array, it is often possible to compute antenna gains by performing redundant-baseline calibration on just the shortest baselines. Since only a subset of the full visibility matrix is generated for the purpose of calibration, this technique is called subset redundant calibration. Depending on the baseline-types chosen for redundant-baseline calibration, the computational resources required by such a calibrator can also be limited to an $\mathcal{O}(N\log{N})$ scaling.

\section{Metrics for Evaluating Calibration Methods}

Redundant-baseline calibration computes per-antenna complex gains by minimising the scatter in the visibilities of redundant baselines, which makes it a suitable calibration scheme for FFT-correlators. It relies on the fact that pairs of antennas with the same beam patterns, spaced at equal distances, measure the same visibility. If $V_{ij}$ is the visibility product of two antennas spaced a distance $d$ apart in the East-West direction, then the visibility measured by two different antennas, $V_{lm}$, is the same as $V_{ij}$ if they are also spaced a distance $d$ apart in the same direction. In practice, this is often not true because of variations in amplifier gain, timing differences originating in the correlator, cable delays etc., that need to be calibrated. By comparing visibilities that are theoretically identical, it is possible to infer the calibration parameters for the antennas involved.

In the case of highly redundant arrays such as HERA, PAPER, CHIME, and the MWA Phase-II hexes, there are many more visibility measurements than unique baselines. This allows one to build a system of equations, which can be solved to estimate all the antenna calibration parameters. For the array layout shown in Figure~\ref{fig:hexarray}, the system of equations can be constructed as\footnote{In general, all the measurements and variables in this system of equations have a time and frequency dependence. We have omitted writing this explicitly for notation convenience.}:

\begin{align}
    \label{eq:redcal}
    V_{01}^{\mathrm{meas}} &= g_0 g_1^* V_{\alpha}^{\mathrm{true}} + n_{01} \nonumber \\
    V_{12}^{\mathrm{meas}} &= g_1 g_2^* V_{\alpha}^{\mathrm{true}} + n_{12} \nonumber \\
&\vdots \quad (\text{baselines with }\mathbf{ d_{\alpha}} = \mathbf{r_0} - \mathbf{r_1}) \nonumber \\
    V_{04}^{\mathrm{meas}} &= g_0 g_4^* V_{\beta}^{\mathrm{true}}  + n_{04} \nonumber \\
&\vdots \quad (\text{baselines with }\mathbf{d_{\beta}} = \mathbf{r_0} - \mathbf{r_4}) \nonumber \\
    V_{05}^{\mathrm{meas}} &= g_0 g_5^* V_{\gamma}^{\mathrm{true}} + n_{05} \nonumber \\
&\vdots \quad (\text{baselines with }\mathbf{d_{\gamma}} = \mathbf{r_0} - \mathbf{r_5}) \nonumber \\
    V_{02}^{\mathrm{meas}} &= g_0 g_2^* V_{\delta}^{\mathrm{true}} + n_{02} \nonumber \\
&\vdots \nonumber \\ 
\text{baseli}&\text{nes separated by one or more antennas}
\end{align}
\noindent
where $V_{\alpha}^{\mathrm{true}}$ is the unknown, model true visibility of all the baselines with a displacement vector $\mathbf{d_{\alpha}}$, $V_{\beta}^{\mathrm{true}}$ is the unknown model true visibility of baselines with displacement vector $\mathbf{d_{\beta}}$ and so on. Baselines with the same displacement vector are said to be of the same baseline-type. $V_{ij}^{\mathrm{meas}}$ is the visibility measured by the pair of antennas $(i,j)$ in the field, and $n_{ij}$ is the noise in that measurement. The per-antenna, complex gains denoted by $g_i$ represent the calibration parameters of the antennas involved in measuring that visibility. 

The redundant-baseline calibration process estimates the gains and models true visibilities that best describe the measured visibilities. When the full visibility matrix $\mathbf{V^{full}}$ is used for the set of measured visibilities $V_{ij}^{\mathrm{meas}}$, the $V_{\alpha}^{\mathrm{true}}$ returned by the redundant-baseline calibration process represents the minimum-scatter average visibility for that unique baseline-type.

The system of equations in Equation~\ref{eq:redcal} can also be built using the visibility matrix computed by the calibrator $\mathbf{V^{reduced}}$. In the case of low-cadence calibration, this set of visibilities may have a lower SNR than the full visibility matrix due to smaller integration times in the calibrator. In case of subset redundant calibration, this set of visibilities is smaller than the full visibility matrix (but sufficient to determine and over-constrain all the variables in the system of equations) due to fewer cross-correlations computed by the calibrator. The $V_{\alpha}$s estimated by either reduced redundant-baseline calibration schemes are discarded and only the antenna gains are used to calibrate voltages for the spatial Fourier transform. The redundant-baseline averaged visibilities for all unique baseline-types are computed by the FFT-correlator. 

For the purpose of this paper, it is useful to consider an intermediate hypothetical step in the spatial Fourier transform where the redundant baselines have not yet been averaged. At this stage, the visibilities in the FFT-correlator would be equivalent to the calibrated visibilities of an FX-correlator. The baseline averaging stage within the FFT-correlator, that generates $\mathbf{V^{unique}}$, can be written in terms of the full visibility matrix as:

\begin{equation}\label{eq:modelvis}
    V_{\alpha}^{\mathrm{unique}} = \frac{1}{N_{\alpha}}\sum_{(i,j) \in \alpha} \frac{V_{ij}^{\mathrm{full}}}{g_ig_j^*}
\end{equation}
\noindent
where $N_{\alpha}$ is the number of redundant baselines that contribute to that baseline-type. Since it is easier to quantify the effect of calibration on visibilities than on voltages, we use this equation to represent the process of calibration in the FFT-correlator.

Another difference between traditional redundant-baseline calibration and 
reduced redundant-baseline calibration, that is evident from Equation~\ref{eq:modelvis}, is that the latter estimates antenna gains from a different set of visibilities ($\mathbf{V^{reduced}}$) than what they are finally applied to ($\mathbf{V^{full}}$). In this section we discuss two metrics that will help in evaluating the effect of this: (a) the uncertainty in estimated antenna gains and (b) scatter in visibilities calibrated by a reduced redundant-baseline calibration process.

The uncertainty in antenna gains has to decrease or remain constant with increase in array size, for the reduced redundant-baseline calibration scheme, and consequently the calibrator design, to be scalable to large arrays. Moreover, the uncertainty in the estimated gains can be expressed in terms of the SNR of the measured visibilities and the number of baselines used in the calibration process, providing us with a convenient metric to directly compare low-cadence calibration and subset redundant calibration. The overall uncertainty in the redundant-baseline averaged calibrated visibilities comes from both the noise in the measured visibilities and the uncertainty in the estimated gains. A quantitative measure of the antenna gain uncertainty will help us estimate the contribution of gain errors to the overall error in the calibrated visibilities.

The scatter in visibilities post-calibration is a direct probe of the effectiveness of the reduced redundant-baseline calibration process in estimating gains that can calibrate antenna voltages for the FFT-correlation. The spatial Fourier transform averages the visibilities of redundant baselines (Equation~\ref{eq:modelvis}) converting any residual scatter into noise in the estimated visibilities. Hence, quantifying the post-calibration scatter will help  us evaluate the performance of either reduced redundant-baseline calibration scheme with respect to traditional redundant-baseline calibration.

An important mathematical detail, before delving into the metrics that assess reduced redundant-baseline calibration schemes, is that the system of equations represented by Equation~\ref{eq:redcal} is not linear. \citet{wieringa_1992} suggests a logarithmic approach to linearizing which can be written as: 

\begin{equation}\label{eq:logcal}
\ln{V_{ij}^{\mathrm{meas}}} = \ln{g_i} + \ln{g_j^*} + \ln{V^{\mathrm{true}}_{\alpha}} + n^{\prime}_{ij}
\end{equation}
\noindent
The noise parameter $n^{\prime}_{ij}$, evaluates to a non-Gaussian error that depends on the SNR of the measured visibilities. \citet{Liu_et_al_2010} discuss the noise-bias in antenna gains due to this non-Gaussianity, and propose another approach based on Taylor expanding the variables around a starting point. This paper employs a widely used third approach, called \texttt{omnical} \citep{Zheng_et_al_2014, Ali_et_al_2015, Li_et_al_2018, Dillon_et_al_2020}, that was originally developed for the MITEoR experiment. We use the logarithmic approach to make theoretical arguments about the nature of gain solutions since constant coefficients make the system of equations easier to analyse. However, simulations run to test these arguments and the plots shown in this paper have been generated by employing the \texttt{omnical} algorithm that is not noise-biased. In general, most of the results presented in this paper are not dependent on the solving technique used.

\subsection{Uncertainty in Antenna Gains}
\label{sec:redredcal:gainvar}

The uncertainty in the antenna gains, estimated by solving a system of equations, is given by the variance-covariance matrix (covariance matrix henceforth). The linearized system of equations shown in Equation~\ref{eq:logcal} can be written in matrix notation as $\matr{A}\matr{x} + \matr{n} = \matr{b}$ where $\matr{A}$ is a constant complex-valued matrix of dimensions $N_m \times N_v$ ($N_m$ is the number of measured visibilities and $N_v$ is the number of variables). $\matr{x}$, $\matr{b}$ and $\matr{n^{\prime}}$ are one-dimensional matrices of the variables (log-gains and log-unique-visibilities), measured quantities (log-visibilities) and noise in each measurement respectively. The covariance matrix $\matr{C}$ (of dimensions $N_v \times N_v$) for the estimated variables is given by:

\begin{equation}\label{eq:cov}
\matr{C} = \left<\matr{x}\matr{x}^{\dagger}\right> = \left(\matr{A}^{\dagger}\matr{N}^{-1}\matr{A}\right)^{-1}
\end{equation}
\noindent
The diagonal of the covariance matrix gives the variance in estimated variables, including antenna gains. The noise covariance matrix ($\matr{N} = \left<\matr{n} \matr{n}^{\dagger}\right>$), is a statistical estimate of the noise in the measurement matrix $\matr{b}$. The matrix $\matr{n}$ cannot be measured and can only be estimated from the thermal noise expected in the measurements.

The covariance matrix $\matr{C}$ reflects the covariance between all variables returned by the reduced redundant-baseline calibration process. Since the calibrator uses only the antenna gains and discards the model visibilities estimated by redundant-baseline calibration, the covariance matrix of interest is the \textit{marginalised} covariance ($\matr{C^{\prime}}$) of just the antenna gain solutions. Assuming all variables are normally distributed, the marginalised covariance matrix for a subset of variables is given by the rows and columns of the variables of interest. Hence, the marginalised covariance matrix of the gain solutions is given by the first $N$ rows and columns of the covariance matrix in Equation~\ref{eq:cov}, where $N$ is the number of antennas in the array.

\citet[\textsection~2.4]{Liu_et_al_2010} derive the noise covariance matrix for the logarithmic approach to linearizing (Equation~\ref{eq:logcal}) under the assumptions that the measured visibilities have a high SNR, and that the noise in the measured visibilities is uncorrelated between baselines, Gaussian in nature and similar across all baseline-types. This noise covariance matrix evaluates to:

\begin{equation}
\label{eq:noisecov}
    \matr{N} \approx (\mathrm{SNR})^{-2} \; \matr{I}
\end{equation}
\noindent
under the additional assumption that all the baselines in the array have the same average SNR. In general, this assumption does not hold when observing a real sky. However, in this paper, we only use SNR in the context of other array-averaged parameters, for which this assumption is justified. Since we have assumed that the noise is uncorrelated between baselines, the noise covariance matrix is just proportional to the identity matrix $\matr{I}$.

Substituting the noise covariance matrix into Equation~\ref{eq:cov}, and taking the first $N$ rows and columns of the covariance matrix, which we denote by $\left(\matr{A}^{\dagger}\matr{A}\right)^{-1}_{(N\times N)}$, we get the covariance of antenna gains. The diagonal of this matrix represents the variance or uncertainty in the estimated antenna gains.

\begin{equation}
    \label{eq:gaincov}
    \sigma_{g}^2 \approx (\mathrm{SNR})^{-2} \; \mathrm{diag}\left[\left(\matr{A}^{\dagger}\matr{A}\right)^{-1}_{(\mathit{N}\times \mathit{N})}\right]
\end{equation}
\noindent
The two reduced redundant-baseline calibration schemes effect the uncertainty in gains according to the above equation. In low-cadence calibration, the lower SNR of visibilities in the calibrator result in higher gain variance as compared to traditional redundant-baseline calibration. In subset redundant calibration, only a sub-matrix of $\matr{A}$ is used for calibration, again resulting in antenna gains with a higher variance.

The matrix $(\matr{A}^{\dagger}\matr{A})$ is nearly diagonal (small off-diagonal terms), with each entry equal to the number of equations in which the corresponding variable is involved. When redundant-baseline calibration is performed using the full visibility matrix, the first $N$ diagonal entries of this matrix are each equal to $N$ since every antenna is involved in $N$ equations. At constant SNR, this results in the following scaling for gain variance:

\begin{equation}\label{eq:gainvar_redcal}
    \sigma_{g}^2 \propto \frac{1}{N}
\end{equation}
\noindent
In Sections~\ref{sec:lowcadcal} and~\ref{sec:subredcal} we derive a scaling relation for the variance in gains estimated using a $\mathcal{O}(N\log{N})$ calibrator that implements low-cadence calibration and subset redundant calibration respectively, and compare it with the above relation.

\subsection{Scatter in Visibilities of Redundant Baselines}
\label{sec:redredcal:scatter}

The spatial Fourier transform in an FFT-correlator, which decreases the computational scaling from $\mathcal{O}(N^2)$ to $\mathcal{O}(N\log{N})$, averages the visibilities of redundant baselines. Hence, the post-calibration residual scatter in redundant visibilities is an important metric for assessing the gains estimated by a reduced redundant-baseline calibration process.

The scatter in redundant visibilities is quantified by the reduced $\chi^2$ of antenna gains and model visibilities ($\chi_r^2$) which is given by:

\begin{equation}\label{eq:redchi}
    \chi_r^2 = \frac{1}{\mathrm{DoF}}\sum_{(i,j) \in \alpha, \forall \alpha} \frac{\left|V_{ij}^{\text{meas}} - g_i g_j^*V_{\alpha}^{\text{true}}\right|^2}{\sigma_{ij}^2}
\end{equation}
\noindent
where $\sigma^2_{ij}$ is the variance of the noise in measured visibilities, $n_{ij}$ in Equation~\ref{eq:redcal}. DoF is the degrees of freedom in this system of equations given by:

\begin{equation}\label{eq:dof}
    \mathrm{DoF} = \mathit{N}_{obs} - \mathit{N} - \mathit{N}_{ubl} + 2
\end{equation}
\noindent
$N_{\mathrm{obs}}$ is the total number of cross-correlations computed (or number of visibility measurements), $N$ is the number of antennas in the array and $N_{\mathrm{ubl}}$ is the number of unique baseline-types in the system of equations. The additional offset by two accounts for the number of degenerate parameters in the system of equations representing a single-polarisation \citep{Dillon_et_al_2018, Dillon_et_al_2020}.

The reduced redundant-baseline calibration process within the calibrator is setup to minimise the $\chi_r^2$ between the visibility matrix computed by the calibrator $\mathbf{V^{reduced}}$ and the estimated gains and model true visibilities. In an array with identical antennas and perfect redundancy, we expect this $\chi_r^2$ to evaluate to one.

The estimated gains, however, apply to antenna voltages prior to the spatial Fourier transform which computes a different visibility matrix $\mathbf{V^{unique}}$ than that used in the reduced redundant-baseline calibration process. Hence, the $\chi_r^2$ evaluated using the gains estimated by the reduced redundant-baseline calibration process, model visibilities computed by the FFT-correlator (estimated using Equation~\ref{eq:modelvis}) against the $V_{ij}^{\mathrm{meas}}$ drawn from the full visibility matrix computed by an FX-correlator, is a useful a metric to assess the effectiveness of the estimated gains in calibrating the whole array. 

In Section~\ref{sec:comparison}, we use the $\chi_r^2$ computed in this way to compare the performance of reduced redundant-baseline calibration to traditional redundant-baseline calibration with the full visibility matrix.

\subsection{Simulation} 
\label{sec:simulation}

\begin{figure}
  \centering \includegraphics[width=0.65\linewidth]{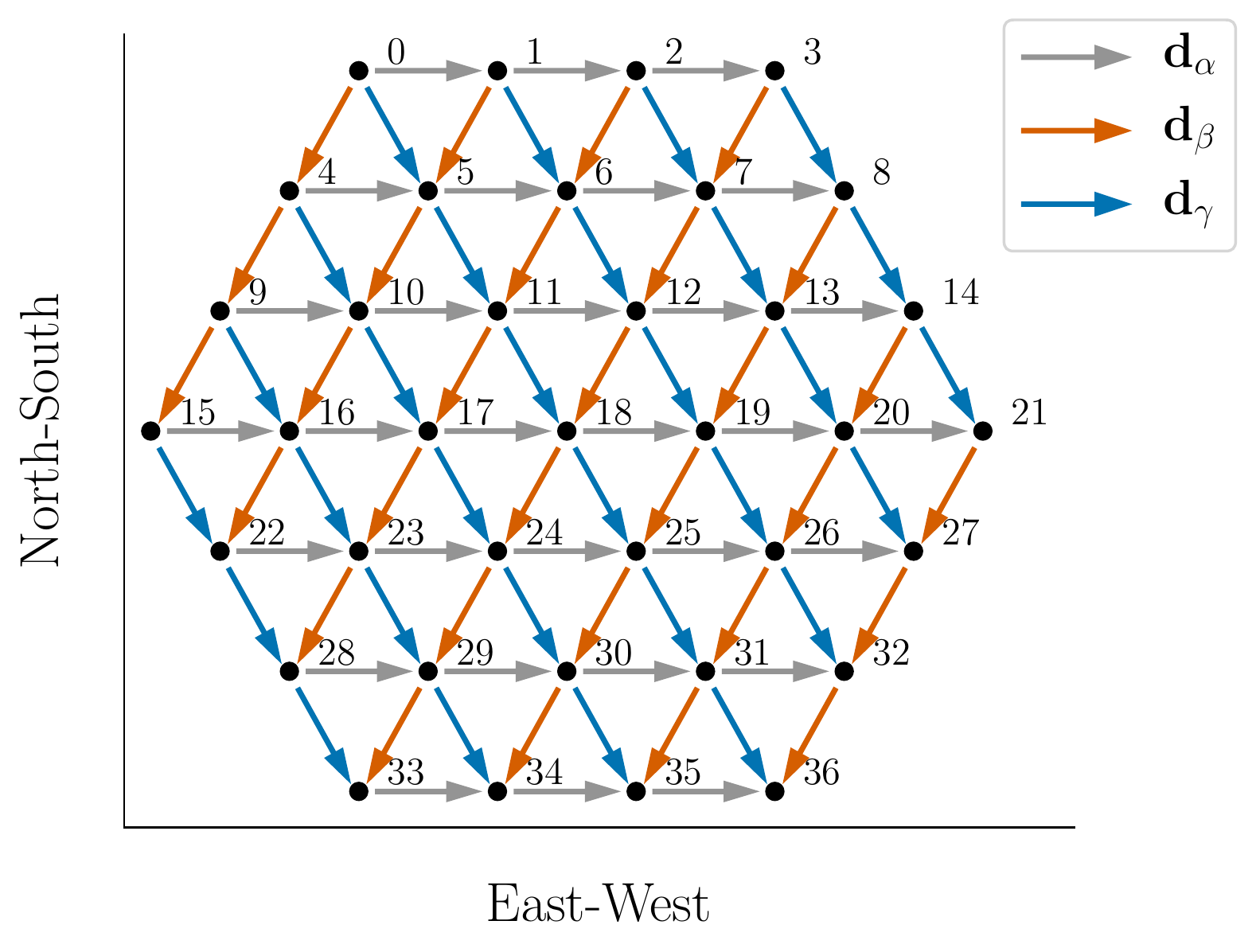}
  \caption{A hexagonal redundant array that is loosely based HERA; is used as a prototype for demonstrating the properties of reduced redundant-baseline calibration in this paper. Three different redundant baseline groups are marked by blue, orange and grey arrows.
    \label{fig:hexarray}}
\end{figure}

In the following sections, we discuss the performance of low-cadence calibration and subset redundant calibration using simulated visibilities and antenna gains. We use hexagonal array layouts, like the one shown in Figure~\ref{fig:hexarray}, with varying number of antennas for the simulations. Though this layout is loosely based on HERA (see \citet{Dillon_and_Parsons_2016}), we expect the derived trends to hold for any two dimensional redundant array layout. We assume perfect redundancy in the array and identical antenna beams.

The simulations start by generating a set of true visibilities for all the unique baseline-types and a set of gains for all the antennas in the array. These simulated true visibilities have a random constant average amplitude across all baseline-types, are constant in time and uncorrelated between baseline-types. While this does not reflect a real sky signal, it is sufficient for the purpose of this paper because redundant-baseline calibration only has a weak dependence on the actual signal observed. The simulated antenna gains are Gaussian-distributed, with an average amplitude of 1. The antenna-to-antenna variation, or gain scatter, is assumed to be at the $\sim$10\% level which was found to be typical for HERA antennas~\citep{Kern_et_al_2020}.

The process of calibration in the FFT-correlator (first step in the yellow boxed region of Figure~\ref{fig:systemlayout}), is simulated by applying the estimated gains to the full visibility matrix. This is equivalent to applying the gains to individual antenna voltages and cross-correlating them. The full visibility matrix computed by an FX-correlator, $\mathbf{V^{full}}$, is generated by multiplying the simulated gains and true visibilities and adding Gaussian random noise. In the simulations where an explicit SNR for the visibilities is not mentioned, we have assumed an SNR of 10 even though a more favourable SNR is expected in practice. The visibilities produced by the FFT-correlator $\mathbf{V^{unique}}$ are generated by applying the gains estimated by one of the reduced redundant-baseline calibration processes to this full visibility matrix (Equation~\ref{eq:modelvis}).

For low-cadence calibration, the visibility matrix computed within the calibrator, $\mathbf{V^{reduced}}$, is generated by adding higher amplitude Gaussian noise to the multiplied gains and true visibilities. For subset redundant calibration, this visibility matrix is generated by choosing the visibilities of the selected antenna pairs from the full visibility matrix. Reduced redundant-baseline calibration is performed on this visibility matrix using the \texttt{omnical} algorithm, with a damping factor of 0.3 and convergence criteria of $10^{-10}$. The amplitude and phase degeneracies of the resulting gains are fixed by comparing with the amplitude and phase of the simulated input gains. Unless otherwise specified, the variance of antenna gains is computed by running 256 simulations with the same underlying gains and different realisations of true visibilities.

The variance of antenna gains is estimated by running 256 simulations (unless otherwise specified) with the same underlying gains and different sets of true visibilities. The variance in antenna gains is given by the average variance of the gains estimated for each antenna in each data set. In all the plots shown in this paper, the errorbars associated with the gain variance show the antenna-to-antenna variation in the average variance.
 
\section{Low-Cadence Calibration}
\label{sec:lowcadcal}

Low-cadence calibration is a reduced redundant-baseline calibration scheme that estimates antenna gains from visibilities that have been computed in a round-robin fashion. The calibrator in Figure~\ref{fig:systemlayout} cross-correlates the baselines required for calibration, but the computational resources allocated to it cannot scale faster than $\mathcal{O}(N\log{N})$. The computational resources required to compute visibilities is determined by the number of baselines that need to be cross-correlated \textit{simultaneously}.
By decreasing the number of antenna pairs that need to be correlated at a time, the computational resources required by the calibrator can be reduced. The full visibility matrix is populated after a few cycles and this is used to redundantly calibrate the array. Since redundant-baseline calibration can only be performed once in a given number of cross-correlation cycles, this calibration scheme is called low-cadence calibration.

The size of a low-cadence calibrator is determined by two parameters-- the time period available for generating the full visibility matrix within the calibrator ($t_{\mathrm{cal}}$) and the integration time allotted to each cycle ($t_{\mathrm{int}}$). These are related by the equation:
\begin{equation}
    t_{\mathrm{cal}} = N_{\mathrm{cycle}} \times t_{\mathrm{int}}
\end{equation}
\noindent
where $N_{\mathrm{cycle}}$ is the number of integration cycles taken by the calibrator to populate the full visibility matrix. The size of the low-cadence calibrator is inversely proportional to $N_{\mathrm{cycle}}$ i.e., for a small calibrator size we require a large $t_{\mathrm{cal}}$ and a small $t_{\mathrm{int}}$. 

Redundant-baseline calibration operates under the assumption that antenna gains and true visibilities are constant during the time period required to compute all the visibilities involved in the system of equations. For an FX-correlator, this is equal to the integration time of the full visibility matrix which is usually smaller than the inherent gain variability, and necessarily smaller than the time period over which the visibilities evolve due to a constantly rotating sky. For a low-cadence calibrator, the constancy of antenna gains and true visibilities within a calibration cycle has to be manually enforced.

The upper limit of $t_{\mathrm{cal}}$ is set by the inherent gain variability of the array, which could depend on numerous factors including the analog signal chain, the radio frequency interference environment and the precision of antenna gains required for the science application. If the time taken to generate the full visibility matrix is larger than the interval within which gains can be assumed to be constant, redundant-baseline calibration can result in erroneous gain solutions. If the time period of gain variability is large, it is possible that the true visibilities change within this period. However, to preserve redundancy we only require that all pairs of antennas with the same baseline be correlated simultaneously. Since this is necessarily always less than $N$ visibilities, a calibrator which can cross-correlate at least $N$ baselines can accommodate the largest redundant-baseline group in the array and satisfy this condition. 

A realistic lower limit for $t_{\mathrm{cal}}$ is the integration time of visibilities in the FFT-correlator. Within this period, the assumption of constant antenna gains and true visibilities holds and redundant-baseline calibration can be solved using the algorithms currently available. While $t_{\mathrm{cal}}$ can theoretically be set to a smaller value, it could unnecessarily increase the size of the calibrator by decreasing $N_{\mathrm{cycle}}$ for a given $t_{\mathrm{int}}$.

\subsection{Scaling in Gain Variance with Integration Time}

The relationship between integration time and SNR of a measured visibility is given by the radiometer equation \citep[see][Appendix 1.1]{tms2017}. Substituting the radiometer equation into the variance of antenna gains in Equation~\ref{eq:gaincov} we get:

\begin{align}
    \label{eq:lowcadcal_gaincov}
        \sigma_{g}^2 &\approx (\mathrm{SNR})^{-2} \; \mathrm{diag}\left[\left(\matr{A}^{\dagger}\matr{A}\right)^{-1}_{(\mathit{N}\times \mathit{N})}\right] \nonumber \\
        &\propto \left(\sqrt{t_{\mathrm{int}}}\right)^{-2} \;
        \mathrm{diag}\left[\left(\matr{A}^{\dagger}\matr{A}\right)^{-1}_{(\mathit{N}\times \mathit{N})}\right] \; \propto \frac{1}{\mathit{t}_{\mathrm{int}}}
\end{align}
\noindent
which quantifies the variance of the antenna gains estimated by performing redundant-baseline calibration on visibilities integrated for a given duration. A shorter integration time leads to lower SNR in the measured visibilities and consequently, a higher variance in the antenna gains.

\begin{figure}
    \centering
    \includegraphics[width=0.75\linewidth]{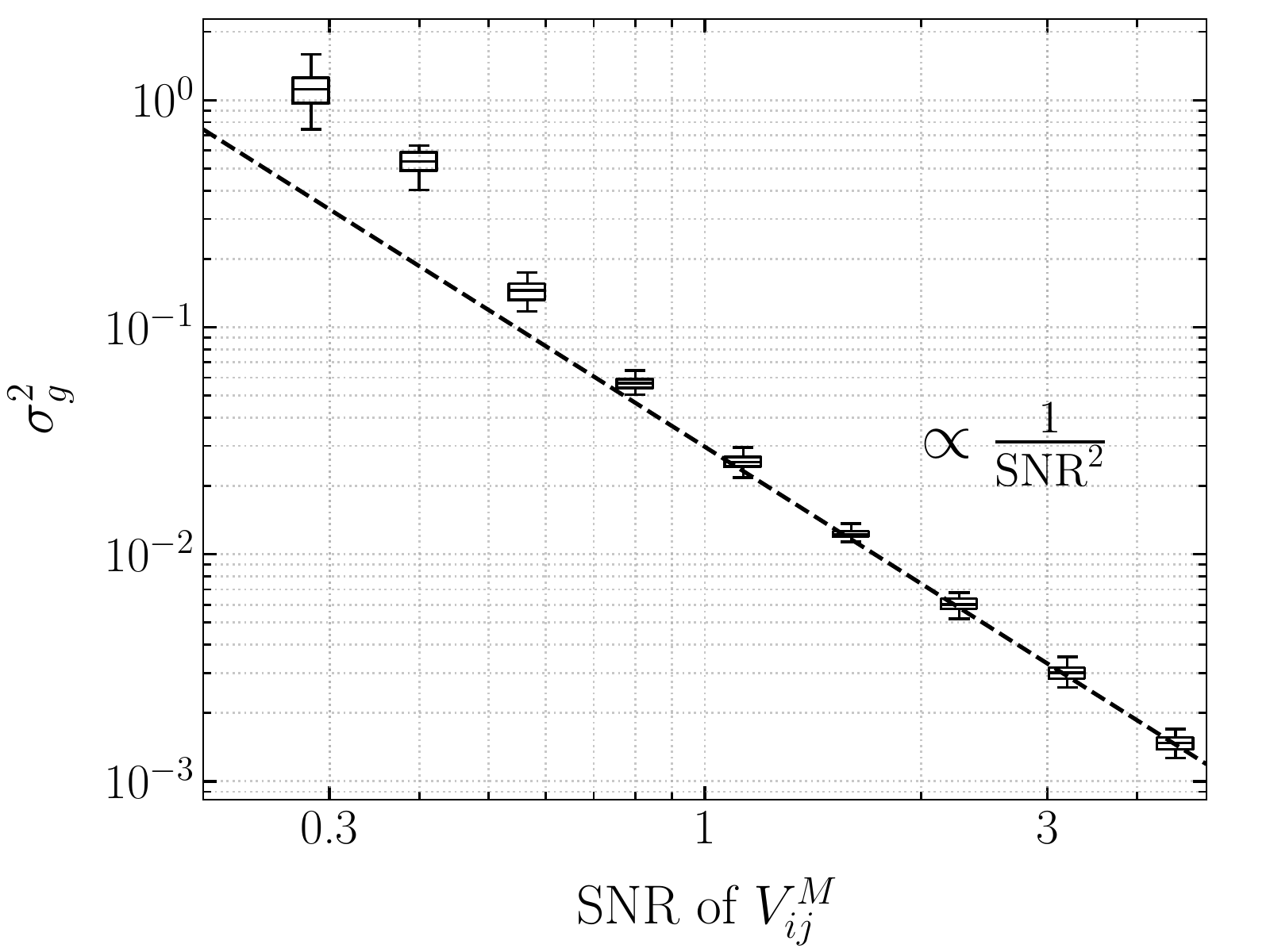}
    \caption{Relationship between the variance of antenna gains and the SNR of visibilities used for low-cadence calibration. The dashed line represents the inverse square relationship predicted by Equation~\ref{eq:gaincov} and the boxes show the results of simulation. The estimated variance of antenna gains in each simulation, is a distribution of $N$ points, where $N$ is the number of antennas in the array. The upper and lower limits of the `box' represent the upper and lower quartiles of the distribution. The horizontal bar within the box represents the mean of the distribution, and the whiskers show the total range of the estimated gain variances. At high SNR the gain variance follows the theoretically expected trend. At low SNR the variance is higher than the predicted value and becomes solver dependent because the $\chi^2$ is not effectively minimised by the solver.
    \label{fig:lowcadcal_wsnr}}
\end{figure}

Figure~\ref{fig:lowcadcal_wsnr} shows the trend in estimated gain variance with the average SNR of all the visibilities in an array. Each box in the figure represents the distribution of gain variance over all antennas in the array and has been generated assuming that all the baseline-types in the array have the same specified SNR. When the average SNR of visibilities is high, the estimated antenna gain variance follows the expected inverse square relationship. At low SNR, when the theoretical gain variance estimated using Equation~\ref{eq:gaincov} becomes comparable to the antenna-to-antenna scatter in gains squared ($\sim$0.01 for this simulation), redundant-baseline calibration fails at estimating antenna gains. Below a threshold SNR, that is set by the gain scatter, the inverse square relationship breaks down and the variance of estimated antenna gains becomes dependent on the solver. That is, the logarithmic approach, Taylor expansion approach, and \texttt{omnical} algorithms of linearizing the system of equations, result in different deviations from the given trend. This is because each algorithm minimises $\chi^2$ in a different way and none of them are effective at converging to the solution.

\begin{figure}
    \centering
    \includegraphics[width=0.75\linewidth]{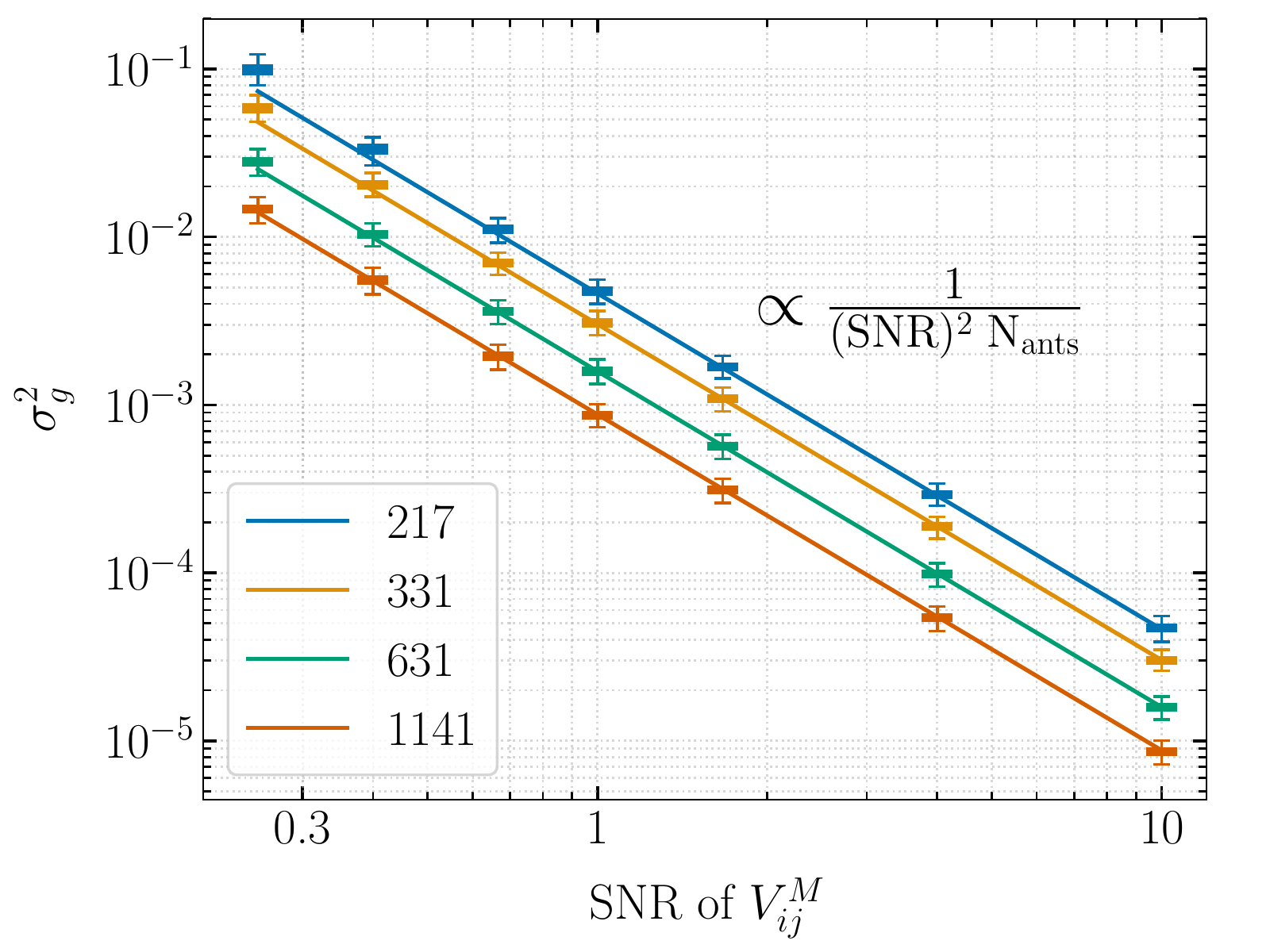}
    \caption{Relationship between distribution of gain variance and SNR of visibilities used in low-cadence calibration. The four different colours
    represent hexagonal arrays of various sizes; legend shows the number of 
    antennas in each layout. The solid lines represent the theoretical trend predicted by Equation~\ref{eq:gaincov} and the boxes show the distribution of gain variance obtained from simulation (see Figure~\ref{fig:lowcadcal_wsnr} for explanation). The threshold SNR, below which the variance deviates from the theoretical trend, is dependent on the number of antennas in the array.
    \label{fig:lowcadcal_wants}}
\end{figure}

\subsection{Scaling in Gain Variance with an \texorpdfstring{$\mathcal{O}(N\log{N})$}{O(NlogN)} Calibrator}

Figure~\ref{fig:lowcadcal_wants} shows the trend in gain variance with SNR of measured visibilities for hexagonal layouts with different number of antennas. The variance is suppressed by a factor of $N$ as the number of antennas in the array increase, because of the $\left(\matr{A}^{\dagger}\matr{A}\right)_{(\mathrm{N}\times \mathrm{N})}^{-1}$ term in Equation~\ref{eq:gaincov}. Notice that in larger arrays, the gain variance follows the theoretical trend even when the SNR of measured visibilities is less than one.

The threshold SNR below which the gain variance diverges from the theoretical prediction, changes with the number of antennas in the array. When the theoretical gain variance is less than the square of the expected gain scatter, the gain variance follows the inverse square relationship \textit{even when the SNR of measured visibilities is less than one}. This result is important because it allows low-cadence calibration to be scaled to extremely large arrays. 

Say, the computational resources allocated to the calibrator are restricted to scale similarly as the FFT-correlator. The calibrator cross-correlates $pN\log{N}$ baselines in each integration cycle, where $p$ is a pre-factor (like a proportionality constant) to convert the $\mathcal{O}(N\log{N})$ scaling into number of baselines. A larger pre-factor would result in a larger calibrator size. For a fixed interval of calibration $t_{\mathrm{cal}}$, the integration time is smaller for larger arrays according to the scaling: 

\begin{equation}
\label{eq:lowcadcal_inttime}
    t_{\mathrm{int}} = \left(\frac{pN\log{N}}{N^2/2}\right) t_{\mathrm{cal}} 
                 = \left(\frac{2p\log{N}}{N}\right) t_{\mathrm{cal}}
\end{equation}
\noindent
Hence, even if the interval of calibration is large, the multiplying factor might become small enough to push the SNR of measured visibilities to less than one. Substituting the result of Equation~\ref{eq:lowcadcal_inttime} into Equation~\ref{eq:lowcadcal_gaincov}, we get:

\begin{equation}
    \label{eq:gainvar_lowcadcal}
        \sigma_{g}^2 \propto \left(\frac{N}{2p\log{N}} \cdot  \frac{1}{t_{\mathrm{cal}}} \right) \frac{1}{N} \propto \left(\frac{1}{p\log{N}}\right) \cdot \frac{1}{t_{\mathrm{cal}}}
\end{equation}
\noindent
which shows that the antenna gain variance improves with array size even at constant $t_{\mathrm{cal}}$. That is, even though the SNR of measured visibilities might decrease to a value less than one, the antenna gain variance decreases. The price that one pays for not using $\mathcal{O}(N^2)$ resources for calibration is that the precision in antenna gains scales more slowly as compared to that of redundant-baseline calibration with the full visibility matrix which is given by Equation~\ref{eq:gainvar_redcal}.

Low-cadence calibration is a calculated way of trading computational resources for precision in the antenna gain solutions. As long as the size of the calibrator scales faster than $\mathcal{O}(N)$ with the size of the array, the variance in antenna gains decreases with increase in the number of antennas. A potential drawback of low-cadence calibration, especially when applied to arrays with over 10,000 antennas, is the time taken by a linearized solver to result in convergent gains. \citet{Dillon_et_al_2020} show that the time taken by the \texttt{omnical} algorithm scales as $\mathcal{O}(N^2)$ when the solver has to optimise $N^2$ baselines. If the time interval between calibration cycles $t_{\mathrm{cal}}$ can be proportionally decreased, it might still be possible to obtain real-time solutions. However, $t_{\mathrm{cal}}$ is usually set by the inherent gain variability in the array which might not be scalable with array size. One way of addressing this issue, is to look at redundant-baseline calibration with a limited set of baselines.

\section{Subset Redundant Calibration}
\label{sec:subredcal}

The spatial Fourier transform in an FFT-correlator averages visibilities of redundant baselines. Traditional redundant-baseline calibration assumes that all the $\sim$$N^2/2$ cross-correlation products are available for calibration, which could be non-viable to compute for large-N arrays. Subset redundant calibration is a reduced redundant-baseline calibration scheme that attempts to estimate antenna gains from visibilities of only a limited set of baseline-types. This section examines the effect of not using all baselines for redundant-baseline calibration on the variance of estimated antenna gains. 

While considering baselines for subset redundant calibration, it is useful to distinguish between baseline-types (or unique baselines) and redundant baseline groups. A baseline-type that a particular antenna pair belongs to is specified by the displacement vector between the two antennas. A redundant baseline group consists of all the antenna-pairs that have the same baseline vector. For instance, Figure~\ref{fig:hexarray} shows three different baseline-types with displacement vectors pointing East (grey), Southeast (blue) and Southwest (orange). Each baseline-type has 30 different antenna pairs in its redundant baseline group, marked in arrows of the same colour. 

\subsection{Brief Discussion on Using Short Baselines}

The short baseline-types are important in subset redundant calibration for two main reasons: (a) they involve all the antennas in the array, allowing redundant-baseline calibration on just these visibilities to estimate all the antenna gains and (b) the redundant baseline groups of these baseline-types are larger than the longer baseline-types. For example, in the layout shown in Figure~\ref{fig:hexarray}, there are 30 baselines in each of the redundant baseline groups that belong to the shortest three baseline-types, while there are only 4 baselines that belong to the group formed by the baseline-type like $(0,31)$. This is important for subset redundant calibration because every new baseline-type added to the system of equations requires a new variable in the form of the unique visibility for that baseline-type. Since the short baselines have a higher ratio of redundant baselines (measurements) to unique visibilities (variables), they contribute more to constraining the gain solutions. 

In addition to this, at low radio frequencies, the short baselines pick up the bright diffuse emission from our galaxy and have high SNR visibilities. As shown in Equation~\ref{eq:gaincov}, this suppresses gain variance and results in higher precision gain solutions. \citet{Orosz_et_al_2018} discuss other advantages of using only short baselines from the perspective of non-redundancies in a realistic array layout. They argue that calibration errors affect the inferred power spectrum, and contamination worsens when longer baselines are included in the redundant-baseline calibration process. \citet{Li_et_al_2018} point out that redundant-baseline calibration performs better than sky-based calibration at low radio frequencies, partly because short baselines have to be ignored for sky-based calibration due to poor diffuse sky models. On the other hand, the shorter baselines are more susceptible to systematic errors like antenna cross-coupling~\citep{Kern_et_al_2019} and may be more non-redundant than the longer baselines \citep{Dillon_et_al_2020} in a realistic array layout. 

A practical subset redundant calibrator would cross-correlate a combination of short and long baselines that produces the best estimate of antenna gains for the array. Since the voltages from all the antennas are available to the subset redundant calibrator, the combination of baselines that it needs to compute can also dynamically change with time/day of observation. In this paper, the baselines used to perform subset redundant calibration are considered in the order of baseline length from shortest to longest. That is, a smaller calibrator preferentially cross-correlates only the shorter baselines. However, the results presented in this section apply to combinations of short and long baselines as well.

\subsection{Degeneracy Criterion}

When using a limited set of baselines for redundant-baseline calibration, it is important to ensure that there are sufficient number of measurements to determine all the variables. The solution space of Equation~\ref{eq:redcal} has a null space with four degenerate parameters \citep{wieringa_1992, Liu_et_al_2010, Dillon_et_al_2018}:
\begin{enumerate}[(a)]
\item the absolute amplitude of the gains (or the sum of all gains), 
\item the absolute phase of the gains (or the sum of all gain phases), 
\item the phase slope of the gains in the $x$ direction and 
\item the phase slope of the gains in the $y$ direction. 
\end{enumerate} 

When selecting baselines for subset redundant calibration, it is important to verify that the null space of the solution set is restricted to these four degenerate parameters. Introducing more degeneracies allows the gain solutions to vary in that dimension and could require future corrections. If the additional degeneracy does not have a physical interpretation this may not even be possible. For a hexagonal layout like the one shown in Figure~\ref{fig:hexarray}, a minimum of three unique baseline-types, with displacement vectors in the directions marked in the figure, are required to satisfy the degeneracy requirement.

\subsection{Scaling in Gain Variance with Number of Baselines}

\begin{figure}
    \centering
    \includegraphics[width=0.75\linewidth]{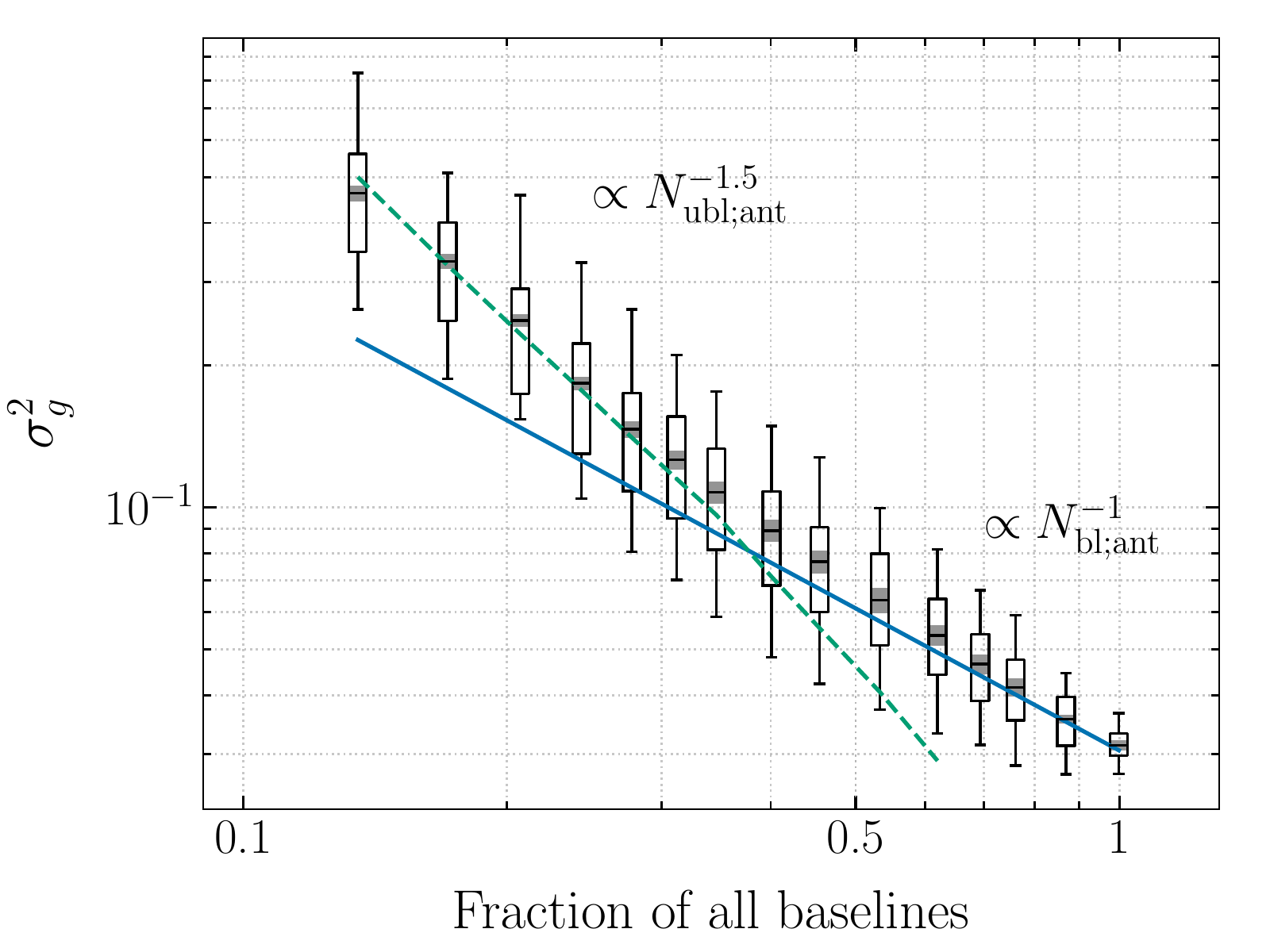}
    \caption{Relationship between fraction of baselines used in subset redundant calibration and the resulting variance in estimated gains. The boxes show the distribution of gain variances in simulation (see Figure~\ref{fig:lowcadcal_wsnr} for explanation), where increasingly longer baselines are included in the subset redundant calibration system of the simulation. When the fraction of baselines used in estimating antenna gains is small, the gain variance depends on the number of unique visibilities per antenna used in calibration ($N_{\mathrm{ubl;ant}}$); this trend is shown by the dashed green line. When a large fraction of baselines are used for calibration, the gain variance depends on the total number of measurements per antenna ($N_{\mathrm{bl;ant}}$); shown by the solid blue line. A second set of boxes in grey show the antenna-to-antenna variation with all the edge antennas excluded.
    \label{fig:subredcal_wbls}}
\end{figure}

Figure~\ref{fig:subredcal_wbls} shows the relationship between number of baselines used in subset redundant calibration and the corresponding variance in the estimated gains. The number of baselines are shown in terms of a fraction of the total baselines in the array. The black boxes show the results of simulation (see Section~\ref{sec:simulation}). The x-axis from left to right, represents baselines added to the subset redundant calibration system in ascending order of baseline length (starting with the minimum required to satisfy the degeneracy criterion). For each baseline-type added to the system, it is assumed that all the redundant baseline pairs that contribute to that baseline-type are used for calibration. The SNR of visibilities is assumed to be similar for all baseline-types (unlike for a real sky), and constant through the simulation. The vertical range of the boxes, which represents distribution of gain variance over antennas in the array, is larger than the case of low-cadence calibration. However, most of this variation comes from the edge antennas in the array. A second set of solid boxes in grey show the distribution of gain variances with the edge antennas excluded. The reason for this decrease in variation is discussed in more detail in Section~\ref{sec:subredcal_cov}.

The solid line, in blue, represents the inverse of the total number of baselines per antenna ($N_{\mathrm{bl;ant}}$) that are used to perform subset redundant calibration.

\begin{equation}
    \label{eq:subredcal_n_bl}
    N_{\mathrm{bl;ant}} = \frac{N_{\mathrm{obs}}}{N} \approx \frac{f \; N^2/2}{N} = f \; \frac{N}{2}
\end{equation}
\noindent
Here $f$ represents the fraction of all baselines used in subset redundant calibration. When the fraction of baselines used in subset redundant calibration is high, the gain variance asymptotes to the inverse of the total number of measurements-per-antenna in the system of equations. This trend is expected from the Gaussian noise in visibility measurements; each new measurement added to the system contributes to decreasing the noise in the estimated gains. 

When the fraction of baselines used in estimating antenna gains is small, gain variance depends on two factors: (a) the number of baseline-types included in subset redundant calibration, and (b) the number of antennas that are involved in forming redundant baselines for these baseline-types. A combination of these two factors is captured by the variable $N_{\mathrm{ubl;ant}}$ which represents the average number of unique visibilities per antenna.

\begin{equation}
    \label{eq:subredcal_n_ubl}
    N_{\mathrm{ubl;ant}} = \frac{1}{N} \sum\limits_{\alpha} N_{\mathrm{ant}} \left[\in V_{\alpha} \right]
\end{equation}
\noindent
In the above equation, the variable being summed is the number of antennas that are involved in forming redundant baselines of the baseline-type $V_{\alpha}$. The summation runs over all the baseline-types that are used in subset redundant calibration and $N$ is the total number of antennas in the array. Consider the case where all the baseline-types used in subset redundant calibration have redundant baselines involving all the antennas in the array (for instance, when only the shortest 3-6 baseline-types are used for calibration). In such a system, $N_{\mathrm{ubl;ant}}$ simply evaluates to the total number of baseline-types (or unique visibilities) used in the calibration process. If some of the baseline-types used in subset redundant calibration involve only a couple of antennas, $N_{\mathrm{ubl;ant}}$ is smaller than the total number of unique visibilities in the system of equations. 

Empirically, we find that the relationship between gain variance and the average number of unique visibilities per antenna is a power law with a slope around $-1.5$ for hexagonal and square layouts. This power law trend is shown by the dashed green line in Figure~\ref{fig:subredcal_wbls}. The large antenna-to-antenna variation in this regime makes it difficult to determine the exact slope or understand the origin of this power law. We suspect that it originates in the way gain error propagates from antenna to antenna. 

In summary, for a subset redundant calibration system that uses only a small fraction of the total baselines in the array, gain variance improves when baseline-types with larger redundant groups are used for calibration.

\subsection{Covariance in Estimated Gains}
\label{sec:subredcal_cov}

The improvement in gain variance obtained when using a higher number of unique baselines per antenna, can also be explained through the gain covariance. When the fraction of baselines used in calibration is small, in addition to high variance, the gains also have a relatively high covariance. The covariance in gains is given by the off-diagonal terms of the marginalised covariance matrix:

\begin{equation}
    \label{eq:gaincov_v1}
    \matr{C^{\prime}} \approx \left(\mathrm{SNR}\right)^{-2} \left(\matr{A}^{\dagger}\matr{A}\right)_{(\mathit{N} \times \mathit{N})}^{-1}
\end{equation}
\noindent
Both the diagonal and off-diagonal terms of the matrix $\left(\matr{A}^{\dagger}\matr{A}\right)$ change when a subset of baseline-types are used for redundant calibration. Inverting this matrix changes the covariance in the resulting gains.

\begin{figure}
    \centering
    \includegraphics[width=0.75\linewidth]{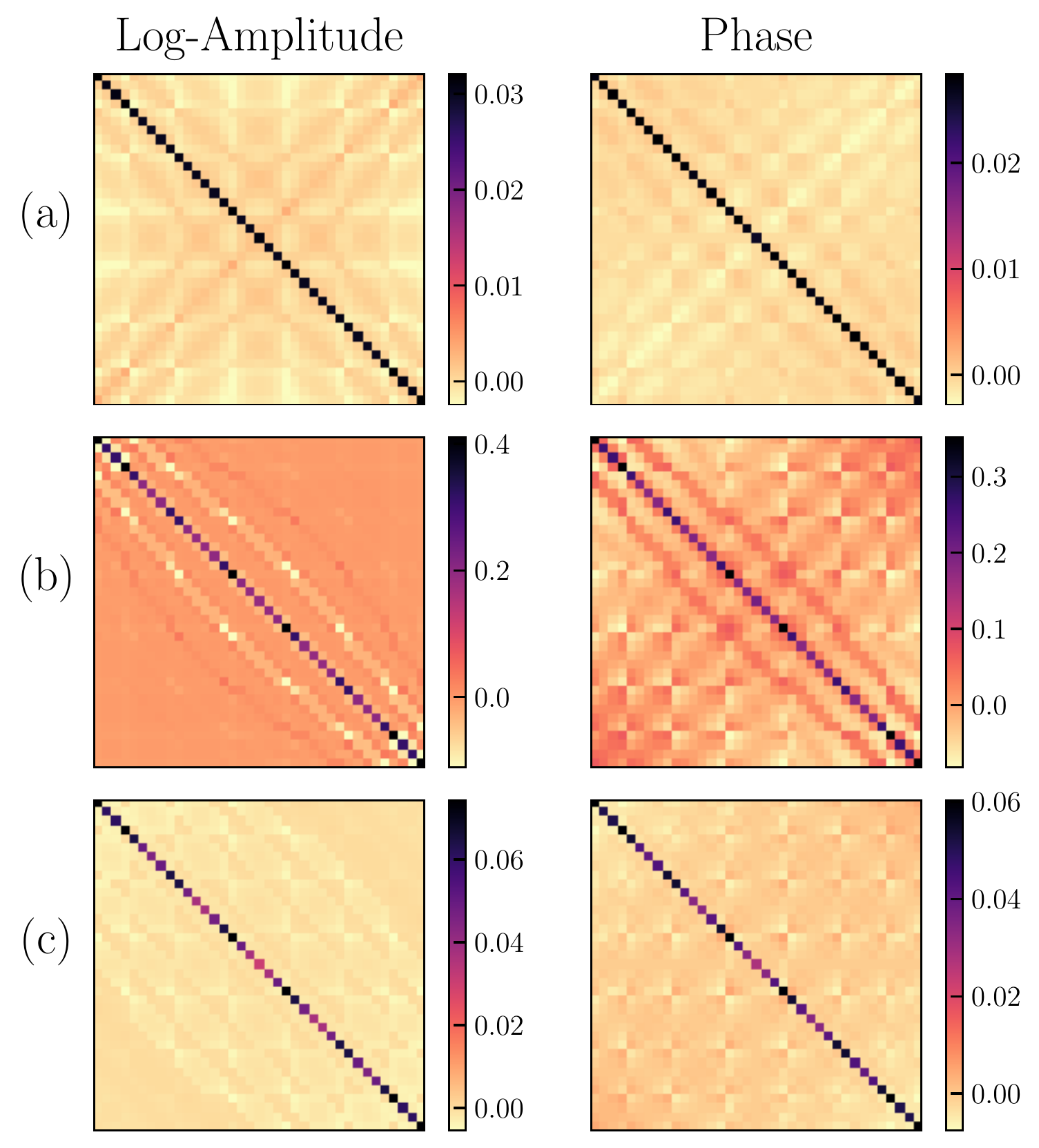}
    \caption{Covariance in the amplitude and phase of antenna gains for subset redundant calibration performed on the hexagonal layout of Figure~\ref{fig:hexarray}. Panel (a) shows the marginalised covariance for the case where redundant-baseline calibration is performed using the full visibility matrix. Panel (b) shows the covariance matrix when subset redundant calibration is performed with just the shortest three baselines required to satisfy the degeneracy criterion. In addition to higher gain variance, the antennas have also have a high covariance. Panel (c) is the covariance matrix for subset redundant calibration performed using more than half of the total baselines in the array. Both the variance and the covariance are comparable to that of redundant-baseline calibration performed with the full visibility matrix even though the covariance has a different structure compared to panel (a), and the variance is still dependent on antenna location.
    \label{fig:subredcal_cov}}
\end{figure}

Figure~\ref{fig:subredcal_cov} shows the marginalised covariance matrices for three different subset redundant calibration systems (for the hexagonal array layout in Figure~\ref{fig:hexarray}). The logarithmic approach to linearizing, shown in Equation~\ref{eq:logcal}, naturally results in separating the amplitude and phase of gains into the real and imaginary parts of the logarithm respectively. Hence, using the real (imaginary) part of the matrix $\matr{A}$ gives the gain covariance in the amplitude (phase) of gains. All the covariance matrices are normalised by the thermal noise in the visibilities used for redundant-baseline calibration. 

Panel (a) of Figure~\ref{fig:subredcal_cov} shows the covariance matrices for the case where redundant-baseline calibration is performed using the full visibility matrix. The variance in antenna gains, given by the diagonal of the matrix, has an average value of $1/N$ as predicted by Equation~\ref{eq:gainvar_redcal}. Though not evident in the figure, this variance is weakly dependent on antenna location as shown by \citet{Dillon_and_Parsons_2016}. There a small but non-zero covariance in the antenna gains. 

Panel (b) of Figure~\ref{fig:subredcal_cov} shows the covariance matrices for the extreme case where only the shortest three baselines (the minimum baseline-types required to satisfy the degeneracy criterion) are used in redundant-baseline calibration. Notice that the variance is nearly an order of magnitude higher than the first case and clearly dependent on antenna location. The covariance between antennas is non-negligible and higher between antennas that have a high variance. 

Panel (c) of Figure~\ref{fig:subredcal_cov} shows the covariance matrices for the case where more than half the baselines are used in subset redundant calibration. This is the threshold at which gain variance starts following the inverse measurements per antenna trend in Figure~\ref{fig:subredcal_wbls}. Even though the variance is still antenna location dependent and the covariance has a different structure, they are comparable to the case of redundant-baseline calibration with the full visibility matrix. 

When the subset redundant calibration system involves a larger number of unique baselines, there are more independent constraints on the gain of each antenna. This decreases the average gain variance, covariance between antennas and also the antenna-location dependence of the variance. In Figure~\ref{fig:subredcal_wbls}, the errorbars associated with the gain variance estimated in simulation, represent the antenna-to-antenna variation within the simulation. Hence, the errorbars are larger for the gains estimated using a smaller fraction of baselines.

\begin{figure}
    \centering
    \includegraphics[width=0.75\linewidth]{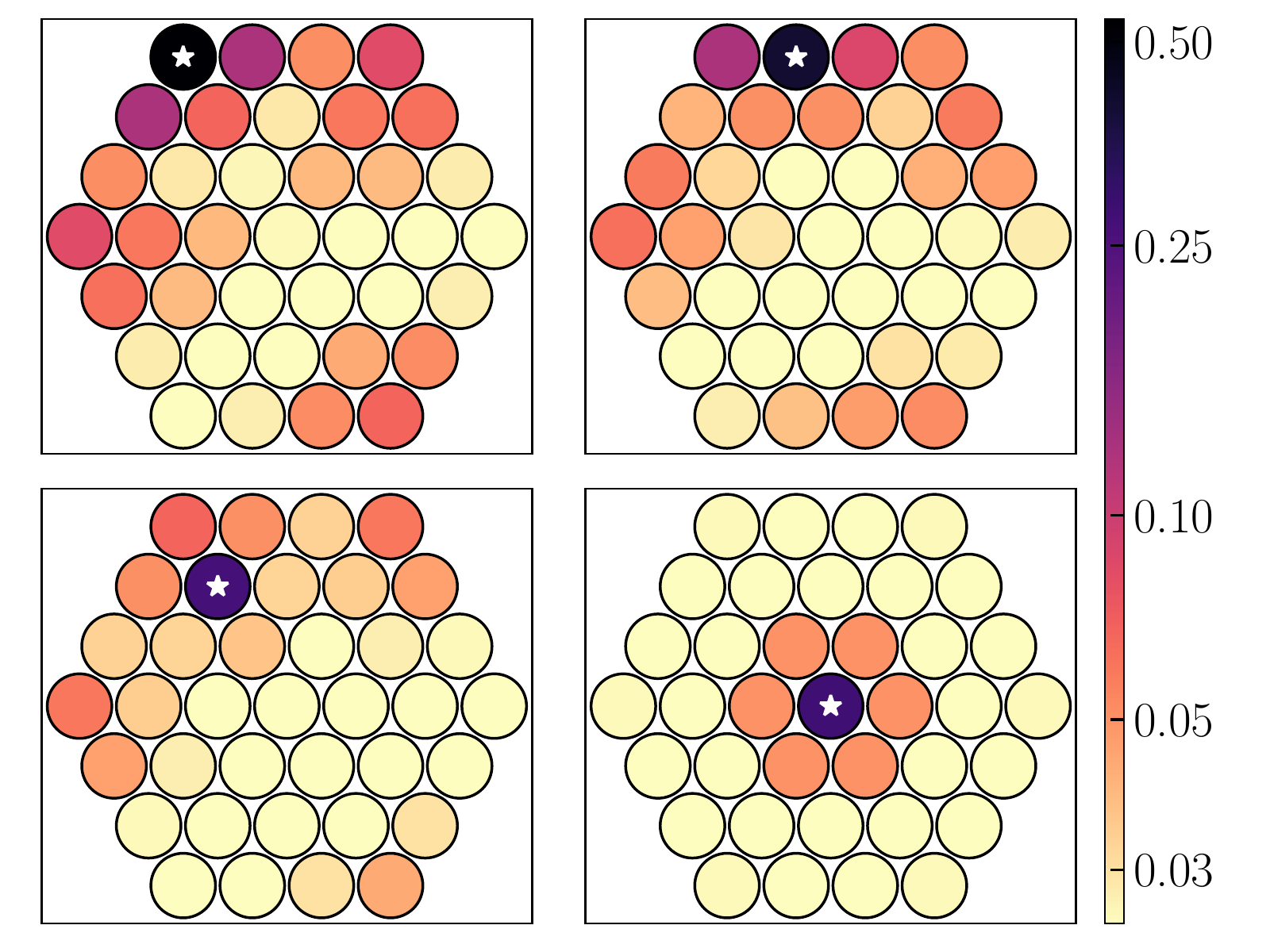}
    \caption{Four rows of the covariance matrix in Panel (b) of Figure~\ref{fig:subredcal_cov}. The colour-bar has been log-normalised to make the covariance more evident. The covariance shown in each panel is the sum of covariance in amplitude and phase for the antenna that is marked with a star. For all antennas, the covariance is highest with adjacent antennas because only those visibility measurements are used for constraining gain solutions. The variance and covariance of corner and edge antennas (upper two panels) are higher than that of antennas placed centrally (lower two panels) because the edge antennas participate in fewer visibility measurements than central antennas.
    \label{fig:subredcal_gaincov}}
\end{figure}

This dependence of antenna gain variance and covariance on the number of independent constraints per antenna is more evident in Figure~\ref{fig:subredcal_gaincov} which shows four different rows of the covariance matrix in Panel (b) of Figure~\ref{fig:subredcal_cov} in an exaggerated manner. In this case, subset redundant calibration is performed with just the shortest three baseline-types required to satisfy the degeneracy criterion. The covariance and variance of edge antennas is higher than that of centrally placed antennas because the corner antennas participate in fewer cross-correlations (three for the antenna in the top left panel, four for the antenna in the top right) than centrally placed antennas (six each for the antennas in the bottom two panels). Hence, there are fewer independent constraints for the edge antennas leading to a higher variance in their estimated gains. Ignoring the edge antennas after subset redundant calibration leads to an 80-30\% decrease in the antenna-to-antenna variation depending on the fraction of baselines used in the calibration process. The distribution of antenna gain variances with the edge antennas excluded is shown using solid grey boxes in Figure~\ref{fig:subredcal_wbls}.

\subsection{Scaling in Gain Variance with an \texorpdfstring{$\mathcal{O}(N\log{N})$}{O(NlogN)} Calibrator}

\begin{figure}
    \centering
    \includegraphics[width=0.75\linewidth]{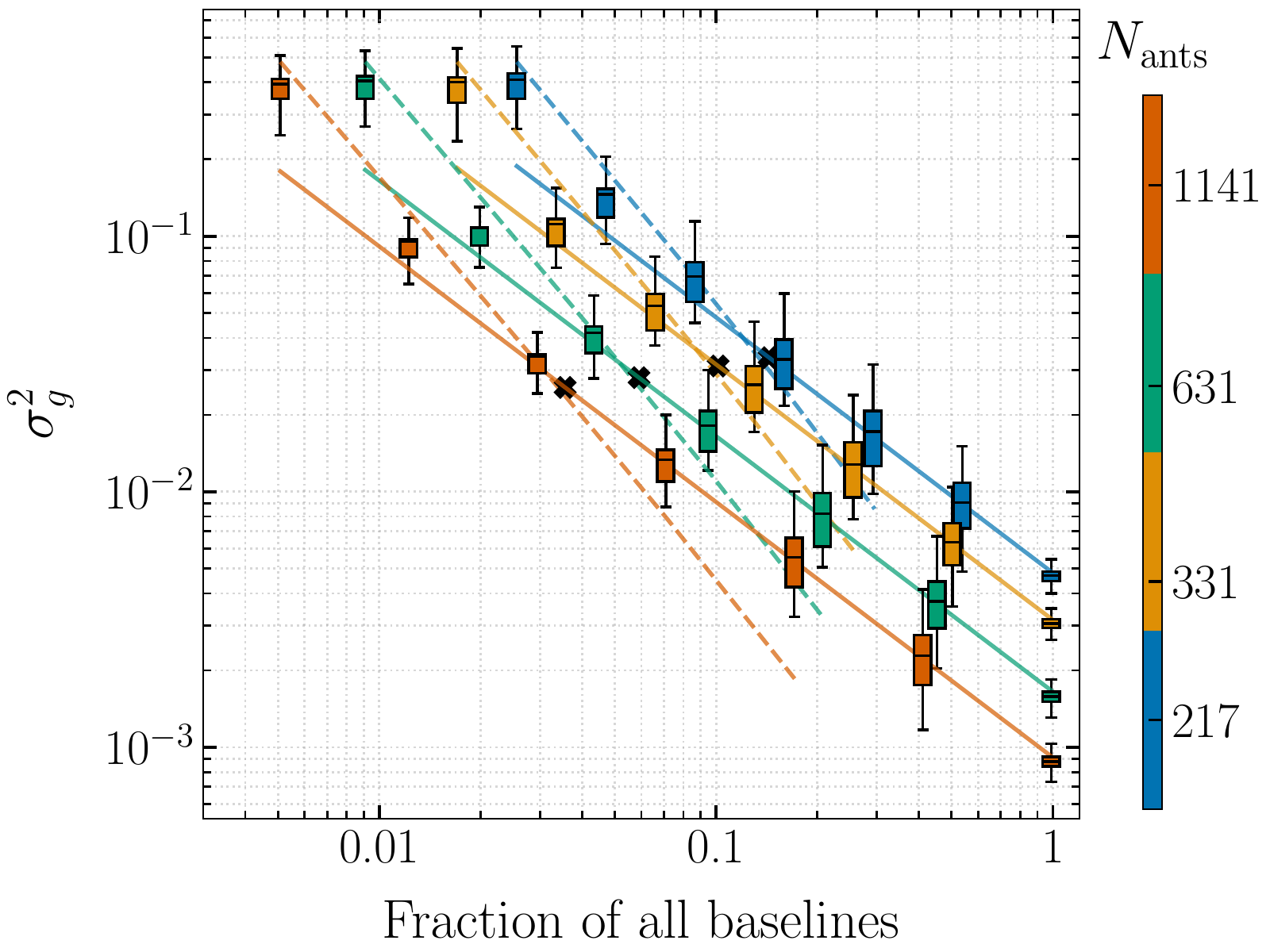}
    \caption{Relationship between variance of antenna gains and the fraction of baselines used in subset redundant calibration, for hexagonal array layouts of various sizes (shown in different colours). The boxes represent the distribution of gain variance obtained from simulation (see Figure~\ref{fig:lowcadcal_wsnr} for explanation). The solid lines represent an inverse trend of baselines per antenna $\left(N_{\mathrm{bl;ant}}^{-1}\right)$. The dashed lines show a power law dependence on the number of different baseline-types per antenna $\left(N_{\mathrm{ubl;ant}}^{-1.5}\right)$ used for calibration. The small black crosses represent the fraction of baselines that can be cross-correlated by a calibrator that can process $N\log{N}$ baselines.
    \label{fig:subredcal_wants}}
\end{figure}

Figure~\ref{fig:subredcal_wants} extends the relationship between gain variance and fraction of baselines used in subset redundant calibration to hexagonal layouts with larger number of antennas. The four different colours represent four different array sizes. Within each colour, the boxes represent the distribution of gain variance over antennas in the array (see Section~\ref{sec:simulation}). The solid lines represent an the inverse measurements per antenna trend $\left(N_{\mathrm{bl;ant}}^{-1}\right)$ and the dashed lines show a power law relationship between gain variance and number of unique visibilities per antenna $\left(N_{\mathrm{ubl;ant}}^{-1.5}\right)$. It is evident that the two asymptotes to the gain variance, shown in Figure~\ref{fig:subredcal_wbls}, hold with changing array size. 

When a large fraction of baselines are used for subset redundant calibration, the gain variance depends on the total number of baselines used for calibration. At a constant fraction of baselines, the number of baselines formed by an antenna is proportional to the number of antennas in the array as shown in Equation~\ref{eq:subredcal_n_bl}. The four solid lines show a scaling in gain variance by this factor. When the fraction of baselines used for calibration is small, the gain variance does not necessarily decrease with increase in array size. The leftmost point within each array size represents the variance in the estimated gains when subset redundant calibration is performed using just the shortest three baseline-types (required for the degeneracy criterion). As evident from the figure, the gain variance is nearly constant despite the larger number of redundant baselines in an array with more antennas. This is because, within increasing array size, there are as many new variables (in the form of antenna gains) as measurements. For improvement in variance, the size of the subset redundant calibrator should increase with the array size.

Say the computational resources allocated to a subset redundant calibrator are restricted to scale similarly to the FFT-correlator. The calibrator can process $pN\log{N}$ baselines for a given array size. The fraction of baselines that can be processed by such a calibrator is given by:

\begin{equation}
    \label{eq:subredcal_frac_bls}
    f = \frac{pN \log{N}}{N^2/2} = \frac{2p\log{N}}{N}
\end{equation}
\noindent
Figure~\ref{fig:subredcal_wants} shows this fraction of baselines, for a pre-factor of one, in small black crosses. When the calibrator cross-correlates exactly $N\log{N}$ baselines, this fraction falls in the transition region between the two asymptotes. However, assuming that the $N_{\mathrm{bl;ant}}^{-1}$ approximation holds at this fraction of baselines, we can substitute Equation~\ref{eq:subredcal_frac_bls} into the number of baselines per antenna in Equation~\ref{eq:subredcal_n_bl} to get the overall scaling in gain variance.

\begin{align}
    \label{eq:gainvar_subredcal}
        \sigma_{g}^2 &\propto N_{\mathrm{bl;ant}}^{-1} \nonumber \\
        &\propto \displaystyle \frac{1}{f \; \frac{N}{2}} \propto \frac{1}{p\log{N}}
\end{align}
\noindent
The scaling in gain variance with array size, using subset redundant calibration, is similar to that obtained using low-cadence calibration (Equation~\ref{eq:gainvar_lowcadcal}). In both cases, the price one pays for not using $\mathcal{O}(N^2)$ baselines for calibration is that the gain variance scales slower than the case where redundant calibration is performed using the full visibility matrix (Equation~\ref{eq:gainvar_redcal}).

Subset redundant calibration leverages the higher constraining power of some baseline-types, by allocating computational resources of the calibrator to preferentially cross-correlating those redundant baselines. The least number of baselines that need to be considered is set by the null space of the solution to the redundant-baseline calibration equations. However, using a small fraction of baselines can result in antenna gains that have a non-negligible covariance and location-dependent variances. If the fraction of baselines cross-correlated by a subset redundant calibrator can scale as $\mathcal{O}(N\log{N})$ or faster, the gain variance decreases with the increase in array size and scaling in gain variance is similar to that of low-cadence calibration.

\section{Comparing the two Calibration Schemes}
\label{sec:comparison}

Low-cadence calibration and subset redundant calibration are two potential solutions to calibrating FFT-correlators for redundant array layouts without computing $\mathcal{O}(N^2)$ cross-correlations. In low-cadence calibration, the calibrator computes the visibilities of all baselines through a round-robin of antenna pairs and spends a shorter amount of time on each visibility measurement. As a consequence, the SNR of measured visibilities is lower and leads to a higher variance in the estimated gains. In subset redundant calibration, the calibrator computes the correlations of only a few baselines (preferentially the shorter baseline pairs) and uses these visibilities to estimate the antenna gains. In this case, having fewer measurements leads to higher variance in the estimated gains.

\subsection{Scaling in Gain Variance}

\begin{figure}
    \centering
    \includegraphics[width=0.75\linewidth]{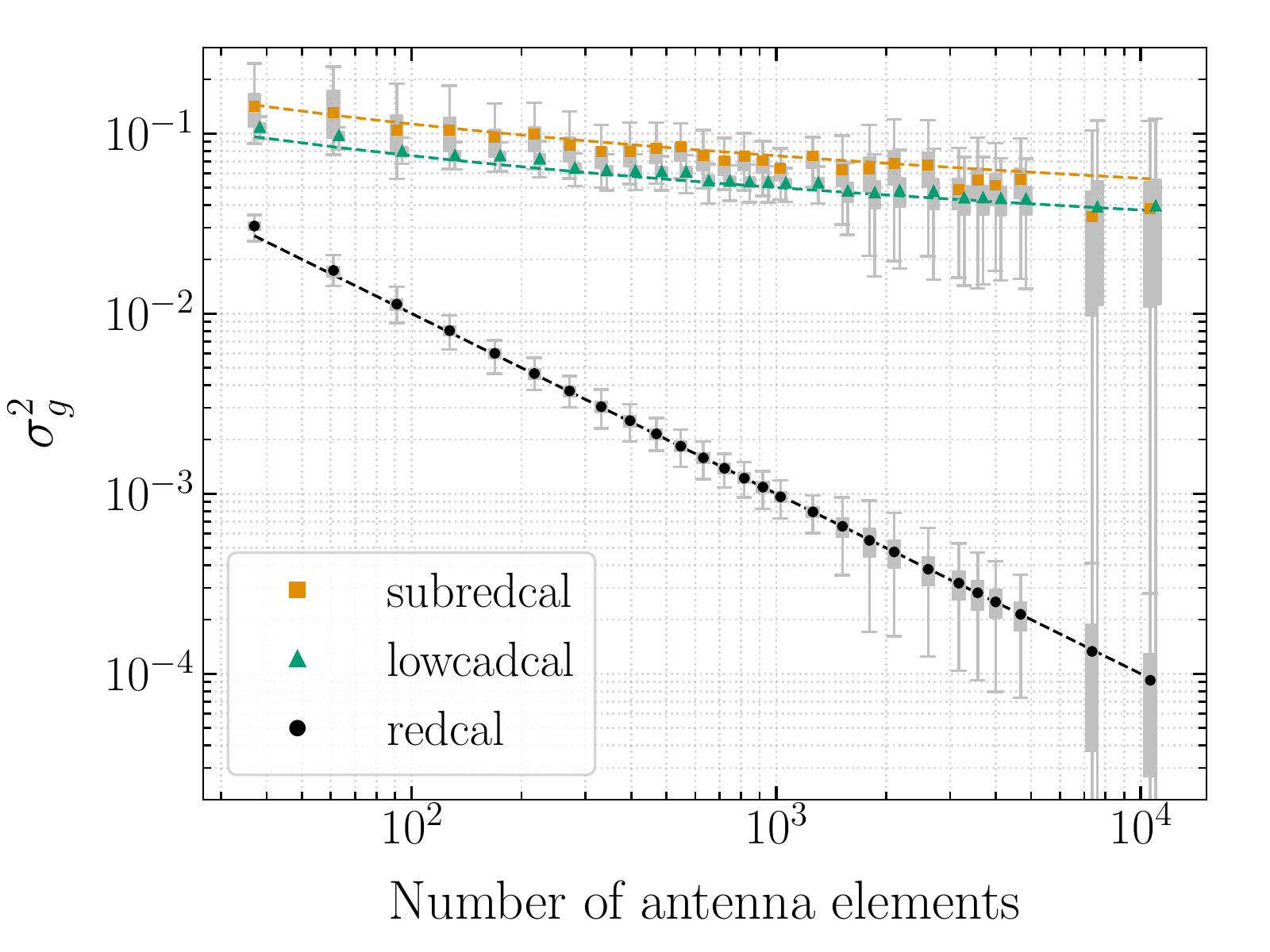}
    \caption{Distributions of antenna gain variance, estimated using low-cadence calibration (\texttt{lowcadcal}), subset redundant calibration (\texttt{subredcal}) and traditional redundant-baseline calibration (\texttt{redcal}) for hexagonal layouts of various sizes. The markers show the mean of the distribution of gain variance for each array size, and the distribution itself is shown using grey solid boxes around each marker (see Figure~\ref{fig:lowcadcal_wsnr} for explanation). The larger vertical extent of boxes for arrays with $N>1500$ elements are due to fewer simulations. The dashed lines represent the theoretically derived trends (Equations~\ref{eq:gainvar_redcal},~\ref{eq:gainvar_lowcadcal},~\ref{eq:gainvar_subredcal}). Low-cadence calibration consistently yields lower gain variance than subset redundant calibration for similar calibrator sizes. While both the reduced redundant-baseline calibration techniques cross-correlate $\sim$$N\log{N}$ baselines for each array size, traditional redundant-baseline calibration assumes that the complete visibility matrix with $\sim$$N^2/2$ baselines is available for calibration. The additional reduction in gain variance for the latter comes at a high computational cost and may not be necessary when scatter in redundant visibilities is dominated by thermal noise (see Equations~\ref{eq:visvar_lowcadcal} and~\ref{eq:visvar_subredcal}).
    \label{fig:scale_ants}}
\end{figure}

Figure~\ref{fig:scale_ants} compares the scaling in the average variance of gains estimated using either reduced redundant-baseline calibration method to the scaling in the average variance of gains estimated using redundant-baseline calibration on the full visibility matrix measured at high SNR. For both low-cadence calibration and subset redundant calibration, we assume that the pre-factor $p=1$ and use only $N\log{N}$ visibility measurements to estimate antenna gains. The distribution of gain variances, for arrays with number of antennas in the range 1500-5000, is computed using only 16 simulations (instead of 256) due to long simulation times. For the same reason, the distributions for the two largest array sizes with $N>5000$ antennas are computed using only 2 simulations each.

For low-cadence calibration, the interval between calibration cycles, $t_{\mathrm{cal}}$ is assumed to be constant for all array sizes. The integration time $t_{\mathrm{int}}$ has been scaled down according to Equation~\ref{eq:lowcadcal_inttime} to keep the size of the calibrator at an $\mathcal{O}(N\log{N})$ scaling. As predicted, the variance in the estimated antenna gains scales according to Equation~\ref{eq:gainvar_lowcadcal} and is shown by the dashed green line in Figure~\ref{fig:scale_ants}.

For subset redundant calibration, the SNR of measured visibilities is assumed to be constant for all array sizes. The number of baselines used in the calibration process for a given array size is the closest whole number, to the fraction given by Equation~\ref{eq:subredcal_frac_bls}, that accounts for an integer number of redundant baseline groups in ascending order of their baseline length. That is, baseline-types are considered in the order of their baseline length and added to the subset redundant calibration system only if all the redundant baselines that contribute to that baseline-type can be considered. This causes a non-uniform increase in the number of baseline-types used for calibration as the array size increases. The overall trend in gain variance follows the scaling predicted by Equation~\ref{eq:gainvar_subredcal} and is shown by the dashed orange line in Figure~\ref{fig:scale_ants}. 

Subset redundant calibration results in a higher gain variance than low-cadence calibration because, as shown in Figure~\ref{fig:subredcal_wants}, the approximation of inverse measurements per antenna (Equation~\ref{eq:gainvar_subredcal}) underestimates the gain variance when the pre-factor is unity. However, the scaling predicted by that approximation holds true for all the simulated array sizes. For two largest array sizes in Figure~\ref{fig:scale_ants}, the estimated gain variance using subset redundant calibration seems comparable to that of low-cadence calibration but is not an exception to this trend. This is because the two simulations used to compute these estimates generated a favourable set of visibilities for subset redundant calibration with the short baselines having larger amplitudes than the long baselines. In general, a subset redundant calibrator would have to cross-correlate more baselines than a low-cadence calibrator to achieve the same gain variance.

Redundant-baseline calibration on the full visibility matrix, measured at high SNR results in a $1/N$ scaling as predicted by Equation~\ref{eq:gainvar_redcal}. and shown by the black dashed line in Figure~\ref{fig:scale_ants}. Such a calibration method requires a calibrator with $\mathcal{O}(N^2)$ computational resources that may not be viable for large arrays. Moreover, the high precision in gains obtained using the full visibility matrix might not be necessary for decreasing the scatter in calibrated redundant visibilities.

\subsection{Variance in Calibrated Redundant Visibilities}

The visibilities of redundant baselines, that have been calibrated using a reduced redundant-baseline calibration processes, are averaged by the spatial Fourier transform. This converts any post-calibration residual scatter in redundant visibilities into additional noise on the unique visibilities returned by the FFT-correlator.

If the estimated antenna gains are exactly the true gains, the scatter in redundant-baseline averaged visibilities is given by $\sigma^2/N_{\alpha}$, assuming that the variance in the thermal noise of visibilities is similar for all baselines and represented by $\sigma^2$. However, gains estimated using redundant-baseline calibration diverge from the true gains with an average scatter that is given by Equation~\ref{eq:gaincov}. Hence, the calibrated redundant visibilities have a residual scatter that comes from both the thermal noise in the measurements and the variance in the estimated gains. 

The variance in calibrated visibilities can be derived using Equation~\ref{eq:modelvis} and the first order approximation for the variance of non-linear functions. In this derivation, we have assumed that the calibrated visibilities are not correlated with each other. Moreover, since the gains estimated using reduced redundant-baseline calibration are applied to a different set of visibilities that those used to estimate them, we can also ignore the covariance between visibilities and gains. The multiplying antenna gains, however, have a non-negligible covariance that is represented by the terms $\rho_{g_ig_j}$ in the following equation. This represents the off-diagonal components in the covariance matrix $\matr{C^{\prime}}$ in Equation~\ref{eq:gaincov_v1}.

\begin{equation}
    \sigma_{V_{\alpha}^{\mathrm{unique}}}^2 \approx \frac{\left|V_{\alpha}^{\mathrm{unique}}\right|^2}{N_{\alpha}^2} \;
    \sum\limits_{(i,j) \in \alpha} \left[\frac{\sigma_{ij}^2}{\left|V_{ij}\right|^2} + \frac{\sigma_{g_i}^2}{\left|g_i\right|^2} + \frac{\sigma_{g_j}^2}{\left|g_j\right|^2} + 2\frac{\rho_{g_i g_j}}{\left|g_ig_j\right|}\right]
\end{equation}
\noindent
We can simplify this further under the assumptions that the average amplitude of all antenna gains is close to one ($\left|g_i\right|^2$$\sim$1) and that the gain variance of all antennas is similar and given by $\sigma_g^2$. Note that the relative variance of visibilities is just the inverse squared SNR. In the equation below, $\mathrm{SNR}_{\alpha; \; \mathrm{unique}}$ is the SNR of the visibility computed by the FFT-correlator for the baseline-type represented by $\alpha$ and $\mathrm{SNR}_{\mathrm{full}}$ is the average SNR of the $N_{\alpha}$ number of redundant visibilities that belong to that baseline-type.

\begin{equation}
    \label{eq:visvar}
    (\mathrm{SNR})_{\alpha; \; \mathrm{unique}}^{-2} \approx \frac{1}{\mathit{N}_{\alpha}} \left[(\mathrm{SNR})_{\mathrm{full}}^{-2} + 2\sigma_\mathit{g}^2 + \frac{2}{\mathit{N}_{\alpha}} \sum\limits_{\mathit{(i,j)} \in \alpha} \rho_{\mathit{g}_\mathit{i}\mathit{g}_\mathit{j}}\right]
\end{equation}
\noindent
This equation gives the acceptable range of gain variance and covariance for the gains estimated using redundant-baseline calibration. When the gain variance and covariance is much smaller than the thermal noise in visibilities, the first term dominates the residual scatter. If this is satisfied, lowering gain variance by using a larger calibrator will not improve the variance in calibrated visibilities. 

\subsubsection{Variance in Redundant Visibilities Calibrated using Low-cadence Calibration}

The variance in antenna gains estimated using low-cadence calibration depends on the SNR of the visibilities computed in the calibrator which in turn depends on the integration time (Equation~\ref{eq:lowcadcal_gaincov}) available for each cycle of computation. The relationship between the SNR of the full visibility matrix and that of the reduced visibility matrix computed in by calibrator, that scales as $\mathcal{O}(N\log{N})$, can be written using Equation~\ref{eq:lowcadcal_inttime} as:

\begin{equation}
    \label{eq:SNR_lowcadcal}
    \displaystyle \frac{\left(\mathrm{SNR}\right)^{-2}_{\mathrm{full}}}{\left(\mathrm{SNR}\right)^{-2}_{\mathrm{reduced}}} = \frac{t_{\mathrm{int;reduced}}}{t_{\mathrm{int;full}}} = \left(\frac{2p\log{N}}{N}\right)\; \frac{t_{\mathrm{cal}}}{t_{\mathrm{int;full}}}
\end{equation}
\noindent
where $t_{\mathrm{int;full}}$ is the integration time used in an FX- or FFT-correlator. Substituting the above relation into Equation~\ref{eq:gaincov}, we can write the variance in estimated gains in terms of the SNR of the visibility matrix that the gains calibrate. 

\begin{align}
    \sigma_g^2 &= (\mathrm{SNR})_{\mathrm{reduced}}^{-2} \; \left(\matr{A}^{\dagger}\matr{A}\right)^{-1}_{(\mathit{N} \times \mathit{N})} \nonumber \\
    &= (\mathrm{SNR})_{\mathrm{full}}^{-2} \; \left(\displaystyle \frac{\mathit{N}}{2\mathit{p}\log{\mathit{N}}}\right) \; \frac{\mathit{t}_{\mathrm{int;full}}}{\mathit{t}_{\mathrm{cal}}} \; \frac{1}{\mathit{N}/2} \nonumber \\
    &= (\mathrm{SNR})_{\mathrm{full}}^{-2} \; \left(\displaystyle \frac{1}{\mathit{p}\log{\mathit{N}}}\right) \; \frac{\mathit{t}_{\mathrm{int;full}}}{\mathit{t}_{\mathrm{cal}}}
\end{align}
\noindent
When $t_{\mathrm{cal}}$ is large, the SNR of the reduced visibility matrix is larger and the corresponding gain variance is smaller. Assuming the case where the interval of calibration is same as the integration time in the FFT correlator and substituting the above equation into Equation~\ref{eq:visvar} we get the following for the variance in redundant visibilities:

\begin{equation}
    \label{eq:visvar_lowcadcal}
    (\mathrm{SNR})_{\mathrm{unique}}^{-2} \approx \displaystyle \frac{(\mathrm{SNR})_{\mathrm{full}}^{-2}}{\mathit{N}_{\alpha}} \left[1 + \left(\frac{2}{\mathit{p}\log{\mathit{N}}}\right)\right]
\end{equation}
\noindent
When redundant-baseline calibration is performed with the full visibility matrix, the gain covariance terms in Equation~\ref{eq:visvar} are around an order of magnitude smaller than the variance term so we drop the third term for clarity. In large arrays, where an FFT-correlator architecture would be preferable to an FX-correlator, the contribution of gain variance (second term) to the variance in calibrated redundant visibilities is smaller than the thermal noise in the measure visibilities (first term). Hence, the precision in gain variance obtained from using an $\mathcal{O}(N\log{N})$ calibrator is sufficient for the purpose of calibrating voltages for an FFT-correlator.

\subsubsection{Variance in Redundant Visibilities Calibrated using Subset Redundant Calibration}

In subset redundant calibration, the integration time does not change between the calibrator and the FFT-correlator. Consequently, the SNR of measured visibilities is the same for both data sets. However, the number of baselines used in the calibration process is lower, resulting in a higher variance in the estimated gains. For a calibrator that scales as $\mathcal{O}(N\log{N})$ with array size, the relationship between gain variance and number of baselines is given by Equation~\ref{eq:gainvar_subredcal} where the proportionality constant is the SNR of visibilities. Substituting Equation~\ref{eq:gainvar_subredcal} into Equation~\ref{eq:visvar} we get:

\begin{equation}
    \label{eq:visvar_subredcal}
    (\mathrm{SNR})_{\mathrm{unique}}^{-2} \approx \displaystyle \frac{(\mathrm{SNR})_{\mathrm{full}}^{-2}}{\mathit{N}_{\alpha}} \left[1 + \frac{2}{\mathit{p}\log{\mathit{N}}} + \frac{2}{\mathit{N}_{\alpha}} \sum\limits_{\mathit{(i,j)} \in \alpha} \rho_{\mathit{g}_\mathit{i}\mathit{g}_\mathit{j}}\right]
\end{equation}
\noindent
The gain covariance terms $\rho_{g_ig_j}$, for gains estimated using subset redundant calibration, are sometimes comparable to the gain variance terms. When the pre-factor (p) is small, the amplitude of the covariance scales similarly to the gain variance with increase in size of the array. This effectively doubles the contribution of the variance term in the above equation. When the pre-factor is large, the covariance is only a small fraction of the gain variance and can be ignored. Overall, the contribution of the second and third terms in the above equation is much smaller than the thermal noise in measured visibilities for any reasonably large array. Hence, the precision in gains estimated using subset redundant calibration is also sufficient for the purpose of calibrating voltages for an FFT-correlator. 

Figure~\ref{fig:scale_ants} might give a misleading impression that traditional redundant-baseline calibration is superior to either of the reduced redundant-baseline schemes, by providing gains that have orders-of-magnitude lower variance. However, this additional gain precision comes at a high computational cost and might not be necessary for large arrays where the contribution of gain variance to the overall scatter in redundant visibilities is only a small fraction of the thermal noise.

\subsection{Bias in Estimated Variables}

\begin{figure}
    \centering
    \includegraphics[width=0.75\linewidth]{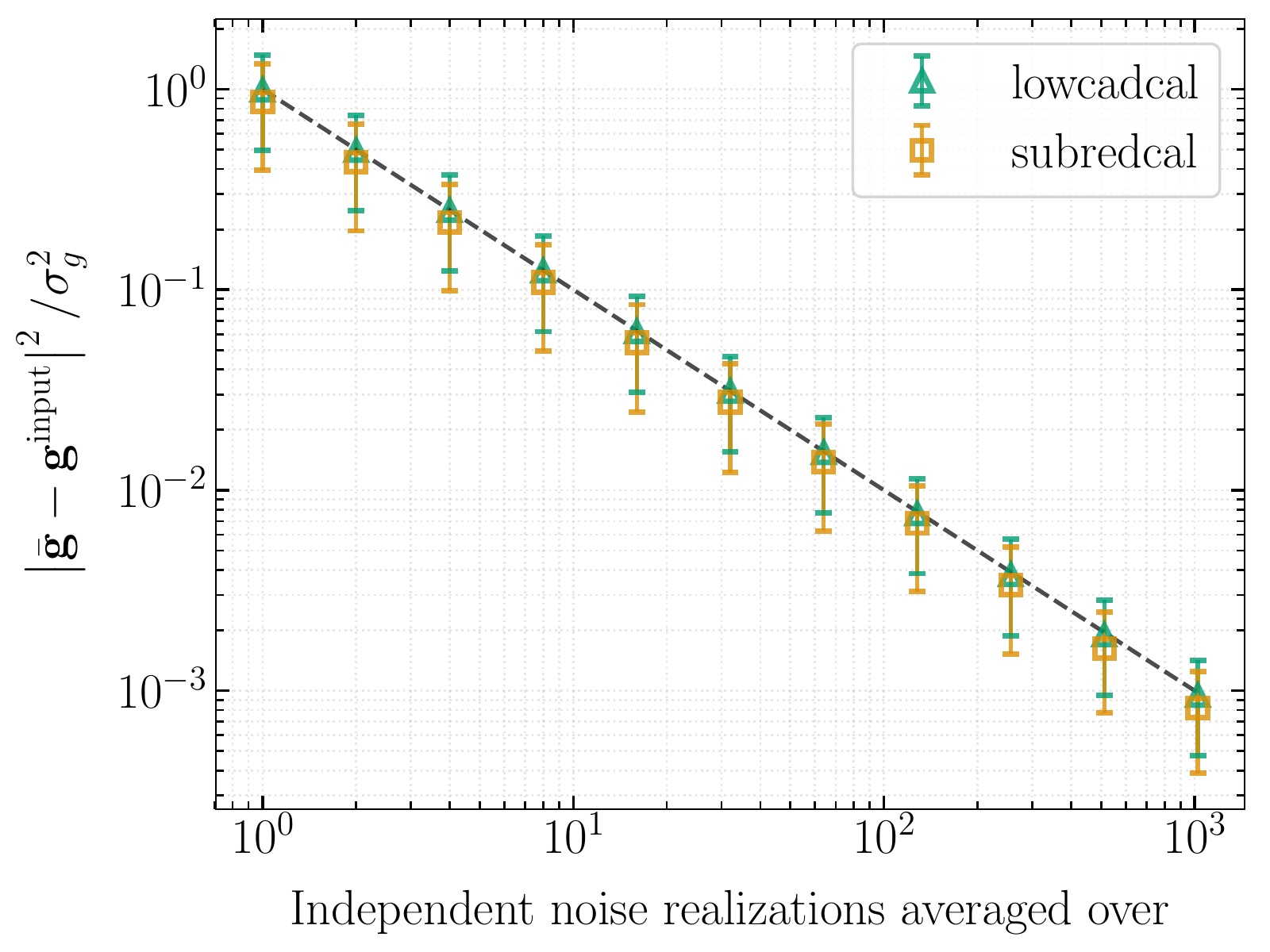}
    \caption{Deviation of averaged gains from their true value for low-cadence calibration (\texttt{lowcadcal}; green triangles) and subset redundant calibration (\texttt{subredcal}; orange squares). The markers are the result of a simulation with the same underlying input variables (both gains and unique visibilities) but different noise realisations of the measured visibilities. The x-axis shows the number of such simulations that the variables have been averaged over. The errorbars reflect the antenna-to-antenna variation. The dashed line represents the trend expected when the estimated gains differ from truth only within Gaussian noise. Both low-cadence calibration and subset redundant calibration yield unbiased gain solutions.
    \label{fig:gain_vis_bias}}
\end{figure}

The gains estimated using either reduced redundant-baseline calibration process are unbiased estimates of the true value. This can be verified through simulations that have constant underlying gains and visibilities, and different realisations of the noise in the measured visibilities. Averaging the solutions obtained over multiple such simulations decreases the noise in the estimated parameters and can expose an underlying bias, if any. Figure~\ref{fig:gain_vis_bias} shows the deviation in averaged gains from the input true gains, normalised by the variance expected in the gains. The errorbars represent the antenna-to-antenna variation which also averages down. Gains that have been averaged over $N_{\mathrm{sim}}$ independent noise realisations have a factor of $1/N_{\mathrm{sim}}$ smaller deviation, which is expected when the estimated gains differ from truth only within the Gaussian random noise in the measurement. This trend is marked by the dashed black line in the figure. A bias in gains, within the precision exposed by averaging down noise, would have resulted in a deviation from this trend.

%The visibilities computed by the FFT-correlator $\mathbf{V^{unique}}$ have a low bias that is exposed when averaging visibilities that have been calibrated by gains with a high variance. This bias can be explained by the gain covariance term $\rho_{g_ig_j^*}$ in Equations~\ref{eq:visvar_lowcadcal} and~\ref{eq:visvar_subredcal}. The first and second terms in the variance of calibrated visibilities are Gaussian in nature and average down over multiple noise realisations. However, the covariance in gains is a function of the baselines used in redundant-baseline calibration and does not average down over noise realisations. This non-Gaussian term becomes evident as a bias in the calibrated visibilities when the contribution of the first two terms is lowered by averaging over noise realisations.

The visibilities computed by the FFT-correlator $\mathbf{V^{unique}}$ are also unbiased when the visibility matrix used for reduced redundant-baseline has a high SNR. The deviation of the calibrated and redundant-baseline averaged visibilities from the simulated input, averages down according to the trend expected for Gaussian random noise. However, empirically, we find that the calibrated visibilities are sometimes biased when the variance in estimated gains is larger than $\sim$$10^{-5}$. It is possible that there is a low level of bias, that is not exposed by averaging over $N_{\mathrm{sim}}=4096$ simulations, in the visibilities calibrated with lower variance gains. To clarify, the gain variance plotted in all the figures in this paper has been normalised by the thermal noise in the visibilities used for calibration. The absolute gain variance in all these simulations is less than the empirical value above which we find a bias in the calibrated visibilities.

Assuming this threshold in gain variance is real, it translates to a requirement of an SNR $\gtrsim 20$ in the measured visibilities for a 300 antenna array when implementing low-cadence calibration, and higher when implementing subset redundant calibration. This minimum SNR requirement decreases as $1/\sqrt{N}$ with increase in array size and may not be an issue for large-N arrays.

This requirement of a minimum SNR in the reduced visibility matrix could be due two possible reasons-- (a) if the assumption that the non-linear equations of redundant-baseline calibration can be optimised by solving the linearized system of equations, does not hold at this limit or (b) if the product of gains form an asymmetric distribution about their mean value, and do not average down. If the former is true, a linearized solver of Equation~\ref{eq:redcal} would result in a biased solution when the noise in measured visibilities is high. However, we find that linearization based on Taylor expansion of variables and \texttt{omnical} always result in unbiased gains and visibilities irrespective of the SNR in the measured visibilities.

The more favourable explanation seems to be latter. As shown by \citet{odonoughue_and_moura_2012}, the probability distribution function (PDF) of the product of two complex Gaussian random variables is not a simple Gaussian distribution. Moreover, the resulting distribution can be asymmetric if two complex random variables are drawn from a non-zero mean Gaussian, as is the case with antenna gains. 

\subsection{\texorpdfstring{$\chi_r^2$}{Chisq} of Estimated Gains and Visibilities}

\begin{figure}
    \centering
    \includegraphics[width=0.75\linewidth]{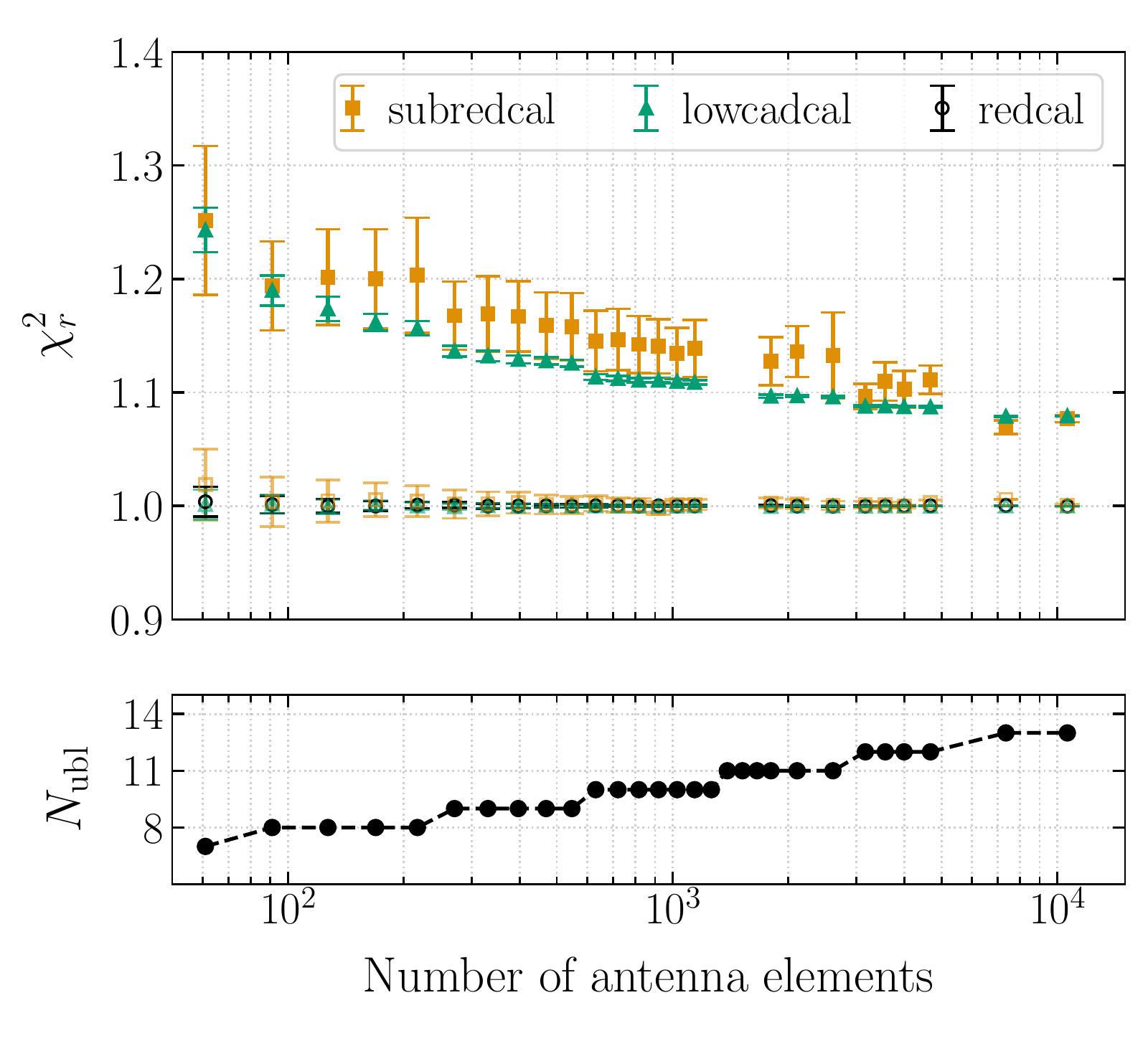}
    \caption{The upper panel shows the goodness-of-fit of the gains and unique visibilities that have been estimated using either reduced redundant-baseline calibration scheme. The hollow points (orange and green) show the fit of the estimated gains to the visibilities computed by the calibrator $\mathbf{V^{reduced}}$ and is close to one as expected. The hollow black points show the fit of gains estimated using full redundant-baseline calibration to $\mathbf{V^{full}}$. The solid orange and green points show the $\chi_r^2$ of the fit to the full visibility matrix $\mathbf{V^{full}}$ and these are also not far from one. The associated errorbars show the range of $\chi_r^2$ obtained over multiple simulations. The jump in $\chi_r^2$ for low-cadence calibration arises from a non-uniform increase in the calibrator size. The number of unique baseline-types that are cross-correlated by the calibrator are shown in the lower panel as a proxy for the size of the calibrator.
    \label{fig:chisq_scale_ants}}
\end{figure}

As was discussed in Section~\ref{sec:redredcal:scatter}, the gains computed using a reduced redundant-baseline calibration scheme, are estimated from the visibility matrix $\mathbf{V^{reduced}}$ but applied to a different visibility matrix $\mathbf{V^{unique}}$. The redundant-baseline calibration process is designed to minimise the $\chi_r^2$ between the visibility matrix used for calibration and the estimated variables. In the case of reduced redundant-baseline calibration, this is the $\chi_r^2$ evaluated between the estimated gains, model visibilities that are discarded by the calibrator and the reduced visibility matrix. In Figure~\ref{fig:chisq_scale_ants}, this $\chi_r^2$ is represented by hollow orange and green markers for subset redundant calibration and low-cadence calibration respectively, and has the expected value of one.

The $\chi_r^2$ estimated using gains computed by the reduced redundant-baseline calibration process, unique visibilities computed by the FFT-correlator and the full visibility matrix is a better metric to assess the effectiveness of the calibration process. This $\chi_r^2$ is represented by the solid orange and green points in Figure~\ref{fig:chisq_scale_ants}, for subset redundant calibration and low-cadence calibration respectively. The $\chi_r^2$ for array sizes with number of antennas in the range 1500--5000 is estimated over only 16 simulations, rather than 256, due to long simulation times. For the same reason, the $\chi_r^2$ for the two largest array sizes with $N>5000$ antennas is estimated using only 2 simulations.

Both the reduced redundant-baseline calibration schemes yield a $\chi^2_r$ that is close to one, indicating that the estimated parameters are a reasonable fit to the full visibility matrix. The larger $\chi^2_r$ of subset redundant calibration as compared to low-cadence calibration, and the larger range of $\chi_r^2$ obtained over multiple simulations (represented by the errorbars) is a direct consequence of the higher gain variance in the former compared to the latter. The non-smooth trend in the $\chi_r^2$ of low-cadence calibration is due to a non-uniform increase in the size of the calibrator used in simulation. The jump in $\chi_r^2$ of low-cadence calibration is correlated with the increase in number of unique-baseline types processed by the calibrator, because this leads to a jump in the integration time available to each cycle of correlation in the calibrator. 

\begin{figure}
    \centering
    \includegraphics[width=0.75\linewidth]{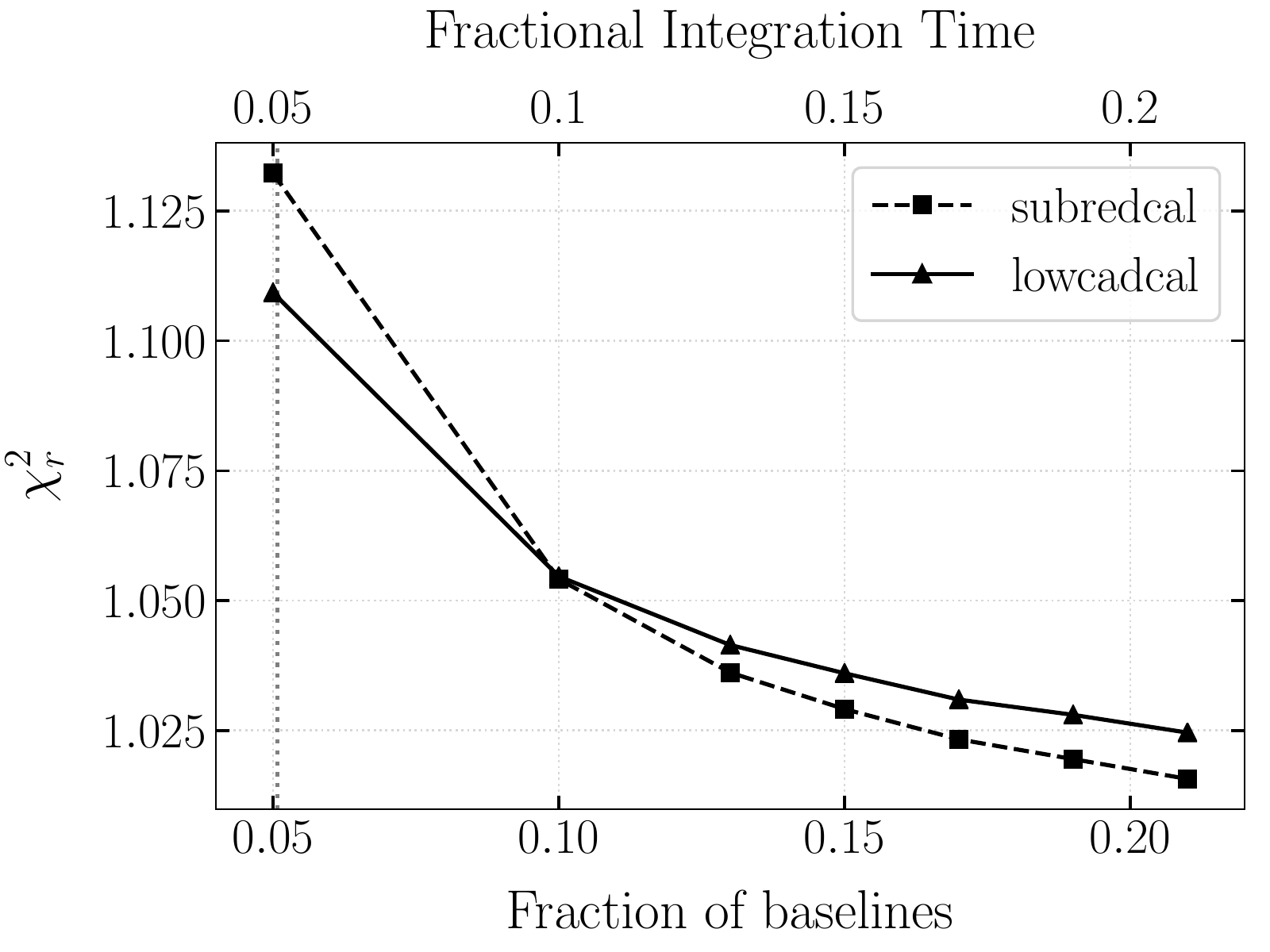}
    \caption{Reduced chi-squared as a function of the computational resources allocated to a reduced redundant calibrator for a 331\,antenna array. The size of the calibrator is marked in the fraction of baselines that can be processed for subset redundant calibration (bottom axis) or the fraction of integration time that can be spent on each baseline for low-cadence calibration (top axis). The vertical dotted line marks the size of a $\mathcal{O}(N \log{N})$ calibrator. At larger calibrator sizes, subset redundant calibration yields better gain estimates than low-cadence calibration.
    \label{fig:low_v_sub}}
\end{figure}

Figure~\ref{fig:low_v_sub} compares the performance of low-cadence calibration to subset redundant calibration for calibrators of various sizes operating a fixed array size. When the resources allocated to the calibrator are sufficient to cross-correlate exactly $N\log{N}$ baselines, which was the assumption throughout this paper, low-cadence calibration yields better gain estimates. This calibrator size is marked by a vertical dotted line in the figure. However, at larger calibrator sizes, subset redundant calibration performs better than low-cadence calibration. This is because the subset redundant calibrator preferentially spends time on baselines that have a higher constraining power and integrates them to a higher SNR than a low-cadence calibrator of a similar size. 

% HERA Correlator System
\chapter{HERA Correlator System}
\label{chp:HERACorr}

The Hydrogen Epoch of Reionization Array (HERA; Section~\ref{chp:Intro:sec:HERA}; \citealt{Deboer_et_al_2017}) is being built to detect and characterize the power spectrum of neutral hydrogen between redshifts $5 < z < 28$ when the first stars, galaxies and quasars started forming in the universe. After Recombination and emission of the Cosmic Microwave Background Radiation, the universe was transparent and predominantly composed of neutral hydrogen that can undergo a spontaneous spin-flip transition and emit a weak signal at a wavelength of 21\,cm. The first luminous structures that formed during Cosmic Dawn ionized this neutral hydrogen, creating spatial and temporal structure in the otherwise homogeneous signal. By characterizing the evolution of the redshifted 21\,cm signal from Cosmic Dawn through the Epoch of Reionization (EoR), HERA aims to provide key insights into the astrophysical phenomena that governed the early universe.

Foreground sources, that emit in the same spectral range as the redshifted 21\,cm signal, are around 5\,orders of magnitude brighter than the EoR signal itself \citep{deOliveiraCosta_et_al_2008}. The methodology for statistical separation of these foregrounds from the EoR signal, which is crucial for any EoR experiment, defines the design, layout and signal processing of HERA. \citet{Parsons_et_al_2012} outline a per-baseline, delay transform based approach for foreground avoidance that has already been used by the Donald C. Backer Precision Array for Probing the Epoch of Reionization (PAPER; \citealt{Parsons_et_al_2010}) with success \citep{Ali_et_al_2015, Kolopanis_et_al_2019}. This approach relies on spectral and spatial smoothness of the instrumental response, and is the defining metric by which the layout, antenna feeds, analog signal chain and the digital correlator have been designed.

HERA is currently under construction in the South African Karoo Radio Astronomy Reserve, sharing premises with the MeerKAT Telescope~\citep{Jonas_2009_meerkat} and C-Band All Sky Survey (CBASS; \citealt{King_2010}) projects. Fully constructed, HERA will have 350 non-tracking parabolic dishes built in the split-core hexagonal close packed manner laid out in~\citet{Dillon_and_Parsons_2016}. The layout has been designed to maximize sensitivity on the short baselines which contribute most to the EoR power spectrum when using the delay transform approach. In addition, the split-core configuration and the outrigger antennas provide higher uv-coverage for imaging and map-making. The parabolic dish, the Vivaldi feeds~\citep{Fagnoni_et_al_2019} and the analog signal path have been designed to minimize reflections within the delay window where the EoR is expected to be detected. Figure~\ref{fig:HERA} shows the current state of the array, with 200 of the dishes built and commissioned.

Generally, using an interferometer for imaging or building a power spectrum involves an intermediate step of computing antenna cross-correlation products or \textit{visibilities}. For a given pair of antennas, the visibility function encodes the interference fringe pattern due to a geometric delay in signal propagation from one antenna to the other, and represents the key difference between observing with a single antenna and an interferometer. The complex visibility, which is a function of the baseline vector between an antenna pair ($\mathbf{b}$) and spectral frequency ($\nu$) can be theoretically modelled as:

\begin{equation}
    \text{V}(\mathbf{b},\nu) = \int\limits_{\text{sky}} I(\mathbf{\hat{s}},\nu) \; B(\mathbf{\hat{s}},\nu) \; \text{exp}\left[-2\pi i \frac{\nu}{c} \mathbf{b}.\mathbf{\hat{s}}\right] \mathrm{d}\Omega
\end{equation}
\noindent
where $I(\hat{s},\nu)$ is the intensity of a source in the sky as function of direction and $B(\hat{s}, \nu)$ is the representative primary beam of all antennas in the array. The exponential term represents the interferometric fringe pattern which is analogous to interference fringes in classical wave optics. As the Earth rotates under a constant radio source, the projection of the baseline vector in the direction of the source, $\mathbf{b}.\mathbf{\hat{s}}/\lambda$ changes, causing the fringe pattern.

Practically, the visibility of a given pair of antennas at a particular frequency, is proportional to the level of signal correlation between those two antennas and is encoded by the cross-correlation product of the time-dependent voltage signal at that frequency. The digital \textit{correlator} system computes the complex visibility for every pair of antennas in the array within narrow spectral bins. The set of cross-correlation products that comprise of all antenna pairs in the array and all frequency channels, spanning the bandwidth of observation, is called the \textit{visibility matrix}. In addition to computing the visibility matrix, the correlator system for HERA also provides functionality for phase switching at antenna feeds (to minimize cross-talk), fringe stopping (to minimize decorrelation due to long integration times) and baseline dependent averaging (for managing the otherwise large datarate).

\section{Architecture Overview}

A correlator system performs all the functions necessary to build the visibility matrix from raw analog voltages of each antenna. The design of the HERA correlator follows the ``FX-architecture" which has become a standard for many modern correlators \citep{Ord_et_al_2015, Kocz_et_al_2015, Foley_et_al_2016, Bandura_et_al_2016, Nita_et_al_2016, Primiani_et_al_2016, Hickish_et_al_2018}. A first stage converts time-domain digitized antenna voltages to frequency-domain using an FFT-based algorithm and forms the ``F-stage" or F-engine of the correlator. The second stage computes cross-correlation products of all antenna pairs in the array within each spectral bin from the first stage, forming the ``X-stage" or X-engine. For arrays with a large number of antennas, the FX architecture is computationally more efficient than the older XF architecture in which the order of the two stages are reversed. The computational efficiency comes from performing only $\mathcal{O}(N^2)$ operations in the X-stage of the FX architecture as compared to $\mathcal{O}(FN^2)$ operations in the X-stage of the XF architecture. 

The large number of redundant baselines in the layout of HERA also support the recently proposed FFT architecture \citep{Tegmark_and_Zaldarriaga_2009, Tegmark_and_Zaldarriaga_2010, Foster_et_al_2014, Thyagarajan_et_al_2017} where the X-stage is implemented using an FFT algorithm. This decreases the number of computations in the X-stage to $\mathcal{O}(N\log_2{N})$, making the FFT architecture more computationally efficient than the FX architecture for large arrays. However, an FFT based X-stage requires calibrated antenna voltages for building visibilities without signal loss~\citep{Beardsley_2017, gorthi_et_al_2021}. This requirement could make the data analysis for detecting the EoR signal more complex, since extraction of the signal from bright foreground sources demands precision calibration~\citep{Liu_et_al_2010,barry_et_al_2016}. For this reason, the implementation-wise simpler FX architecture was chosen for the correlator design with the FFT implementation left as a research exercise.

\begin{figure}[htbp]
    \centering
    \includegraphics[width=\linewidth]{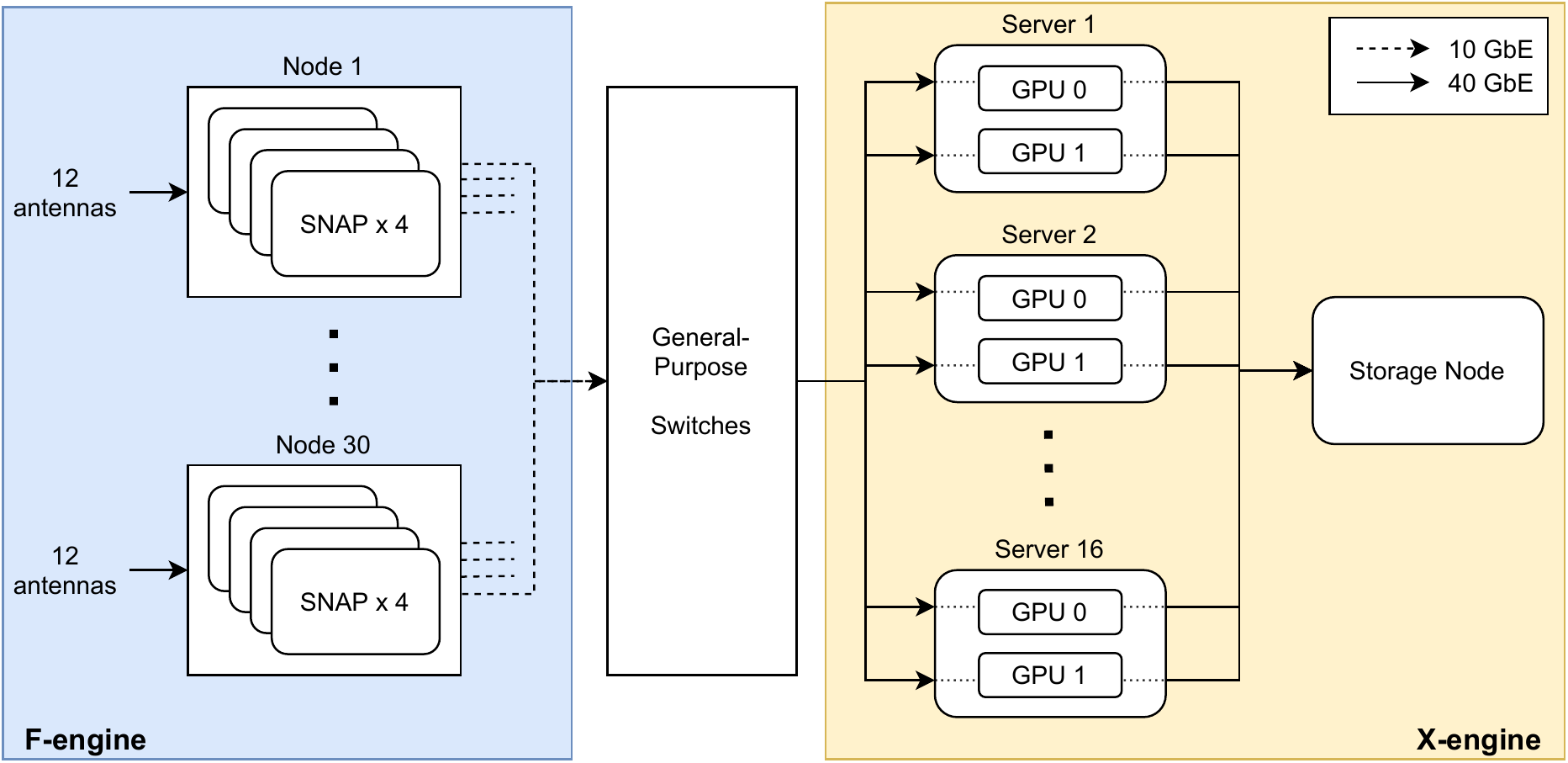}
    \caption{Overview of the HERA correlator pipeline built in the FX-architecture. The F-engine is implemented on custom-designed FPGA boards called the Smart Network ADC Processor (SNAP). A set of 4 SNAP boards serve 12 antennas and reside in a node enclosure. A total of 120 SNAP boards distributed across 30 nodes form the F-engine of the correlator. 32 Nvidia GPU cards installed on 16 servers perform the antenna cross-correlation and act as the X-engine. A separate server is used for file writing and storage. All components are connected via off-the-shelf 10\,Gb and 40\,Gb Ethernet switches.}
    \label{fig:hera_corr}
\end{figure}

Figure~\ref{fig:hera_corr} shows an overview of the layout of the FX correlator for HERA. The F-stage is implemented on custom-designed FPGA boards called the Smart Network ADC Processor (SNAP). Each SNAP board has enough resources to process three dual-polarization antennas or six inputs. A group of four SNAP boards, serving a total of 12 antennas, are placed within a node enclosure for cooling and radio frequency interference (RFI) shielding. Node enclosures, that are scattered across the array, also house Post-Amplification Modules (PAMs) that provide analog input to the SNAP boards, power distribution units for remote powering off/on of each node and temperature and humidity sensors for monitoring. Each node is placed at the center of the 12 antennas that it serves, to minimize signal attenuation in the cable carrying analog RF signal from the antenna feed to the receivers. Even though only 117 SNAP boards are required to support the 350 antennas of the array, three additional boards with dummy inputs are used to maintain symmetry in the network layout. A total of 120 SNAP boards, distributed across 30 nodes, serve as the platform for the F-stage of the HERA correlator.

The X-stage of the correlator, responsible for computing the cross-correlation products, is built on 16 servers hosting 2 GPU cards each. Each server processes $1/8^{\mathrm{th}}$ the observing bandwidth by splitting the non-lapping frequency channels. The 16 servers are split into two sets of 8 servers and the data from the SNAP boards is time-multiplexed between the two sets. The rearrangement from a per-antenna distribution of the signal on the F-engine side, to a per-frequency distribution on the X-engine side, is performed by general-purpose, off-the-shelf Ethernet switches. A fibre optic patch panel connects the SNAP boards on-site to the X-engine servers located in the Karoo Array Processor Building (KAPB). Section~\ref{sec:network} decribes the network configuration in more detail. A storage node hosting a 50TB hard-drive is used to write the complete visibility matrices to files in a custom HDF5-based file format~\citep{Hazelton_et_al_2017} and store them for calibration and post-processing.

\section{Design Specifications}

The design specifications for the HERA digital correlator system are shown in Table~\ref{tab:corr_specs}. The frequency range for observation was chosen to be 50--250\,MHz to probe the large redshift range of $4.7 \leq z \leq 27.4$ \citep{Deboer_et_al_2017} where the EoR is expected to be detected. Rather than using a local oscillator to down-convert the signal, the analog voltages are digitized by Nyquist oversampling at 500\,MHz and the portion of the signal corresponding to 0--50\,MHz is dropped by the F-engine after channelizing. 

\begin{table}[htbp]
\centering
\setlength{\extrarowheight}{3pt}
\begin{tabular}{c|c}
    \textbf{Property}            & \textbf{Specification} \\ 
    \hline
    Receiver Bandpass            & 30 -- 250\,MHz         \\
    Sampling Frequency           & 500\,MHz               \\
    Number of channels           & 8192                   \\
    Frequency resolution         & 30.5\,kHz              \\
    Channels retained            & 6144                   \\
    Effective Bandpass           & 47 -- 234\,MHz         \\
    Shortest Integration time    & 2 seconds              \\
    Longest Integration time     & 16 seconds             \\
    \hline
\end{tabular}
\caption{HERA Digital Correlator Specifications}
\label{tab:corr_specs}
\end{table}

\subsection{Frequency Resolution}

The number of channels computed by the F-engine of the correlator, which dictates the frequency resolution of cross-correlated data, was chosen to limit decorrelation in interferometric fringes due to a finite bandwidth of each frequency channel. The Fourier transform of the frequency response of each channel forms an envelope to the interferometric response. When the channel width is large, a narrow envelope could result in decorrelation of fringes. The decorrelation factor due to a finite bandwidth of each frequency channel can be computed as:
\begin{equation}
\label{eq:decorr_freq}
    f_{D} = \Delta \nu \; \frac{b}{c} \sin{\theta}
\end{equation}
\noindent
where $\Delta \nu$ is the channel width, $b$ is the East-West distance between antennas and $\theta$ is the angle of observation, measured from zenith. For the static antennas that make up HERA, the limiting angle is determined by the field-of-view of antennas which is $90^{\circ}$. To limit the decorrelation on the longest baselines to 10\%, the maximum channel width that can accommodated is $\sim$35\,kHz. This motivation drives the design specification of 8192 channels across 0--250\,MHz which results in a frequency resolution of 30.5\,kHz.

\subsection{Integration time}

Similarly, the integration time or the time period for which cross-correlation products computed in the X-engine are averaged, is also motivated by decorrelation in fringes. The time period for which a visibility can be averaged depends on the resolution of the telescope and the rate of Earth's rotation. Any integration time results in a small amount of decorrelation in the fringes of the interferometric response. The decorrelation factor due to a given integration time can be estimated as:
\begin{equation}
\label{eq:decorr_time}
f_{D} = \frac{\omega_e \; t_{\textrm{int}}}{\sin^{-1}\left(\lambda/d\right)}
\end{equation}
\noindent
where $\omega_e$ is the Earth's rotation velocity and $t_{\textrm{int}}$ is the integration time. A shorter integration time is, hence, preferable for limiting the signal decorrelation, especially on long baselines. However, a small integration time also implies a large data rate and more data stored to disk. For example, with an integration time of 100\,ms, the total data rate into the storage node(s) is approximately 7 Tbps resulting in nearly 40\,PB of data for a single observing session (12 hours). To limit the data rate and the data volume, it is preferable to integrate the computed visibilities over a longer duration. HERA uses two techniques to increase the integration time without compromising too much on signal decorrelation: (a) fringe rotation or beam phasing and (b) baseline dependent averaging.

\subsection{Fringe Rotation}

Fringe rotation is a mechanism to reduce the fringe oscillations in the visibilities computed by the correlator. In general, fringe oscillations occur due to a changing component of the baseline vector between a pair of antennas in the direction of the source, caused by the rotation of the Earth. The rate at which the fringe oscillates, called the fringe frequency, depends on various factors like the location of the source being observed in the sky, the location of the telescope, the distance between the pair of antennas, their orientation and the Earth's rotation velocity. For a given baseline vector and latitude, the fringe frequency is highest for a source at zenith and lowest for a source on the horizon. Baseline vectors oriented in the East-West direction have larger fringe rates than baselines oriented in the North-South direction. 

Fringe oscillations that prevent large integration times, can be minimized by mixing the observed voltage signal with a predicted fringe frequency. If the predicted value is close to the actual fringe frequency, the integration time can be increased without affecting decorrelation too much. (Note that the integration time still cannot be indefinitely large for telescopes located at non-zero latitudes because the North-South component of Earth's rotation still causes fringe oscillations.) For HERA, the fringe frequency is predicted for each antenna by using the baseline vector between that antenna and the array center and fixing the \textit{fringe stopping phase center} to the local zenith where the fringe rate is the highest. The fringe frequency is predicted for the full length of the integration time window, with samples spaced 30\,ms apart. This predicted signal is used to multiply the digitized voltage signal in the F-engine prior to cross-correlation in the X-engine. This mechanism of fringe rotation allows the integration time to be increased to up to 32 seconds for the shortest baselines in the array.

\subsection{Baseline Dependent Averaging}

\begin{table}[htbp]
    \centering
    \begin{tabular}{c|c}
    Integration time & Number of baselines \\
    \hline
    2 sec    &   385    \\
    4 sec    &   1533   \\
    8 sec    &   7168   \\
    16 sec   &   52339  \\
    \hline
    \end{tabular}
    \caption{Distribution of baselines across the five different integration time bins used for implementing baseline dependent averaging as part of the HERA correlator system.}
    \label{tab:bda_dist}
\end{table}

Baseline Dependent Averaging, as the name suggests, involves using different integration times for baselines of different lengths. As can be inferred from Equation~\ref{eq:decorr_time}, a given integration time creates larger decorrelation in the visibilities of long baselines as compared to shorter baselines. This can also be understood in terms of the fringe frequency; long baselines have a higher fringe frequency and hence require shorter integration periods to avoid decorrelation. On the other hand, short baselines have relatively low fringe rates and can be integrated for a longer period of time, decreasing the data rate and the data volume stored. By adjusting the integration time of a visibility product, according to the length of the baseline vector between the pair of antennas forming that visibility, the data rate can be kept low while preventing decorrelation on long baselines.

For maximally redundant arrays like HERA, the number of short baselines always exceeds the number of long baselines. By allowing a large integration time of 16 seconds for these numerous short baselines, the effective data rate into the storage node can be brought down to nearly 15\,Gbps. The X-engine of the HERA correlator performs baseline dependent averaging by sorting antenna pairs into one of four integration time bins: 2 seconds, 4 seconds, 8 seconds and 16 seconds. The motivation for this integration time quantization, the exact distribution of baselines into bins, and the practical effect of baseline dependent averaging on EoR detection will be discussed in an upcoming paper. For the purpose of designing the X-engine, the tentative distribution of baselines used is shown in Table~\ref{tab:bda_dist}. Only the longest baselines of 870\,m, formed by outrigger antenna pairs, are integrated for only 2 seconds while short baselines formed by antenna pairs that are adjacent to each other are integrated for 16\,seconds. As evident, most of the baselines fall into the longer integration time bins and drives the motivation for implementing baseline dependent averaging.

\subsection{Phase Switching}

The HERA correlator enables phase switching at the feed of antennas to minimize signal crosstalk in the receiver signal chain and analog-to-digital converters (ADCs). Due to the large number of antennas and a large observing duration, even a small signal contamination below the detection threshold, in the analog voltages, can lead to large unwanted components in the correlator output. A common solution for minimizing signal crosstalk and other spurious signals in the analog receivers, is multiplying the signal with a periodic function whose phase oscillates between $0^{\circ}$ and $180^{\circ}$. If the modulating functions for a pair of antennas are orthogonal within the time period of integration, the crosstalk introduced after switching is significantly minimized when the cross-correlation products are averaged. To minimize the pre-switching signal path, HERA performs phase switching at the feed of each antenna. The F-engine of the correlator sets the orthogonal modulation function for each antenna by generating Walsh functions \citep{Walsh_1923} from Hadamard matrices, and communicates this modulation pattern to the feed. After accounting for the appropriate time delay due to signal propagation, the digitized signal in the F-engine is demodulated using the same pattern.

\section{F-engine}

The F-engine of the HERA correlator is primarily responsible for digitizing and channelizing the analog voltage signal from each antenna, and building UDP packets that can be transported via 10\,Gb Ethernet links to one of the 16 servers in the X-engine. In addition to this functionality, the SNAP F-engine design enables communication with the FEMs via the I2C protocol and allows coarse or fine delay settings on each antenna for phasing the primary beam to zenith (fringe stopping). The different modules of the FPGA design that implement these features are described below.

\subsection{SNAP Boards}

The F-stage for HERA is built on custom-designed FPGA boards called the Smart Network ADC Processor (SNAP), which has been designed on similar lines as the other FPGA-based hardware boards developed and maintained by the Collaboration for Astronomy Signal Processing and Electronics Research\footnote{https://casper.berkeley.edu} (CASPER; \citealt{Hickish_et_al_2016}). The SNAP board was designed to be a low-cost solution to digitization and channelization close to the antenna feed, which is useful for minimizing signal attenuation in long RF cables.

\begin{figure}[htbp]
    \centering
    \includegraphics[width=0.6\linewidth]{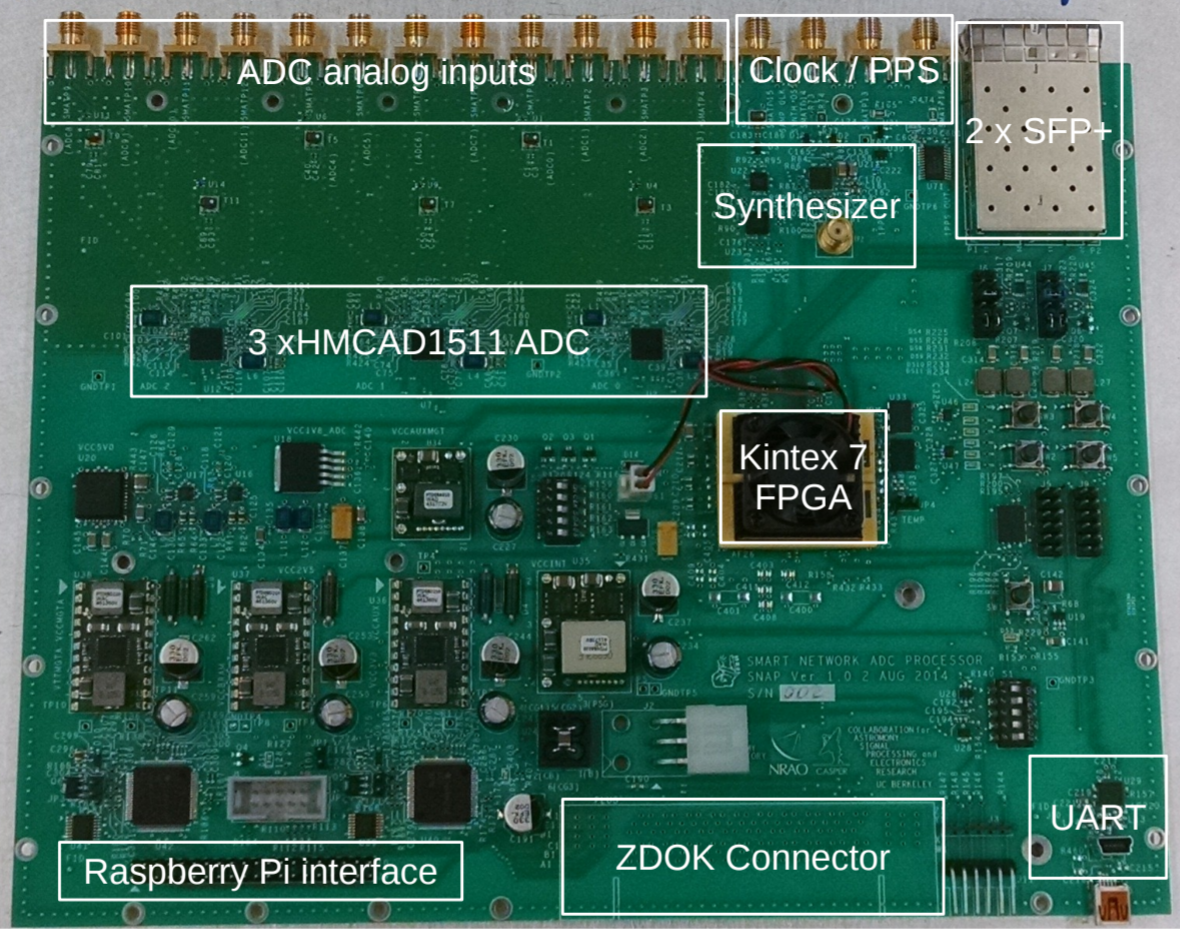}
    \caption{The Smart Network ADC Processor (SNAP) is a custom-designed FPGA board for low-cost digitization and channelization close to the feed of an antenna. It hosts three ADCs that can sample at up to 1000\,Msps, two 10\,Gb Ethernet ports and the Xilinx Kintex-7 FPGA among other components.}
    \label{fig:SNAP}
\end{figure}

The SNAP board (Figure~\ref{fig:SNAP}) hosts three HMCAD1511 ADC chips that can each operate in single, dual or quad modes with sampling frequencies up to 1000, 500 and 250\,Msps respectively, producing an 8\,bit output in each mode. The three ADC chips can support a maximum of 12 inputs when sampling at 250\,Msps, 6 inputs when sampling at 500\,Msps and 3 inputs when sampling at 1000\,Msps. For HERA, since the operating bandwidth is 250\,MHz, the ADCs are used in dual mode and sample at the Nyquist rate of 500\,Msps.

In addition to the analog inputs, the SNAP also supports the input of an external reference clock signal and an external pulse-per-second (PPS) that is useful for maintaining clock synchronization across multiple boards. An on-board synthesizer chip, the Texas Instruments LMX2581, uses the reference clock to generate clock signals for the all the components on-board. The output from the FPGA can be routed to two 10\,Gb Ethernet ports in the form of SFP+ connectors. 

The central compute engine on the SNAP is a Xilinx Kintex-7 FPGA (model XC7K160T, with modifications available to support XC7K325T or XC7K410T). With 600\,DSP slices and 2.14\,MB RAM, the Kintex-7 FPGA is smaller than the Virtex-6 that the older CASPER hardware board called the Reconfigurable Open Architecture Computing Hardware (ROACH-2) housed. The choice of decreasing the compute capacity available on-board was intentionally made to keep the cost of the board low. A higher performance successor board, called the SNAP-2, hosts the Kintex Ultrascale FPGA which offers higher compute and memory bandwidth.

A bitfile that programs the FPGA can be uploaded via the Raspberry-Pi\footnote{https://www.raspberrypi.org/} interface on the SNAP. A 40\,pin ribbon cable connects the SNAP board to the GPIO pins on the Raspberry-Pi, and enables the CPU on the Raspberry-Pi to act as the communication interface between a user and the FPGA. The Kintex-7 FPGA also supports a Microblaze\footnote{https://www.xilinx.com/products/design-tools/microblaze.html} soft-core processor, which repurposes some FPGA resources to act as a CPU interface between the rest of the FPGA fabric and the user. The Microblaze interface is used as the primary means of programming and communicating with the SNAP boards for HERA. For debugging, communication with the FPGA can also be established via a JTAG connector or the UART interface.

\subsection{Clocking}

The SNAP board ADCs sample the input analog voltage signal at 500\,MHz and the FPGA is clocked at 250\,MHz. Both these clock signals are generated on the SNAP board by a synthesizer chip, the LMX2581, by using a reference clock signal at 10\,MHz. The reference clock is supplied by a White Rabbit\footnote{https://ohwr.org/project/white-rabbit} clocking network, which is a 1\,Gbps system that is independent from the 10\,Gbps data network. The White Rabbit also generates a pulse-per-second (PPS) signal which synchronizes the 120 SNAP boards across the array.

\subsection{FPGA Design}

\begin{sidewaysfigure}
    \centering
    \includegraphics[width=\linewidth]{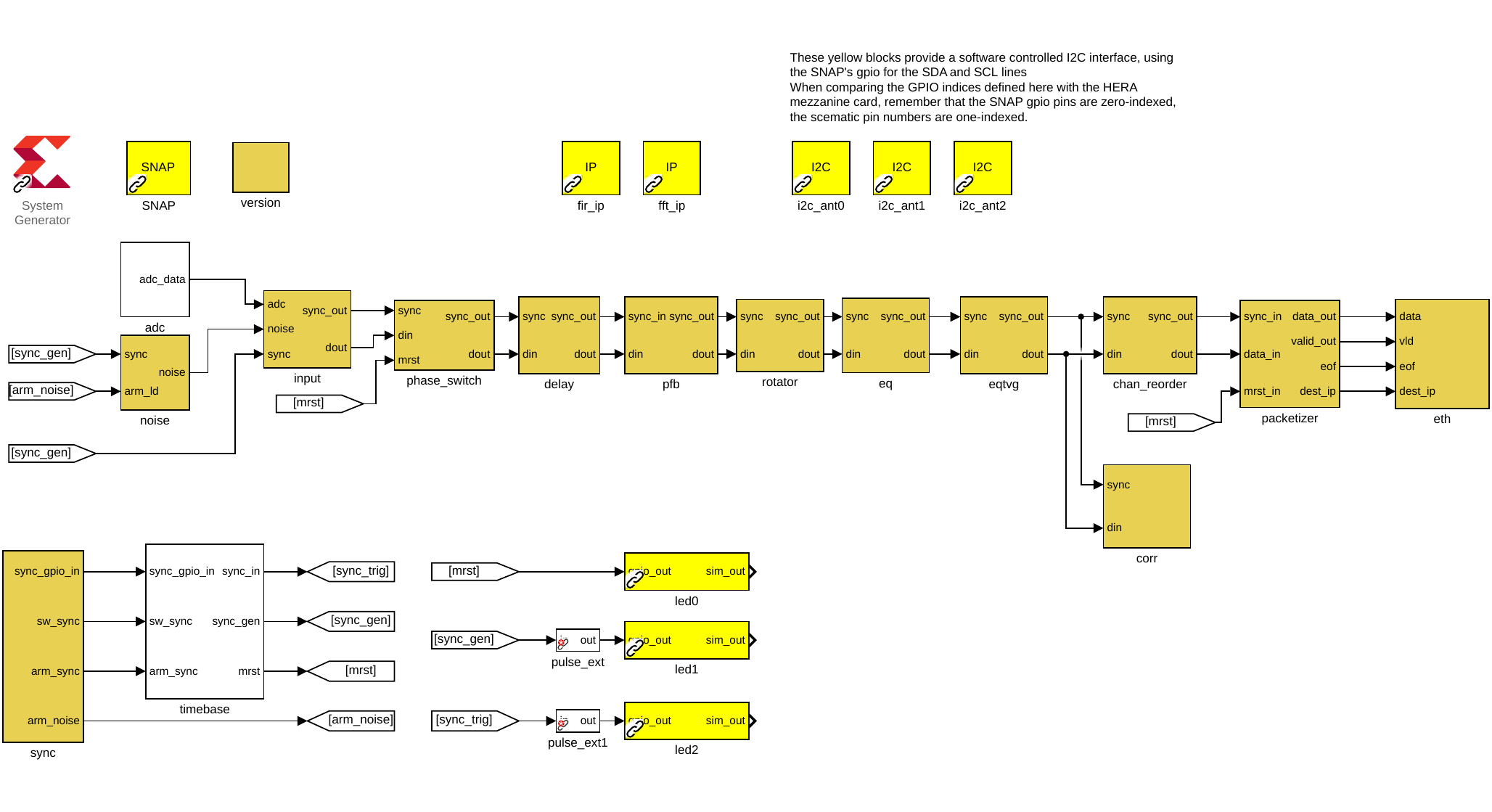}
    \caption{The MATLAB$^{\tiny \text{\textregistered}}$ Simulink$^{\tiny \text{\textregistered}}$ design that programs the Xilinx$^{\tiny \text{\textregistered}}$ Kintex-7 FPGA on SNAP boards, forming the F-stage of the HERA correlator. See text for details about the functionality of each block.}
    \label{fig:fpga_design}
\end{sidewaysfigure}

Figure~\ref{fig:fpga_design} shows the MATLAB$^{\tiny \text{\textregistered}}$ Simulink$^{\tiny \text{\textregistered}}$ design used to program the Kintex-7 FPGAs on the SNAP board. The System Generator for DSP$^{\text{\copyright}}$ module (red colored block in the top left corner), provided by Xilinx, enables the Simulink design to be directly compiled into a Vivado project to generate a bitfile that can be used to program the FPGA. The integration of the hardwired pins of ADCs, GPIOs, 10\,GbE etc. on the FPGA into the Simulink environment was done by members of the CASPER collaboration. These hardware interface modules are available as yellow blocks in the CASPER hardware library. The top-level SNAP block (beside the System Generator) informs the backend compiler that the hardware pin configuration of this design should map to that of a SNAP board.

\subsubsection{\texttt{adc}}

This block encapsulates the ADC CASPER block and generates a 96\,bit bus, corresponding to the twelve 8\,bit inputs. The SNAP ADCs Nyquist sample the input analog signal at 500\,MHz using inter-channel time-multiplexing to achieve the higher sampling rate. Each SNAP FPGA processes data from three dual-pol antennas or a total of six inputs. Two dummy inputs, set to constant zero, are added to the digitized inputs to make the output bus width to a power of two. 

\subsubsection{\texttt{noise}}

This block generates fixed-point Gaussian white noise that can be used in place of the antenna voltage signal, which is useful for debugging other blocks in the design.

\subsubsection{\texttt{input}}

This block allows the user to select between the on-board noise generator, the ADCs and a constant zero value for each of the 6 parallel data streams, corresponding to the three dual-polarization antennas connected to each SNAP board. This is useful for testing that each input is processed independently by the design.

\subsubsection{\texttt{phase\_switch}}

This block enables phase switching at the feed of the antenna for minimizing signal cross-talk in the Post Amplification Modules (PAMs) and in SNAP ADCs. Walsh functions for modulating/ demodulating the signal are stored in BRAMs on the FPGA and sent via dedicated GPIO pins to the Front-End Module (FEM) at the feed of the antenna, where the modulation takes place. The demodulation circuit accounts for the propagation delay in sending the modulation pattern via 30\,m CAT7 cables and receiving the RF signal back via 500\,m optic fiber cables. 

\subsubsection{\texttt{delay}}

This block allows the user to delay the time-domain voltage signal of each antenna by up to 7 clock cycles. At an FPGA clock frequency of 250\,MHz this allows a time delay of up to 28\,ns. Since the HERA antennas are always phased to zenith, this block does not have to correct for large geometric delays. This functionality is available only for diagnostic and debugging purposes.

\subsubsection{\texttt{pfb}}

This block performs a primary task of the F-stage of the correlator, i.e. converting the time-domain voltage signal into frequency-domain. A straight-forward Fourier transform on a time window of 8192 samples creates a frequency response where the individual channels taper off like a sinc function. To create a steeper roll-off, time domain signals are often conditioned by window functions like the Hann, Hamming or Blackman-Harris window etc. This can be achieved by a applying a Polyphase Filter Bank (PFB) to the input signal, prior to the Fourier transform.

In this design, a 4-tap Finite Impulse Response (FIR) filter with a Hamming window function is used for the PFB design. The CASPER \texttt{pfb\_fir\_real}\footnote{\url{https://github.com/casper-astro/mlib\_devel/blob/master/casper\_library/pfb\_fir\_real\_init.m}} module has been modified for this implementation. Due to RAM constraints on the FPGA, the PFB coefficients are set to 12\,bit fixed-point numbers rather than the standard 18\,bit coefficients. The bitwidth of the signal within the 4-tap PFB is allowed to grow from a 8\_7 fixed-point number at the input to a 12\_11 fixed-point number at the output to avoid excess quantization.

The 14-point Fourier transform is performed using a modified version of the CASPER \texttt{fft\_wideband\_real} block. The FFT is performed in 10 stages with bit-growth in the first six stages to an 18\_17 fixed-point number and dynamic shifting of bits in the last four stages with a round-to-even quantization rule. Incidentally, a previous version of the design with round-to-infinity quantization rule in the last four stages led to sharp downward spikes, that we termed ``icicles", in the auto-correlation spectra. These icicles are not evident in Gaussian white noise input but become very pronounced in presence of high-amplitude RFI.

\subsubsection{\texttt{rotator}}

Short baselines of 14.6\,m and a low observing frequency allow relatively long integration times on the visibilities of some baselines. However, since the sky-drift in this interval can be non-negligible, the primary beam of each antenna has to be phased to zenith periodically in the integration time window. This is achieved by multiplying the input signal with appropriate phase offsets, using the Xilinx CORDIC 6.0 blocks that are capable of rotating a fixed point complex number by a specified phase angle within the interval $(-\pi, \pi]$. A BRAM is used to store the phase angles for $2^{19}$ spectra or a duration of 17.17\,seconds with a resolution of $2^{10}$ spectra or 33\,ms. A master reset signal ensures that the first spectrum sent over Ethernet, to the X-engine, is at the boundary of a phasing window.

\subsubsection{\texttt{eq}}

This block, labelled short for equalization, re-quantizes the signal from a 18\_17 fixed-point number to 4\_3 fixed-point number. This is done to limit the output data rate at the Ethernet ports of the SNAP to be less than 10\,Gbps. To ensure that the 4\,bit representation does not result in signal-loss, the data can be optionally multiplied by per-antenna, per-channel coefficients before the requantization. These multiplicative constants can be set by the user at run-time through a BRAM. In practice, an automated script sets the value of these coefficients once at startup, by measuring the power level at the input of the SNAP. 

\subsubsection{\texttt{eq\_tvg}}

This block generates optional test-vectors which can be used to test the design circuitry downstream that is responsible for generating UDP packets and sending them to the IP address of the right server in the X-engine. The X-engine functionality can also be tested using this block, by setting all the 120 SNAP boards in the F-stage to transmit test vectors. A software BRAM can be programmed with 8\,bit (4\,bit real and 4\,bit imaginary) complex numbers for each of the four (one dummy) dual-polarization inputs and 8192 frequency channels.

\subsubsection{\texttt{chan\_reorder}}

This block changes the ordering of data into a format that is more convenient for constructing UDP packets. The data ordering at the input to this block, from the slowest to the fastest varying indices is 2 time samples, 4 antennas (one dummy), 2 polarizations, 16 blocks per spectra and 64 channels per block with 8 channels forming the data bus. The data is buffered and reordered to the format: 16 blocks per spectra, 4 antennas, 64 channels per block, 2 times and 2 polarizations. The slowest varying index of blocks per spectra allows the distribution of UDP packets to X-engine servers per frequency sub-band.

\subsubsection{\texttt{packetizer}}

This block creates UDP packets with 384 channels, 3 antennas, 2 time samples and 2 polarizations in a single packet. Including a 64\,bit header, which can be used to infer the timestamp of the packet and its contents, the UDP packets are 4616 bytes wide. These jumbo UDP packets are transmitted via the 10\,GbE SFP+ connectors to one of the 16 X-engine servers.

Of every block of 512 channels (64 consecutive clock cycles of 8 parallel channels), only 384 channels are sent to an X-engine. Effectively, this results in sending only $3/4^{\text{th}}$ of the bandwidth to the X-stage of the correlator. The \texttt{reorder} block also ensures that the channels that are not transmitted include only the bandpass rollover regions of the spectrum. The dummy fourth antenna in the design so far is also discarded at this stage.

For customization at a later stage, the header bits which specify the antenna numbers and channel numbers in a given UDP packet and the destination IP address for that packet are all read from software BRAMs that can be programmed by the user. A configuration file, that is read during startup, can be used to specify this information for each SNAP.

\subsubsection{\texttt{eth}}

This block is a wrapper around the CASPER \texttt{ten\_GbE} block. Data \textit{valid} and \textit{end-of-file} signals generated by the \texttt{packetizer} block define the clock boundaries of each UDP packet. The CASPER block wraps the 4616 bytes in each packet with a UDP frame for transmission.

\subsubsection{\texttt{corr}}

This block functions as an on-board correlator that can cross-correlator a single antenna pair at a time. A vector accumulator module is used to integrate the cross-correlation spectrum for a user-specified amount of time before writing the output to a software BRAM that can be read during run-time. This feature of the design is useful in monitoring observation runs during the commissioning phase of the array. 

An additional feature of this block is recording the maximum amplitude spectrum within a specified integration window. This is useful for monitoring sporadic time-dependent RFI pulses on-site which might not be captured by autocorrelation spectra which are time-averaged. To conserve RAM resources on the FPGA, this feature is only made available for autocorrelation spectra which are theoretically always real-valued. The BRAM which otherwise stores the imaginary part of the cross-correlation spectrum is re-purposed to store the max-hold spectrum in the case of autocorrelations.

\subsubsection{\texttt{sync} and \texttt{timebase}}

These two blocks contain the logic to generate the signals that synchronize the data signal across all the blocks in the design and, indirectly, across all the SNAP boards on the network. As evident in Figure~\ref{fig:fpga_design}, every block in the design, barring the ADCs, accepts a sync signal. The sync signal is internally used to reset counters, FIFO memory segments, reorder cycles etc. to their initial values at the time when a logical start-point of data is expected to arrive. After initialization of all blocks, a single sync-pulse (signal that is high for exactly one FPGA clock cycle) propagating serially through the design is used to bring the functionality of all the blocks into synchronization. This sync-pulse is generated on the arrival of the external PPS signal from the White Rabbit network. The distributed PPS signal, which has a nanosecond accuracy, ensures that all the SNAP boards across the array start observing at exactly the same time.

\subsection{Firmware}

The FPGA design in the MATLAB$^{\tiny \text{\textregistered}}$ Simulink$^{\tiny \text{\textregistered}}$ environment is compiled into a Vivado project and from there into a programmable bitfile by the CASPER toolflow\footnote{https://casper-toolflow.readthedocs.io/en/latest/index.html} that is developed and maintained by the CASPER collaboration. The compiled bitfile is in a \texttt{.fpg} file format and can be directly uploaded to the FPGA.

Communication with the FPGA is setup via a Trivial File Transfer Protocol (TFTP) server running on the Microblaze softcore processor, on the Xilinx Kintex-7 chip. The \texttt{casperfpga} software package\footnote{https://casper-toolflow.readthedocs.io/projects/casperfpga/en/latest/}, developed by the CASPER collaboration, is used to communicate with the TFTP server for uploading the bitfile, programming the FPGA, reading/writing various software registers etc.

\subsection{Control Software}

The 120 SNAP boards distributed across the array, can be programmed and controlled from a single server via the same 10\,Gb Ethernet network that output data is routed through (see Figure~\ref{fig:network}). An additional server that is connected in parallel to the X-engine servers, located within the KDRA, serves as the \texttt{hera-snap-head} node. 

The control software consists of three main scripts: (a) \texttt{blocks.py} which defines the classes and methods for each block in the FPGA design. This script exposes the functionality of each block to an end-user by abstracting the bit-level implementation. (b) \texttt{hera\_snap\_feng\_init.py} is the initialization script that performs all the tasks necessary to bring-up a single SNAP board, program it with the relevant bitfile and initialize all the blocks in preparation for observing. (c) \texttt{hera\_corr.py} contains classes and methods that perform a given operation on all the 120 SNAP boards across the array. It also implements monitoring functionality, periodically logging the status of all the SNAPs and uploading status variables to a Redis database.

\subsubsection{FPGA Design Control (\texttt{blocks.py})}

This script uses \texttt{casperfpga} commands to communicate with the Microblaze software processor for reading, writing and manipulating bits of software registers, BRAMs and GPIO pins. The functionality of each block in the FPGA design is encapsulated in its own class. Objects of these classes are created on the instantiation of a object of class \texttt{SnapFengine}, and can be accessed as members of this class object. This organization makes it highly modular and efficient to access all the functionality of the FPGA design from a single object declaration.

An abstract parent class \texttt{Block} defines communication methods like \texttt{read\_int}, \texttt{read}, \texttt{write\_int}, \texttt{write}, \texttt{blindwrite}, \texttt{change\_reg\_bits} etc. for manipulating 32-bit software registers and BRAMs, using \texttt{casperfpga} commands. The class definitions for each block are derived classes of the \texttt{Block} parent block and include functionality that is specific to that particular block.

\paragraph{\texttt{Adc}} Calls the SNAP ADC calibration script from the  \texttt{casperfpga} software package and sets the internal gain of the ADC chips.

\paragraph{\texttt{NoiseGen}} Sets the seeds of the noise generator modules for each of the six inputs.

\paragraph{\texttt{Input}} Toggles the appropriates bits of a software register to select between a constant zero input, the noise generator or the ADC inputs when methods like \texttt{use\_zero}, \texttt{use\_noise} or \texttt{use\_adc} are called.

\paragraph{\texttt{Delay}} Toggles bits of a single software register to set the delay in clock cycles for each of the inputs.

\paragraph{\texttt{Pfb}} Allows the user to set a bit shift schedule for the FFT, i.e. which stages of the FFT will result in a bit-shift and hence halving of the signal amplitude at the cost of loss in precision. For Gaussian white noise, shifting every other stage would suffice. Correspondingly, the right bits of a software register are read to report if an FFT overflow occured with the given bit-shift schedule.

\paragraph{\texttt{PhaseSwitch}} Generates and writes the Hadamard matrices to the BRAM which are used for modulation and demodulation of the input signal. This class also includes methods to modify the time delay between modulation and demodulation, toggling GPIO pins and enabling and disabling the block entirely.

\paragraph{\texttt{Eq}} Allows the user to write an array of short-integers to the BRAM that multiples the input signal before equalization. This array can be user-defined but, in practice, is set by reading the input power level of the signal just before the ADC. It provides the user a count of the number of times a signal was clipped using the given coefficients.

\paragraph{\texttt{EqTvg}} Provides the functionality to write two kinds of test vectors to the BRAM, even though, in principle, any array of 8\,bit numbers can be written to it. The end user can choose between a frequency ramp that is unique for each antenna (or polarization) or a value that is a unique constant for each antenna (or polarization). 

\paragraph{\texttt{ChanReorder}} Writes the BRAM that contains the reorder map for channels. The reorder block rejects 128 channels of every 512 channel block that it receives, or 1/4th of the total frequency band. The reorder map is set such that the rejected frequency bins correspond to the bandpass roll-over regions which are unimportant for the scientific goal of the experiment.

\paragraph{\texttt{Packetizer}} Contains methods that can write an array of integers to the BRAMs that contain the UDP packet header information of antennas and channels contained in each packet. These allocation, along with BRAM that contains destination IP addresses, perform the important task of assigning each frequency sub-band to a specific X-engine. 

\paragraph{\texttt{Rotator}} Provides methods that allow the end-user to set an array of phase values that the signal of each antenna is multiplied by. To minimize decorrelation due to sky movement, within the relatively long integration period of 16\,seconds, the antennas are phased to a central zenith. The time delay in arrival of an astronomical signal between two antennas in the interferometer is given by: Equation~\ref{eq:time_delay}, which depends on the baseline vector $\bar{d}$ and the wavelength or frequency of observation. To phase all antennas to a central zenith, an arbitrary location like the center of the array is chosen to compute relative time delays. The phase correction is given by:

\begin{equation}
    \Delta \phi = 2 \pi i {\tiny \Delta} \nu \tau_g
\end{equation}
\noindent
where $\tau_g$ is computed with respect to the center of the array.

\paragraph{\texttt{Eth}} This class populates the ARP (Address Resolution Protocol) table for the Ethernet block, with the MAC addresses of the destination X-engine servers. It also allows the user to probe statistics like the number of packets sent, the number of times the output FIFO buffer became full, overrun or faced errors etc.

\paragraph{\texttt{Corr}} This class provides functionality to set the antenna pair whose cross-correlation needs to be computed and provides the output of the operation by reading the BRAM. The integration time for computing the cross-correlation can be manually set, but care should be taken to ensure that the output can be represented by the 8\,bits that is allocated to each frequency channel.

\subsubsection{Initialization Script (\texttt{hera\_snap\_feng\_init.py})}

This script is called at the startup of the correlator at the beginning of each day in the observing season. It performs all the tasks necessary to get the SNAP boards ready for observing. Following is the list of operations executed in time order:

\begin{enumerate}
    \item Attempt to establish TFTP communication with all SNAPs listed in a configuration file and report number of ``dead" SNAPs.
    \item Check the number of SNAPs that need to be programmed or initialized. This is done by checking the presence of a particular software register in the design, and two specific bits of that software register which are set when the SNAPs have been initialized.
    \item Check if the configuration file itself is valid. This performs logical checks like the number of antennas and frequency channels assigned to each SNAP etc.
    \item Prevent monitoring scripts from sending TFTP traffic to the SNAP boards. This is a simple attempt at preventing broadcast storms on the network which could occur on ARP requests from uninitialized SNAP boards.
    \item Program all the SNAP boards that have not been programmed with the correct bitfile (optionally, program all of them).
    \item Initialize all the SNAP boards. This involves running the initialization method of all the blocks in the FPGA design.
    \item Set the BRAM that allocates a sub-band of the frequency range to a particular X-engine server and configure the BRAMs that set UDP header content on each SNAP.
    \item Set the SNAPs in test-vector mode or noise generator mode if requested.
    \item Disable the Ethernet output and arm the sync block for generating the sync pulse on reception of the PPS clock.
    \item Enable Ethernet output from the SNAPs and monitoring scripts.
\end{enumerate}

The entire F-engine codebase including the FPGA design in Simulink, the control software and other configuration and test files is available publicly at \url{https://github.com/HERA-Team/hera_corr_f}. In summary, the F-engine of the correlator primarily digitizes and channelizes the input analog voltage signal, along with implementing functionality that enhances the data quality during observing like phase switching and fringe rotation. At the output of the F-engine, 6144 frequency channels across a bandwidth of 187.5\,MHz are sent to the X-engine servers via a 10\,Gb Ethernet network as UDP packets. The output data rate from each SNAP board is close to 9\,Gbps bringing the total data rate into the X-engine to a little more than 1\,Tbps.

\section{Networking}
\label{sec:network}

\begin{figure}
    \centering
    \includegraphics[width=0.85\textwidth]{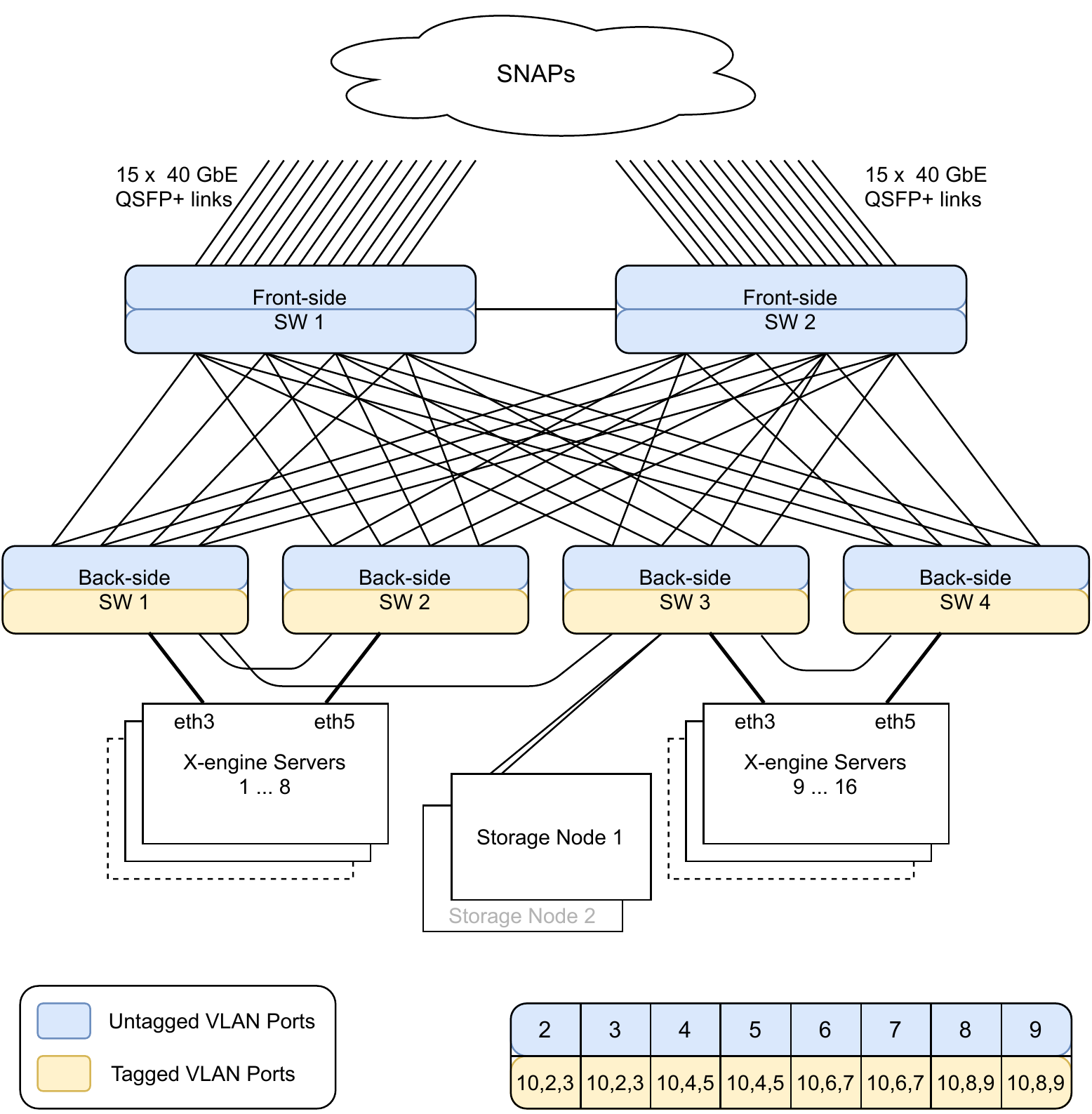}
    \caption{HERA network configuration that connects SNAP boards on-site to the X-engine servers in the KDRA. The network performs a corner-turn on the data, routing UDP packets from each SNAP board to the appropriate X-engine server while preventing closed-loops that lead to broadcast storms. Color indicated if the switch ports are configured as tagged or untagged VLAN. The VLAN configuration for each back-side switch is shown in the inset on the bottom right.}
    \label{fig:network}
\end{figure}

The HERA correlator network, shown in Figure~\ref{fig:network}, performs the functionality of connecting the SNAP boards on-site to the X-engine servers located in the Karoo Data Rack Area (KDRA). A fibre-optic patch panel (not shown in the figure) routes the data traffic via optic fibres that run underground from site to the KDRA. In addition to this, the network configuration also performs a corner-turn on the data. In a general FX-architecture correlator, the data on the F-engine side is distributed in a per-antenna manner. That is, each SNAP board handles the data of a given set of antennas. However, the data on the X-engine side needs to be distributed in a per-frequency manner. For computing cross-correlation products of all antenna pairs, a given X-engine server needs the data from all antennas. However, since data volumes tend to be large, the processing is often distributed among multiple servers, with each server receiving data in a narrow frequency range. The network configuration performs this corner-turn by routing UDP packets from each SNAP to the appropriate X-engine server.

The HERA network performs this corner-turn using six 32-port QSFP+ switches, split into ``front-side" and ``back-side" switches, as shown in Figure~\ref{fig:network}. Each of the two front-side switches services 60 SNAP boards via 15 40\,GbE links which can be split into 60 10\,GbE links. The four back-side switches service a subset of the bandwidth from the front-side and route traffic to the X-engine servers. Each back-side switch uses 40\,GbE links to connect to Ethernet ports on an X-engine server, serving a total of 8 X-engine servers each. The four back-side switches have plenty of spare ports available for back-up X-engine servers, control nodes and re-routing data acquisition traffic between the X-engine servers and storage nodes (not shown in the figure).

The interconnect between a pair of front-side and back-side switches consists of 4 40\,GbE links. Each front-side switch uses 16 ports to connect to the back-side switches while each back-side switch uses 8 ports for the interconnect. This system, unfortunately, creates many closed loops in the network which can lead to broadcast storms when the network encounters broadcast packets like DHCP or ARP requests. To prevent closed loops, each switch segregates different parts of the network by using VLANs. The front-side switches are configured to be access ports (or untagged VLAN ports) with default VLAN ID 1, since they receive untagged UDP packets from the SNAP boards. The back-side switches which connect to the front-side switches are also configured as access ports, but with a unique VLAN ID for each port as shown in the inset on the bottom right of Figure~\ref{fig:network}. All even numbered VLAN IDs are connected to one front-side switch and odd numbered VLAN IDs are connected to the other. The previously closed loop is broken by configuring the ports connecting to the X-engine servers as trunk ports (tagged VLAN ports) with VLAN IDs that match the unique ID from the front-side connection. Each trunk port is configured with two VLAN IDs corresponding to the two front-side switch connections that it needs to receive data from. 

The numerous spare sports on the back-side switches are theoretically capable of re-routing the data acquisition traffic, which consists of the UDP packets with cross-correlated data that the X-engine servers transmit to the storage node. However, the servers send these untagged UDP packets to trunk ports with the configuration described above. To handle this untagged traffic, all the trunk ports are configured with a native VLAN ID 10. This configuration also allows control nodes to communicate with each other and with the SNAP boards.

\section{X-engine}

The X-engine of the correlator computes the visibility matrix, or the cross-correlation product of all antenna pairs in the array within each frequency channel of the observing bandwidth. This compute intensive process is performed on 32 GPU cards spread across 16 servers. Each server hosts two Intel Xeon E5-2620-v4 CPUs (8 cores each), two Nividia GTX-1080 GPU cards and two 40\,Gb Ethernet ports. Two separate servers with the same hardware specifications have also been setup in preparation for unexpected server failures during observing runs. A separate storage node with 10 data drives, forming a 50\,TB RAID system, is used to write the visibility matrices to disk.

\subsection{Hashpipe}

The X-engine pipeline is built using \texttt{hashpipe}\footnote{https://github.com/HERA-Team/hashpipe}, a general-purpose code to write multithreaded processes with shared memory. The \texttt{hashpipe} software codebase, written in C language, was developed at Berkeley for the PAPER correlator. It provides functionality for allocating shared memory, moving ownership of memory segments between threads using semaphores, polling the status of threads or memory and monitoring thread-specific statistics like timing, packet loss, timestamp of data being processed etc. \texttt{Hashpipe} also provides an interface to some thread variables via a central Redis\footnote{\url{https://redis.io/}} database which can be used to set the start-time of file writing or data cross-correlation etc., and for monitoring thread behavior. 

\begin{figure}
    \centering
    \includegraphics[width=\linewidth]{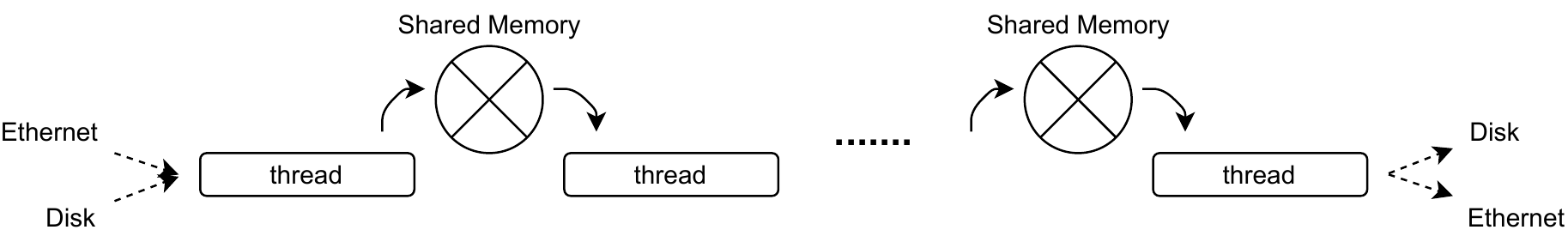}
    \caption{The general setup of a \texttt{hashpipe}-line. Input data, read from a disk or captured at an Ethernet port, is processed and stored on a shared ring buffer memory segment or the output buffer to that thread. The same memory segment acts as the input buffer to another thread, which processes this piece of data some more and loads to its own output buffer. The final data product is then output via an Ethernet port or written to disk.}
    \label{fig:hashpipe}
\end{figure}

Figure~\ref{fig:hashpipe} shows the general layout of a \texttt{hashpipe}-line. The input data to the pipeline is usually in the form of UDP packets or data read of a disk system. \texttt{Hashpipe} has the functionality to deal with raw UDP packets that have not been processed by the kernel stack, by creating packet sockets at the device driver level. This allows data capture at a much higher datarate than permitted by a Linux kernel. An input thread reads in the data by processing UDP packets or files stored on disk and stores it in a shared ring buffer memory segment. A ring buffer is a piece of memory, allocated by the program (either during compile-time or run-time), consisting of two or more logical segments that are represented by the same data structure. At any given point of time, only a single thread is allowed to read/write to each segment in the ring buffer. When the data in that segment has been processed, it is marked as complete and released to another thread.

\texttt{Hashpipe} restricts the access of a ring buffer to only two threads, as shown in Figure~\ref{fig:hashpipe}. Hence each ring buffer acts as an input buffer to exactly one thread and as an output buffer to exactly one thread. The user of \texttt{hashpipe} can define the data layout within each memory segment, including the datatype used to represent the data and the various metadata variables that need to be stored for proper functioning of the pipeline. The number of memory segments in each ring buffer, the total number of threads and the functionality of each thread can be customized for the specific application for which \texttt{hashpipe} is being used. Like the input to the pipeline, the output of the pipeline can be either written to disk or sent out of the server via Ethernet.

\subsection{Cross-correlation Pipeline}

Each server in the X-engine runs two parallel, completely independent \texttt{hashpipe} processes. Each \texttt{hashpipe}-line uses one CPU (with 8 cores) and one GPU to cross-correlate all antenna pairs within 384 channels (or 1/16th of 1644 channels). Half the servers process the `even' time samples and the other half process the `odd' time samples to limit the datarate at the input of each server to be less than 40\,Gbps. This time multiplexing also enables the computation of the sum and difference of consecutive time samples, which is useful for rejecting short period RFI in the data. Each X-engine server outputs UDP packets containing the cross-correlation products of a sub-band of the observing frequency. The storage node or the `catcher' machine runs an independent \texttt{hashpipe}-line which captures the UDP packets from all the X-engine servers and writes the data to disk. 

\begin{figure}
    \centering
    \includegraphics[width=0.55\linewidth]{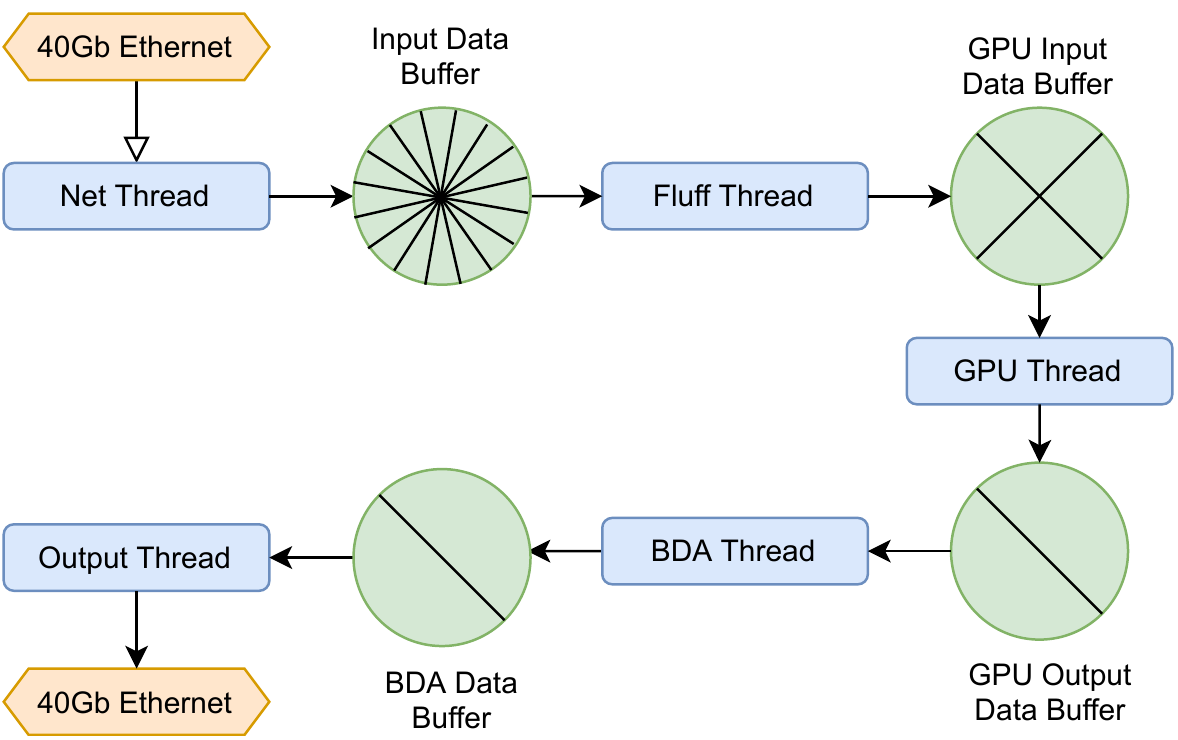}
    \caption{Structure of the \texttt{hashpipe}-line that runs on each X-engine GPU server. The pipeline consists of five threads (blue blocks) and four shared ring buffers (green circles) to perform task of cross-correlating antenna pairs and integrating each baseline pair for a custom integration period. The output of the process is transmitted to the storage node via Ethernet. See text for details.}
    \label{fig:px_pipeline}
\end{figure}

The \texttt{hashpipe}-line running on each of the GPU servers (labelled `px1' through `px16') consists of five threads and four shared ring buffers as shown in Figure~\ref{fig:px_pipeline}. The functionality of each thread is below:

\subsubsection{Net Thread}

The net thread sets up a packet socket (a software device that is capable of capturing raw packets) to the 40\,Gb Ethernet port where the UDP packets from SNAP boards are expected. On receiving a packet, its header is decoded and the contents of the packet are placed in an appropriate location within the memory allocated to the shared ring buffer `input data buffer'.

\subsubsection{Input Data Buffer}

This shared ring buffer consists of 16 memory segments, each 520\,MB wide, occupying a total of 8.25\,GB on the RAM. This buffer acts as the output to the net thread and the input to the fluffing thread. Each segment of this ring buffer contains 384 channels of all the dual-polarization antennas, or 704 inputs, and 2048 (even/odd) time samples. The memory is structured in the following order from the slowest to fastest varying indices: mcount, antenna number, channel number, time, polarization, complexity (real/imag). The order is dictated by the structure of a single UDP packet.

\subsubsection{Fluff Thread}

The output data from the SNAP consist of 8 bit complex numbers with 4 bit real and 4 bit imaginary components. The minimum datatype width that the Nvidia GPUs can handle is 8 bit numbers. This requires ``fluffing" the input with sign-extension to 16\,bit numbers with 8 bit real and 8 bit imaginary components. A second requirement of the fluff thread is to transpose the input data buffer to change the order from mcount, antenna, channel, time, polarization, complexity to time//4, channel, antenna, polarization, complexity and time\%4. These two tasks are performed using Intel IPP commands on 256\,bit words which optimize bit manipulation at the CPU hardware architecture level, and helps keep up with the input datarate.

\subsubsection{GPU Input Data Buffer}

This ring buffer acts as the output to the fluff thread and the input to the GPU thread. It consists of 4 memory segments, each 1\,GB wide, occupying a total of 4\,GB of the available RAM. Within each memory segment the data is ordered in the following fashion, from the slowest to fastest varying indices: time, channel, antenna, polarization, complexity and time. The first time index increments in units of 4 and the last time index contains 4 time samples. That is, each group of 8 bytes contains four times samples of a single antenna, polarization and channel (both real and imaginary components).

\subsubsection{GPU Thread}

This thread uses the package \texttt{xGPU}\footnote{\url{https://github.com/GPU-correlators/xGPU}} \citep{Clark_LaPlante_Greenhill}, which is a library to perform antenna cross-correlations on a GPU card. The dp4a architecture on Nvidia GTX-1080 cards supports 8-bit integers, which allows \texttt{xGPU} to perform pure integer correlations which involves a 8\,bit integer multiply and a 32\,bit integer accumulation. This thread sets the start-time of observation to microsecond precision by adjusting the mcount at which cross-correlation and integration begin, based on a second-precision start time set by the observer. The computation of the mcount at which observation starts takes into account phasing boundaries set by the SNAP, the GPU buffer size and any time delay specified by the user. The integration time period for the cross-correlation products, within the GPU, is set to a fixed value of 2\,seconds to enable baseline dependent averaging at the next stage.

\subsubsection{GPU Output Data Buffer}

The output of the GPU cross-correlation process, or the visibility matrix with 384 channels, is copied into this buffer as-is without modifying the data format. The visibility matrix for each channel is stored in a register tile order, where each unique antenna-polarization pair, or input pair, corresponds to a unique index in the output memory. The real and imaginary components of the cross-correlation products are stored in two independent visibility matrices. This buffer consists of two memory segments, each storing the full visibility matrix of all antenna cross-correlation products (real and imaginary components) and 384 channels, occupying a total of 1.5\,GB of RAM.

\subsubsection{Baseline Dependent Averaging Thread}

The BDA thread uses a user-defined configuration file to determine the integration time for each antenna pair in the array, and accumulates the GPU output until the specified integration period is achieved. Once a GPU Output Data Buffer is available for processing, the BDA thread reads the data corresponding to each pair of antennas serially and determines if that antenna pair needs to be added to a previous result or buffered in a new location. The period of time for which data is accumulated informs the RAM required by the output buffer of this thread and subsequently, the data rate out of the X-engine servers and the total data rate into the storage node. These specifications for various accumulation times are shown in Table~\ref{tab:bda_resource_specs}.

\begin{table}[htbp]
    \centering
    \begin{tabular}{c|c|c}
        \hline
    \textbf{Buffer Accumulation Time}  & \textbf{Data Rate (Gbps)} 
    & \textbf{RAM (GB)} \\ \hline
    2 sec  &  96.6  &  1.50  \\
    4 sec  &  48.6  &  1.52  \\
    8 sec  &  25.2  &  1.58  \\
    16 sec &  15    &  1.86  \\
    32 sec &  12    &  2.97  \\
    \hline
    \end{tabular}
    \caption{Resources required to support a given total accumulation time in baseline dependent averaging. Buffer accumulation time is the amount of time for which baselines are buffered on the server before sending them to the storage node, assuming that the remainder of the baseline averaging occurs on the storage node. Column 2 shows the net data rate into the storage node for the given buffer time and Column 3 shows the RAM required on each X-engine server.}
    \label{tab:bda_resource_specs}
\end{table}

The RAM usage on each X-engine is computed as:
\begin{equation}
    \label{eq:bda_ram}
    \text{RAM} = \# \text{ baselines accumulated} * 8 \text{ bytes} * 4 \text{ stokes} * 2 \text{ samples} * 384 \text{ channels}
\end{equation}
\noindent
where the number of baselines accumulated depends on the distribution of baselines into each integration time bin and the total amount of time for which data is accumulated. For example, if the BDA buffer accumulated data for 8 seconds, 4 samples of all the visibility products with a 2 second integration time, 2 samples of visibility products with 4 second integration time and one sample each of everything with an integration time of 8 seconds or greater would have to be stored in the buffer (see Figure~\ref{fig:bda_buffer}). The RAM usage in Table~\ref{tab:bda_resource_specs} has been computed by assuming the distribution of baselines shown in Table~\ref{tab:bda_dist}.

Similar to RAM usage, the total data rate into the storage node can be computed as:
\begin{equation}
    \label{eq:bda_datarate}
    \text{Data Rate} = \frac{\# \text{ baselines accumulated} * 8 \text{ bytes} * 4 \text{ stokes} * 2 \text{ samples} * 6144 \text{ channels}}{\text{accumulation time}}
\end{equation}
\noindent
As expected, a smaller accumulation time leads to a larger data rate into the storage node; because it necessarily implies that only a portion of the baseline dependent averaging, which can decrease the data rate, is occurring on the X-engine servers. The remainder of the averaging should be performed on the storage node before data is written to disk.

Hence, the accumulation time of the BDA buffer introduces a trade-off between RAM usage on each server and the total data rate into the storage node. Theoretically, the shortest baselines in the array, formed by pairs of antennas that are adjacent to each other, can be integrated up to 32 seconds. However, the RAM required to buffer a single set of BDA data is nearly 3\,GB leading to 6\,GB of memory usage for a ring buffer with just two blocks. To limit the RAM usage, the accumulation time for BDA is set to be 16 seconds. The last step of averaging the short baselines to 32 seconds is omitted for convenience since the data rate reduction obtained is not too significant.

\subsubsection{BDA Data Buffer}

\begin{figure}[htbp]
    \centering
    \includegraphics[width=0.5\linewidth]{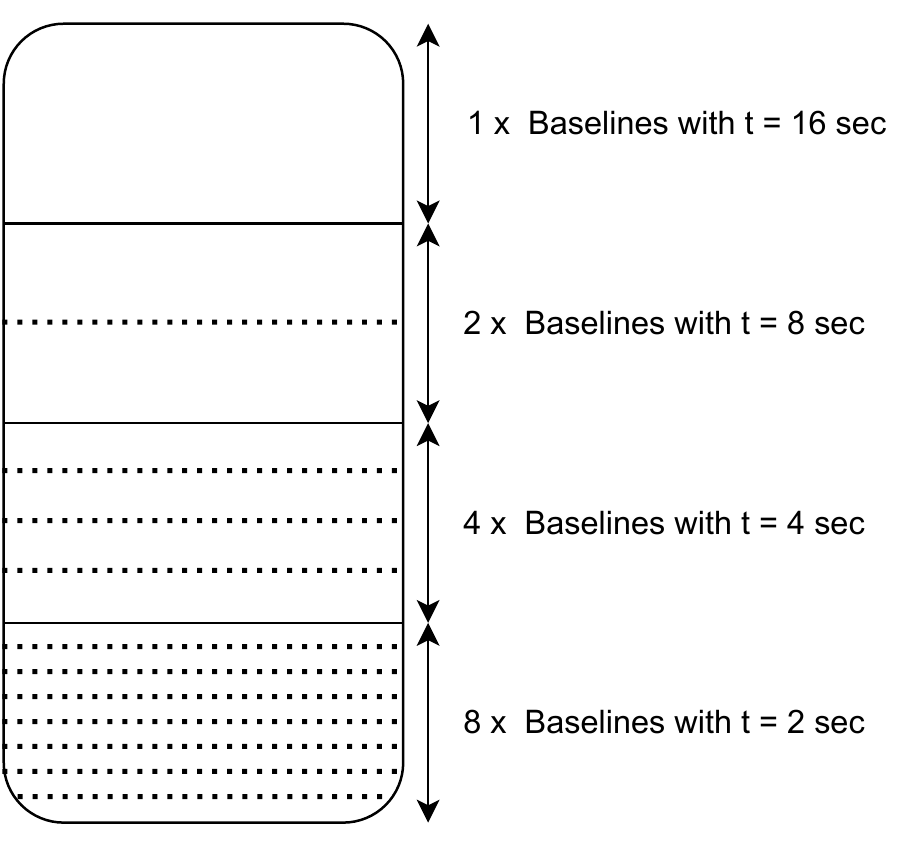}
    \caption{Structure of the data buffer allocated during run-time for storing the output of baseline dependent averaging.}
    \label{fig:bda_buffer}
\end{figure}

The BDA ring buffer consists of two memory blocks, each of which can accommodate all the different baselines that comprise 16\,seconds of observing. Unlike the other data buffers which are allocated on the stack at compile-time, this ring buffer is allocated during run-time on the heap. Dynamic allocation is necessary for supporting the user-defined configuration file which can change from start-up to start-up. Figure~\ref{fig:bda_buffer} shows the structure of a single block of the ring buffer allocated for holding the BDA data.

\subsubsection{Output Thread}

The output thread reads the data in the BDA buffer and packetizes it into UDP packets that are transmitted over the Ethernet to the storage node. It appends a useful header to each UDP packet, labelling the antenna pair, channel range, and time stamp of each packet. The UDP packets are 4120 bytes wide including the header, carrying a payload of 4096 bytes corresponding to 64 channels of data for a given visibility product. The output data rate from each X-engine server nearly 1\,Gbps.

\subsection{Data Acquisition}

The storage node receiving UDP traffic from GPU servers, also runs a \texttt{hashpipe} process for collecting the UDP packets and writing them to disk. The format of the \texttt{hashpipe}-line running on the storage node is very similar in structure to that of the X-engine servers. It primarily consists of two threads, one for collecting UDP packets and time-ordering them into a ring buffer and another for writing the data to disk. A third process periodically uploads the auto-correlation data into a Redis server for commissioning and debugging.

The data format that is used to write the cross-correlation products to disk is based on HDF5 and adapted specifically for radio astronomy~\citep{Hazelton_et_al_2017}. The recorded files are processed by the real-time processing (RTP) system \citep{LaPlante_et_al_2021} which performs calibration, RFI-excision and automatic flagging of bad data. The processed data is uploaded to servers at the National Radio Astronomy Observatory (NRAO) in New Mexico using the HERA Librarian\footnote{\url{https://github.com/HERA-Team/librarian}} software package.

\section{Preliminary Results}

\begin{figure}
    \centering
    \includegraphics[width=0.65\textwidth]{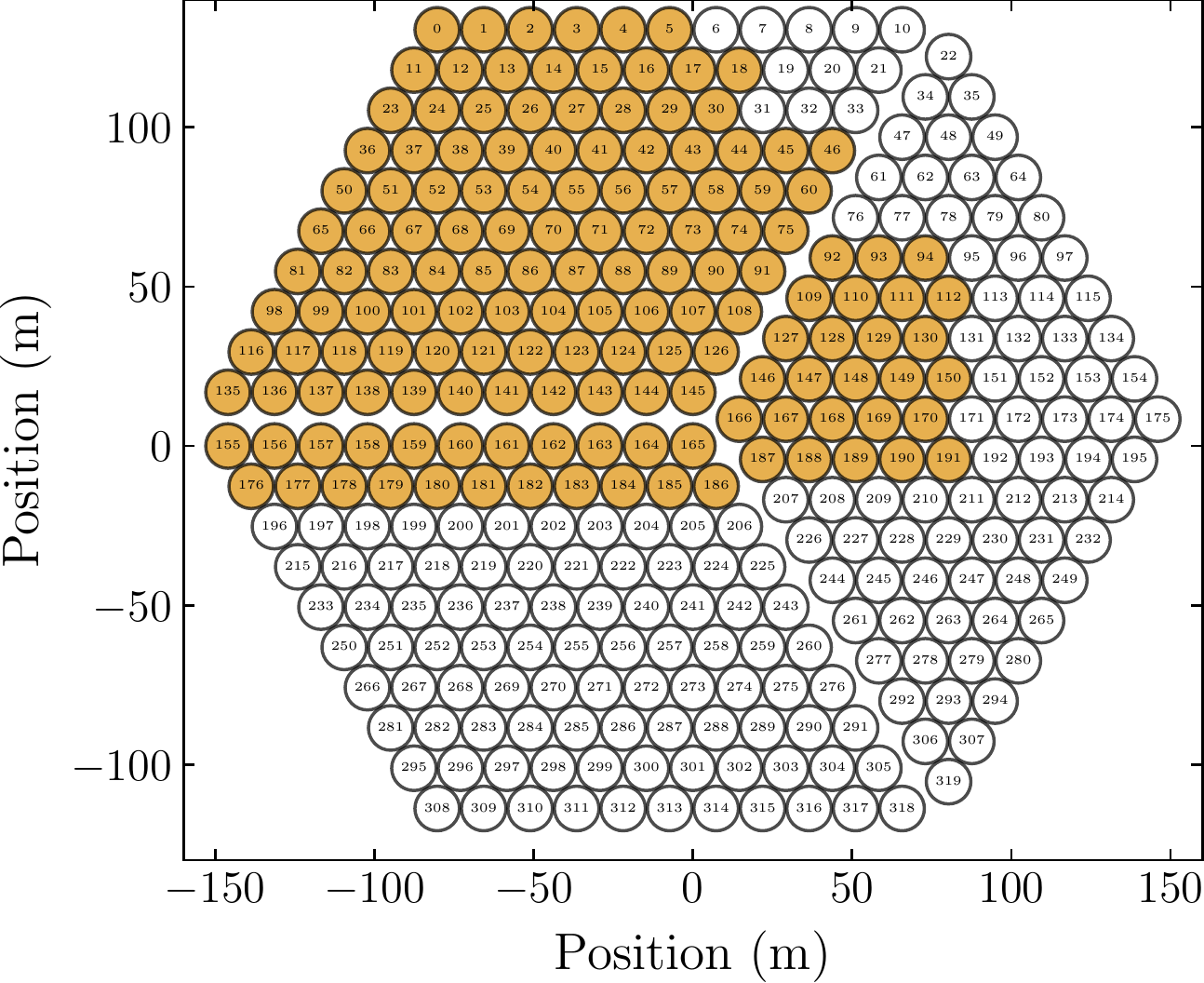}
    \caption{Antennas online (filled circles) during commissioning operations between September--December 2020. The rest of the antennas currently under-construction.}
    \label{fig:comm_ants}
\end{figure}

Figure~\ref{fig:comm_ants} shows the status of the array in September 2020 when commissioning operations began. The filled circles are antennas that have been constructed and equipped with feeds and receiver electronics. The BDA portion of the \texttt{hashpipe}-line allows files to be written with just the data from the antennas being commissioned, rather than the full visibility matrix. This flexibility proved key to commencing observing early since the storage nodes are not equipped to handle the data rate from the full array yet. 

In addition to the primary data product of antenna cross-correlations files, the HERA correlator periodically uploads metadata to a Redis database for monitoring. This metadata is tracked by control scripts that upload this information to a Grafana Dashboard\footnote{\url{https://heranow.reionization.org}} that can be used by an observer for real-time monitoring. The metadata contains various useful statistics like the temperature of nodes on-site, PAM and FEM power levels, number of operating SNAPs, ADC histograms from the SNAP board, packet statistics from the correlator \texttt{hashpipe}-line, and the auto-correlations uploaded by the data acquisition pipeline to name a few. These statistics are monitored by the observer during commissioning for problems or errors in the correlator operation. 

\begin{figure}
    \centering
    \includegraphics[width=\textwidth]{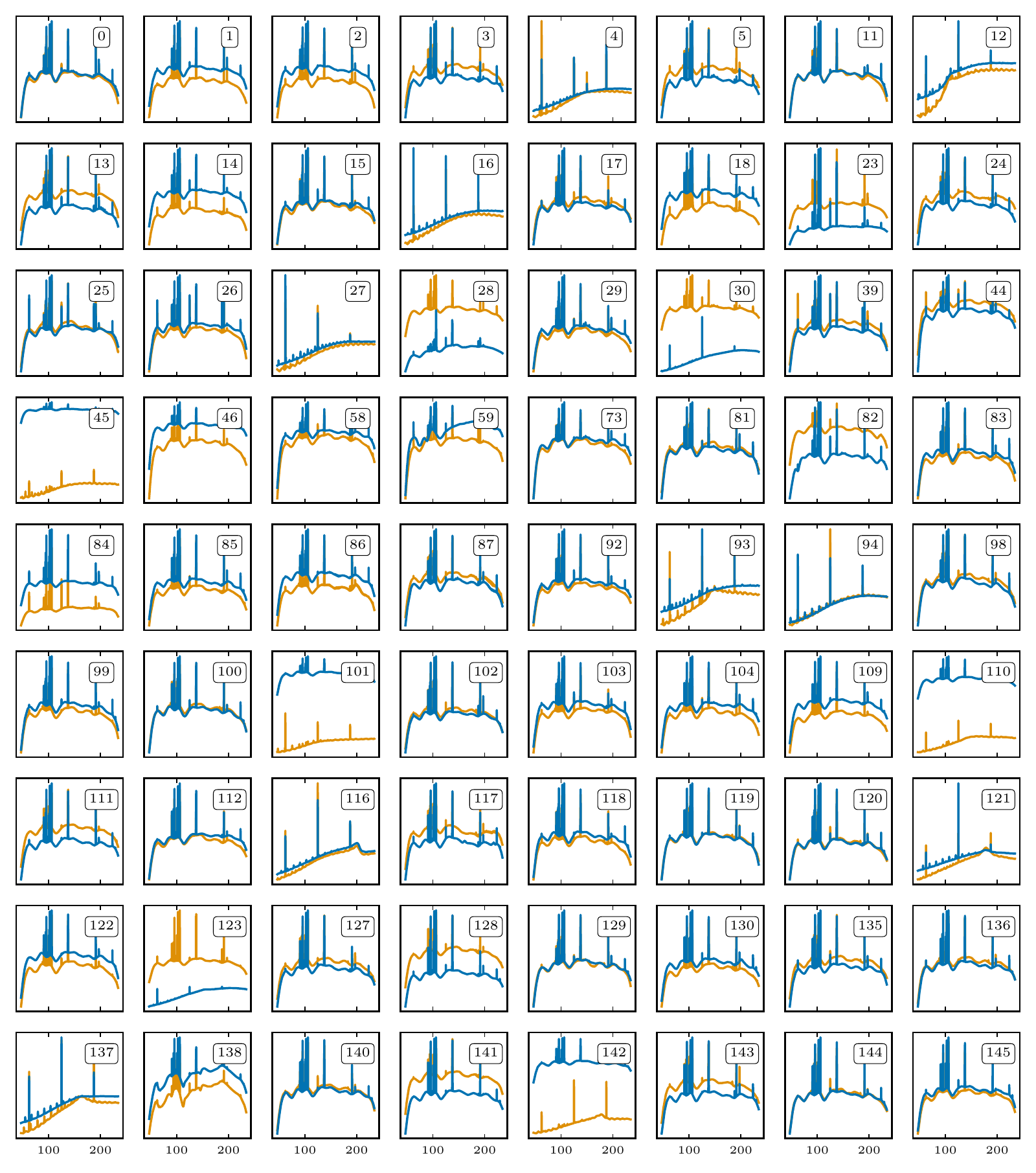}
    \caption{Antenna auto-correlation spectra taken on JD 2459139 (Oct 20, 2020), shown for a single integration time picked arbitrarily from the middle of the night.}
    \label{fig:autocorrs}
\end{figure}

Figure~\ref{fig:autocorrs} shows the auto-correlated power spectra for an arbitrarily picked set of 72 antennas, uploaded by the \texttt{hashpipe}-line to the Redis database. The auto-correlation spectra of an antenna are used as a metric to flag `bad' antennas which are classified by either a consistently low power, high RFI contamination, high temporal variability, inconsistent bandpass shape, or low correlation with other antennas.

The data files written by the correlator are processed daily by the commissioning team, who generate Jupyter Notebooks\footnote{\url{https://github.com/HERA-Team/H4C_Notebooks}} to assess the health of the array and data quality. The Notebooks are used to monitor antenna auto-correlations as a function of time and frequency, on-site and internal RFI contaminating each antenna, phase correlation between antennas, redundancy in various baseline groups and their corresponding delay spectra. 

The primary scientific goal of HERA is to detect and characterize the power spectrum of redshifted neutral hydrogen from the epoch of reionization. The foreground avoidance technique that will be employed on HERA data for extracting this weak signal, relies on the redundancy of measured visibilities to build sensitivity. That is, antenna pairs separated by the same baseline vector, theoretically expected to measure the same power spectrum mode, should measure the same visibility to within noise limits. Additionally, redundancy is also important for obtaining accurate calibration solutions from the redundant-calibration calibration process that is used to compute antenna gains.

\begin{figure}
    \centering
    \includegraphics[width=\textwidth]{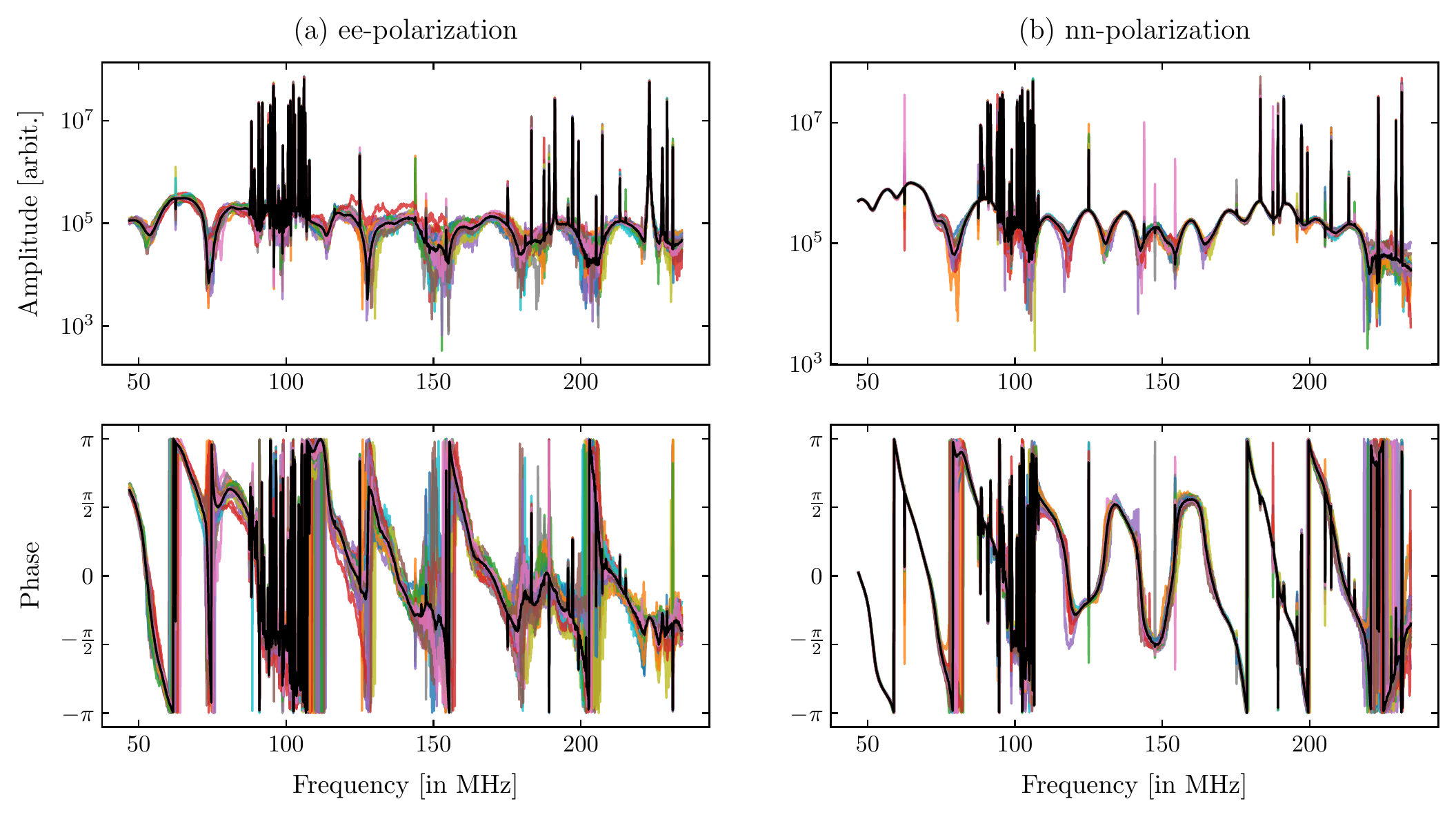}
    \caption{Results from the commissioning of HERA in September 2020. Plot shows the amplitude and phase of the visibility functions (colored lines) measured by the antenna pairs separated by 14.6\,m in the East-West direction. Panels (a) show the East-West polarization result and Panels (b) show the North-South polarization. The solid black line in all the panels is the redundant-baseline calibration solution obtained for this set.}
    \label{fig:commissioning_results}
\end{figure}

Figure~\ref{fig:commissioning_results} shows the visibility functions measured by the set of antenna pairs that share the baseline vector of 14.6\,m separation in the East-West direction. The visibilities have been averaged for one integration period of 9.66\,seconds, arbitrarily chosen from a full night of observing in October 2020. The data has been calibrated in real-time using redundant-baseline calibration, and the calibration result is marked by a solid black line in the figure. The multiple colored lines show the calibrated visibilities measured by the different antenna pairs in this particular redundant set.  While redundancy can be qualitatively defined in multiple ways, as presented by \citet{Dillon_et_al_2020} for HERA Phase-I, this plot shows that the shortest baseline group is fairly redundant to eye. That is, the correlator does not introduce digital errors to the data path and accurately reflects the level of signal cross-correlation in the analog voltages of different antennas.

%A peculiarity of the HERA correlator is that it needs to be restarted at the beginning of every night of observation and shutdown at the end of it. The SNAP boards deployed on-site, within node enclosures, are not sufficiently cooled during the day and overheat in the hot desert if they are not turned off. Hence, a better metric of the performance of the HERA correlator is its reliability. Early on in the commissioning of the correlator, we found that TFTP errors in the communication between the SNAP boards and the control node led to some SNAP boards not being programmed and the corresponding three antennas being dropped out of observations for the night. 

% Conclusion
\chapter{Conclusion}
\label{chp:Conclusion}

The formation of the first stars, about $\sim$400,000 years after the Big Bang, lead to the reionization of neutral hydrogen in the inter-galactic medium. This event, called the Epoch of Reionization (EoR), is considerably unexplored so far. Indirect probes like the Ly-$\alpha$ forest of high redshift quasars, the Cosmic Microwave Background and galaxy surveys have placed coarse limits on the time, duration, amount of ionizing radiation, X-ray heating, etc. but are model dependent and uncertain. A direct probe of this period is 21\,cm emission from cosmological neutral hydrogen that dominated the baryonic content of the universe until the EoR. 

The 21\,cm emission line is extremely faint compared to foregrounds at the same frequency, like the Galactic synchrotron emission, making it hard to detect. Moreover, the chromatic beam response of radio interferometers, that are often used to probe this signal, couples the foregrounds to the EoR in a frequency-dependent way. This coupling is hard to model and filter, posing a formidable challenge to the detection of the EoR signal. Over the last decade, a new idea that has come up is to exploit the containment of foregrounds into a wedge-like region in cylindrical power spectrum coordinates, and probe the EoR signal outside this region. 

Using this foreground avoidance technique requires radio interferometers with numerous short baselines and large bandwidths. Combined with the large collecting area required to build sensitivity towards the EoR signal, this has given rise to a new class of radio interferometers with $N > 100$ antennas. Such arrays require large correlators in the backend to cross-correlate antenna pairs and generate visibility matrices in real-time. A limit is being reached in the cost of such correlators, making them equivalent to the cost of the array itself. Some authors have suggested FFT correlators as a viable solution to this problem, but the spatial Fourier transform requires calibrated antenna voltages to avoid signal loss.

In this thesis, I propose a $\mathcal{O}(N\log{N})$ calibrator design that can operate in parallel with the FFT-correlator, forming a self-contained correlator system that can be scaled to large-N arrays built on a regular grid. I discuss two calibration algorithms that can be employed by such a calibrator and show that the resulting calibration solutions are unbiased and converge to the true values when noise averaged. I compare both the calibration schemes and show that low cadence calibration produces superior results for small calibrator sizes, but that subset redundant-baseline calibration is better for larger calibrator sizes. Ultimately, the calibration method that is suitable for a particular experiment depends on the computational resources available and the science goal at hand.

To verify this theoretical proposition, I built a voltage recorder system that could digitize and store raw voltage data from 12 antennas, each with a bandwidth of 100\,MHz. The system was built on CASPER-designed FPGA boards called the Smart Network ADC Processor (SNAP) which uses 250\,MHz ADCs to sample the input voltage signal and a 10\,Gb Ethernet port to send the digitized data to a CPU system. A program that uses shared ring buffers to time order the UDP packets from the SNAP board, runs on the CPU and writes the voltage data to disk. This system was deployed on two different sets of antennas, PAPER antennas at the Green Bank site in West Virginia, and HERA in South Africa, over two consecutive years. Unfortunately, the data collected both times was insufficient to verify calibration methods. However, the data from PAPER antennas was useful to confirm that post-calibration output of FFT correlators matches the baseline-averaged visibility matrix of a traditional correlator.

Lastly, I detail the correlator system built for HERA, which is a case demonstration of the computational resources required by a large-N radio interferometer. The digital correlator requirements, set by the array size and science goals, drive the hardware and software design. The correlator is built in the traditional FX-architecture on a hybrid platform. The F-stage of the design is implemented on aforementioned SNAP boards, which are installed close to antenna feeds to prevent signal loss in long RF cables. The X-stage of the correlator uses the large computational capacity of GPU cards to perform cross-correlation and generate visibility matrices. These two stages are connected by a network of 40\,Gb Ethernet switches which use VLANs to prevent broadcast storms. The correlator is currently operational and being commissioned on roughly a third of the full array.

Overall, this thesis lays out the instrumentation required to enable current- and future-generation large-N radio interferometers built on a regular grid. The rapidly increasing compute capacity of GPUs and FPGAs, driven by upcoming industries like autonomous vehicles, bitcoin mining, gaming, etc. might mean that the FX-correlator architecture is here to stay. However, if the trend in decreasing cost per compute does not continue, FFT correlators will be inevitable to realize the sensitivity required for the next-generation EoR power spectrum experiments. The broader story that I've tried to address, is that the instrument working at the backend of a telescope, which is often taken for granted and expected to deliver sky-accurate data products, is complicated and requires careful science-oriented considerations. 

\printbibliography

% \appendix
% \chapter{More Monticello Candidates}

\end{document}